\documentclass[12pt,twoside,openany]{book}

\usepackage{color}
\definecolor{indigo}{RGB}{0,0,120}
 
\usepackage[top = 2.5cm, bottom = 2.5cm, left = 2.5cm, right = 2.5cm]{geometry}

\usepackage[nocompress]{cite}

\usepackage[pdftex]{graphicx}

\usepackage{calligra} 

\usepackage{subcaption}
\usepackage{wrapfig}

\usepackage{amsmath}

\usepackage[colorlinks=true, linkcolor=indigo, citecolor=blue, urlcolor=indigo]{hyperref}

\usepackage{amssymb}

\newcommand\abstractname{Abstract}

\makeatletter
\if@titlepage
  \newenvironment{abstract}{%
      \titlepage
      \null\vfil
      \@beginparpenalty\@lowpenalty
      \begin{center}%
        \bfseries \abstractname
        \@endparpenalty\@M
      \end{center}}%
     {\par\vfil\null\endtitlepage}
\else
  
\fi
\makeatother

\newcount\colveccount  
\newcommand*\colvec[1]{\global\colveccount#1  \begin{pmatrix} \colvecnext} \def\colvecnext#1{#1 \global\advance\colveccount-1
        \ifnum\colveccount>0 \\ \expandafter\colvecnext
        \else \end{pmatrix} \fi}
        
\DeclareMathAlphabet{\mathcalligra}{T1}{calligra}{m}{n}
\DeclareFontShape{T1}{calligra}{m}{n}{<->s*[2.2]callig15}{}
\newcommand{\scripty}[1]{\ensuremath{\mathcalligra{#1}}}

\def\tr{\;{\rm tr}\;}

\def\imply{\Rightarrow}

\def\fl{\noindent}

\newcommand{\Bra}{\left \langle}
\newcommand{\Ket}{\right \rangle}
\newcommand{\bra}{\langle}
\newcommand{\ket}{\rangle}
\newcommand{\lv}{\left\vert}
\newcommand{\rv}{\right\vert}
\newcommand{\tl}[1]{\tilde{#1}}
\newcommand{\dd}[2]{\frac {\partial #1}{\partial #2}}

\newcommand{\pdr}{\partial}
\newcommand{\DD}[2]{\frac {d #1}{d #2}}
\newcommand{\grad}{{\bf \nabla}}

\newcommand{\beq}{\begin{equation}}
\newcommand{\eeq}{\end{equation}}
\newcommand{\beqs}{\begin{eqnarray}}
\newcommand{\eeqs}{\end{eqnarray}}

\newcommand{\half}{\frac{1}{2}}
\newcommand{\ov}[1]{\frac{1}{#1}}
\newcommand{\fr}[2]{\frac{#1}{#2}}

\def\al{\alpha}		\def\g{\gamma} 		\def\G{\Gamma} 
\def\del{\delta}	\def\D{\Delta}		\def\eps{\epsilon} 
			 
\def\la{\lambda}		\def\sig{\sigma}		
		\def\tht{\theta}	
		\def\om{\omega}		\def\Om{\Omega}  

\def\vf{\varphi}

\newcommand{\bfr}{{\bf r}}

\newcommand{\bfp}{{\bf p}}
\newcommand{\bfq}{{\bf q}}

\newcommand{\bfR}{{\bf R}}
\newcommand{\bfL}{{\bf L}}
\newcommand{\bfP}{{\bf P}}

\newcommand{\bfJ}{{\bf J}}

\newcommand{\err}{\scripty{r}}
\newcommand{\mC}{{\mathbb{C}}}
\newcommand{\mR}{{\mathbb{R}}}
\newcommand{\mS}{{\mathbb{S}}}

\def\span{\text{span}}

\def\T{{\cal T}}
\def\V{{\cal V}}

\begin{document}

\frontmatter

\thispagestyle{empty}

\begin{center}

\vspace*{5mm}

{\LARGE \bf Instabilities and chaos in the classical }

\vspace{.3cm}

{\LARGE \bf three-body and three-rotor problems}
\vspace{1cm}
\\{\Large by}\\ 
\vspace{1cm}
{\Large \textbf{Himalaya Senapati}} 
\large 

\vspace{2cm}

{\it A thesis submitted in partial fulfillment of the requirements for \\ the degree of Doctor of Philosophy in Physics} \\
\vspace*{0.8cm}
to  
\vspace*{0.5cm} 

Chennai Mathematical Institute

\vspace{0.5cm}

Submitted: April 2020 

Defended: July 23, 2020

\vspace{1cm}

\begin{figure}[h]
\begin{center}
 \includegraphics[width = 7cm]{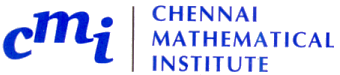}
 \end{center}
\end{figure}

\vspace{0.1cm}

Plot H1, SIPCOT IT Park, Siruseri,\\
Kelambakkam, Tamil Nadu 603103, \\
India

\end{center}



\newpage
\thispagestyle{empty}

\begin{flushleft}
Advisor:

Prof.~Govind~S.~Krishnaswami, \emph{Chennai Mathematical Institute.}\\ 
\vspace*{1cm}
Doctoral Committee Members: 
\begin{enumerate}
\item Prof.~K.~G.~Arun, \emph{Chennai Mathematical Institute.}\\
\item Prof.~Arul~Lakshminarayan, \emph{Indian Institute of Technology Madras.}
\end{enumerate}
\end{flushleft}



\newpage

\chapter*{\center Declaration}

This thesis is a presentation of my original research work, carried out under the guidance of Prof.~Govind~S.~Krishnaswami at the Chennai Mathematical Institute. This work has not formed the basis for the award of any degree, diploma, associateship, fellowship or other titles in Chennai Mathematical Institute or any other university or institution of higher education.

\vspace*{5mm}
\begin{flushright}
Himalaya Senapati\\
April, 2020
\end{flushright}

\vspace*{30mm}

\fl In my capacity as the supervisor of the candidate's thesis, I certify that the above statements are true to the best of my knowledge.
\vspace*{5mm}
\begin{flushright}
Prof.~Govind~S.~Krishnaswami\\
April, 2020
\end{flushright}


\newpage

\chapter*{\center Acknowledgments}
\addcontentsline{toc}{chapter}{Acknowledgments}

I would like to start by thanking my supervisor, Professor Govind S. Krishnaswami for his support and guidance, it has been a pleasure to be his student and learn from him. I am grateful to him for suggesting interesting research directions while giving me necessary space to pursue my own reasoning. His office was always open and he was willing to carefully listen to arguments, even when it was a different line of thought than his own, and offer corrections and insights. His persistent effort to improve the presentation of our results so that they are self-contained and appeal to a wide audience is something to strive for. Aside from research, he has also been very kind in other aspects of life, ranging from helping preparing applications for workshops and conferences to taking care of administrative matters. 

I thank Prof. K. G. Arun who is on my doctoral committee for his encouragement and for organizing departmental seminars with Prof. Alok Laddha where Research Scholars could present their work. I  thank Prof. Arul Lakshminarayan for his insightful comments and references to the literature during our doctoral committee meetings. I also extend my thanks to Prof. Athanase Papadopoulos for encouraging me to contribute articles on non-Euclidean geometries to a book and for inviting me to conferences at MFO, Oberwolfach and BHU, Varanasi. I thank Prof. Sudhir Jain for valuable discussions. I would also like to thank Professors S G Rajeev and MS Santhanam for carefully reading this thesis and for their questions and suggestions.

I am indebted to Prof. Swadheenananda Pattanayak and Prof. Chandra Kishore Mohapatra for instilling in me a love for mathematical sciences from a young age via Olympiad training camps at the state level. I am grateful to Banamali Mishra, Gokulananda Das and Ramachandra Hota for their teaching. I also thank Sandip Dasverma for his support.

I thank Kedar Kolekar, Sonakshi Sachdev and T R Vishnu for many coffee time discussions. I also thank my officemates Abhishek Bharadwaj, Dharm Veer and Sarjick Bakshi as well as A Manu, Anbu Arjunan, Aneesh P B, Anudhyan Boral, Athira P V, Gaurav Patil, Keerthan Ravi, Krishnendu N V, Miheer Dewaskar, Naveen K, Pratik Roy, Praveen Roy, Priyanka Rao, Rajit Datta, Ramadas N, Sachin Phatak, Shanmugapriya P, Shraddha Srivastava and Swati Gupta for their help and friendship. A special thanks to Apolline Louvet for hosting me in France for a part of my stay during both my visits to Europe. 

I thank the faculty at the Chennai Mathematical Institute including Professors H S Mani, G Rajasekaran, N D Hari Dass, R Jagannathan, A Laddha, K Narayan, V V Sreedhar, R Parthasarathy, P B Chakraborty, G Date and A Virmani for their time and help. Thanks are also due to the administrative staff at CMI including S Sripathy, Rajeshwari Nair, G Ranjini and V Vijayalakshmi as well as the mess, security and housekeeping staff.

I gratefully acknowledge support from the Science and Engineering Research Board, Govt. of India in the form of an International Travel Support grant to attend the Berlin Mathematical Summer School in 2017, for sponsoring a school on nonlinear dynamics at SPPU, Pune in 2018 and for travel support to attend a CIMPA school at BHU, Varanasi and CNSD 2019 at IIT Kanpur. I also acknowledge MFO, Oberwolfach for awarding me an Oberwolfach Leibniz Graduate Students travel grant to attend a conference held there. I thank CMI for my research fellowship and for supporting my travel to attend schools and workshops at ICTS Bengaluru, MFO Oberwolfach, RKMVERI Belur Math, IISER Tirupati and IIT Kanpur and to meet Professors A Chenciner and L Zdeborova in Paris. I am also grateful to the Infosys Foundation and J N Tata trust for financial support.

Finally, I would like to thank my parents Niranjan Senapati and Annapurna Senapati and brother Meghasan Senapati for their love and support.


\newpage

\chapter*{\center Abstract}
\addcontentsline{toc}{chapter}{Abstract}

This thesis studies instabilities and singularities in a geometrical approach to the planar three-body problem as well as instabilities, chaos and ergodicity in the three-rotor problem.

Trajectories of the three-body problem are expressed as geodesics of the Jacobi-Maupertuis (JM) metric on the configuration space. Translation, rotation and scaling isometries lead to reduced dynamics on quotients of the configuration space, which encode information on the full dynamics. Riemannian submersions are used to find  quotient metrics and to show that the geodesic formulation regularizes collisions for the $1/r^2$, but not for the $1/r$ potential. Extending work of Montgomery, we show the negativity of the scalar curvature on the center of mass configuration space and certain quotients for equal masses and zero energy. Sectional curvatures are also found to be largely negative, indicating widespread geodesic instabilities.

In the three-rotor problem, three equal masses move on a circle subject to attractive cosine inter-particle potentials. This problem arises as the classical limit of a model of coupled Josephson junctions. The energy serves as a control parameter. We find analogues of the Euler-Lagrange family of periodic solutions: pendula and breathers at all energies and choreographies up to moderate energies. The model displays order-chaos-order behavior and undergoes a fairly sharp transition to chaos at a critical energy with several manifestations: (a) a dramatic rise in the fraction of Poincar\'e surfaces occupied by chaotic sections, (b) spontaneous breaking of discrete symmetries, (c) a geometric cascade of stability transitions in pendula and (d) a change in sign of the JM curvature. Poincar\'e sections indicate global chaos in a band of energies slightly above this transition where we provide numerical evidence for ergodicity and mixing with respect to the Liouville measure and study the statistics of recurrence times.


\chapter*{List of publications}
\addcontentsline{toc}{chapter}{List of publications}

This thesis is based on the following publications.

\begin{enumerate}
	
	\item G. S. Krishnaswami  and H. Senapati, {\it Curvature and geodesic instabilities in a geometrical approach to the planar three-body problem}, \href{https://doi.org/10.1063/1.4964340}{J. Math. Phys.}, {\bf 57}, 102901 (2016). \href{https://arxiv.org/abs/1606.05091}{arXiv:1606.05091}. [Featured Article]	
	
	\item G. S. Krishnaswami and H. Senapati, {\it An introduction to the classical three-body problem: From periodic solutions to instabilities and chaos},  \href{https://doi.org/10.1007/s12045-019-0760-1}{Resonance}, {\bf 24}, 87-114 (2019). \href{https://arxiv.org/abs/1901.07289}{arXiv:1901.07289}. 
	
	\item G. S. Krishnaswami and H. Senapati, {\it Stability and chaos in the classical three rotor problem}, \href{https://www.ias.ac.in/article/fulltext/conf/002/0020}{Indian Academy of Sciences Conference Series}, {\bf 2} (1), 139 (2019). \\ \href{https://arxiv.org/abs/1810.01317}{arXiv:1810.01317}. 
	
	\item  G. S. Krishnaswami and H. Senapati, {\it Classical three rotor problem: periodic solutions, stability and chaos}, \href{https://doi.org/10.1063/1.5110032}{Chaos}, {\bf 29} (12), 123121 (2019). \href{https://arxiv.org/abs/1811.05807}{arXiv:1811.05807}. [Editor's pick, Featured article]
	
	\item G. S. Krishnaswami  and H. Senapati, {\it Ergodicity, mixing and recurrence in the three rotor problem}, \href{https://doi.org/10.1063/1.5141067}{Chaos}, {\bf 30} (4), 043112 (2020). \href{https://arxiv.org/abs/1910.04455}{arXiv:1910.04455}. [Editor's pick]
	
\end{enumerate}


\small

\tableofcontents

\newpage
\thispagestyle{empty}

\normalsize

\mainmatter

\chapter{Introduction}
\label{c:introduction}

The classical three-body problem arose in an attempt to understand the effect of the Sun on the Moon's Keplerian orbit around the Earth. It has attracted the attention of some of the best physicists and mathematicians and led to the discovery of chaos. In the first part of this thesis (Chapter \ref{chapter:three-body}), we study a geometrical approach to the planar three-body problem subject to Newtonian or inverse-square potentials and describe results on instabilities and near collision dynamics by treating trajectories as geodesics of an appropriate metric on the configuration space. The second part (Chapter \ref{chapter:three-rotor}) concerns instabilities, chaos and ergodicity in the classical three-rotor problem which we propose as an interesting variant of the three-body problem. It also arises as the classical limit of a model for chains of coupled Josephson junctions. Despite the close connections, the two parts are reasonably self-contained and may be read independently.

\section[Geometrical approach to the planar three-body problem]{Geometrical approach to the planar three-body \\problem \sectionmark{Geometrical approach to the planar three--body problem}}
\label{s:intro-3body}
\sectionmark{Geometrical approach to the planar three--body problem}

The classical gravitational three-body problem \cite{gutzwiller-book,gutzwiller-three-body} is one of the oldest problems in dynamics\footnote{A survey of some landmarks in the history of the three-body problem is presented in Appendix \ref{a:history-three-body}.} and was the place where Poincar\'e discovered chaos \cite{chenciner-poincare-three-body}. It continues to be a fertile area of research with discovery of new phenomena such as choreographies \cite{chenciner-montgomery} and Arnold diffusion \cite{xia-arnold-diffusion}. Associated questions of stability have stimulated much work in mechanics and nonlinear and chaotic dynamics \cite{laskar,routh}. Quantum and fluid mechanical variants with potentials other than Newtonian are also of interest, e.g., (a) the dynamics of two-electron atoms and the water molecule \cite{gutzwiller-book}, (b) the $N$-vortex problem with logarithmic potentials \cite{newton-n-vortex}, (c) the problem of three identical bosons with inverse-square potentials (Efimov effect in cold atoms \cite{efimov, efimov-cold-atom}) and (d) the Calogero-Moser system, also with inverse-square potentials \cite{calogero}. 

The inverse-square potential has some simplifying features over the Newtonian one, due in part to the nature of the scaling symmetry of the Hamiltonian,	
	\beq
	H(\la \bfr_1, \la \bfr_2, \la \bfr_3, \la^{-1} \bfp_1, \la^{-1} \bfp_2, \la^{-1} \bfp_3) = \la^{-2} H(\bfr_1, \bfr_2, \bfr_3, \bfp_1, \bfp_2, \bfp_3).
	\eeq
Here, for $a = 1,2$ and $3$, $\bfr_a$ and $\bfp_a$ are position and momentum vectors of the three bodies and $\la$ is a positive real number. As a consequence, the sign of the energy $E$ controls asymptotic behavior: bodies fly apart or suffer a triple collision according as $E$ is positive or negative, leaving open the special case $E=0$ \cite{Rajeev}. Indeed, if $m_{1,2,3}$ are the masses of the three bodies, the time evolution of the moment of inertia 
	\beq
	I = \sum_a m_a {\bfr_a}^2 = \sum_{a,i} m_a r_a^i r_a^i
	\eeq
for the inverse square potential is easily obtained from the canonical Poisson brackets $\{ {r_a}^i , {p_b}_j \}$ $= \del_{ab} \del^i_j$:
	\beq
	\{\dot I, r_a^i\} = \left\{ \sum_{b,j} 2 m_b \: r_b^j \: \dot r_b^j, r_a^i \right\} 
	= \left\{ \sum_{b,j} 2 r_b^j \: p_{bj}, r_a^i \right\}
	= -2 r_a^i \quad \text{and} \quad \{\dot I, p_{aj}\} = 2p_{aj}
	\eeq
implying $\{\dot I, T\} = 4T$ and $\{\dot I, V\} = 4V$ where $T = (1/2)\sum \bfp_a^2/m_a$ is the kinetic energy and $V$ is the potential energy. Thus, one obtains the Lagrange-Jacobi identity $\ddot I = \{\dot I, H\} = 4E$ where $E = T+V$ is the total conserved energy of the 3 bodies. Consequently, if $E > 0$ then $I \to \infty$ with bodies flying apart while if $E < 0$ then $I \to 0$ and the bodies suffer a triple collision. The intermediate case where $I$ remains non-zero and bounded for all time is particularly interesting. This happens when initial conditions are chosen so that $E = 0$ and $\dot I = 0$. By contrast, for the Newtonian potential, 
	\beq
	H(\la^{-2/3} \bfr_{1,2,3}, \la^{1/3} \bfp_{1,2,3}) = \la^{2/3} H(\bfr_{1,2,3}, \bfp_{1,2,3})
	\eeq
leads to $\ddot I = 4E - 2V$, which is not sufficient to determine the long-time behavior of $I$ when $E < 0$.

Here, we adopt a geometrical approach to  the planar three-body problem with Newtonian and attractive inverse-square potentials. It is well known that trajectories of a free particle moving on a Riemannian manifold are geodesics of a mass/kinetic metric $m_{ij}$ defined by the kinetic energy $\half m_{ij}(x) \dot x^i \dot x^j$. Indeed, geodesic flow on a compact Riemann surface of constant negative curvature is a prototypical model for chaos \cite{gutzwiller-book}. In the presence of a potential $V$, trajectories are reparametrized geodesics of the conformally related Jacobi-Maupertuis (JM) metric $g_{ij} = (E-V(x))m_{ij}$ (see \cite{arnold,lanczos} and \S \ref{s:traj-as-geodesics}). The linear stability of geodesics to perturbations is then controlled by sectional curvatures of the JM metric.

Several authors have tried to relate the geometry of the JM metric to chaos. For systems with many degrees of freedom, Pettini et al. \cite{pettini-2000-phys-rpts,pettini-2008, pettini-book-2007} obtain an approximate expression for the largest Lyapunov exponent in terms of curvatures. In \cite{pettini-1996}, the geometric framework is applied to investigate chaos in the H\'enon-Heiles system and a suitable average sectional curvature proposed as an indicator of chaos for systems with few degrees of freedom (see also \cite{Ramasubramanian-Sriram}). While negativity of curvature need not imply chaos, as the Kepler problem shows for $E > 0$, these works suggest that chaos could arise both from negativity of curvature and from fluctuations in curvature. Interestingly, the system of three coupled rotors studied in Chapter \ref{chapter:three-rotor} provides a striking connection  between a change in sign of the curvature and the onset of widespread chaos.

For the {\it planar} gravitational three-body problem (i.e. with pairwise Newtonian potentials), the JM metric on the full configuration space $\mR^6 \cong \mC^3$ has isometries corresponding to translation and rotation invariance groups $\bf C$ and U$(1)$ (\S \ref{s:jm-metric-config-space-hopf-coords}). This allows one to study the reduced dynamics on the quotients: the center-of-mass configuration space $\mC^2 \cong \mC^3/{\bf C}$ and shape space $\mR^3 \cong \mC^2/{\rm U}(1)$ \cite{montgomery-american-monthly}. Here, collision configurations are excluded from $\mC^3$ and its quotients. When the Newtonian potential is replaced with the inverse-square potential, the zero-energy JM metric  acquires a scaling isometry leading to additional quotients: $\mS^3 \cong \mC^2/{\rm scaling}$ and the shape sphere $\mS^2 \cong \mR^3/{\rm scaling}$ (see Fig. \ref{f:flow-chart}). Since the collision configurations have been removed, the (non-compact) shape sphere $\mS^2$ has the topology of a pair of pants and fundamental group given by the free group on two generators. As part of a series of works on the planar three-body problem, Montgomery \cite{montgomery-pants} shows that for three equal masses with inverse-square potentials\footnote{The $1/r^3$ force corresponding to the inverse-square potential is sometimes called a `strong' force.}, the curvature of the JM metric on $\mS^2$ is negative except at the two Lagrange points, where it vanishes. As a corollary, he shows the uniqueness of the analogue of Moore's `figure $8$' choreography solution (see Fig.~\ref{f:figure-8} and \cite{c-moore-braid-group}) up to isometries and establishes that collision solutions are dense within bound ones. In \cite{montgomery-syzygy,montgomery-2007}, he uses the geometry of the shape sphere to show that  zero angular momentum negative energy solutions (other than the Lagrange homotheties\footnote{In a Lagrange homothety, three bodies always occupy vertices of an equilateral triangle which shrinks to a triple collision at the center of mass without rotation.}) of the gravitational three-body problem have at least one syzygy\footnote{A syzygy is an instantaneous configuration where the three bodies are collinear.}.

We begin by extending some of Montgomery's results on the geometry of the shape sphere to the center-of-mass configuration space $\mC^2$ (without any restriction on angular momentum) and its quotients. In \S \ref{s:quotient-metrics} and \S \ref{s:curvature-newtonian-potential}, metrics on the quotients are obtained explicitly via Riemannian submersions \cite{Carmo} which simplify in `Hopf' coordinates \cite{nakahara}, as the Killing vector fields point along coordinate vector fields. These coordinates also facilitate our explicit computation of metrics and curvatures near binary and triple collisions. We interpret Lagrange and Euler homotheties (`central configurations' \cite{chenciner-scholarpedia}) as radial geodesics at global and local minima of the conformal factor in the JM metric for the inverse-square potential (\S \ref{s:geodesic-completeness}) and thereby deduce geodesic completeness of $\mC^2$ and its quotients $\mR^3$ and $\mS^3$ for arbitrary masses and allowed energies. The estimates showing completeness on $\mC^2$ are similar to those showing that the classical action (integral of Lagrangian) diverges for collisional trajectories. In a private communication, Montgomery points out that this was known to Poincar\'e and has been rediscovered several times (see for example \cite{c-moore-braid-group, montgomery-braid-groups, chenciner-icm-notes}). Completeness establishes that the geodesic reformulation `regularizes' pairwise and triple collisions by reparametrizing time so that any collision occurs at $t = \infty$. In contrast with other regularizations \cite{celletti, yeomans}, this does not involve an extrapolation of the dynamics past a collision nor a change in dependent variables. Unlike for the inverse-square potential, we show that geodesics for the Newtonian potential {\it can} reach curvature singularities (binary/triple collisions) in finite geodesic time (\S \ref{s:geodesic-incompleteness-newtonian-pot}). This may come as a surprise, since the Newtonian potential is {\it less} singular than the inverse-square potential and masses collide sooner under Newtonian evolution in the inverse-square potential. However, due to the reparametrization of time in going from trajectories to geodesics, masses can collide in finite {\it geodesic} time in the Newtonian potential while taking infinitely long to do so in the inverse-square potential. Indeed, for the attractive $1/r^n$ potential, the JM line-element leads to estimates $\propto \int_0^{\eta_0} \frac{d \eta}{\eta^{n/2}}$ and $\int_0^{r_0} \frac{d r}{r^{n/2}}$ for the distances to binary and triple collisions from a nearby location (\S \ref{s:geodesic-completeness}). These diverge for $n \geq 2$ and are finite for $n < 2$.

To examine stability of geodesics, we evaluate scalar and sectional curvatures of the zero-energy, equal-mass JM metrics on $\mC^2$ and its quotients. For the inverse-square potential, we obtain strictly negative upper bounds for scalar curvatures on $\mC^2$, $\mR^3$ and $\mS^3$ (\S \ref{s:scalar-curvature-inv-sq-pot-c2-r3-s3-s2}), indicating widespread linear geodesic instabilities. Moreover, scalar curvatures are shown to be bounded below. In particular, they remain finite and negative at binary and triple collisions. O'Neill's theorem is used to determine or bound various sectional curvatures on $\mC^2$ using the more easily determined ones on its Riemannian quotients; they are found to be largely negative (\S \ref{s:sectional-curvature-inv-sq-pot}). On the other hand, for the Newtonian potential, we find that the scalar curvature on $\mC^2$ is strictly negative, though it can have either sign on shape space $\mR^3$ (\S \ref{s:curvature-newtonian-potential}). Unlike for the inverse-square potential, scalar curvatures $\to - \infty$ at collision points. We also discuss the geodesic instability of Lagrange rotation and homothety solutions for equal masses (\S \ref{s:stability-tensor}). We end the chapter with a cautionary remark  comparing stability of geodesics to that of corresponding trajectories: simple examples are used to illustrate that the two notions of stability need not always coincide. 

While it is still a challenge to relate the above geodesic instabilities in the planar three-body problem to medium- and long-time behavior as well as to chaos, the problem we now turn to, i.e., the classical three-rotor problem, provides an arena to study this connection without the added complications of collisions.

\section[Classical three-rotor problem]{Classical three-rotor problem \sectionmark{Classical three--rotor problem}}
\sectionmark{Classical three--rotor problem}
\label{s:intro-3rotor}

\begin{figure}
 	\centering
		\begin{subfigure}[t]{5cm}
		\centering
		\includegraphics[width=5cm]{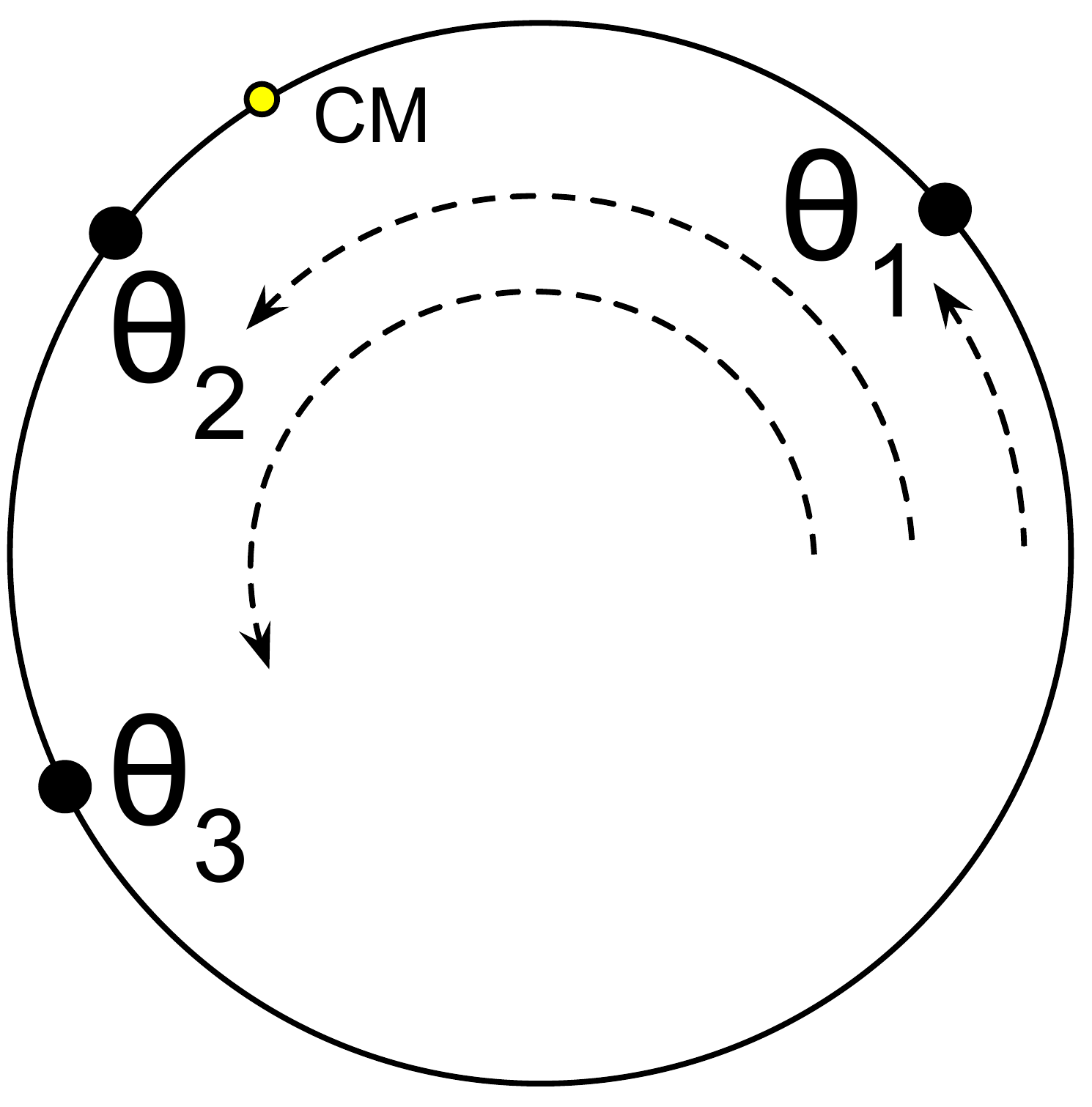}
		\caption{\small Three coupled rotors}
		\label{f:3rotor-config}
		\end{subfigure}
		\qquad
		\begin{subfigure}[t]{9cm}
		\centering
		\raisebox{1.7cm}{\includegraphics[width=9cm]{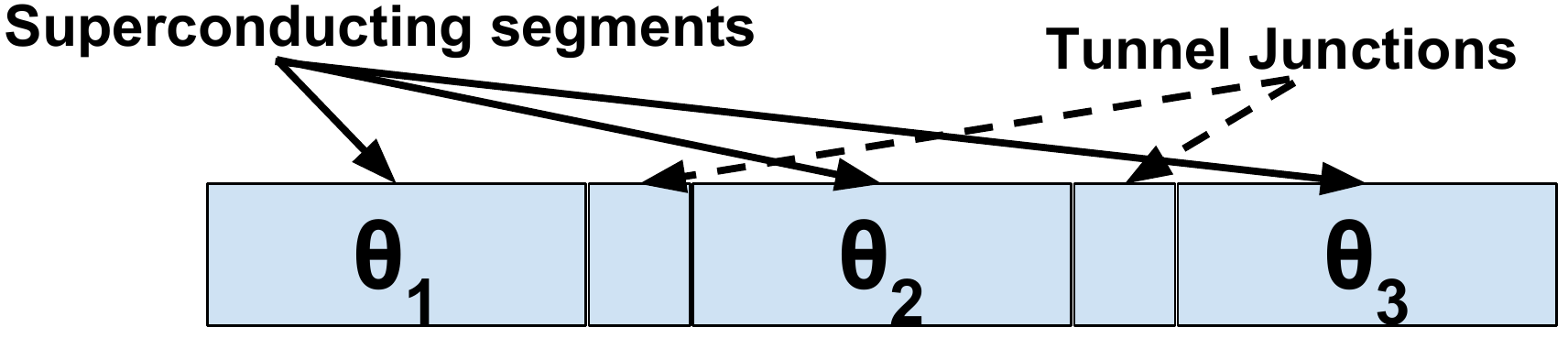}}
		\caption{\small Chain of Josephson junctions}
		\label{f:3rotor-JJchain}
		\end{subfigure}
	\caption{\small (a) Three coupled classical rotors with angular positions $\tht_{1,2,3}$ and center of mass CM. (b) An open chain of three coupled Josephson junctions. A closed chain obtained by connecting the first and third segments via a junction may be modeled by the quantum three-rotor problem.}
	\label{f:3rotor}		
	\end{figure}

 In the classical three-rotor problem, three point particles of equal mass $m$ move on a circle subject to attractive cosine inter-particle potentials of strength $g$ (see Fig.~\ref{f:3rotor-config}). The problem of two rotors reduces to that of a simple pendulum while the three-rotor system bears some resemblance to a double pendulum as well as to the planar restricted three-body problem. However, unlike in the gravitational three-body problem, the rotors can pass through each other\footnote{As we will soon see, this is physically reasonable since the rotors occupy distinct sites when the three-rotor problem is viewed as the classical limit of a chain of coupled Josephson junctions.} so that there are no collisional singularities. In fact, the boundedness of the potential also ensures the absence of non-collisional singularities leading to global existence and uniqueness of solutions. Despite these simplifications, the dynamics of three (or more) rotors is rich and displays novel signatures of the transition from regular to chaotic motion as the coupling (or energy) is varied.
 
The quantum version of the $n$-rotor problem is also of interest as it is used to model chains of coupled Josephson junctions \cite{sondhi-girvin} (see Fig. \ref{f:3rotor-JJchain}). Here, the rotor angles are the phases of the superconducting order parameters associated to the segments between junctions. It is well-known that this model for chains of coupled Josephson junctions is related to the XY model of classical statistical mechanics \cite{sondhi-girvin,wallin-1994} (see  also Appendix \ref{a:n-rotor-from-xy} where we obtain the quantum $n$-rotor problem from the XY model via a partial continuum limit and a Wick rotation). While in the application to the insulator-to-superconductor transition in arrays of Josephson junctions, one is typically interested in the limit of large $n$, here we focus on the classical dynamics of the $n=3$ case. 

The classical $n$-rotor problem also bears some resemblance to the Frenkel-Kontorova (FK) model \cite{braun-kivshar-FK}. The latter describes a chain of particles subject to nearest neighbor harmonic and onsite cosine potentials. Despite having different potentials and `target spaces' ($\mR^1$ vs $S^1$), the FK and $n$-rotor problems both admit continuum limits described by the sine-Gordon field \cite{braun-kivshar-FK,sachdev}. The $n$-rotor problem also bears some superficial resemblance to the Kuramoto oscillator model \cite{kuramoto}: though the interactions are similar, the equations of motion are of second and first order respectively.

Though quite different from our model, certain variants of the three-rotor problem have also been studied, e.g., (a) chaos in the dynamics of three masses moving on a line segment with periodic boundary conditions subject to harmonic and 1d-Coulombic inter-particle potentials \cite{kumar-miller}, (b) three free but colliding masses moving on a circle and indications of a lack of ergodicity therein \cite{Rabouw-Ruijgrok}, (c) coupled rotors with periodic driving and damping, in connection with mode-locking phenomena \cite{Thouless-Choi} and (d) an open chain of three coupled rotors with pinning potentials and ends coupled to stochastic heat baths, in connection to ergodicity \cite{Eckmann}.

In the center of mass frame of the three-rotor problem, we discover three classes of periodic solutions: choreographies up to moderate relative energies $E$ and pendula and breathers at all $E$. The system is integrable at $E=0$ and $\infty$ but displays a fairly sharp transition to chaos around $E \approx 4g$, thus providing an instance of an order-chaos-order transition. We find several manifestations of this transition: (a) a geometric cascade of stable to unstable transition energies in pendula as $E \to 4g^{\pm}$; (b) a transition in the curvature of the Jacobi-Maupertuis metric from being positive to having both signs as $E$ exceeds four, implying widespread onset of instabilities; (c) a dramatic rise in the fraction of the area of Poincar\'e surfaces occupied by chaotic trajectories and (d) a breakdown of discrete symmetries in Poincar\'e sections present at lower energies. Slightly above this transition, we find evidence for a band of global chaos where we conjecture ergodic behavior. This is in contrast with the model of three free but colliding masses moving on a circle \cite{Rabouw-Ruijgrok} discussed above where numerical investigations indicated a lack of ergodicity. 

There are several few degrees of freedom models that display global chaos as well as ergodicity and mixing. Geodesic flow on a constant negative curvature compact Riemann surface is a well-known example \cite{sinai-geodesic-flow,sinai-central-limit}. Ballistic motion on billiard tables of certain types including Sinai billiards \cite{sinai-billiard} and its generalization to the Lorentz gas \cite{lenci} provide other canonical examples. Kicked rotors and the corresponding Chirikov standard map \cite{chirikov-phys-rept} are also conjectured to display global chaos and ergodicity for certain sufficiently large parameter values \cite{stdmap-ergodicity}. An attractive feature of the three-rotor system is that, in contrast to these canonical examples, it offers the possibility of studying ergodicity in a continuous time autonomous Hamiltonian system of particles without boundaries or specular reflections (rotors can `pass through' each other without colliding). Interestingly, the center of mass dynamics of three rotors may also be regarded as geodesic flow on a 2-torus with non-constant curvature (of both signs) of an appropriate Jacobi-Maupertuis metric (see \S \ref{s:JM-approach}).

The statistics of recurrence times provides another window into chaotic dynamics \cite{kac, zaslavsky}. It is well-known that the distribution of recurrence times to small volumes in phase space approaches an exponential law for sufficiently mixing dynamics (e.g. Axiom-A systems \cite{hirata-axiomA} and some uniformly hyperbolic systems \cite{hirata-uniformly-hyperbolic}). Moreover, successive recurrence times are independently distributed so that the sequence of recurrence times is Poissonian. 

In the three-rotor problem, we provide evidence for ergodicity in the band of global chaos by showing that numerically determined time averages approach the corresponding  ensemble averages. Evidence for mixing in the same band is obtained by showing that trajectories with a common energy from a small volume approach a uniform distribution on the energy hypersurface. Finally, we show that the distribution of recurrence times to finite size cells on such energy hypersurfaces follows an exponential law. Moreover, the mean recurrence time obeys a scaling law with exponent as expected from global chaos and ergodicity.

We now summarize our results on the three-rotor problem described in Chapter \ref{chapter:three-rotor}. We begin by formulating the classical three-rotor problem in \S \ref{s:three-rotor-setup}. We show absence of singularities and eliminate the center of mass motion to arrive at dynamics on a $2$ dimensional configuration torus parametrized by the relative angles $\vf_1$ and $\vf_2$. In \S \ref{s:dynamics-on-2torus}, we discuss the dynamics on the $\vf_1$-$\vf_2$ torus, find all static solutions for the relative motion and discuss their stability (see Fig. \ref{f:static-solutions-3rotors}). The system is also shown to be integrable at zero and infinitely high relative energies $E$ (compared to the coupling $g$) due to the emergence of additional conserved quantities. Furthermore, using Morse theory, we discover changes in the topology of the Hill region of the configuration space at $E = 0$, $4g$ and $4.5g$ (see Fig. \ref{f:topology-hill-region}). 

 In \S \ref{s:reduction-one-dof}, we use consistent reductions of the equations of motion to one degree of freedom to find two families of periodic solutions at all energies (pendula and isosceles breathers, see Fig. \ref{f:periodic-soln}). This is analogous to how the Euler and Lagrange solutions of the three-body problem arise from suitable Keplerian orbits. We investigate the stability of the pendula and breathers by computing their monodromies. Notably, we find that the stability index of pendula becomes periodic on a log scale as $E \to 4g^\pm$ and shows an accumulation of stable to unstable transition energies at $E = 4g$ (see Fig. \ref{f:monodromy-evals}). In other words, the largest Lyapunov exponent switches from positive to zero infinitely often with the widths of the (un)stable windows asymptotically approaching a geometric sequence as the pendulum energy approaches $4g$. This accumulation bears an interesting resemblance to the Efimov effect \cite{efimov} as discussed in \S \ref{c:discussion} and to the cascade of period doubling bifurcations in unimodal maps \cite{logistic-map}.

In \S \ref{s:JM-approach}, we reformulate the dynamics on the $\vf_1$-$\vf_2$ torus as geodesic flow with respect to the Jacobi-Maupertuis metric. We prove in Appendix \ref{a:positivity-of-JM-curvature} that the scalar curvature is strictly positive on the Hill region for $0 \leq E \leq 4g$ but acquires both signs above $E = 4g$ (see Fig. \ref{f:curvature-3rotors-phi1-phi2-torus}) indicating widespread geodesic instabilities as $E$ crosses $4g$. In \S \ref{s:poincare-section}, we examine Poincar\'e sections and observe a marked transition to chaos in the neighborhood of $E = 4g$ as manifested in a rapid rise of the fraction of the area of the energetically allowed `Hill' region occupied by chaotic sections (see Fig. \ref{f:chaos-vs-egy}). This is accompanied by a spontaneous breaking of two discrete symmetries present in Poincar\'e sections below this energy (see Figs. \ref{f:psec-egy=2-3} and \ref{f:psec-egy-near-4}). This transition also coincides with the accumulation of stable to unstable transition energies of the pendulum family of periodic solutions at $E = 4g$. Slightly above this energy, we find a band of global chaos $5.33g \lesssim E \lesssim 5.6g$, where the chaotic sections fill up the entire Hill region on all Poincar\'e surfaces, suggesting ergodic behavior (see Fig. \ref{f:global-chaos}). Appendix \ref{a:estimate-f} summarizes the numerical method employed to estimate the fraction of chaos on Poincar\'e surfaces.

In \S \ref{s:choreographies}, we derive a system of delay differential and algebraic equations for periodic choreography solutions of the three-rotor problem. We discover three families of choreographies. The first pair are uniformly rotating versions of two of the static solutions for the relative motion. The third family is non-rotating, stable and exists for all relative energies up to the onset of global chaos (see Fig. \ref{f:choreo-time-vs-egy-phi1-phi2-plot}). It is found by a careful examination of Poincar\'e sections. What is more, we  prove that choreographies cannot exist for arbitrarily high relative energies.

In \S \ref{s:ergodicity}, we present evidence for ergodicity in the band of global chaos by showing that numerically determined time averages agree with ensemble averages.  In particular, we find the distributions of relative angles ($\vf_{1,2}$) and momenta ($p_{1,2}$) over constant energy hypersurfaces weighted by the Liouville measure. While the joint distribution function of $\vf_{1,2}$ is uniform on the Hill region of the configuration torus at all energies, the distribution of $p_1$ (and of $p_2$) shows interesting transitions from the Wigner semi-circular distribution when $E \ll g$ to a bimodal distribution for $E > 4.5g$ (see Fig.~\ref{f:collage-ensemble-time-avg-dist}).  In the band of global chaos, we find that distributions of $\vf_{1,2}$ and $p_{1,2}$ along generic (chaotic) trajectories are independent of the chosen trajectory and agree with the corresponding distributions over constant energy hypersurfaces, indicating ergodicity. This agreement fails for energies outside this band. In \S \ref{s:approach-to-ergodicity}, we investigate the rate of approach to ergodicity. We find that time averages such as $\bra \cos^2 \vf_1 \ket_{\rm t}$ and $\bra p_1^2 \ket_{\rm t}$ along a generic trajectory over the time interval $[0,T]$  approach the corresponding ensemble averages as a power law $\sim T^{-1/2}$ (see Fig. \ref{f:mean-vs-time}). This is expected of an ergodic system where correlations decay sufficiently fast in time as shown in Appendix \ref{a:approach-to-ergodicity} (see also \cite{prl-dechant} for a stochastic formulation).

In \S \ref{s:mixing}, we show that the dynamics is mixing (with respect to the Liouville measure) in the band of global chaos. This is done by showing that the histogram of number of trajectories in various cells partitioning the energy hypersurface approaches a distribution strongly peaked at the expected value with increasing time (see Fig. \ref{f:approach-to-mixing-in-time}). We also observe characteristic departures from mixing {\it even in chaotic regions} of the phase space at energies just outside this band (see Fig.~\ref{f:approach-to-mixing-in-energy}). 

In \S \ref{s:recurrenc-time-dist}, we study the distribution of recurrence times to a finite size cell \cite{altmann} in a given energy hypersurface. Within the band of global chaos, we find that the normalized distribution of recurrence times $\tau$ follows the exponential law $(1/\bar \tau) \exp(-\tau/\bar \tau)$ with possible deviations at small recurrence times (see Fig.~\ref{f:recurrence-time-distribution}). Though the mean recurrence/relaxation time $\bar \tau$ varies with the Liouville volume ${\rm v}$ of the cell, we find that it obeys the scaling law $\bar \tau \times {\rm v}^{2/3} = \tau^*$. This scaling law is similar to the ones discussed in \cite{balakrishnan-scaling-law, gao} with the scaling exponent $2/3$ consistent with global chaos and ergodicity. The rescaled mean recurrence time $\tau^*$ can vary with the location of the cell center, but does {\it not} vary significantly with energy in the band of global chaos. Finally, we demonstrate a loss of memory by showing that the gaps between successive recurrence times are uncorrelated.

We conclude the thesis with a discussion in Chapter \ref{c:discussion}.

\chapter[Instabilities in the planar three-body problem]{Instabilities in the planar three-body problem:  A geometrical approach}
\chaptermark{Instabilities in the planar three--body problem}
\label{chapter:three-body}

This chapter is based on \cite{gskhs-three-body} and \cite{gskhs-three-body-resonance}. Here, we study the planar three-body problem via a geometrical approach. To set the stage, in \S \ref{s:traj-as-geodesics} we introduce a reformulation of trajectories of Newtonian mechanics as geodesics of the Jacobi-Maupertuis metric on the configuration space.

\section[Trajectories as geodesics of the Jacobi-Maupertuis metric]{Trajectories as geodesics of the Jacobi-Maupertuis metric \sectionmark{Trajectories as geodesics of the Jacobi--Maupertuis metric}}
\sectionmark{Trajectories as geodesics of the Jacobi--Maupertuis metric}
\label{s:traj-as-geodesics}

Fermat's principle in optics states that light rays extremize the optical path length $\int n(\bfr(\tau)) \: d\tau$ where $n(\bfr)$ is the (position dependent) refractive index and $\tau$ a parameter along the path\footnote{The optical path length $\int n(\bfr) \, d\tau$ is proportional to $\int d\tau/\la$, which is the geometric length in units of the local wavelength $\la(\bfr) = c/n(\bfr) \nu$. Here, $c$ is the speed of light in vacuum and $\nu$ the constant frequency.}. The variational principle of Euler and Maupertuis (1744) is a mechanical analogue of Fermat's principle \cite{arnold,lanczos}. It states that the curve that extremizes the abbreviated action $\int_{\bfq_1}^{\bfq_2} {\bf p}\cdot d{\bf q}$ holding energy $E$ and the end-points $\bfq_1$ and $\bfq_2$ fixed has the same shape as the Newtonian trajectory. By contrast, Hamilton's principle of extremal action (1835) states that a trajectory going from $\bfq_1$ at time $t_1$ to $\bfq_2$ at time $t_2$ is a curve that extremizes the action. 

It is well-known that the trajectory of a free particle (i.e., subject to no forces) moving on a plane is a straight line. Similarly, trajectories of a free particle  moving on the surface of a sphere are great circles. More generally, for a mechanical system with configuration space $M$ and Lagrangian $L = \half m_{ij}(\bfq) \dot q^i \dot q^j$, Lagrange's equations $\DD{p_i}{t} = \dd{L}{q^i}$ are equivalent to the geodesic equations with respect to the `mass' or `kinetic metric' $m_{ij}$ on $M$:
	\beq
	 m_{ij} \: \ddot q^j(t) = - \half \left(m_{ji,k} + m_{ki,j} - m_{jk,i} \right) \dot q^j(t) \: \dot q^k(t).
	 \eeq
Here, $m_{ij,k} = \pdr m_{ij}/\pdr q^k$ and $p_i = \dd{L}{\dot q^i} = m_{ij}\dot q^j$ is the momentum conjugate to coordinate $q^i$. For instance, the kinetic metric ($m_{rr} = m$, $m_{\tht \tht} = m r^2$, $m_{r \tht} = m_{\tht r} = 0$) for a free particle moving on a plane may be read off from the Lagrangian $L = \half m (\dot r^2 + r^2 \dot \tht^2)$ in polar coordinates, and the geodesic equations shown to reduce to Lagrange's equations of motion $\ddot r = r \dot \tht^2$ and $d(m r^2 \dot \tht)/dt = 0$.

Remarkably, this correspondence between trajectories and geodesics continues to hold even in the presence of conservative forces derived from a potential $V$ and follows from a refinement of the Euler-Maupertuis principle due to Jacobi. The shapes of trajectories and geodesics coincide but the Newtonian time along trajectories is not the same as the arc-length parameter along geodesics. Precisely, the equations of motion (EOM)
	\beq 
	 m_{ki} \ddot x^i(t) = - \pdr_k V - \half \left(m_{ik,j} + m_{jk,i} - m_{ij,k} \right) \dot x^i(t) \: \dot x^j(t)
	 \label{e:Lagrange-eqns-kin-metric-and-V}
	 \eeq
may be regarded as reparametrized geodesic equations for the Jacobi-Maupertuis (JM) metric,
	\beq
	ds^2 = g_{ij} dx^i dx^j = (E-V) m_{ij} dx^i dx^j
	\eeq
on the classically allowed `Hill' region $E - V \geq 0$. Notice that $\sqrt{2} \int ds = \int p dq = \int (L+E) dt$ so that the length of a geodesic is related to the classical action of the trajectory. The formula for the inverse JM metric $g^{ij} = m^{ij}/(E-V)$ may also be read off from the time-independent Hamilton-Jacobi (HJ) equation $(m^{ij}/2(E-V))\; \pdr_i W \pdr_j W = 1$ by analogy with the rescaled kinetic metric $m^{ij}/2E$ appearing in the free particle HJ equation $(m^{ij}/2E) \pdr_i W \pdr_j W = 1$ (see p.74 of \cite{Rajeev}). The JM metric is conformal to the kinetic metric and depends parametrically on the conserved energy $E = \half m_{ij} \dot x^i \dot x^j + V$. The geodesic equations 
	\beq
	\label{e:jm-geodesic-equation}
	\ddot x^l(\la) = -  \ov{2}g^{lk}\left(g_{ki,j}+g_{kj,i}-g_{ij,k}\right)\dot x^i(\la) \dot x^j(\la)
	\eeq 
for the JM metric reduce to (\ref{e:Lagrange-eqns-kin-metric-and-V}) under the reparametrization 
	\beq
	\fr{d}{d\la} = \fr{1}{\sigma} \fr{d}{dt} \quad \text{where} \quad \sigma = \fr{(E-V)}{\sqrt{\cal{T}}}.
	\eeq
Here, ${\cal T} = \half g_{ij} \dot x^i \dot x^j$ is the conserved `kinetic energy' along geodesics and equals one-half for arc-length parametrization. To obtain $\sigma$, suppose $y^i(t)$ is a trajectory and $z^i(\la)$ the corresponding geodesic. Then at a point $x^i = z^i(\la) = y^i(t)$, the velocities are related by $\sigma \dot z^i = \dot y^i$ leading to
	\beq
	{\cal T} = \half g_{ij} \dot z^i \dot z^j =\fr{E-V}{2}m_{ij} \dot{z}^i\dot{z}^j = \fr{E-V}{2 \sigma^2} m_{ij} \dot{y}^i\dot{y}^j  = \left(\fr{E-V}{ \sigma}\right)^2.
	\eeq
This reparametrization of time may be inconsequential in some cases [e.g. Lagrange rotational solutions where $\sigma$ is a constant since $V$ is constant along the trajectory (see \S \ref{s:stability-tensor})] but may have significant effects in others [e.g. Lagrange homothety solutions where the exponential time-reparametrization regularizes triple collisions (see \S \ref{s:triple-collision-inv-sq-near-collision-geom})] and could even lead to a difference between linear stability of trajectories and corresponding geodesics (see \S \ref{s:stability-tensor}). 

The curvature of the JM metric encodes information on linear stability of geodesics (see \S \ref{s:sectional-curvature-inv-sq-pot}). For example, in the planar isotropic harmonic oscillator with potential $k r^2/2$ in plane polar coordinates, the gaussian curvature $R = 16Ek/(2E-kr^2)^3$ of the JM metric on configuration space is non-negative everywhere indicating stability. In the planar Kepler problem with Hamiltonian $\bfp^2/2m - k/r$, the gaussian curvature of the JM metric $ds^2 = m(E+k/r)(dr^2+r^2d\theta^2)$ is $R =  -{Ek}/ ({m(k+Er)^3})$. $R$ is  everywhere negative/positive for $E$ positive/negative and vanishes identically for $E = 0$. This reflects the divergence of nearby hyperbolic orbits and oscillation of nearby elliptical orbits. Negativity of curvature could lead to chaos, though not always, as the hyperbolic orbits of the Kepler problem show. As noted, chaos could also arise from curvature fluctuations \cite{pettini-2000-phys-rpts}.

\section[Planar three-body problem with inverse-square potential]{Planar three-body problem with inverse-square \\potential \sectionmark{Planar three--body problem with inverse--square potential}}
\sectionmark{Planar three--body problem with inverse--square potential}

\subsection{Jacobi-Maupertuis metric on the configuration space}
\label{s:jm-metric-config-space-hopf-coords}

\begin{figure}	
 	\centering
	\includegraphics[height=5cm]{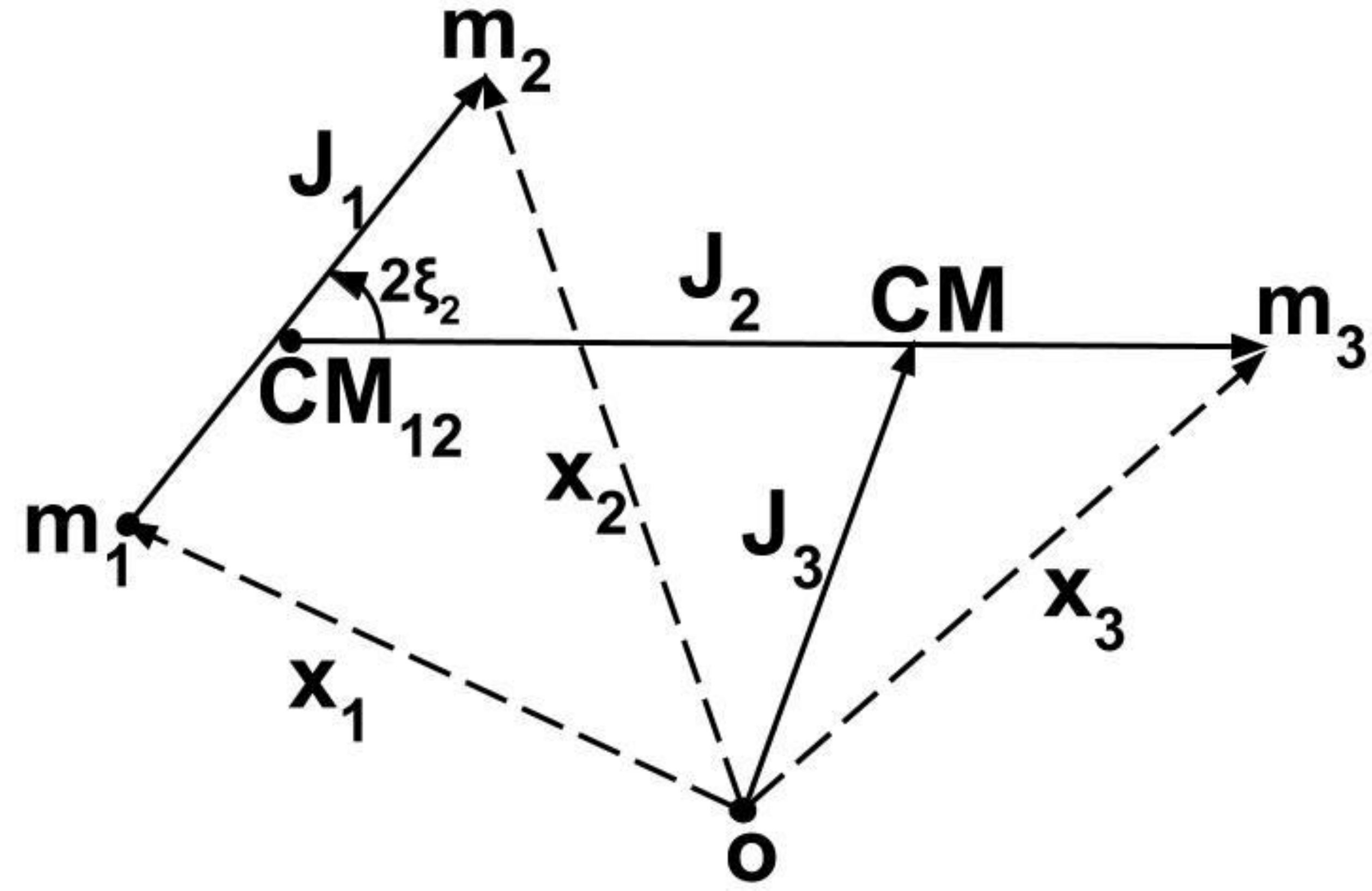}
	\caption{\small Position vectors $x_{1,2,3}$ of masses relative to origin and Jacobi vectors $J_{1,2,3}$.}
	\label{f:jacobi-vectors}		
\end{figure}

We consider the three-body problem  with masses moving on a plane regarded as the complex plane $\bf{C}$. Its $6$D configuration space (with collision points excluded) is identified with ${\mC^3}$. A point on $\mC^3$ represents a triangle on the complex plane with the masses $m_{1,2,3}$ at its vertices $x_{1,2,3} \in \mathbf{C}$. It is convenient to work in Jacobi coordinates (Fig. \ref{f:jacobi-vectors})
	\beq
	\label{e:jacobi-coordinates-on-c3} 
J_1=x_2-x_1, \quad J_2=x_3-\fr{m_1x_1+m_2x_2}{m_1+m_2} \quad \text{and} \quad J_3=\fr{m_1x_1+m_2x_2+m_3x_3}{M_3},
	\eeq
in which the kinetic energy $KE = (1/2) \sum_i m_i |\dot x_i|^2$ remains diagonal:
	\beq
	\label{e:ke-in-jacobi-coordinates-on-c3}
	KE = \half \sum_i M_i | \dot J_i|^2 \;\; \text{where} \;\;
	\ov{M_1}=\ov{m_1}+\ov{m_2}, \;\; \ov{M_2}=\ov{m_3}+\ov{m_1+m_2}
	\eeq
and $M_3 = \sum_i m_i$. The KE for motion about the center of mass (CM) is $\half (M_1 |\dot J_1|^2 + M_2 |\dot J_2|^2)$. The moment of inertia about the origin $I = \sum_{i=1}^3 m_i |x_i|^2$ too remains diagonal in Jacobi coordinates ($I = \sum_{i=1}^3 M_i |J_i|^2$), while about the CM we have $I_{\rm CM} = M_1 |J_1|^2 + M_2 |J_2|^2$. With 
	\beq
	U = -V = \sum_{i < j} \fr{G m_i m_j}{|x_i - x_j|^2}
	\eeq
denoting the (negative) potential energy, the JM metric for energy $E$ on $\mC^3$ is
	\beq
	\label{e:jm-metric-in-jacobi-coordinates-on-c3}
	ds^2 = \left( E + U  \right) \sum_{i=1}^3 M_i |dJ_i|^2 \quad \text{where} \quad
	 U = \fr{G m_1 m_2}{|J_1|^2} + \fr{G m_2 m_3}{|J_2 - \mu_1 J_1|^2} + \fr{G m_3 m_1}{|J_2+\mu_2 J_1|^2}
	\eeq
and $\mu_i= m_i/(m_1 + m_2)$. Due to the inverse-square potential, $G$ {\it does not} have the usual dimensions. The metric is independent of the CM coordinates $J_3$ and $\bar J_3$, while $J_1,\bar J_1, J_2$ and $\bar J_2$ are invariant under translations $x_i \to x_i + a$ for $a \in \mathbf{C}$. Thus translations act as isometries of (\ref{e:jm-metric-in-jacobi-coordinates-on-c3}). Similarly, we will see that scalings (for $E =0$) and rotations also act as isometries. These isometries also act as symmetries of the Hamiltonian. For instance the dilatation $D = \sum_i \vec x_i \cdot \vec p_i = \sum_i \Re (x_i \bar p_i)$ generates scale transformations $x_i \to \la x_i$ and $p_i \to \la^{-1} p_i$ via Poisson brackets: $\{x_i, D \} = x_i$ and $\{p_i, D \} = - p_i$. Since $\{ H, D \} = - 2 H$, scaling is a symmetry of the Hamiltonian only when energy vanishes.

\subsubsection{Isometries and Riemannian submersions}

The study of the geometry of the JM metric is greatly facilitated by first considering the geometry of its quotients by isometries (for instance, geodesics on a quotient lift to horizontal geodesics). Riemannian submersions \cite{Carmo, oneill} provide a framework to define and compute metrics on these quotients. Suppose $(M,g)$ and $(N,h)$ are two Riemannian manifolds and $f: M \rightarrow N$ a surjection (an onto map). Then the linearization $df(p): T_p M \to T_{f(p)} N$ is a surjection between tangent spaces. The vertical subspace $V(p) \subseteq T_p M$ is defined to be the kernel of $df$ while its orthogonal complement $\ker(df)^\perp$ with respect to the metric $g$ is the horizontal subspace $H(p)$. $f$ is a Riemannian submersion if it preserves lengths of horizontal vectors, i.e., if the isomorphism $df(p) \colon \ker(df(p))^{\perp} \rightarrow T_{f(p)} N$ is an isometry at each point. The Riemannian submersions we are interested in are associated to quotients of a Riemannian manifold ($M,g$) by the action of a suitable group of isometries $G$. There is a natural surjection $f$ from $M$ to the quotient $M/G$. Requiring $f$ to be a Riemannian submersion defines the quotient metric on $M/G$: the inner product of a pair of tangent vectors $(u,v)$ to $M/G$ is defined as the inner product of {\it any} pair of horizontal preimages under the map $df$.

The surjection $\left( J_1,\bar J_1, J_2,\bar J_2, J_3,\bar J_3\right) \mapsto\left( J_1,\bar J_1, J_2,\bar J_2\right)$ defines a submersion from configuration space $\mC^3$ to its quotient $\mC^2$ by translations. Linearization of this map $d{\cal J}( J) : T_ J {\mC^3} \to T_{{\cal J}( J)} \mC^2$ is the Jacobian matrix 
	\beq
	d{\cal J}=\colvec{4}{1&0&0&0&0&0&0}{0&1&0&0&0&0&0}{0&0&1&0&0&0&0}{0&0&0&1&0&0&0}.
	\eeq
$T_J {\mC^3}$ is the span of  $\pdr_{ J_1}, \pdr_{\bar J_1}, \pdr_{ J_2}, \pdr_{\bar J_2}, \pdr_{ J_3}, \pdr_{\bar J_3}$ and a typical tangent vector $a_1\pdr_{ J_1}+a_2 \pdr_{\bar J_1}+a_3 \pdr_{ J_2}+a_4 \pdr_{\bar J_2}+a_5 \pdr_{ J_3}+a_6 \pdr_{\bar J_3}$ is represented by the column vector $\colvec{1}{a_1 & a_2 & a_3 & a_4 & a_5 & a_6}^t$. The vertical subspace $V(J)$ of the submersion is defined to be the kernel of $d{\cal J}( J)$ i.e. the span of $ \pdr_{ J_3}$ and $\pdr_{\bar J_3}$. The orthogonal complement of $V( J)$ in $T_ J {\mC^3}$ is the horizontal subspace $H( J)$. $H( J)$ is spanned by the four orthogonal vectors  $\pdr_{ J_1}, \pdr_{\bar J_1}, \pdr_{ J_2}$ and  $\pdr_{\bar J_2}$. For the map ${\cal J}$ to be a riemannian submersion, lengths of horizontal vectors must be preserved. A typical horizontal vector is  of the form $a_1\pdr_{ J_1}+ \bar a_1 \pdr_{\bar J_1}+a_2 \pdr_{ J_2}+\bar a_2 \pdr_{\bar J_2}$ with norm-square $( E + U ) \sum M_i \; a_i \; \bar a_i$. This defines the quotient metric on $\mC^2$ in coordinates $J_1,\bar J_1, J_2$ and $\bar J_2$:
	\beq
	\label{e:jm-metric-in-jacobi-coordinates-on-c2}
	ds^2=( E + U ) (M_1 \; |d J_1|^2 + M_2 \; |d J_2|^2).
	\eeq 
It is convenient to define rescaled coordinates on $\mC^2$, $z_i = \sqrt{M_i} \:  J_i$, in terms of which (\ref{e:jm-metric-in-jacobi-coordinates-on-c2}) becomes $ds^2 = (E + U) (|dz_1|^2 + |dz_2|^2)$. The kinetic energy in the CM frame is $KE = (1/2) (|\dot z_1|^2 + |\dot z_2|^2 )$ while $I_{\rm CM} = |z_1|^2 + |z_2|^2$.

\subsubsection{Hopf coordinates on \texorpdfstring{$\mC^2$}{C2} and quotient spaces \texorpdfstring{$\mR^3$, $\mS^3$ and $\mS^2$}{R3, S3 and S2}}

\begin{figure}[!ht]	
	\centering
	\begin{subfigure}[t]{4.4cm}
		\centering
		\includegraphics[height=4.5cm]{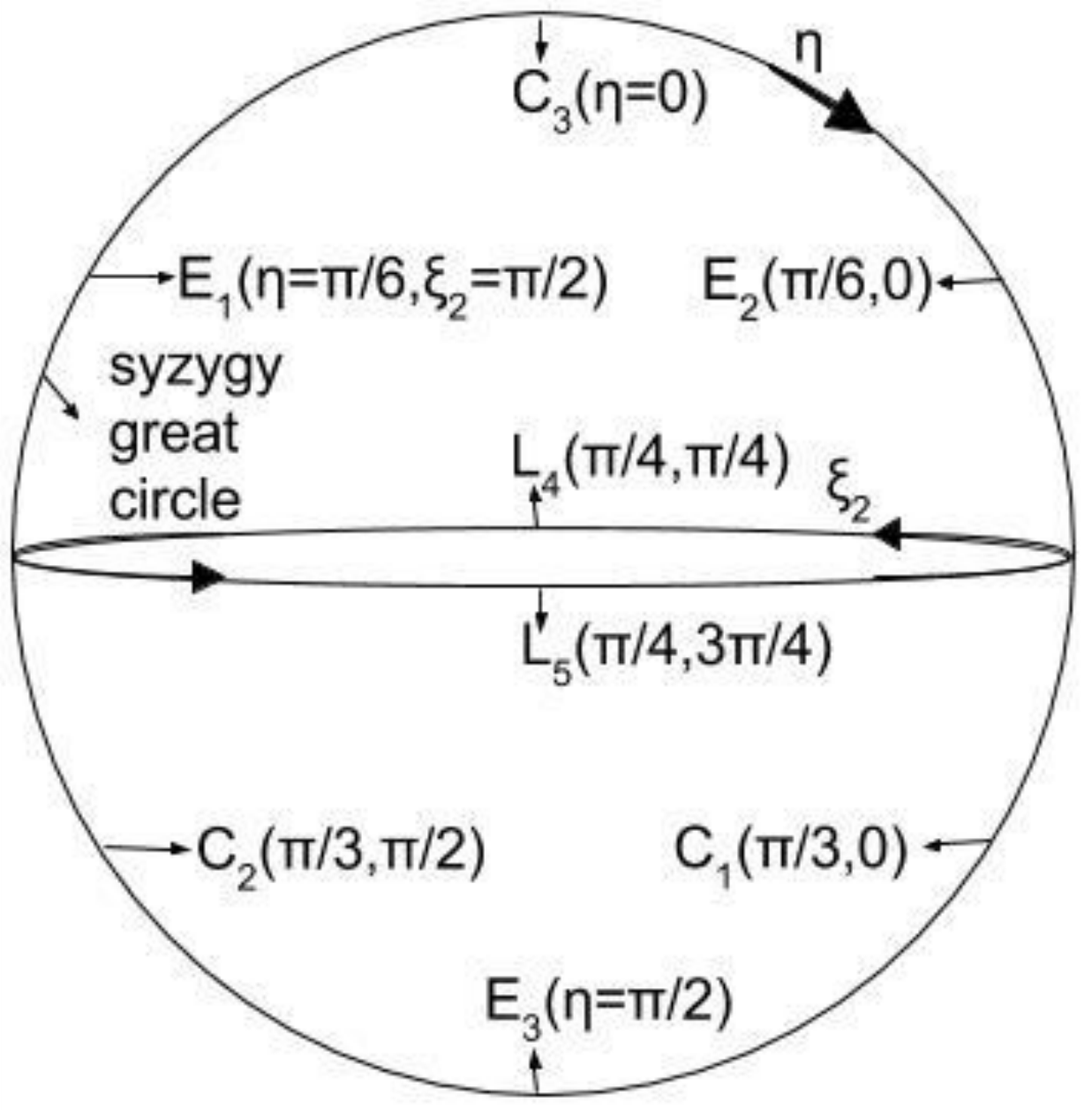}
		\caption{\small }
		\label{f:shape-sphere-xi2-eta}
	\end{subfigure}
	\begin{subfigure}[t]{3in}
		\centering
		\includegraphics[width=5cm]{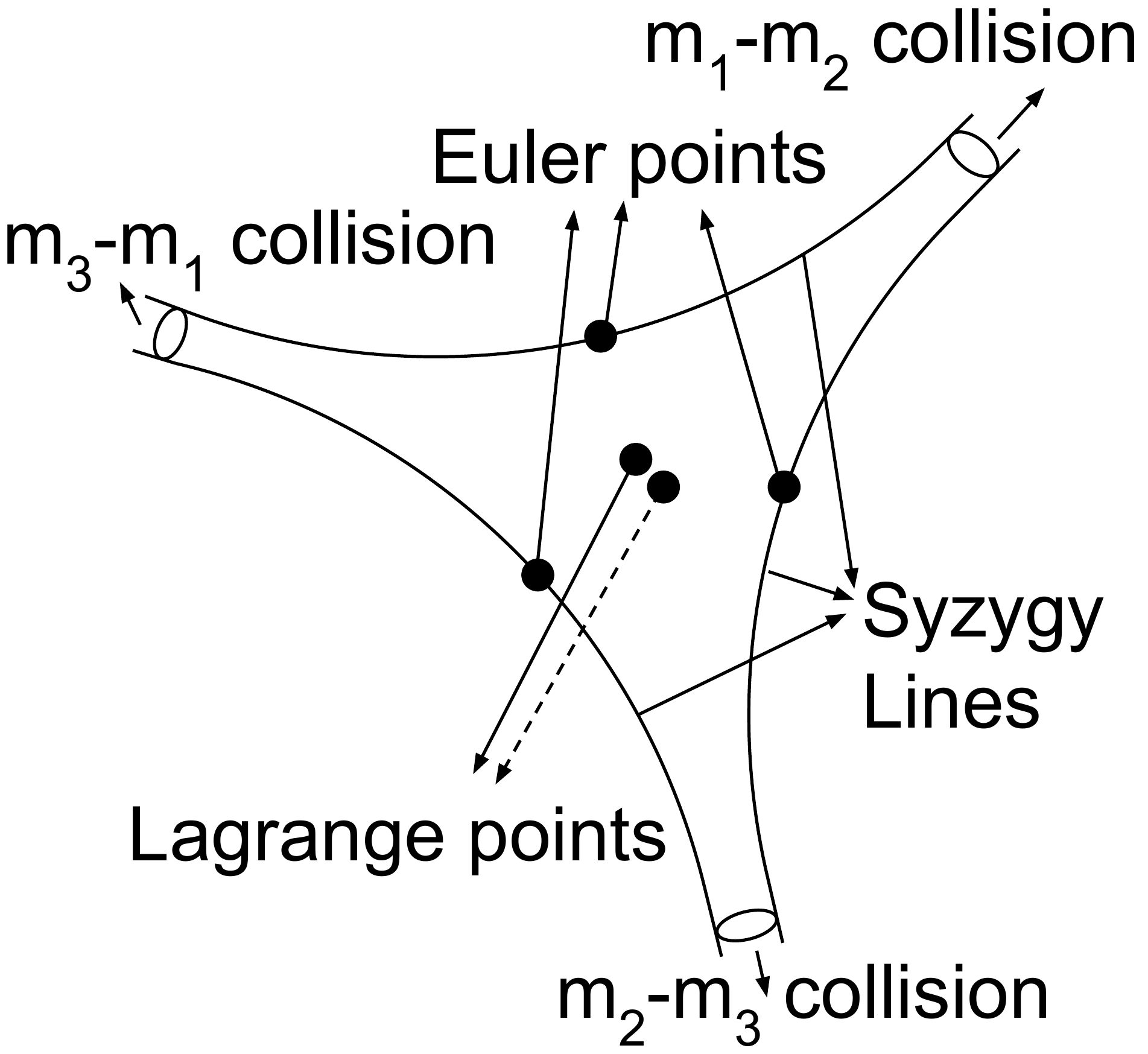}
		\caption{\small }
		\label{f:horn-shape-sphere}		
	\end{subfigure}	
	\begin{subfigure}[t]{3cm}
		\centering
		\includegraphics[height=5cm]{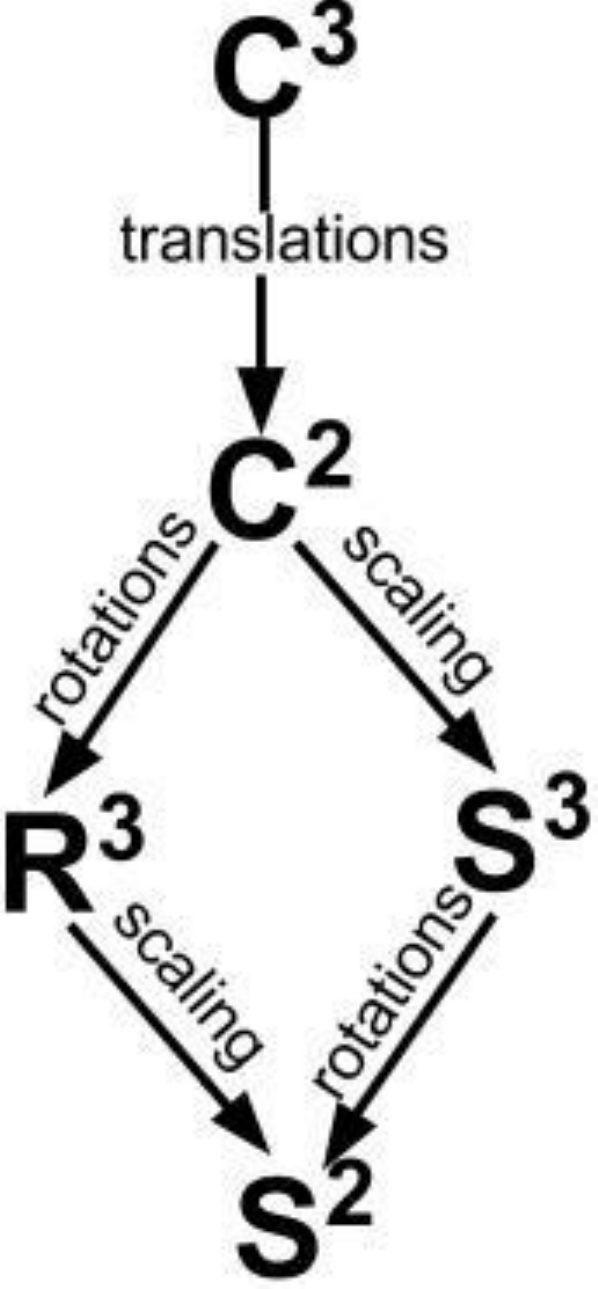}
		\caption{\small }
		\label{f:flow-chart}
	\end{subfigure}
	\caption{\small (a) The shape sphere is topologically a 2-sphere with the three collision points $C_{1,2,3}$ removed, endowed with the quotient JM metric of {\it negative} gaussian curvature. Coordinates and physical locations on the shape sphere are illustrated. 2$\eta$ is the polar angle ($0 \leq \eta \leq \pi/2$). 2$\xi_2$ is the azimuthal angle ($0\leq \xi_2 \leq \pi$). The `great circle' composed of the two longitudes $\xi_2=0$ and $\xi_2=\pi/2$ consists of collinear configurations (syzygies) which include $C_{1,2,3}$ and the Euler points $E_{1,2,3}$. Lagrange points $L_{4,5}$ lie on the equator $\eta = \pi/4$. The shape space $\mR^3$ is a cone on the shape sphere. The origin $r=0$ of shape space is the triple collision point. (b) The negatively curved `pair of pants' metric on the shape sphere $\mS^2$. (c) Flowchart of Riemannian submersions.}
	\label{f:shape-sphere-xi2-eta-and-jacobi-vectors}
\end{figure}

We now specialize to equal masses ($m_i = m$) so that $M_1 = m/2$,$M_2={2m}/{3}$ and $\mu_i = 1/2$. The metric on $\mC^2$ is seen to be conformal to the flat Euclidean metric  via the conformal factor $E + U$:
	\beq
	ds^2 = \left(E+\fr{G m^3}{2|z_1|^2}+\fr{2 G m^3}{3|z_2-\ov{\sqrt{3}}z_1|^2}+\fr{2 G m^3}{3|z_2+\ov{\sqrt{3}}z_1|^2}\right) \left(|dz_1|^2+|dz_2|^2 \right) .
	\label{e:jm-metric-in-rescaled-jacobi-coordinates-for-eq-mass-on-c2}
	\eeq
Rotations U$(1)$ act as a group of isometries of $\mC^2$, taking $\left(z_1,z_2\right) \mapsto\left(e^{i \tht}z_1,e^{i \tht}z_2\right)$ and leaving the conformal factor invariant. Moreover if $E = 0$, then scaling $z_i \mapsto \la z_i$ for $\la \in {\bf R}^+$ is also an isometry. Thus we may quotient the center-of-mass configuration manifold $\mC^2$ successively by its isometries. We will see that $\mC^2/$U$(1)$ is the shape space $\mR^3$ and $\mC^2$/scaling is $\mS^3$. Furthermore the quotient of $\mC^2$ by both scaling and rotations leads to the shape sphere $\mS^2$ (see Fig. \ref{f:flow-chart}, note that collision points are excluded from $\mC^2, \mR^3, \mS^3$ and $\mS^2$). Points on shape space $\mR^3$ represent oriented congruence classes of  triangles while those on the shape sphere $\mS^2$ represent oriented similarity classes of triangles. Each of these quotient spaces may be given a JM metric by requiring the projection maps to be Riemannian submersions. The geodesic dynamics on $\mC^2$ is clarified by studying its projections to these quotient manifolds. We will now describe these Riemannian submersions explicitly in local coordinates. This is greatly facilitated by choosing coordinates (unlike $z_1, z_2$) on $\mC^2$ in which the Killing vector fields (KVFs) corresponding to the isometries point along coordinate vector fields. As we will see, this ensures that the vertical subspaces in the associated Riemannian submersions are spanned by coordinate vector fields. Thus we introduce the Hopf coordinates $(r, \eta, \xi_1, \xi_2)$ on $\mC^2$ \cite{nakahara} via the transformation
	\beq
	z_1=r e^{i ( \xi_1+ \xi_2)} \sin\eta \quad \text{and} \quad z_2=r e^{i ( \xi_1- \xi_2)} \cos\eta.
	\label{e:Hopf-coords}
	\eeq
Here the radial coordinate $r = \sqrt{|z_1|^2 + |z_2|^2} = \sqrt{I_{\rm CM}} \geq 0$ is a measure of the size of the triangle with masses at its vertices. $\xi_2$ determines the relative orientation of $z_1$ and $z_2$ while $\xi_1$ fixes the orientation of the triangle as a whole. More precisely, $2 \xi_2$ is the angle from the rescaled Jacobi vector $z_2$ to $z_1$ while $2 \xi_1$ is the sum of the angles subtended by $z_1$ and $z_2$ with the horizontal axis in Fig \ref{f:jacobi-vectors}. Thus we may take $0 \leq \xi_1 + \xi_2 \leq 2\pi$ and $0 \leq \xi_1 -\xi_2 \le 2 \pi$ or equivalently, $-\pi\leq\xi_2\leq\pi$ and $|\xi_2| \leq \xi_1 \leq 2\pi-|\xi_2|$. Finally, $0 \leq \eta \leq \pi/2$ measures the relative magnitudes of $z_1$ and $z_2$, indeed $\tan \eta = |z_1|/|z_2|$. When $r$ is held fixed, $\eta, \xi_1$ and $\xi_2$ furnish the standard Hopf coordinates parametrizing the three sphere $|z_1|^2 + |z_2|^2 = r^2$. For fixed $r$ and $\eta$, $\xi_1 + \xi_2$ and $\xi_1 - \xi_2$ are periodic coordinates on tori. These tori foliate the above three-sphere as $\eta$ ranges between $0$ and $\pi/2$. Furthermore, as shown in \S \ref{s:quotient-metrics}, $2 \eta$ and $2 \xi_2$ are polar and azimuthal angles on the two-sphere obtained as the quotient of $\mS^3$ by rotations via the Hopf map.

Let us briefly motivate these coordinates and the identification of the above quotient spaces. We begin by noting that the JM metric (\ref{e:jm-metric-in-rescaled-jacobi-coordinates-for-eq-mass-on-c2}) on $\mC^2$ is conformal to the flat Euclidean metric $|dz_1|^2 + |dz_2|^2$. Recall that the cone on a Riemannian manifold $(M, ds^2_M)$ is the Cartesian product $\mathbf{R}^+ \times M$ with metric $dr^2 + r^2 ds_M^2$ where $r > 0$ parameterizes $\mathbf{R}^+$. In particular, Euclidean $\mathbf{C}^2$ may be viewed as a cone on the round three sphere $\mathbf{S}^3$. If $\bf{S}^3$ is parameterized by Hopf coordinates $\eta, \xi_1$ and $\xi_2$, then this cone structure allows us to use $r, \eta, \xi_1$ and $\xi_2$ as coordinates on $\bf{C}^2$. Moreover, the Hopf map\footnote{ The Hopf map $\mathbf{S}^3 \to \mathbf{S}^2$ is often expressed in Cartesian coordinates. If  $|z_1|^2 + |z_2|^2 = 1$ defines the unit-$\mathbf{S}^3 \subset \mathbf{C}^2$ and $w_1^2 + w_2^2 + w_3^2 = 1/4$ defines a 2-sphere of radius $1/2$ in $\mathbf{R}^3$, then $w_3 = \left(|z_2|^2-|z_1|^2\right)/2$ and $w_1+ i w_2 = z_1 \bar z_2$. Using Eq. \ref{e:Hopf-coords}, we may express the Cartesian coordinates $w_i$ in terms of Hopf coordinates:
	\beq \nonumber
	2 w_3 = {r^2}\cos2\eta,
\quad 2 w_1 = r^2 \sin(2\eta)\cos(2\xi_2)\quad \text{and} \quad 2 w_2 = r^2 \sin(2\eta)\sin(2\xi_2).
	\eeq
} defines a Riemannian submersion from the round $\mathbf{S}^3$ to the round two sphere $\mathbf{S}^2$. Indeed, if we use Hopf coordinates $\eta, \xi_1, \xi_2$ on $\bf{S}^3$, then the Hopf map takes $(\eta, \xi_1, \xi_2) \mapsto (\eta, \xi_2) \in \bf{S}^2$. In general, if $M \to N$ is a Riemannian submersion, then there is a natural submersion\footnote{Let $f:(M,g)\mapsto(N,h)$ be a Riemannian submersion with local coordinates $m^i$ and $n^j$. Let $(r, m^i)$ and $(r,n^j)$ be local coordinates on the cones $C(M)$ and $C(N)$. Then $\tl f: (r, m) \mapsto (r, n)$ defines a submersion from $C(M)$ to $C(N)$. Consider a horizontal vector $a \pdr_r + b_i \pdr_{m_i}$ in $T_{(r,m)}C(M)$. We will show that $d\tl f$ preserves its length. Now, if $df(b_i \pdr_{m_i}) = c_i \pdr_{n_i}$ then $d\tilde f(a \pdr_r + b_i \pdr_{m_i}) = a \pdr_r + c_i \pdr_{n_i}$. Since $\pdr_r \perp \pdr_{m^i}$, $||a \pdr_r+ b_i \pdr_{m_i}||^2 = a^2 + r^2  \|b_i \pdr_{m_i}\|^2$ $= a^2 + r^2\|c_i \pdr_{n_i}\|^2$ as $f$ is a Riemannian submersion. Moreover $a^2 + r^2 \|c_i \pdr_{n_i}\|^2 = \| a \pdr_r + c_i \pdr_{n_i} \|^2$ since $\pdr_r \perp \pdr_{n^i}$. Thus $\tilde f$ is a Riemannian submersion.} from the cone on $M$ to the cone on $N$. In particular, the Hopf map extends to a Riemannian submersion from the cone on the round $\bf{S}^3$ to the cone on the round $\bf{S}^2$, i.e. from Euclidean $\bf{C}^2$ to Euclidean $\bf{R}^3$ taking $(r, \eta, \xi_1, \xi_2) \mapsto (r, \eta, \xi_2)$. As the conformal factor is independent of rotations, the same map defines a Riemannian submersion from $\mC^2$ with the JM metric to shape space $\mR^3$ with its quotient JM metric. Finally, for $E=0$, scaling ${\vec r} \to \la {\vec r}$ defines an isometry of the quotient JM metric on shape space $\mR^3$. Quotienting by this isometry we arrive at the shape sphere $\mS^2$ with Montgomery's `pair of pants' metric. Alternatively, we may quotient $\mC^2$ first by the scaling isometry of its JM metric to get $\mS^3$ and then by rotations to get $\mS^2$ (see Fig. \ref{f:flow-chart}).

With these motivations, we express the equal-mass JM metric on $\mC^2$ in Hopf  coordinates [generalization to unequal masses is obtained by replacing $Gm^3 h$ below with $\tl h(\eta, \xi_2)$ given in Eq. (\ref{e:unequal-mass-htilde})]:
	\beq
	ds^2 = \left( E + \fr{Gm^3 h(\eta,\xi_2)}{r^2} \right) \left(dr^2 + r^2 \left( d\eta^2 + d\xi_1^2 - 2 \cos 2\eta \; d\xi_1 \; d\xi_2 + d\xi_2^2 \right) \right).
	\label{e:c2-metric-in-r-eta-xi1-xi2-coordinates}
	\eeq
It is convenient to write 
	\beq
	h(\eta,\xi_2) = v_1+ v_2 + v_3 
	\label{e:conformal-prefactor-h}
	\eeq	
where $v_1 = r^2/(m |x_2 - x_3|^2)$ is proportional to the pairwise potential between $m_2$ and $m_3$ and cyclic permutations thereof. The $v_i$ are rotation and scale-invariant, and therefore functions only of $\eta$ and $\xi_2$ in Hopf coordinates:
	\beq
	v_{1,2} = \fr{2}{\left( 2 + \cos2\eta \mp \sqrt{3}   \sin2\eta\; \cos2\xi_2\right)} \quad \text{and} \quad v_3 = \fr{1}{2\sin^2\eta}.
	\eeq 
Notice that $h \to \infty$ at pairwise collisions. The $v_i$'s have the common range $1/2 \leq v_i < \infty$ with $v_3 = 1/2$ when $m_3$ is at the CM of $m_1$ and $m_2$ etc. We also have $h \geq 3$ with equality when $v_1 = v_2 = v_3$, corresponding to Lagrange configurations with masses at vertices of an equilateral triangle. To see this, we compute the moment of inertia $I_{\rm CM}$ in two ways. On the one hand $I_{\rm CM}=|z_1|^2+|z_2|^2=r^2$ . On the other hand, for equal masses the CM lies at the centroid of the triangle defined by masses. Thus $I_{\rm CM}$ is $(4m/9) \times$ the sum of the squares of the medians, which by Apollonius' theorem is equal to $(3/4) \times$ the sum of the squares of the sides. Hence $I_{\rm CM}=\sum_{i=1}^3 r^2/3 v_i$. Comparing, we get  $\sum_{i=1}^3 1/v_i = 3$. Since the arithmetic mean is bounded below by the harmonic mean,
	\beq
	\label{ineq:conformal-factor-h}
	{h}/{3} = {\left( v_1 + v_2 + v_3 \right)}/{3} \geq 3 \left( {v_1}^{-1} + {v_2}^{-1} + {v_3}^{-1} \right)^{-1} = 1.
	\eeq

\subsubsection{Lagrange, Euler, collinear and collision configurations}

The geometry of the JM metric displays interesting behavior at Lagrange and collision configurations on $\mC^2$ and its quotients. We identify their locations in Hopf coordinates for {\it equal} masses. The Jacobi vectors in Hopf coordinates are 
	\beq
	J_1=\sqrt{\fr{2}{m}} r e^{i ( \xi_1+ \xi_2)} \sin\eta \quad \text{and} \quad 
	J_2=\sqrt{\fr{3}{2 m}}r e^{i ( \xi_1- \xi_2)} \cos\eta.
	\eeq
At a Lagrange configuration, $m_{1,2,3}$ are at vertices of an equilateral triangle. So $|J_2| = \sqrt{3}|J_1|/2$ (i.e. $\eta = \pi/4$) and $J_2$ is $\perp$ to $J_1$ (i.e. $\xi_2 = \pm \pi/4$, the sign being fixed by the orientation of the triangle). So Lagrange configurations $L_{4,5}$ on $\mC^2$ occur when $\eta = \pi/4$ and $\xi_2=\pm\pi/4$ with $r$ and $\xi_1$ arbitrary. On quotients of $\mC^2$, $L_{4,5}$ occur at the images under the corresponding projections. Since $2 \eta$ and $2 \xi_2$ are polar and azimuthal angles on the shape sphere, $L_{4,5}$ are at diametrically opposite equatorial locations (see Figs. \ref{f:shape-sphere-xi2-eta} and \ref{f:horn-shape-sphere}). Collinear configurations (syzygies) occur when $J_1$ and $J_2$ are (anti)parallel, i.e. when $\xi_2 = 0$ or $\pi/2$, with other coordinates arbitrary. On the shape sphere, syzygies occur on the `great circle' through the poles corresponding to the longitudes $2\xi_2 = 0$ and $\pi$. Collisions are special collinear configurations. By $C_i$ we denote a collision of particles other than the $i^{\rm th}$ one. So $C_3$ corresponds to $J_1=0$ which lies at the `north pole' ($\eta=0$) on $\mS^2$. $m_2$ and $m_3$ collide when $J_2 =  J_1/2$ so  $\eta = \pi/3$ and $\xi_2 = 0$ at $C_1$. Similarly, at $C_2$, $J_2 = - J_1/2$ which corresponds to $\eta = \pi/3$ and $\xi_2=\pi/2$. The Euler configurations $E_i$ for equal masses are collinear configurations where mass $m_i$ is at the midpoint of the other two. 

Finally, we note that the azimuth and co-latitude ($\tht$ and $\phi$) \cite{montgomery-pants} are often used as coordinates on the shape sphere, so that $L_{4,5}$ are at the poles while $C_{1,2,3}$ and $E_{1,2,3}$ lie on the equator. This coordinate system makes the symmetry under permutations of masses explicit, but is not convenient near any of the collisions (e.g. sectional curvatures can be discontinuous). On the other hand, our coordinates $\eta$ and $\xi_2$, which are related to $\tht$ and $\phi$ by suitable rotations,
	\beqs \nonumber
	\label{e:thtphi-xieta}
	\sin\phi &=&  \cos(2\eta-\pi /2)\sin(2\xi_2), \cr \cos\phi \sin\tht &=& \cos(2\eta-\pi/2)\cos(2\xi_2) \quad \text{and} \cr 
\cos\phi \cos\tht &=& \sin\left(2\eta-\fr{\pi}{2}\right),
	\eeqs
 are convenient near $C_3$ but not near $E_3$ or $C_{1,2}$. For instance, sectional curvatures can be discontinuous, as seen in Fig. \ref{f:shape-space-curvature}. The neighborhoods of the latter configurations may be studied by re-ordering the masses.

\subsection{Quotient JM metrics on shape space, \texorpdfstring{$\mS^3$}{S3} and the shape sphere}
\label{s:quotient-metrics}

\subsubsection{Submersion from \texorpdfstring{$\mC^2$}{C2} to shape space \texorpdfstring{$\mR^3$}{R3}}

Rotations $z_j \to e^{i \tht} z_j$ act as isometries of the JM metric (\ref{e:c2-metric-in-r-eta-xi1-xi2-coordinates}) on $\mC^2$. In the Hopf coordinates of Eq. (\ref{e:Hopf-coords}),
	\beq
	z_1=r e^{i ( \xi_1+ \xi_2)} \sin\eta \quad \text{and} \quad z_2=r e^{i ( \xi_1- \xi_2)} \cos\eta, \quad 
	\eeq
rotations are generated by translations  $\xi_1 \to \xi_1 + \tht$ and a discrete shift $\xi_2 \to \xi_2 + \pi$ (${\rm mod} \; 2\pi$). The shift in $\xi_2$ rotates $z_i \mapsto -z_i$, which is not achievable by a translation in $\xi_1$ due to its restricted range, $|\xi_2| \leq \xi_1 \leq 2\pi - |\xi_2|$ and $-\pi \le \xi_2 \le \pi$. To quotient by this isometry, we define a submersion from $\mC^2 \to \mR^3$ taking 
	\beqs
	(r, \eta, \xi_1,\xi_2) &\mapsto& (r, \eta, \xi_2)\quad \text{if} \quad \xi_2 \ge 0 
	\quad \text{and} \cr
	  (r,\eta, \xi_1,\xi_2) &\mapsto& (r,\eta, \xi_2+ \pi) \quad \text{if} \quad \xi_2 < 0.
	\eeqs
The radial, polar and azimuthal coordinates on $\mR^3$ are given by	$r$, $2\eta$ and $2\xi_2$ with $m_1$-$m_2$ collisions occurring on the ray $\eta =0$. Under the linearization of this submersion at a point $p \in \mC^2$, $V(p)$ is spanned by $\pdr_{\xi_1}$ and $H(p)$ by $\pdr_r$, $\pdr_\eta$ and $\cos 2 \eta \; \pdr_{\xi_1} + \pdr_{\xi_2}$. These horizontal basis vectors are mapped respectively to $\pdr_r$, $\pdr_\eta$ and $\pdr_{\xi_2}$ under the linearization of the map. Requiring lengths of horizontal vectors to be preserved we arrive at the following quotient JM metric on $\mR^3$, conformal to the flat metric on $\mR^3$:
	\beq
	\label{e:r3-metric-in-r-eta-xi2-coordinates}
	ds^2=  \left( E + \fr{Gm^3 h(\eta,\xi_2)}{r^2} \right)\left(dr^2+r^2\left(d\eta^2+\sin^2 2\eta \;d\xi_2^2\right)\right).
	\eeq
This metric may also be viewed as conformal to a cone on a round $2$-sphere of radius one-half, since $0 \le 2 \eta \le \pi$ and $0 \le 2 \xi_2 \le 2 \pi$ are the polar and azimuthal angles.

\subsubsection{Submersion from \texorpdfstring{$\mR^3$}{R3} to the shape sphere \texorpdfstring{$\mS^2$}{S2}}

The group $\bf{R}^+$ of scalings $(r, \eta, \xi_2) \mapsto (\la r, \eta, \xi_2)$ acts as an isometry of the {\it zero-energy} JM metric (\ref{e:r3-metric-in-r-eta-xi2-coordinates}) on shape space $\mR^3$. The orbits are radial rays emanating from the origin (and the triple collision point at the origin, which we exclude). The quotient space $\mR^3/{\rm scaling}$ is the shape sphere $\mS^2$. We define a submersion from shape space to the shape sphere taking $(r, \eta, \xi_2) \mapsto (\eta, \xi_2)$. Under the linearization of this map at $p \in \mR^3$, $V(p) = \text{span}(\pdr_r)$. Its orthogonal complement $H(p)$ is spanned by $\pdr_\eta$ and $\pdr_{\xi_2}$ which project to $\pdr_\eta$ and $\pdr_{\xi_2}$ on $\mS^2$. Requiring the submersion to be Riemannian, we get the quotient `pair of pants' JM metric on the shape sphere which is conformal to the round metric on a $2$-sphere of radius one-half:
	\beq
	\label{e:s2-metric-in-eta-xi2-coordinates}
	ds^2 = Gm^3 h(\eta,\xi_2) \left(d\eta^2+\sin^2 2\eta \;d\xi_2^2\right).
	\eeq

\subsubsection{Submersion from \texorpdfstring{$\mC^2$}{C2} to \texorpdfstring{$\mS^3$}{S3} and then to \texorpdfstring{$\mS^2$}{S2}}

For zero energy, it is also possible to quotient the JM metric (\ref{e:c2-metric-in-r-eta-xi1-xi2-coordinates}) on $\mC^2$, first by its scaling isometries to get $\mS^3$ and then by rotations to arrive at the shape sphere. Interestingly, it follows from the Lagrange-Jacobi identity that when $E$ and $\dot I$ vanish, $r$ is constant and the motion is confined to a $3$-sphere embedded in $\mC^2$. To quotient by the scaling isometries $(r, \eta, \xi_1, \xi_2) \mapsto (\la r, \eta, \xi_1, \xi_2)$ of $\mC^2$, we define the submersion $(r, \eta, \xi_1, \xi_2) \mapsto (\eta, \xi_1, \xi_2)$ to $\mS^3$, with ranges of coordinates as on $\mC^2$. The vertical subspace is spanned by $\pdr_r$ while $\pdr_\eta$, $\pdr_{\xi_1}$ and $\pdr_{\xi_2}$ span the horizontal subspace. The latter are mapped to $\pdr_\eta$, $\pdr_{\xi_1}$ and $\pdr_{\xi_2}$ on $\mS^3$. The submersion is Riemannian provided we endow $\mS^3$ with the following conformally-round metric 
	\beq
ds^2=Gm^3 h\left(\eta,\xi_2\right)\left(d\eta^2+ d\xi_1^2-2\cos2\eta\;d\xi_1\;d\xi_2+ d\xi_2^2\right). 
\label{e:s3-metric-in-eta-xi1-xi2-coordinates}
	\eeq
Rotations generated by $\xi_1 \to \xi_1 + \tht$ and $\xi_2 \to \xi_2 + \pi$ (mod $2\pi$) act as isometries of this metric on $\mS^3$. We quotient by rotations to get the metric (\ref{e:s2-metric-in-eta-xi2-coordinates}) on $\mS^2$ via the Riemannian submersion defined by 
	\beq
	(\eta,\xi_1,\xi_2) \mapsto (\eta,\xi_2) \quad \text{if} \quad \xi_2 \ge 0 \quad \text{and} \quad (\eta,\xi_1,\xi_2) \mapsto (\eta,\xi_2+\pi) \quad \text{if} \quad \xi_2 < 0. 
	\eeq
\subsection{JM metric in the near-collision limit and its completeness}
\label{s:geodesic-completeness}

The equal-mass JM metric components on center-of-mass configuration space $\mC^2$ and its quotients blow up at two- and three-body collisions. However, we study the geometry in the neighborhood of collision configurations and show that the curvature remains finite in the limit. Remarkably, it takes infinite geodesic time for collisions to occur which we show by establishing the geodesic completeness of the JM metric on $\mC^2$ and its quotients. By contrast, collisions can occur in finite time for the Newtonian three-body evolution. The JM geodesic flow avoids finite time collisions by reparametrizing time along Newtonian trajectories (see Eq. \ref{e:jm-geodesic-equation}). Thus the geodesic reformulation of the inverse-square three-body problem `regularizes' pairwise and triple collisions.

\subsubsection{Geometry near pairwise collisions}
\label{s:near-pairwise-collision-geometry}

For equal masses (see \S\ref{s:jm-metric-config-space-hopf-coords}), the first pair of masses collide when $\eta=0$ (with other coordinates arbitrary) while the other two binary collisions occur at $C_1$ and $C_2$ (see Fig. \ref{f:shape-sphere-xi2-eta}). Triple collisions occur when $r = 0$. Unlike for the Newtonian potential, sectional curvatures on coordinate $2$-planes are finite at pairwise and triple collisions, though some JM metric (\ref{e:c2-metric-in-r-eta-xi1-xi2-coordinates}) and Riemann tensor components blow up. It is therefore interesting to study the near-collision geometry of the JM metric.

The geometry of the equal-mass JM metric in the neigborhood of a binary collision is the same irrespective of which pair of bodies collide. Since Hopf coordinates are particularly convenient around $\eta = 0$, we focus on collisions between the first pair of masses. Montgomery (see Eq. 3.10c of \cite{montgomery-pants}) studied the near-collision geometry on $\mS^2$ and showed that it is geodesically complete. Let us briefly recall the argument. Expanding the equal-mass $\mS^2$ metric (\ref{e:s2-metric-in-eta-xi2-coordinates}) around the collision point $\eta = 0$, we get
	\beq
	ds^2 \approx \left(\fr{G m^3}{2\eta^2}\right) \left(d\eta^2+4\eta^2 \;d\xi_2^2\right) = \frac{G m^3}{2\rho^2}(d\rho^2 + \rho^2 d\chi^2)
	\label{e:S2-near-collision-metric-inv-sq}
	\eeq
where $\rho = 2 \eta$ and $\chi = 2 \xi_2$. $\pdr_{\chi}$ is a KVF, so `radial' curves with constant $\chi$ are geodesics. Approaching $\rho = 0$ along a `radial' geodesic shows that the collision point $\rho = 0$ is at an infinite distance $(\sqrt{G m^3/2} \int_{\rho_0}^0 d\rho/\rho)$ from any point $(\rho_0,\chi)$ in its neighborhood $(0< \rho_0 \ll 1)$. The symmetry of the metric under exchange of masses ensures that the same holds for the other two collision points: geodesics may be extended indefinitely. Thus the shape sphere ($\mS^2$ with three collision points excluded) is geodesically complete. To clarify the near-collision geometry let $d \la = -d\rho/\sqrt{2}\rho$ or $\la = - \log(\rho/\rho_0)/\sqrt{2}$. This effectively stretches out the neighborhood of the collision point $\la = \infty$. The asymptotic metric $ds^2 = Gm^3 \left( d \la^2 + d\chi^2/2 \right)$ for $0 \leq \chi \leq 2\pi$ and $\la \ge 0$ is the metric on a semi-infinite right-circular cylinder of radius $\sqrt{Gm^3/2}$ with $\la$ the coordinate along the height and $\chi$ the azimuthal angle. Thus the JM metric looks like that of a semi-infinite cylinder near any of the collision points. 

More generally, for {\it unequal} masses, the near-collision metric (\ref{e:S2-near-collision-metric-inv-sq}) is 	
	\beq
	ds^2 \approx \fr{G m_1 m_2 M_1}{2\eta^2} \left(d\eta^2 + 4\eta^2 d\xi_2^2\right)  \quad {\rm  [see \quad Eq. (\ref{e:jm-metric-in-jacobi-coordinates-on-c3}-\ref{e:Hopf-coords})] }
	\eeq
and essentially the same argument implies that the JM metric on the shape sphere is geodesically complete for arbitrary masses.

Since $\mS^2$ arises as a Riemannian submersion of $\mR^3$, $\mS^3$ and $\mC^2$, the infinite distance to binary collision points on the shape sphere can be used to show that the same holds on each of the higher dimensional manifolds. To see this, consider the submersion from (say) $\mC^2$ to $\mS^2$. Any curve $\tl \g$ on $\mC^2$ maps to a curve $\g$ on $\mS^2$ with $l(\tl \g) \geq l(\g)$ since the lengths of horizontal vectors are preserved. If there was a binary collision point at finite distance on $\mC^2$, there would have to be a geodesic of finite length ending at it. However, such a geodesic would project to a curve on the shape sphere of finite length ending at a collision point, contradicting its completeness. 

Thus we have shown that the JM metrics (necessarily of zero energy) on $\mS^2$ and $\mS^3$ with binary collision points removed, are geodesically complete for arbitrary masses. On the other hand, to examine completeness on $\mC^2$ and $\mR^3$ we must allow for triple collisions as well as non-zero energy. Geodesic completeness in these cases is shown in \S\ref{s:triple-collision-inv-sq-near-collision-geom}. In the sequel we examine the near-collision geometry on $\mR^3$, $\mS^3$ and $\mC^2$ in somewhat greater detail by Laurent expanding the JM metric components around $\eta = 0$ and keeping only leading terms.

{\fl \bf Shape space geometry near binary collisions:} The equal-mass shape space metric around $\eta=0$, in the leading order, becomes
	\beq
	ds^2 \approx \fr{G m^3}{2\eta^2 r^2} \left( dr^2+ r^2  \left(d\eta^2+4\eta^2 \;d\xi_2^2\right)\right) = G m^3 \left( \fr{2 dr^2}{\rho^2 r^2} + \frac{d\rho^2}{2\rho^2}  +\frac{d\chi^2}{2} \right),
	\eeq
where $\rho=2\eta$ and $\chi=2\xi_2$. We define new coordinates $\la$ and $\kappa$ by $d \la = -d\rho/\sqrt{2}\rho$, $d\kappa= dr/r$ so that $\rho = \rho_0 e^{-\sqrt{2}\la}$.  In these coordinates the collision occurs at $\la = \infty$. The asymptotic metric is
	\beq
	ds^2 \approx G m^3 \left( \fr{2}{\rho_0^2}e^{2 \sqrt{2}\la} d\kappa^2 + d \la^2 + \half d\chi^2\right)
	\eeq
where $0 \leq \chi \leq 2 \pi$ (periodic), $\la \geq 0$ and $-\infty < \kappa < \infty$. This metric has a constant scalar curvature of $-4/Gm^3$. The sectional curvature in the $\pdr_\la-\pdr_\kappa$ plane is equal to $-2/Gm^3$, it vanishes in the other two coordinate planes. These values of scalar and sectional curvatures agree with the limiting values at the $1$-$2$ collision point calculated for the full metric on shape space. The near-collision topology of shape space is that of the product manifold $\bf{S}^1_\chi \times \bf{R}^+_\la \times \bf{R}_\kappa$.

{\fl \bf Near-collision geometry on $\mC^2$:} The equal-mass JM metric in leading order around $\eta = 0$ is
	\beq
	ds^2 \approx \fr{G m^3}{2\eta^2 r^2} \left( dr^2+ r^2  \left(d\eta^2+d\xi_1^2-2(1-2\eta^2) d\xi_1 d\xi_2+d\xi_2^2\right)\right).
	\eeq
Let us define new coordinates $\la,\kappa,\xi_\pm$ such that $d \la = -d\eta/\sqrt{2}\eta$, $d\kappa= -dr/r$ and $\xi_\pm=\xi_1\pm\xi_2$. $0 \leq \xi_\pm \leq 2\pi$ are periodic coordinates parametrizing a torus. The asymptotic metric is
	\beq
	ds^2 \approx G m^3\left( \fr{d\kappa^2}{2\eta^2}  + d \la^2 + \ov{2\eta^2} d\xi_-^2 + \ov{2} d\xi_+^2\right)
	\label{e:C2-near-collision-metric}
	\eeq
where $\eta = \eta_0 e^{-\sqrt{2}\la}$.  This metric has a constant scalar curvature $-12/Gm^3$. The sectional curvature of any coordinate plane containing $\pdr_{\xi_+}$ vanishes due to the product form of the metric. The sectional curvatures of the remaining coordinate planes ($\pdr_\kappa-\pdr_\la, \pdr_\kappa - \pdr_{\xi_-}, \pdr_{\xi_-}-\pdr_\la$) are equal to $-2/Gm^3$. The scalar and sectional curvatures (of corresponding planes) of this metric agree with the limiting values computed from the full metric on $\mC^2$. 

{\fl \bf Near-collision geometry on $\mS^3$:} The submersion $\mC^2 \to \mS^3$ takes $(\kappa, \la, \xi_\pm) \mapsto ( \la, \xi_\pm)$. As the coordinate vector fields on $\mC^2$ are orthogonal, from (\ref{e:C2-near-collision-metric}) the asymptotic metric on $\mS^3$ near the $1$-$2$ collision point is 
	\beq
	ds^2 \approx G m^3\left( d \la^2 + \ov{2\eta^2} d\xi_-^2 + \ov{2} d\xi_+^2\right).
	\eeq
This metric has a constant scalar curvature equal to $-4/Gm^3$. The sectional curvatures on the $\la- \xi_-$ coordinate 2-plane is $-2/Gm^3$ while it vanishes on the other two coordinate 2-planes.

\subsubsection{Geometry on  \texorpdfstring{$\mR^3$  and $\mC^2$}{R3 and C2} near triple collisions}

\label{s:triple-collision-inv-sq-near-collision-geom}

We argue that the triple collision configuration (which occurs at $r=0$ on $\mC^2$ or shape space $\mR^3$) is at infinite distance from other configurations with respect to the equal-mass JM metrics (Eqs. (\ref{e:c2-metric-in-r-eta-xi1-xi2-coordinates}) and (\ref{e:r3-metric-in-r-eta-xi2-coordinates})) which may be written in the form:
	\beq
	ds^2 = (G m^3 h/r^2) dr^2 + G m^3 \: h \: g_{i j} \: dx^i dx^j.
	\eeq
$g_{ij}$ is the positive (round) metric on $\mathbf{S}^3$ ($x^i = (\eta, \xi_1, \xi_2)$) or $\mathbf{S}^2$ ($x^i = (\eta, \xi_2)$) of radius one-half:
	\beq
	g^{\mC^2}_{ij} = \colvec{3}{1 & 0 & 0}{0 & 1 & -\cos 2\eta}{0 & -\cos 2\eta & 1}\quad \text{and} \quad g^{\mR^3}_{ij} =  \colvec{2}{1 & 0 }{0 & \sin 2\eta}.
	\label{e:kinetic-metric-s3-s2}
	\eeq
Together with our results on pairwise collisions (\S \ref{s:near-pairwise-collision-geometry}), it will follow that the manifolds are geodesically complete. As a consequence, the geodesic flow reformulation of the three-body problem regularizes triple collisions. To show that triple collision points are at infinite distance we will use the previously obtained lower bound on the conformal factor, $h(\xi_2, \eta) \geq 3$ (see Eq. \ref{ineq:conformal-factor-h}).

Let $\g(t)$ be a curve joining a non-collision point $\g(t_0) \equiv (r_0, x^i_0)$ and the triple collision point $\g(t_1) \equiv (r=0, x^i_1)$. We show that its length $l(\g)$ is infinite. Since $G m^3 h g_{ij}$ is a positive matrix,
	\beq
	l(\g) = \int_{t_0}^{t_1} dt\sqrt{\fr{G m^3 h}{r^2} \dot r^2 + G m^3 h g_{i j} \dot x^i \dot x^j} \;  \geq \; \int_{t_0}^{t_1} dt  \sqrt{\fr{G m^3 h}{r^2} \dot r^2 }.
	\eeq
Now using $|\dot r| \geq - \dot r$ and $h \geq 3$, we get
	\beq
	l(\g) \geq -\sqrt{3 G m^3} \int_{t_0}^{t_1}  \fr{\dot r}{ r} dt = \sqrt{3 G m^3}\int_{0}^{r_0}  \fr{d r}{ r} = \infty .
	\eeq	
In particular, a geodesic from a non-collision point to the triple collision point has infinite length. Despite appearances, the above inequality $l(\g) \geq \sqrt{3 G m^3} \int_0^{r_0} dr/r$ does not imply that radial curves are always geodesics. This is essentially because $h$ along $\g$ may be less than that on the corresponding radial curve. However, if $(\eta, \xi_1, \xi_2)$ is an angular location where $h$ is minimal (locally), then the radial curve with those angular coordinates is indeed a geodesic because a small perturbation to the radial curve increases $h$ and consequently its length. The global minima of $h$ ($h = 3$) occur at the Lagrange configurations $L_{4,5}$ and local minima ($h = 9/2$) are at the Euler configurations $E_{1,2,3}$ indicating that radial curves at these angular locations are geodesics. In fact, the Christoffel symbols $\G^i_{rr}$ vanish for $i = \eta, \xi_1, \xi_2$ at $L_{4,5}$ and at $E_{1,2,3}$ so that radial curves $\g = (r(t), x^i_0)$ satisfying $\ddot r + \G^r_{rr} \dot r^2= 0$ are geodesics.
  
These radial geodesics at minima of $h$ describe Lagrange and Euler homotheties (where the masses move radially inwards/outwards to/from their CM which is the center of similitude). These homotheties take infinite (geodesic) time to reach the triple collision. By contrast, the corresponding Lagrange and Euler homothety solutions to Newton's equations reach the collision point in finite time. This difference is due to an exponential time-reparametrization of geodesics relative to trajectories. In fact, if $t$ is trajectory time and $s$ arc-length along geodesics, then from \S \ref{s:traj-as-geodesics} and \S \ref{s:jm-metric-config-space-hopf-coords}, $\sigma = ds/dt = \sqrt{2} (E + 3 Gm^3/r^2)$ since $h = 3$. Near a triple collision (small $r$), $ds^2 \approx 3 Gm^3 dr^2/r^2$ so that $s \approx - \half \sqrt{3 Gm^3} \log(1-t/t_c) \to \infty$ as $t \to t_c = r(0)^2/2\sqrt{6 G m^3}$ which is the approximate time to collision. In fact, the exact collision time $t_c = \sqrt{6 G m^3} \left( -1 + \sqrt{1+ {\kappa r(0)^2/6 G m^3}} \right)/\kappa$ may be obtained by reducing Newton's equations for Lagrange homotheties to the one body problem $r^3 \ddot r = - 6 G m^3$ whose conserved energy is $\kappa = \dot r^2 - 6 G m^3 / r^2$. These homothety solutions illustrate how the geodesic flow reformulation regularizes the original Newtonian three-body dynamics in the inverse-square potential.

More generally, for unequal masses (\ref{e:jm-metric-in-jacobi-coordinates-on-c3})-(\ref{e:Hopf-coords}) give the JM metric $ds^2 = \tl h dr^2/r^2 + \tl g_{i j} dx^i dx^j$ where 
	\beq
	\tl h = \fr{G m_1 m_2 M_1}{\sin^2 \eta} + \fr{G m_2 m_3 M_2}{\lv \cos \eta - \mu_1 e^{2 i \xi_2} \sqrt{\fr{M_2}{M_1}} \sin \eta \rv^2}
	 + \fr{G m_1 m_3 M_2}{\lv\cos \eta + \mu_2 e^{2 i \xi_2} \sqrt{\fr{M_2}{M_1}} \sin \eta \rv^2}.
	\label{e:unequal-mass-htilde}
	\eeq
Irrespective of the masses, $\tl g_{ij}$ (\ref{e:kinetic-metric-s3-s2}) is positive and $\tl h$ has a strictly positive lower bound (e.g. $G m_1 m_2 M_1$). Thus by the same argument as above, triple collisions are at infinite distance. Combining this with the corresponding results for pairwise collision points (\S\ref{s:near-pairwise-collision-geometry}), we conclude that the zero-energy JM metrics on $\mC^2$ and $\mR^3$ are geodesically complete for arbitrary masses. 

For non-zero energy, $ds^2 = (E + \tl h/r^2) (dr^2 + r^2 \tl g_{ij} dx^i dx^j)$ which can be approximated with the zero-energy JM metrics both near binary (say, $\eta = 0$) and triple ($r = 0$) collisions. If $\g$ is a curve ending at the triple collision, $l(\g) \geq l(\tl \g)$ where $\tl \g$ is a `tail end' of $\g$ lying in a sufficiently small neighborhood of $r=0$ (i.e., $r \ll |\tl h/E|^{1/2}$ which is guaranteed, say, if $r \ll |G m_1 m_2 M_1/E|^{1/2}$). But then, $l(\tl \g)$ may be estimated using the zero-energy JM metric giving $l(\tl \g) = \infty$. Thus $l(\g) = \infty$. A similar argument shows that curves ending at binary collisions have infinite length. Thus we conclude that the JM metrics on $\mC^2$ and $\mR^3$ are geodesically complete for arbitrary  energies and masses.

\subsection{Scalar curvature for equal masses and zero energy}
\label{s:scalar-curvature-inv-sq-pot-c2-r3-s3-s2}

\begin{figure}	
	\centering
	\begin{subfigure}[t]{7cm}
		\centering
	\includegraphics[width=7cm]{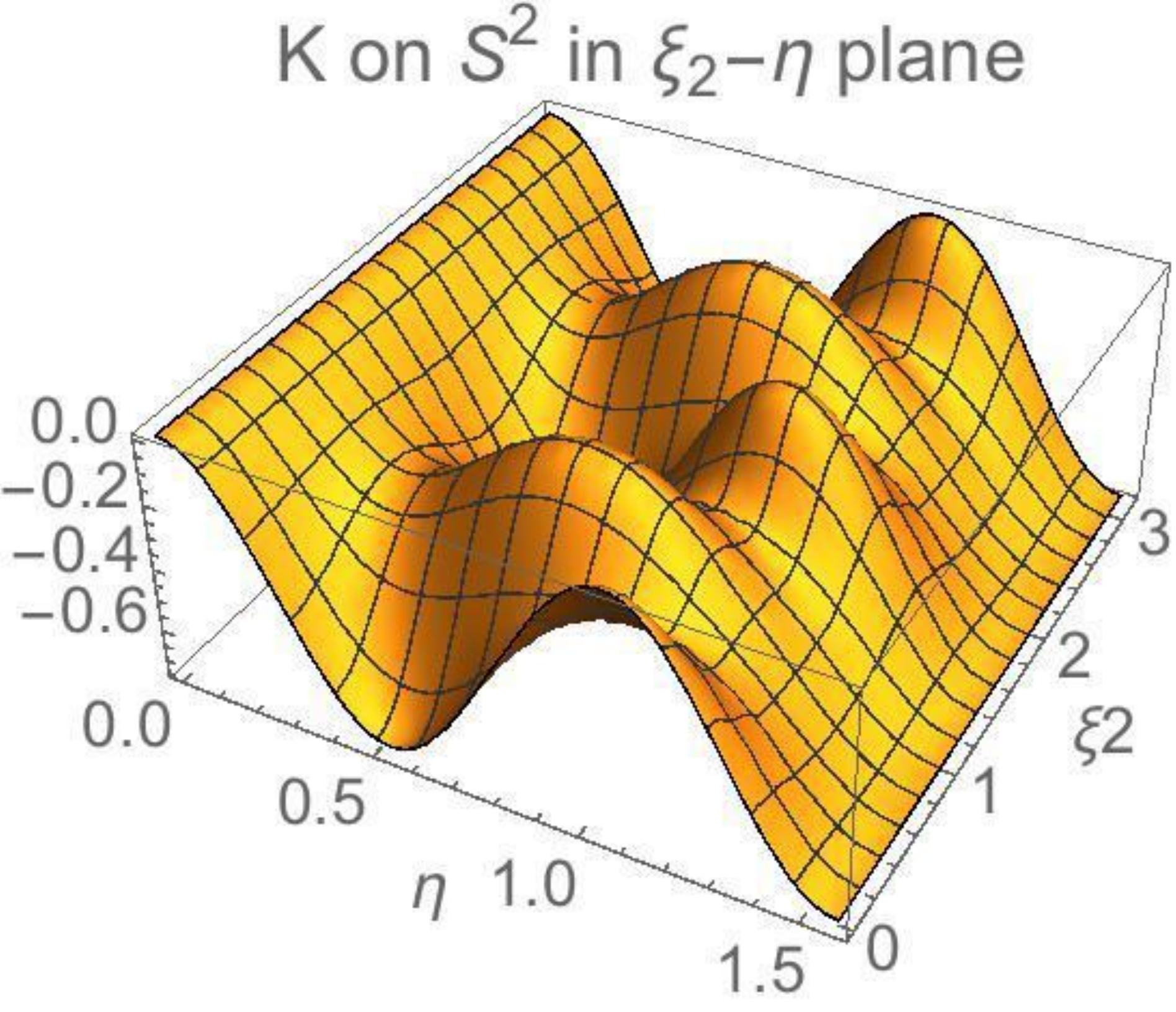}
	\end{subfigure}
	\caption{\small Gaussian curvature $K$ (in units of ${1/Gm^3}$) on $\mS^2$ for equal masses and $E=0$. $K = 0$ at $L_{4,5}$ and $C_{1,2,3}$.}
	\label{f:shape-sphere-curvature}
\end{figure}

\begin{figure}	
	\centering
	\begin{subfigure}[t]{5cm}
		\centering
		\includegraphics[height=4.5cm]{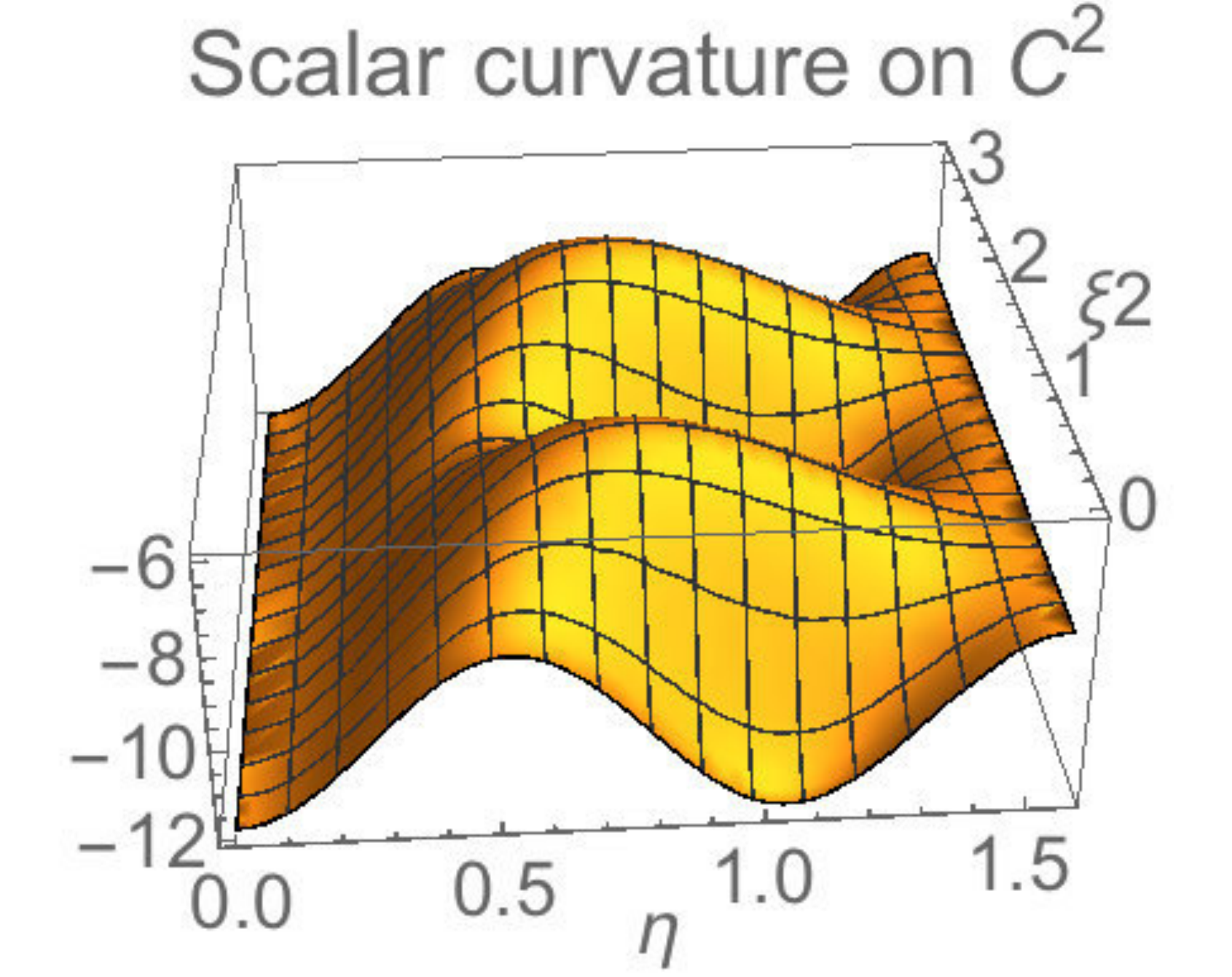}
	\end{subfigure} \quad
	\begin{subfigure}[t]{5cm}
		\centering
		\includegraphics[height=4.5cm]{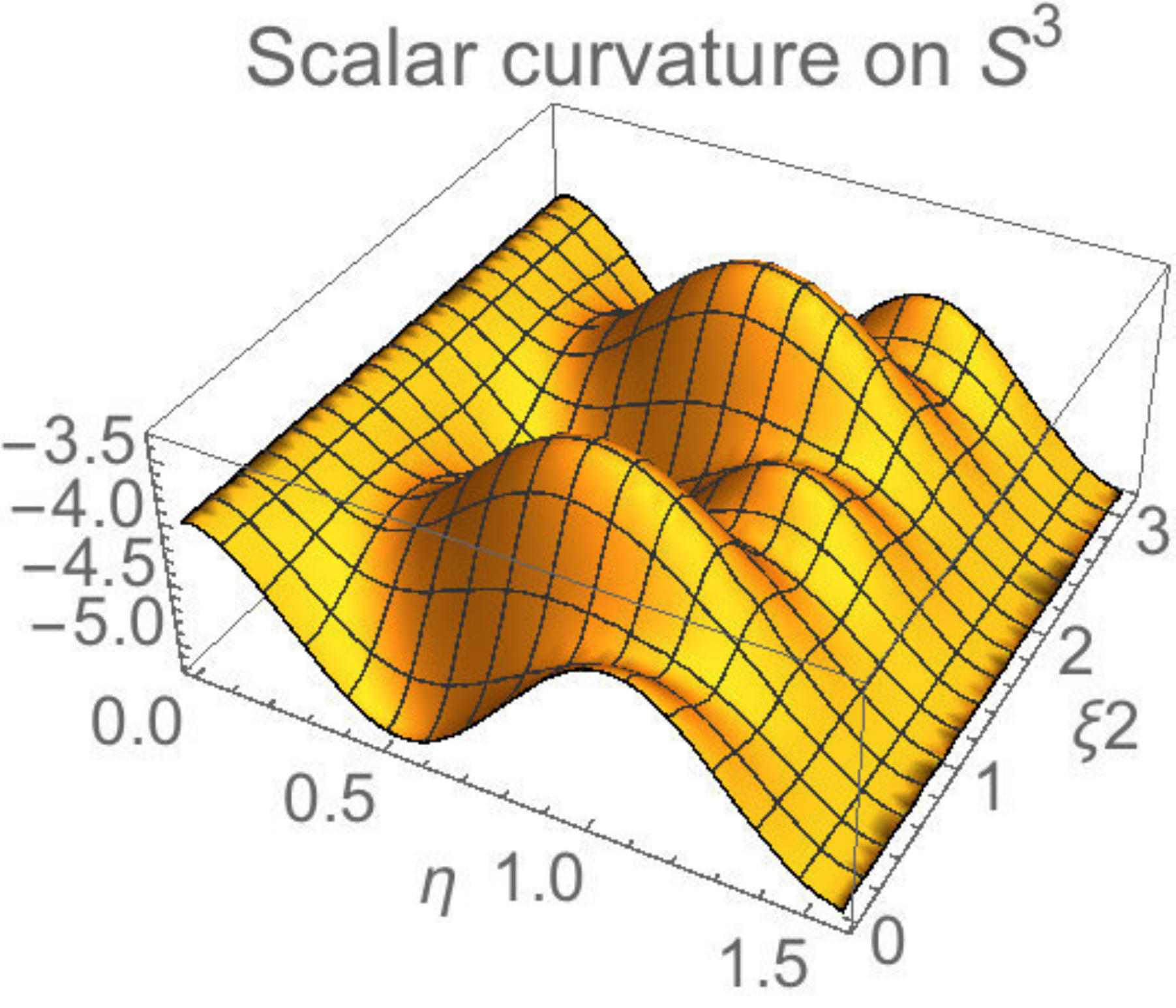}
	\end{subfigure} \quad
	\begin{subfigure}[t]{5cm}
		\centering
		\includegraphics[height=4.5cm]{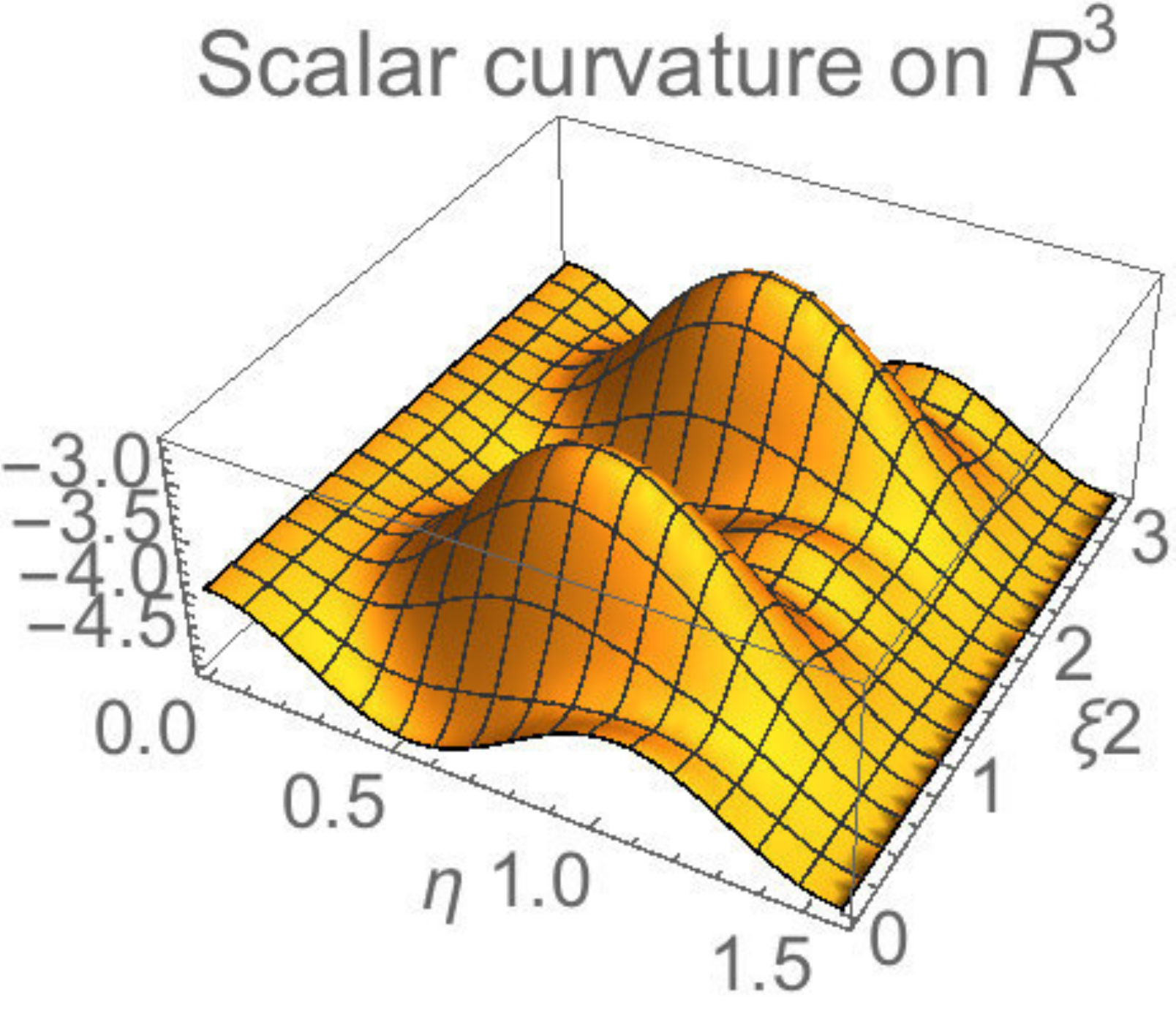}
	\end{subfigure}
\caption{\small Scalar curvatures $R$ on $\mC^2$, $\mS^3$ and $\mR^3$ in units of $1/{G m^3}$. $R$ is strictly negative and has a global maximum at $L_{4,5}$ in all cases. It attains a global minimum at $C_{1,2,3}$ on $\mC^2$ and a local maximum at collisions on $\mR^3$ and $\mS^3$. $E_{1,2,3}$ are saddles on $\mC^2$ and global minima on $\mR^3$ and $\mS^3$.}
\label{f:scalar-curvature}
\end{figure}

A geodesic through $P$ in the direction $u$ perturbed along $v$ is linearly stable/unstable [see \S \ref{s:stability-tensor}] according as the sectional curvature $K_P(u,v)$ is positive/negative. The scalar curvature $R$ at $P$ is proportional to an average of sectional curvatures in planes through $P$ (\S \ref{s:sectional-curvature-inv-sq-pot}). Thus $R$ encodes an average notion of geodesic stability. Here, we evaluate the scalar curvature $R$ of the equal-mass zero-energy JM metric on $\mC^2$ and its submersions to $\mR^3$, $\mS^3$ and $\mS^2$. In each case, due to the rotation and scaling isometries, $R$ is a function only of the coordinates $\eta$ and $\xi_2$ that parametrize the shape sphere. In \cite{montgomery-pants} Montgomery proves that $R_{\mS^2} \leq 0$ with equality at Lagrange and collision points (see Fig. \ref{f:shape-sphere-curvature}). We generalize this result and prove that the scalar curvatures on $\mC^2$, $\mR^3$ and $\mS^3$ are strictly negative and bounded below (see Fig. \ref{f:scalar-curvature}) indicating widespread linear instability of the geodesic dynamics.

\subsubsection{Scalar curvature on the shape sphere \texorpdfstring{$\mS^2$}{S2}}

The quotient JM metric on $\mS^2$ (\ref{e:s2-metric-in-eta-xi2-coordinates}) is conformal to the round (kinetic) metric on a sphere of radius $1/2$:
	\beq
	ds_{\mS^2}^2 = Gm^3 \: h(\eta, \xi_2) \: ds_{\rm kin}^2 \quad \text{where} 
	\quad ds_{\rm kin}^2 =
	d\eta^2+ \sin^2 2\eta \: d\xi_2^2.
	\label{e:jm-metric-s2}
	\eeq
Here the conformal factor ($h = - (r^2/Gm^3) \times$ potential energy) (\ref{e:conformal-prefactor-h}) is a strictly positive function on the shape sphere with double poles at collision points. The scalar curvature of (\ref{e:jm-metric-s2}) is
	\beq
	\label{e:scalar-curvature-on-s2}
	R_{\mS^2} =  \ov{ G m^3 h^3 } \left( 8 h^2 + |\grad h|^2 - h \D h \right),
	\eeq	
where $\D$ is the Laplacian and $\grad^i h = g^{ij} \pdr_j h$ the gradient on $\mS^2$ relative to the kinetic metric:
	\beqs
	\D h &=& \left(\ov{\sin^2 2\eta} \fr{\pdr^2 h}{\pdr \xi_2^2} + 2\cot 2\eta \fr{\pdr h}{\pdr \eta} + \fr{\pdr^2 h}{\pdr \eta^2} \right) \quad \text{and} \cr
	|\grad h|^2 &=& \fr{1}{\sin^2 2\eta} \left(\fr{\pdr h}{\pdr \xi_2}\right)^2 +  \left(\fr{\pdr h}{\pdr \eta}\right)^2 .
	\label{e:grad-laplace-on-s2-kin}
	\eeqs
In fact we have an explicit formula for the scalar curvature, $R_{\mS^2} = AB/C$ where
	\beqs
	A &=& 8 \sin ^2\eta \left((\cos 2 \eta + 2)^2 - 3 \sin ^2 2 \eta  \cos ^2 2\xi_2 \right), \cr
	C &=& {3 \left(2 \sin ^2 2 \eta  \cos 4 \xi_2+\cos 4 \eta -13\right)^3} \;\; \text{and} \cr
	B &=& -8 \sin^4 2 \eta \cos 8 \xi_2 - 16 \sin^2 2 \eta   \cos 4 \xi_2 (\cos 4 \eta - 29) \cr
	&&+ 236 \cos 4 \eta  - 3 \cos 8 \eta + 727.
	\label{e:scalar-curv-shape-sph}
	\eeqs
As shown in \cite{montgomery-pants}, $R_{\mS^2} \leq 0$ with equality only at Lagrange and collision points. Negativity of $R_{\mS^2}$ also follows from (\ref{e:scalar-curv-shape-sph}): each factor in the numerator is $\geq 0$ (the third vanishes at $L_{4,5}$, the second at $C_{1,2}$ and the first at $C_3$) while the denominator is strictly negative. We now use this to show that the scalar curvatures on center-of-mass configuration space $\mC^2$ and its quotients $\mR^3$ and $\mS^3$ are strictly negative.

\subsubsection{Scalar curvature on the center-of-mass configuration space \texorpdfstring{$\mC^2$}{C2}}

The equal-mass zero-energy JM metric on $\mC^2$ from Eq. (\ref{e:c2-metric-in-r-eta-xi1-xi2-coordinates}) is
	\beq
	ds^2_{\mC^2}= \left( {G m^3}/{r^2} \right) h(\eta,\xi_2) \left(dr^2+r^2\left(d\eta^2+ d\xi_1^2-2\cos2\eta\;d\xi_1\;d\xi_2+ d\xi_2^2\right)\right).
	\eeq
The scalar curvature of this metric is expressible as
	\beq
	\label{e:scalar-curvature-on-c2}
	R_{\mC^2}= \left( 3/2G m^3h^3 \right) \left(4h^2+ |\grad h|^2 - 
	2 h \: \D h \right),
	\eeq
where $\D h$ and $\grad h$ are the Laplacian and gradient with respect to the {\it round} metric on $\bf S^2$ of radius one-half (\ref{e:grad-laplace-on-s2-kin}). Due to the scaling and rotation isometries, $R_{\mC^2}$ is in fact a function on the shape sphere. The scalar curvatures on $\mC^2$ (\ref{e:scalar-curvature-on-c2}) and $\mS^2$(\ref{e:scalar-curvature-on-s2}) are simply related:
	\beq
	\label{e:scalar-curv-c2-compare-s2}
	R_{\mC^2}=3 R_{\mS^2} -  \left( 3/2 G m^3 h^3 \right) \left( 12 h^2 +|\grad h|^2\right).
	\eeq
This implies $R_{\mC^2} < 0$ since the second term is strictly negative everywhere as we now show. Notice that the second term can vanish only when $h$ is infinite, i.e., at collisions. Taking advantage of the fact that the geometry (on $\mS^2$ and $\mC^2$) in the neighborhood of all 3 collision points is the same for equal masses, it suffices to check that the second term has a strictly negative limit at $C_3$ $(\eta = 0)$. Near $\eta =0$, $h \sim 1/2 \eta^2$ so that $R_{\mC^2} \to - 12/Gm^3 < 0$. Combining with the $r$-independence of $R_{\mC^2}$, we see that the scalar curvature is non-singular at binary and triple collisions.

With a little more effort, we may obtain a non-zero upper bound for the Ricci scalar on $\mC^2$. Indeed, using $R_{\mS^2} \leq 0$ and the inequality $12 h^2 + |\grad h|^2 \geq \zeta h^3$ proved in Appendix \ref{a:upper-bound-scalar-curvature}, we find
	\beq
	R_{\mC^2} < - 3 \zeta/2 G m^3 \quad \text{where} \quad \zeta = 55/27.
	\eeq
Numerically, we estimate the optimal value of $\zeta$ to be $8/3$.

\subsubsection{Scalar curvatures on shape space \texorpdfstring{$\mR^3$}{R3} and on \texorpdfstring{$\mS^3$}{S3}}

 Recall that the equal-mass zero-energy quotient JM metrics on shape space $\mR^3$ (\ref{e:r3-metric-in-r-eta-xi2-coordinates}) and $\mS^3$ (\ref{e:s3-metric-in-eta-xi1-xi2-coordinates}) are
	\beqs
	ds_{\mR^3}^2 &=&  \left( Gm^3 h/r^2 \right) \left(dr^2+r^2\left(d\eta^2+\sin^22\eta \;d\xi_2^2\right) \right) \quad \text{and} \quad 
	\cr
	ds_{\mS^3}^2 &=& Gm^3 h \: \left(d\eta^2+ d\xi_1^2-2\cos2\eta\;d\xi_1\;d\xi_2+ d\xi_2^2\right). 
	\label{e:r3-s3-jm-metric}
	\eeqs
The corresponding scalar curvatures are
	\beqs
	R_{\mR^3} &=& \left(16h^2+3 |\grad h|^2 - 4 h \D h \right)/2 G m^3h^3
	 \quad \text{and} \cr
	R_{\mS^3} &=& \left(12h^2+3|\grad h|^2-4h \D h\right)/2 G m^3h^3.
	\eeqs
Here $\D h$ and $\grad h$ are as in Eq. (\ref{e:grad-laplace-on-s2-kin}).  The scalar curvatures are related to that on $\mS^2$ as follows
	\beqs
	R_{\mR^3} &=& 2 R_{\mS^2} - \left( 16 h^2 +|\grad h|^2\right)/{2 G m^3 h^3} 
	\quad \text{and} \cr
	R_{\mS^3} &=& 2 R_{\mS^2} - \left( 20 h^2 + |\grad h|^2\right)/{2 G m^3 h^3}.
	\eeqs
As in the case of $\mC^2$ we check that the second terms in both relations are strictly negative. This implies both the scalar curvatures are strictly negative. In fact, using the inequality $12 h^2 + |\grad h|^2 > \zeta h^3$ (see Appendix \ref{a:upper-bound-scalar-curvature}) we find (non-optimal) non-zero upper bounds 
	\beq
	R_{\mS^3, \mR^3} < - \zeta/2 G m^3 \quad \text{where} \quad \zeta = 55/27.
	\eeq
Moreover, we note that
	\beq
	R_{\mC^2} = R_{\mS^3} - \frac{h \D h}{G m^3 h^3} < R_{\mS^3} \quad \text{and} \quad
	R_{\mS^3} = R_{\mR^3} - \frac{4 h^2}{2 G m^3 h^3} \leq R_{\mR^3},
	\eeq
with equality at collision configurations. Recalling that on the shape sphere, the scalar curvature vanishes at collision points (in a limiting sense) and at Lagrange points, we have the following inequalities
	\beq
	0 \geq R_{\mS^2} > R_{\mR^3} \geq R_{\mS^3} > R_{\mC^2}.
	\eeq
Thus we have the remarkable result that the scalar curvatures of the JM metric on $\mC^2$ and its quotients by scaling $(\mS^3)$ and rotations $(\mR^3)$ are strictly negative everywhere and also strictly less than that on $\mS^2$. So the full geodesic flow on $\mC^2$ is in a sense more unstable than the corresponding flow on $\mS^2$. 

In addition to strict negativity, we may also show that the scalar curvatures are bounded below. For instance, from Eq. (\ref{e:scalar-curvature-on-s2}) $R_{\mS^2}$ can go to $- \infty$ only when $\D h \to \infty$ since $h \geq 3$. Now from Eq. (\ref{e:grad-laplace-on-s2-kin}) $\D h$ can diverge only when $\sin 2 \eta = 0$ or when one of the relevant derivatives of $h$ diverges. From Eq. (\ref{e:conformal-prefactor-h}) this can happen only if $\eta= 0$ (C3) or $\eta = \pi/2$ (E3) or when one of the $v_i \to \infty$, i.e., at collisions. However $\D h = 66$ is finite at $\eta = \pi/2$ and we know from \S \ref{s:near-pairwise-collision-geometry} that $R_{\mS^2}$ is finite at collisions so that $R_{\mS^2}$ is bounded below. The same proof shows that scalar curvatures are bounded below on $\mR^3, \mS^3$ and $\mC^2$ as well.

\subsection{Sectional curvature for three equal masses}
\label{s:sectional-curvature-inv-sq-pot}

In \S \ref{s:scalar-curvature-inv-sq-pot-c2-r3-s3-s2}, we showed that the Ricci scalars $R$ on configuration space and its quotients are negative everywhere, save at Lagrange and collision points on the shape sphere where it vanishes. However, $R$ encodes the stability of geodesics only in an average sense. More precisely, a geodesic through $P$ in the direction $u$ subject to a perturbation along $v$ is linearly stable/unstable according as the sectional curvature $K_P(u,v)$ is positive/negative (see \S \ref{s:stability-tensor}). Here, the sectional curvature which is a function only of the 2-plane spanned by $u$ and $v$ generalizes the Gaussian curvature to higher dimensions. It is defined as the ratio of the curvature biquadratic $\scripty{r} = g(R(u,v)v,u)$ to the square of the area ${\rm Ar}(u,v)^2 =  g(u,u) g(v,v) - g(u,v) g(v,u)$ of the parallelogram spanned by $u$ and $v$. Here $g(u,v)$ is the Riemannian inner product and $R(u,v) = [\grad_u, \grad_v] - \grad_{[u,v]}$ the curvature tensor with components $R(e_i, e_j) e_k = R^l_{\; k ij} e_l$ in any basis for vector fields. Furthermore, if $e_1, \ldots, e_n$ are an orthonormal basis for the tangent space at $P$, then  the scalar curvature $R = \sum_{i \ne j} K(e_i,e_j)$ is the sum of sectional curvatures in $\binom{n}{2}$ planes through $P$. It may also be regarded as an average of the curvature biquadratic $R = \iint \scripty{r}(u,v) d\mu_g(u) d\mu_g(v)$ where $d\mu_g(u) = \exp \left(- u^i u^j g_{ij}/2 \right) du$ is the gaussian measure on tangent vectors with mean zero and covariance $g^{ij}$ \cite{Rajeev-geometry-fluid-rigid}. Thus $R$ provides an averaged notion of stability. To get a more precise measure of linear stability of geodesics we find the sectional curvatures in various (coordinate) tangent $2$-planes of the configuration space and its quotients. On account of the isometries, these sectional curvatures are functions only of $\eta$ and $\xi_2$ [explicit expressions are omitted due to their length]. Unlike scalar curvatures which were shown to be non-positive, we find planes in which sectional curvatures are non-positive as well as planes where they can have either sign.

O'Neill's theorem allows us to determine or bound certain sectional curvatures on the center-of-mass configuration space $\mC^2$ in terms of the more easily determined curvatures on its quotients. Roughly, the sectional curvature of a horizontal two-plane increases under a Riemannian submersion. Suppose $f: (M,g) \to (N,\tilde g)$ is a Riemannian submersion. Then O'Neill's theorem \cite{oneill} states that the sectional curvature in any horizontal $2$-plane at $m \in M$ is less than or equal to that on the corresponding $2$-plane at $f(m) \in N$:
	\beq
	\label{e:oneill}
	K_N(df(X),df(Y))=K_M(X,Y)+\frac{3}{4}\fr{|[X,Y]^V|^2}{{\rm Ar}(X,Y)^2}.
	\eeq
Here $X$ and $Y$ are horizontal fields on $M$ spanning a non-degenerate $2$-plane (${\rm Ar}(X,Y)^2 \ne 0$) and $[X,Y]^V$ is the vertical projection of their Lie bracket. In particular, the sectional curvatures are equal everywhere if $X$ and $Y$ are coordinate vector fields.

\begin{figure}	
	\centering
	\begin{subfigure}[t]{5cm}
		\centering
		\includegraphics[height=4.5cm]{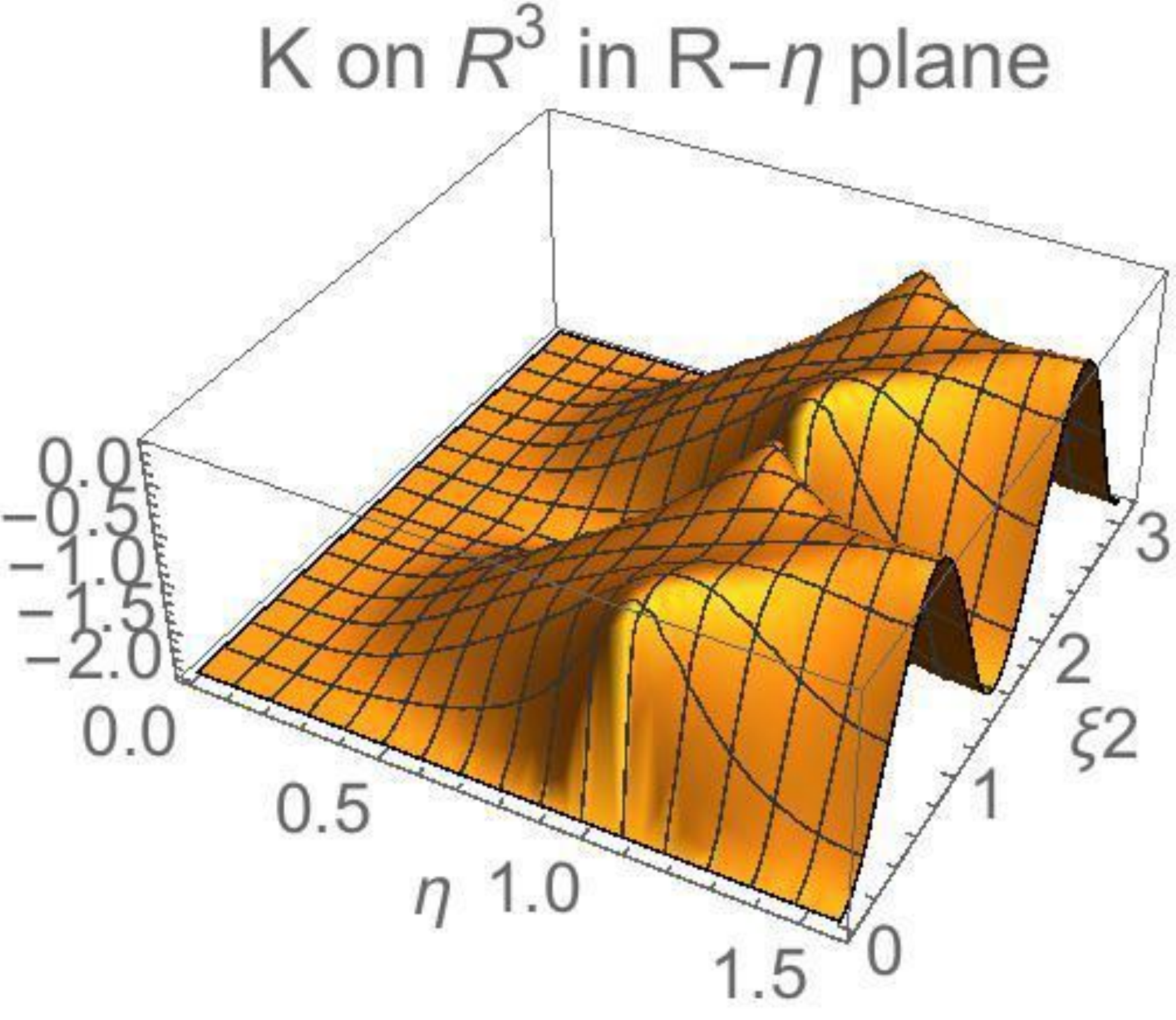}
		\caption{\small }
	\end{subfigure} \quad
	\begin{subfigure}[t]{5cm}
		\centering
		\includegraphics[height=4.5cm]{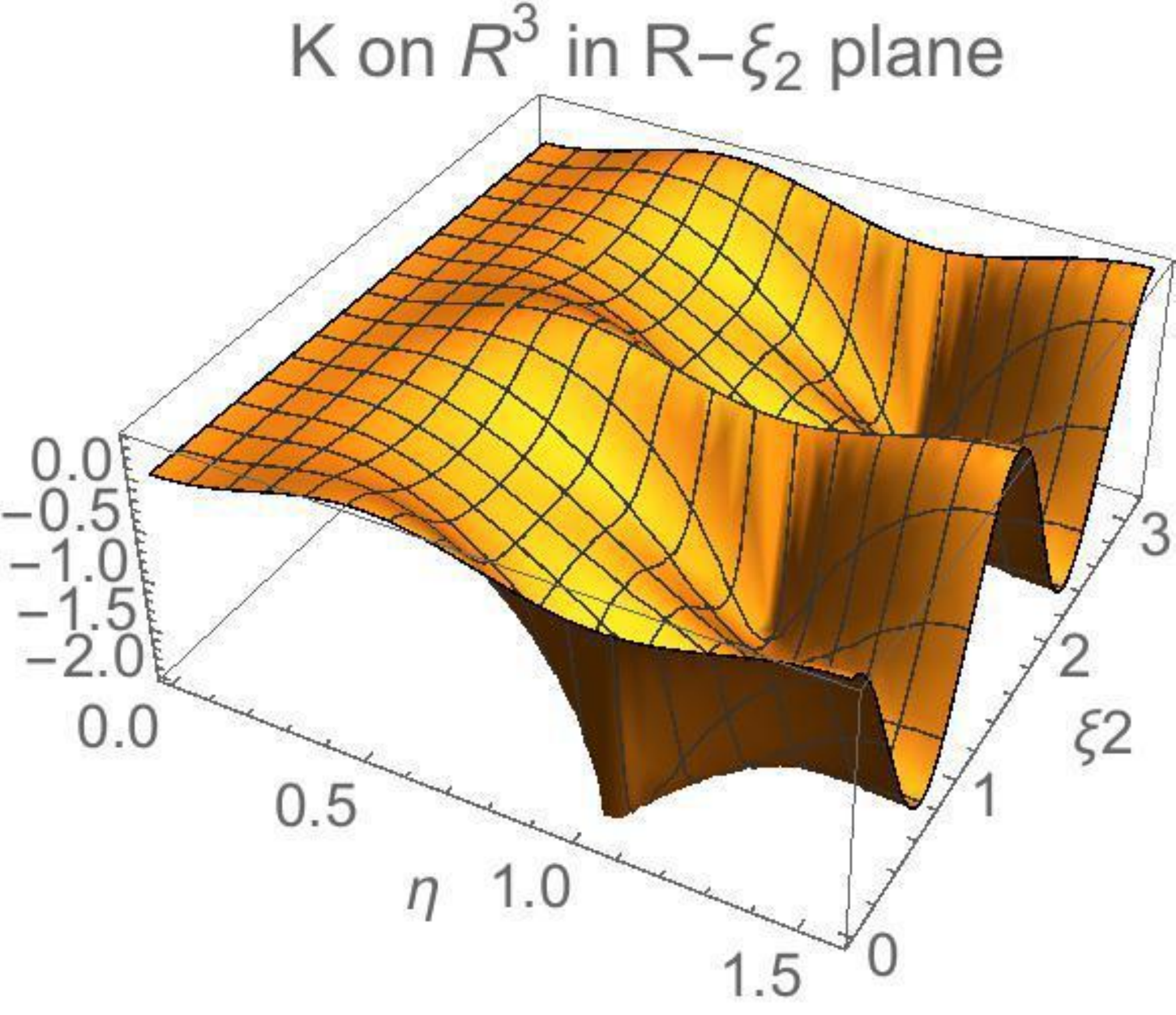}
		\caption{\small }
	\end{subfigure} \quad
	\begin{subfigure}[t]{5cm}
		\centering
		\includegraphics[height=4.5cm]{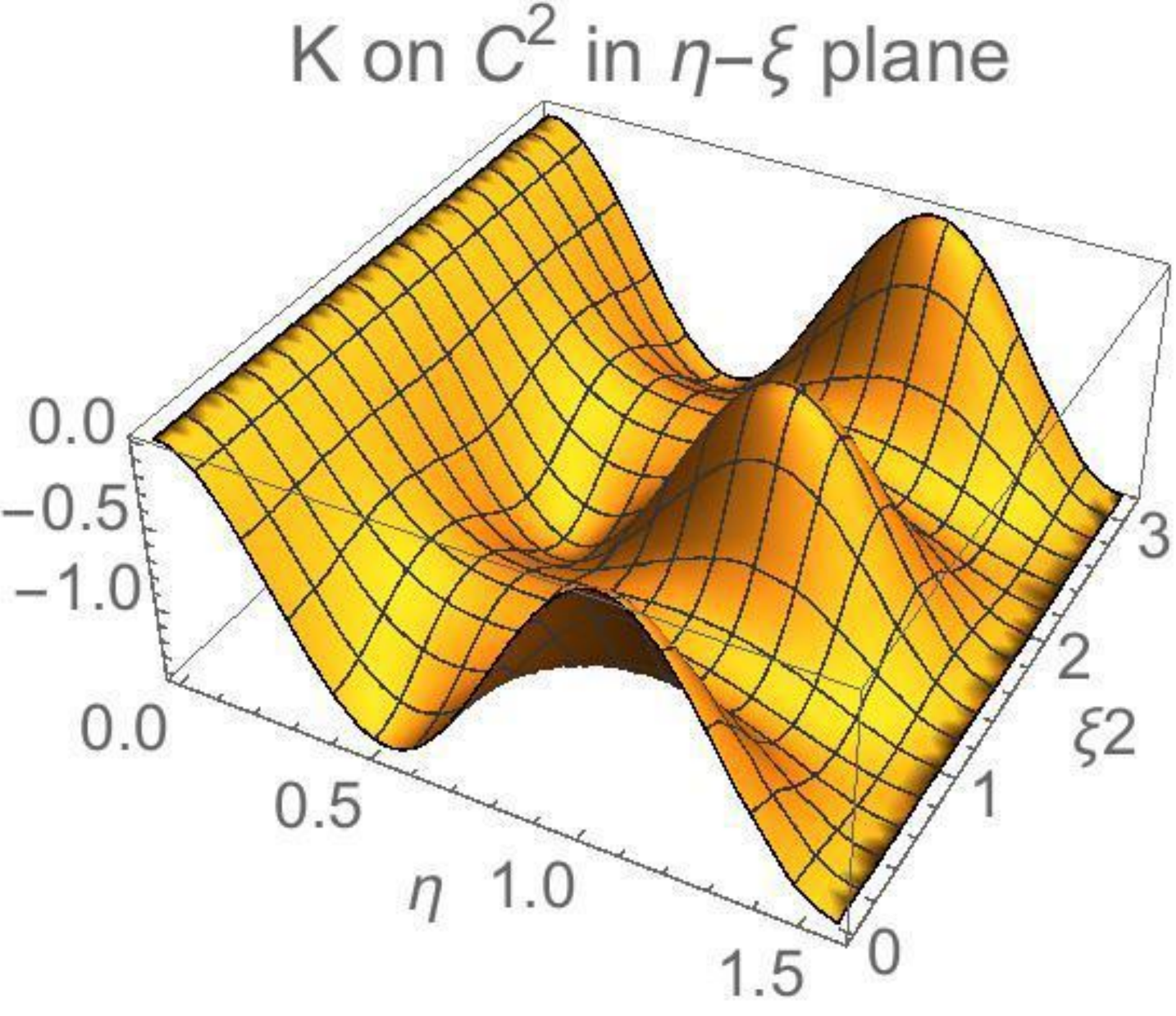}
		\caption{\small }
	\end{subfigure}
\caption{\small Sectional curvatures on horizontal 2-planes of submersion from $\mC^2$ to $\mR^3$ in units of $1/{Gm^3}$. (a) $K_{\mC^2}(\pdr_r,\pdr_\eta) = K_{\mR^3}(\pdr_r,\pdr_\eta) \leq 0$ everywhere except in neighborhoods of $E_3$. $K = -2$ at its global minimum $C_3$ and $K=-2/3$ at $L_{4,5}$. $K \to 0,-2$ when $C_{1,2}$ are approached holding $\eta$ or $\xi_2$ fixed. (b) $K_{\mC^2}(\pdr_r, \pdr_\xi) = K_{\mR^3}(\pdr_r, \pdr_{\xi_2})$ is negative except in neighborhoods of $C_3$ and $E_3$. $K = 0$ at its minimum $C_3$ $(\eta = 0)$ and $K=-2/3$ at $L_{4,5}$. $K \to -2$ or $0$ on approaching $C_{1,2}$ $(\eta = \pi/3, \xi_2 = 0, \pi/2)$ along $\eta$ or $\xi_2$ constant. (c) $K_{\mC^2}(\pdr_\eta, \pdr_{\xi}) \leq K_{\mR^3}(\pdr_\eta, \pdr_{\xi_2})$. $K_{\mC^2}(\pdr_\eta, \pdr_{\xi}) = 0$ at global maxima $C_{1,2,3}$ and is negative elsewhere. $K=-1$ at its local maxima $L_{4,5}$.}
\label{f:shape-space-curvature}
\end{figure}

We consider sectional curvatures in $6$ interesting $2$ planes on $\mC^2$ which are horizontal with respect to submersions to $\mR^3$ and $\mS^3$. Under the submersion from $\mC^2$ to $\mR^3$ (\S \ref{s:quotient-metrics}), the horizontal basis vectors $\pdr_r$, $\pdr_\eta$ and $\pdr_\xi \equiv \cos 2 \eta \pdr_{\xi_1} + \pdr_{\xi_2}$ map respectively to $\pdr_r$, $\pdr_\eta$ and $\pdr_{\xi_2}$ defining three pairs of corresponding $2$-planes. Since $[\pdr_r, \pdr_\eta] $ and $[\pdr_r,\pdr_\xi]$ vanish, we have $K_{\mC^2}(\pdr_r,\pdr_\eta) = K_{\mR^3}(\pdr_r,\pdr_\eta)$ and  $K_{\mC^2}(\pdr_r,\pdr_\xi) = K_{\mR^3}(\pdr_r,\pdr_{\xi_2})$. Fig. \ref{f:shape-space-curvature} shows that $K_{\mC^2}(\pdr_r,\pdr_\eta)$ is mostly negative, though it is not continuous at $E_3$, $C_1$ and $C_2$. On the other hand $K_{\mC^2}(\pdr_r,\pdr_{\xi})$ is largely negative except in a neighborhood of $C_3$. Finally, as $[\pdr_\xi,\pdr_\eta]^V = - 2 \sin 2 \eta \pdr_{\xi_1} \ne 0$, we have $K_{\mC^2}(\pdr_\eta,\pdr_\xi) < K_{\mR^3}(\pdr_\eta,\pdr_{\xi_2})$ with equality at collisions. Moreover the submersion from $\mR^3 \to \mS^2$ (\S \ref{s:quotient-metrics}) implies that $K_{\mR^3}(\pdr_\eta,\pdr_{\xi_2})$ coincides with $K_{\mS^2}(\pdr_\eta,\pdr_{\xi_2})$ which vanishes at Lagrange and collision points and is strictly negative elsewhere (see \S \ref{s:scalar-curvature-inv-sq-pot-c2-r3-s3-s2}). Thus $K_{\mC^2}(\pdr_\eta,\pdr_\xi)$ vanishes at collision points and is strictly negative everywhere else (see Fig. \ref{f:shape-space-curvature}). In particular, Lagrange points are more unstable on the configuration space $\mC^2$ than on the shape sphere.

\begin{figure}	
	\centering
	\begin{subfigure}[t]{5cm}
		\centering
		\includegraphics[height=4.5cm]{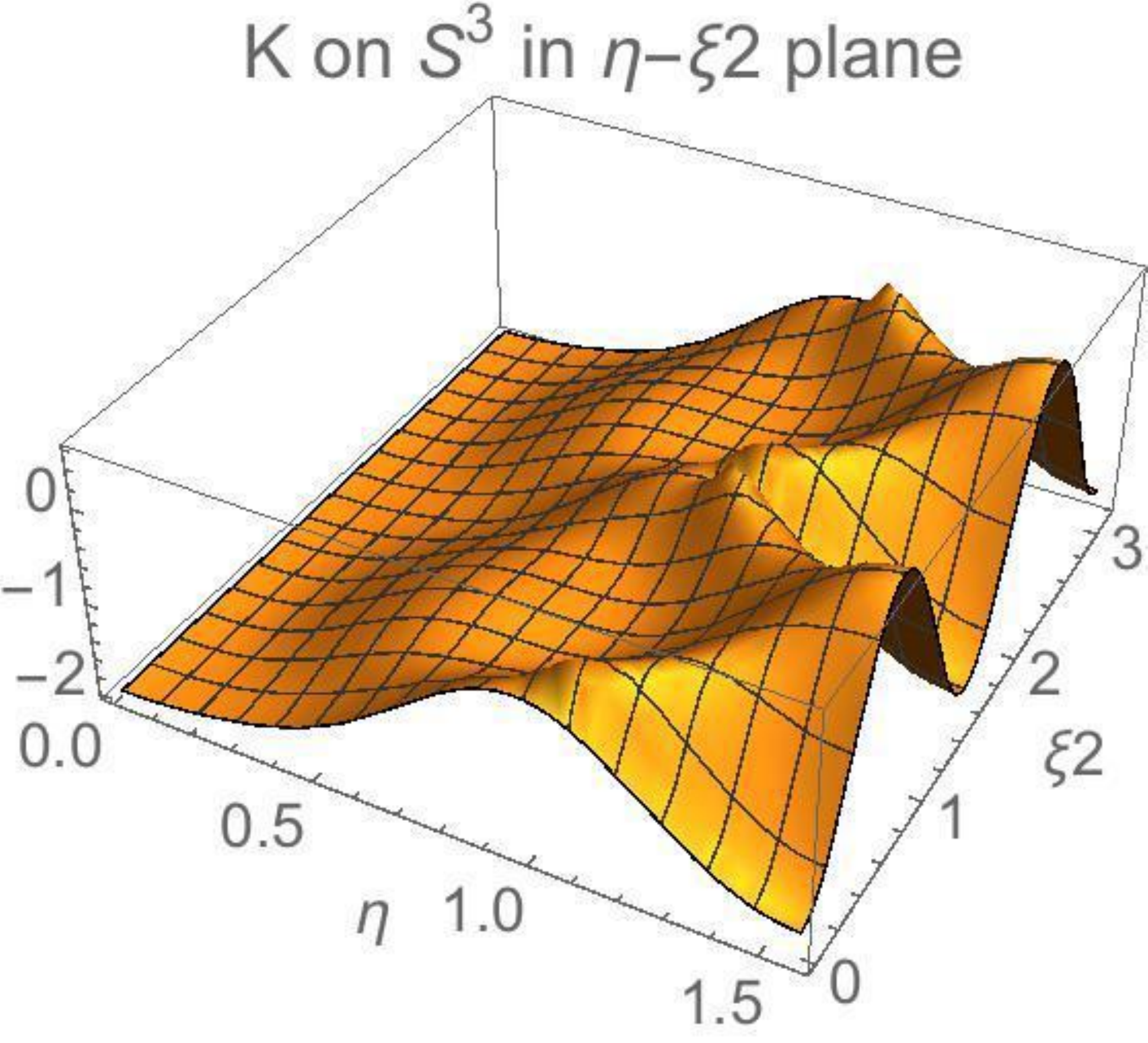}
		\caption{\small }
	\end{subfigure} \quad
	\begin{subfigure}[t]{5cm}
		\centering
		\includegraphics[height=4.5cm]{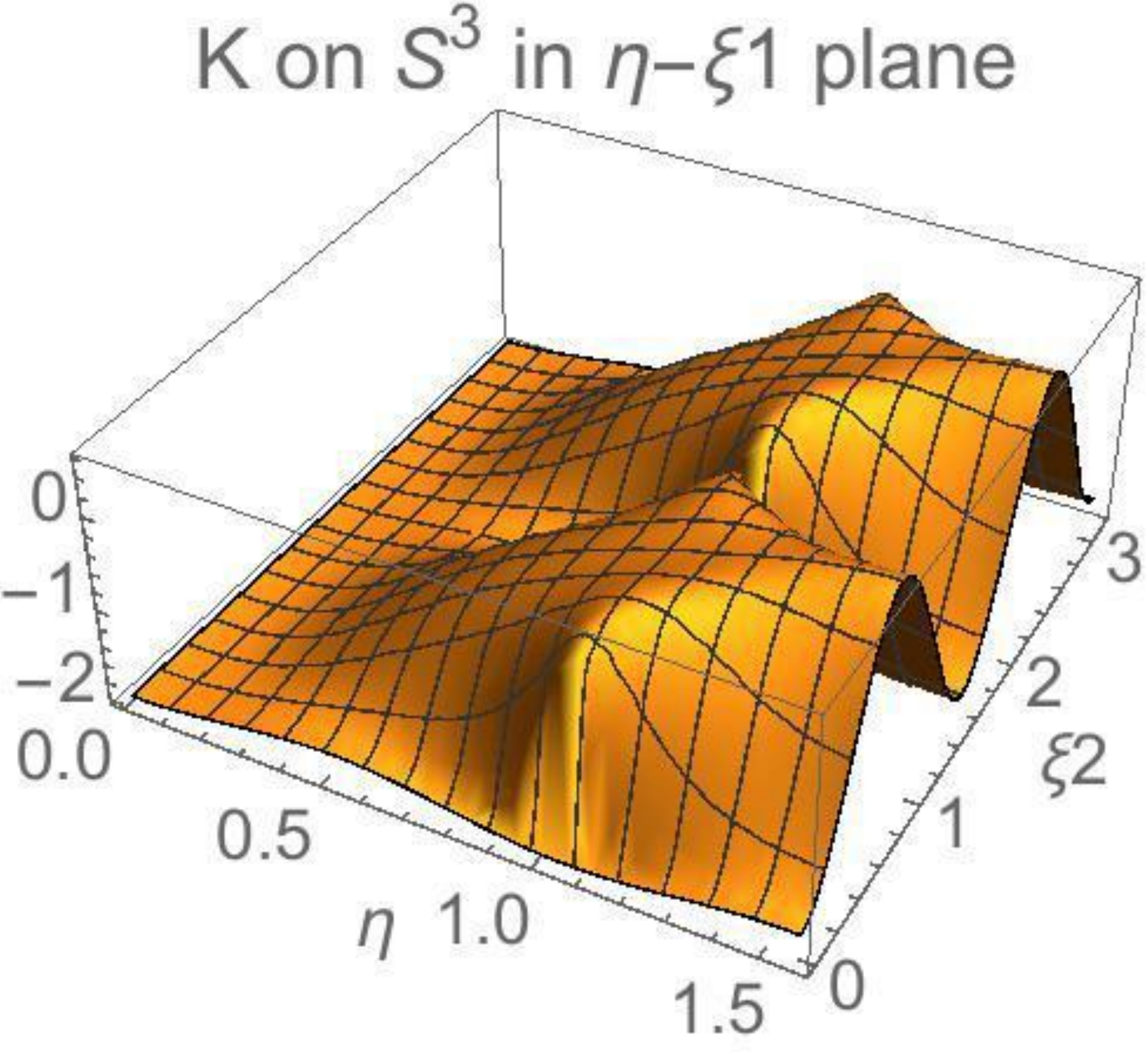}
		\caption{\small }
	\end{subfigure} \quad
	\begin{subfigure}[t]{5cm}
		\centering
		\includegraphics[height=4.5cm]{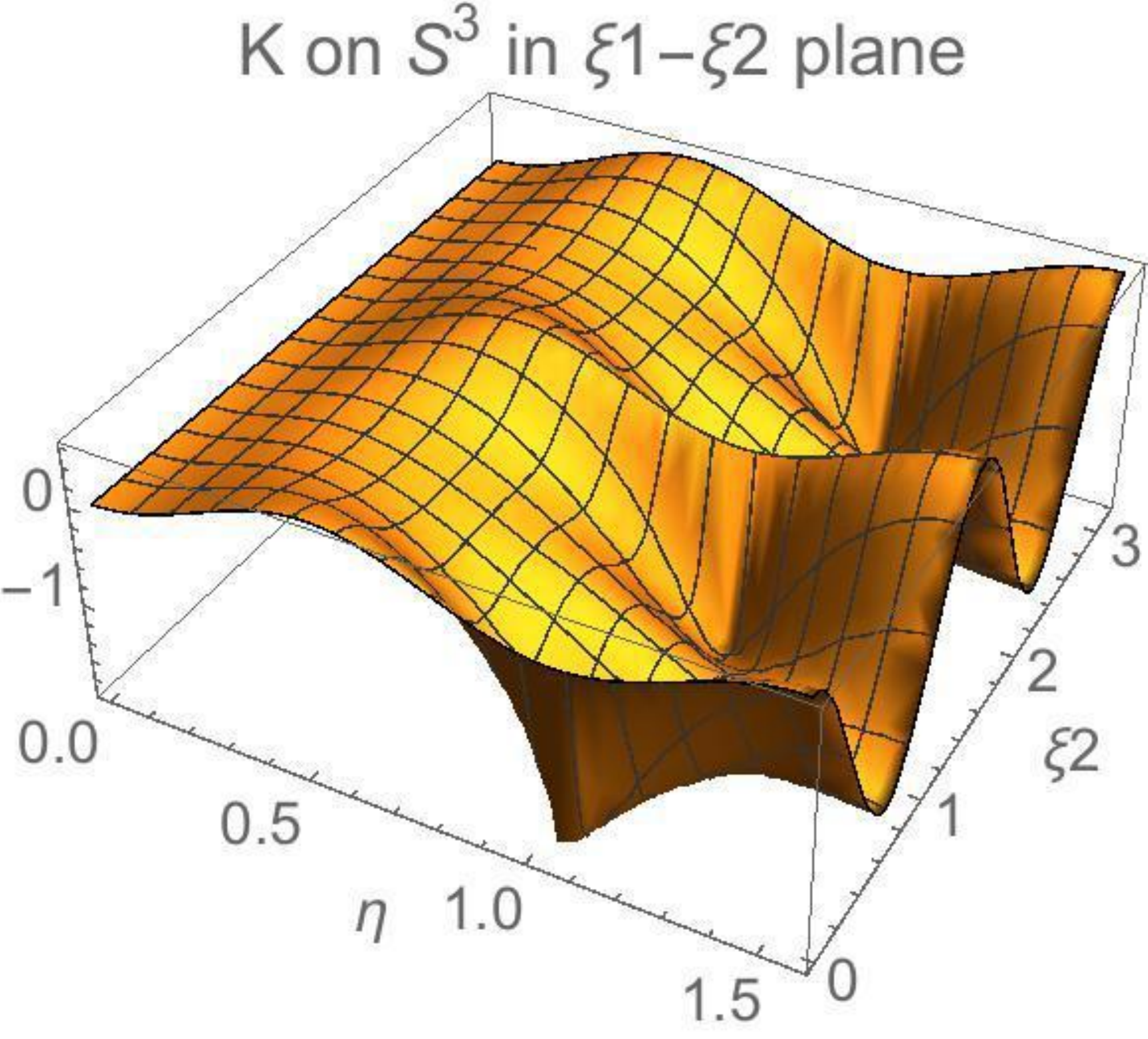}
		\caption{\small }
	\end{subfigure}
\caption{\small Sectional curvatures on horizontal 2-planes of submersion from $\mC^2$ to $\mS^3$ in units of $1/{Gm^3}$. (a) $K_{\mC^2}(\pdr_\eta, \pdr_{\xi_2}) = K_{\mS^3}(\pdr_\eta, \pdr_{\xi_2}) > 0$ in a neighborhood of $E_3$ and negative elsewhere. $K = -2$ at its global minimum $C_3$. $K = -1$ at its local maxima $L_{4,5}$. $K \to$ $0$ or $-1/2$ upon approaching $C_{1,2}$ along constant $\eta$ or $\xi_2$. (b) $K_{\mC^2}(\pdr_\eta, \pdr_{\xi_1}) = K_{\mS^3}(\pdr_\eta, \pdr_{\xi_1}) > 0$ in a neighborhood of $E_3$ and is negative elsewhere. $K= -2$ at its global minimum $C_3$ and $K = -1/3$ at $L_{4,5}$. $K \to 0$ or $-2$ upon approaching $C_{1,2}$ holding $\eta$ or $\xi_2$ fixed. (c) $K_{\mC^2}(\pdr_{\xi_1}, \pdr_{\xi_2}) = K_{\mS^3}(\pdr_{\xi_1}, \pdr_{\xi_2}) > 0$ in some neighborhoods of $C_3$ and $E_{3}$ and negative elsewhere. $K = 0$ at its local minimum $C_3$. $K=-1/3$ at $L_{4,5}$. $K \to -2$ or $0$ upon approaching $C_{1,2}$ while holding $\eta$ or $\xi_2$ fixed.}
\label{f:s3-curvature}
\end{figure}

Under the submersion from $\mC^2$ to $\mS^3$ (\S \ref{s:quotient-metrics}), the horizontal basis vectors $\pdr_\eta$, $\pdr_{\xi_1}$ and $\pdr_{\xi_2}$ map respectively to $\pdr_\eta$, $\pdr_{\xi_1}$ and $\pdr_{\xi_2}$. The sectional curvatures on corresponding pairs of 2-planes are equal, e.g. $K_{\mC^2}(\pdr_\eta,\pdr_{\xi_2}) =K_{\mS^3}(\pdr_\eta,\pdr_{\xi_2})$. 
As shown in Fig. \ref{f:s3-curvature}, $K_{\mC^2}(\pdr_\eta,\pdr_{\xi_2})$ is negative everywhere except in a neighborhood of $E_3$ where it can have either sign. The qualitative behavior of the other two sectional curvatures $K_{\mC^2}(\pdr_{\xi_1},\pdr_{\xi_2})$ and $K_{\mC^2}(\pdr_{\xi_1}, \pdr_\eta)$  is similar to that of $K_{\mC^2}(\pdr_{r},\pdr_{\xi_2})$ and $K_{\mC^2}(\pdr_{r}, \pdr_\eta)$ discussed above. The approximate symmetry under $\pdr_{\xi_1} \leftrightarrow \pdr_r$ is not entirely surprising given that $\pdr_{\xi_1}$ and $\pdr_{r}$ are vertical vectors in the submersions to $\mR^3$ and $\mS^3$ respectively.
   
The remaining two coordinate 2-planes on $\mC^2$ are not horizontal under either submersion. We find that $K_{\mC^2}(\pdr_r, \pdr_{\xi_1})$ is negative everywhere except at $L_{4,5}$ and $K_{\mC^2}(\pdr_r, \pdr_{\xi_2})$ is negative except around $E_{1,2}$.

\subsection{Stability tensor and linear stability of geodesics}
\label{s:stability-tensor}

In this section we use the stability tensor (which provides a criterion for linear geodesic stability) to discuss the stability of Lagrange rotational and homothety solutions. We end with a remark on linear stability of trajectories and geodesics. Consider the $n$-dimensional configuration manifold $M$ with metric $g$. The geodesic deviation equation (GDE) for the evolution of the separating vector (Jacobi field) $y(t)$ between a geodesic $x(t)$ and a neighboring geodesic is \cite{oneill}
	\beq
	\label{e:geodesic-deviation-eqn-jacobi-field}
	\grad_{\dot x}^2 y = R(\dot x, y) \dot x = - R(y, \dot x) \dot x.
	\eeq
We expand the Jacobi field $y=c^k(t) e_k(t)$ in any basis $e_i(t)$ that is parallel transported along the geodesic i.e. $\grad_{\dot x}e_k=0$ [$e_i(0)$ could be taken as coordinate vector fields at $x(0)$]. Taking the inner product of the GDE with $e_m$ and contracting with $g^{im}$, we get $\ddot c^i = - S^i_j c^j$, where the `stability tensor' $S^i_k = R^i_{ jkl}\dot x^j \dot x^l$. As $S$ is real symmetric, its eigenvectors $f_i$ can be chosen to form an orthonormal basis for $T_x M$. Writing $y = d^m f_m$, the GDE becomes $\ddot d^m = -\kappa_m d^m$ (no sum on $m$) where $\kappa_m$ is the eigenvalue of $S$ corresponding to the eigenvector $f_m$. The eigenvalues of $S$ (say at $t = 0$) control the initial evolution of the Jacobi fields in the corresponding eigendirections. Since $\kappa_m = \left( {\rm Area}{\bra f_k , \dot x \ket} \right)^2 K(f_m, \dot x)$ (\S \ref{s:sectional-curvature-inv-sq-pot}), positive (negative) $\kappa$ or $K$ imply local stability (instability) for the initial evolution. We note that calculating $S$ and its eigenvalues at a given instant (say $t=0$) requires no knowledge of the time evolution of $e_i(t)$. So we may simply use the coordinate vector fields as the basis. Notice that the tangent vector to the geodesic $\dot x$ is always an eigendirection of $S$ with eigenvalue zero.

\subsubsection{Rotational Lagrange solutions in Newtonian potential}

 Consider the Lagrange rotational solutions where three equal masses ($m_i = m$) rotate at angular speed $\om = \sqrt{3 G m / a^3}$ around their CM at the vertices of an equilateral triangle of side $a$. The rotational trajectory on $\mC^2$ in $r,\eta,\xi_{1,2}$ coordinates is given by $x(t) = (a/\sqrt{m}, \pi/4, \om t, \pm \pi/4)$ with velocity vector $\om \pdr_{\xi_1}$. Note that trajectory and geodesic times are proportional since $\sigma = ds/dt = (E-V)/\sqrt{\cal T}$ with $V(r, \eta, \xi_2)$ and $\cal T$ constant along $x(t)$. The stability tensor along the geodesic, $S = \om^2 \; \text{diag} (1,-1/2, 0, -1/2 )$ is diagonal in the coordinate basis $r,\eta, \xi_1, \xi_2$. As always, $\dot x$ is a zero-mode. A perturbation along $\pdr_r$ is linearly stable while those directed along $\pdr_\eta$ or $\pdr_{\xi_2}$ are linearly unstable. Note that Routh's criterion $27 (m_1 m_2 + m_2 m_3 + m_3 m_1) < M^2$ \cite{routh} predicts that Lagrange rotational solutions are linearly unstable for equal masses.

\subsubsection{Lagrange homotheties} 

For equal masses, a Lagrange homothety solution is one where the masses move radially (towards/away from their CM) while being at the vertices of equilateral triangles. The geodesic in Hopf coordinates takes the form $(r(t), \eta = \pi/4, \xi_1, \xi_2 = \pm \pi/4)$ where $\xi_1$ is arbitrary and independent of time. Though an explicit expression is not needed here, $r(t)$ is the solution of $\ddot r + \G^r_{rr} \dot r^2 = 0$ where $\G^r_{rr} = - 3 G m^3/(E r^3 + 3 Gm^3 r)$ for the inverse-square potential. The stability tensor is diagonal:
	\beq
	S = \fr{6 G m^3 \dot r^2}{\left( 3 G m^3 r + E r^3 \right)^2}\text{diag} \left(0,- 3 G m^3 - 2 E r^2, - E r^2 , - 3 G m^3 - 2 E r^2 \right).
	\eeq
For a given $r$ and positive energy, perturbations along $\pdr_{\xi_{1,2}}$ and $\pdr_\eta$ are unstable while they are stable when $-3Gm^3/r^2 < E < - 3 Gm^3/2r^2$. For intermediate (negative) energies, $\pdr_{\eta}$ and $\pdr_{\xi_2}$ are unstable directions while $\pdr_{\xi_1}$ is stable. For the Newtonian potential, we have similar conclusions following from the corresponding stability tensor:
	\beq
	S = \fr{3 G m^{5/2} \dot r^2}{4 r^2 \left( 3 G m^{5/2} + E r \right)^2} \text{diag} \left(0,- 9 G m^{5/2} - 5 E  r, - 2 E r, - 9 G m^{5/2} - 5 E  r \right).
	\eeq
We end this section with a cautionary remark. For a system whose trajectories can be regarded as geodesics of the JM metric, linear stability of geodesics may not coincide with linear stability of corresponding trajectories. This may be due to the reparametrization of time (see \S \ref{s:triple-collision-inv-sq-near-collision-geom} for examples) as well as the restriction to energy conserving perturbations in the GDE. We illustrate this with a 2D isotropic oscillator with spring constant $k$. Here the curvature of the JM metric (see \S \ref{s:traj-as-geodesics}) is $R = 2Ek/T^3$ where $T$ is the kinetic energy. Thus for positive $k$, geodesics are always linearly stable while for negative $k$ they are stable/unstable according as energy is negative/positive. By contrast, linearizing the EOM $\ddot \del x_i = - (k/m) \del x_i$ shows that trajectories are linearly stable for positive $k$ and linearly unstable for negative $k$. This (possibly atypical) example illustrates the fact that geodesic stability does not necessarily imply stability of trajectories. 

\section[Planar three-body problem with Newtonian potential]{Planar three-body problem with Newtonian \\potential \sectionmark{Planar three--body problem with Newtonian potential}}
\sectionmark{Planar three--body problem with Newtonian potential}
\label{s:three-body-coulomb}

\subsection{JM metric and its curvature on configuration and shape space}
\label{s:curvature-newtonian-potential}

In analogy with our geometric treatment of the planar motion of three masses subject to inverse-square potentials, we briefly discuss the gravitational analogue with Newtonian potentials. As before, the translation invariance of the Lagrangian 
	\beq
	L = \half \sum_{i=1,2,3} m_i \dot x_i^2 - \sum_{i < j} \fr{G m_i m_j}{|x_i - x_j|}
	\eeq
allows us to go from the configuration space $\mC^3$ to the center-of-mass configuration space $\mC^2$ endowed with the JM metric
	\beq
	ds^2 = \left( E + \fr{G m_1 m_2}{|J_1|} + \fr{G m_2 m_3}{|J_2 - \mu_1 J_1|} + \fr{G m_3 m_1}{|J_2+\mu_2 J_1|} \right) \left( M_1 |dJ_1|^2 + M_2 |dJ_2|^2  \right).
	\eeq
The Jacobi coordinates $J_{1,2}$, mass ratios $\mu_{1,2}$ and reduced masses $M_{1,2}$ are as defined in Eqs. (\ref{e:jacobi-coordinates-on-c3}, \ref{e:ke-in-jacobi-coordinates-on-c3}, \ref{e:jm-metric-in-jacobi-coordinates-on-c3}). In rescaled Jacobi coordinates  $z_i = \sqrt{M_i} \:  J_i$ (\ref{e:jm-metric-in-jacobi-coordinates-on-c2}), the JM metric on $\mC^2$ for {\it equal masses} becomes
	\beq
	ds^2 = \left(E+\fr{G m^{5/2}}{\sqrt{2}|z_1|}+\fr{\sqrt{2} G m^{5/2}}{\sqrt{3}|z_2-\ov{\sqrt{3}}z_1|}+\fr{\sqrt{2} G m^{5/2}}{\sqrt{3}|z_2+\ov{\sqrt{3}}z_1|}\right) \left(|dz_1|^2+|dz_2|^2 \right).
	\eeq
Rotations $z_j \mapsto e^{i \tht} z_j$ continue to act as isometries  corresponding to the KVF $\pdr_{\xi_1}$ in Hopf coordinates (\ref{e:Hopf-coords}), where the JM metric is
	\beqs
	\label{e:c2-jm-metric-coulomb}
	ds^2 &=& \left(E + \fr{Gm^{5/2}U}{r}\right) \left(dr^2+r^2\left(d\eta^2+ d\xi_1^2-2\cos2\eta\;d\xi_1\;d\xi_2+ d\xi_2^2\right)\right)
	\cr \text{with} \quad
	 U &=&\fr{1}{\sqrt{2}\sin\eta} + \fr{\sqrt{2}}{\sqrt{2 + \cos 2 \eta - \sqrt{3} \sin 2 \eta \cos 2 \xi_2}} \cr
	 &&+ \fr{\sqrt{2}}{\sqrt{2 + \cos 2 \eta + \sqrt{3} \sin 2 \eta \cos  2 \xi_2 }}.
	\eeqs
Requiring the submersion $(r,\eta,\xi_1, \xi_2) \mapsto (r,\eta,\xi_2)$ from $\mC^2$ to its quotient by rotations to be Riemannian gives us the JM metric on shape space $\mR^3$:
	\beq
	\label{e:r3-jm-metric-coulomb}
ds^2=  \left(E + {Gm^{5/2}U}/{r}\right) \left(dr^2+r^2\left(d\eta^2+\sin^2 2\eta \;d\xi_2^2\right)\right). 
	\eeq
Unlike for the inverse-square potential, scaling $r \mapsto \la r$ is not an isometry of the JM metric even when $E = 0$. Thus we do not have a further submersion to the shape sphere. However, in what follows, we will consider $E=0$, as it leads to substantially simpler curvature formulae.
 	
\begin{figure}	
	\centering
	\begin{subfigure}[t]{7.5cm}
		\centering
		\includegraphics[height=5.3cm]{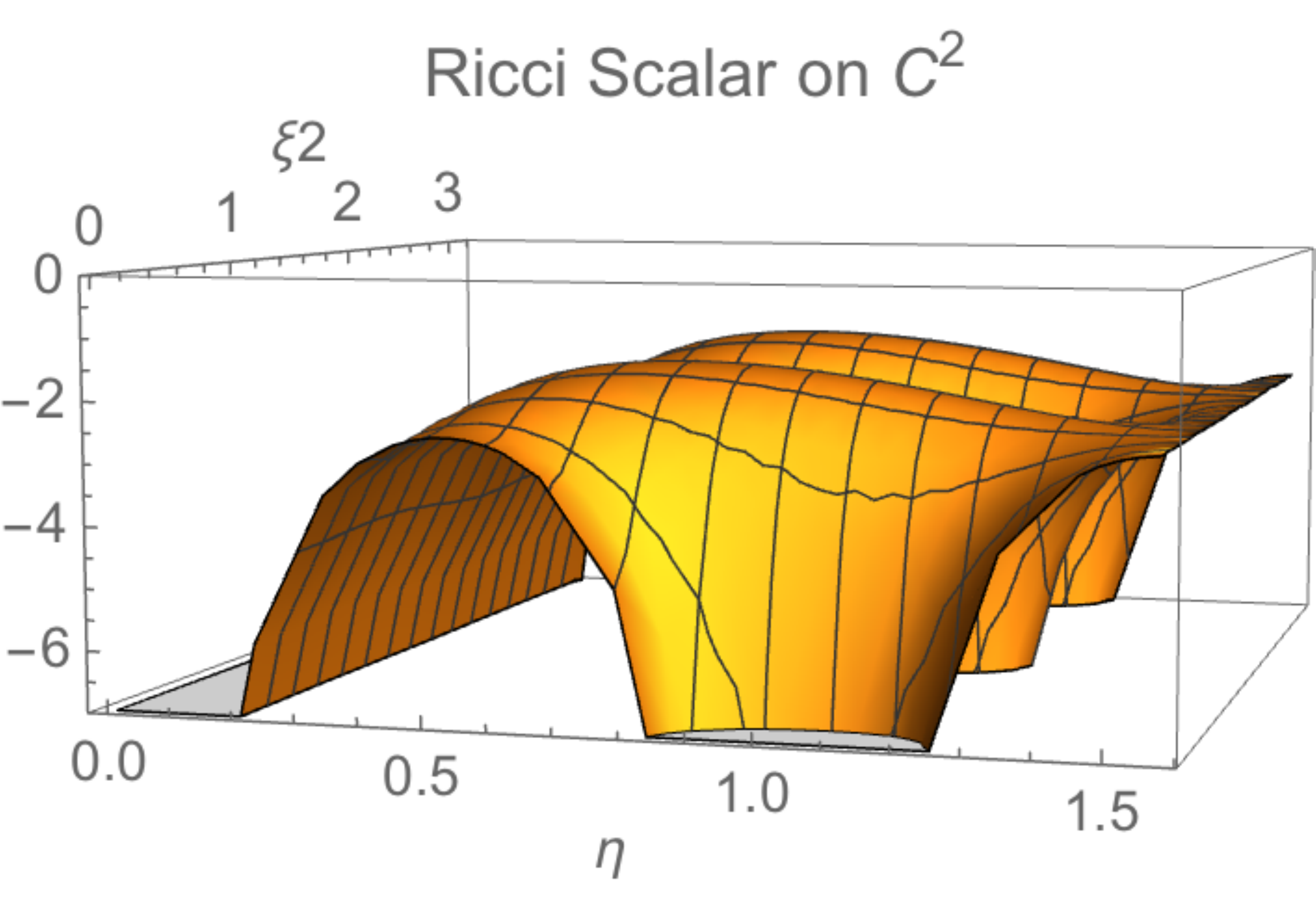}
		\caption{\small }
	\end{subfigure}
	\;
	\begin{subfigure}[t]{7.5cm}
		\centering
		\includegraphics[height=5cm]{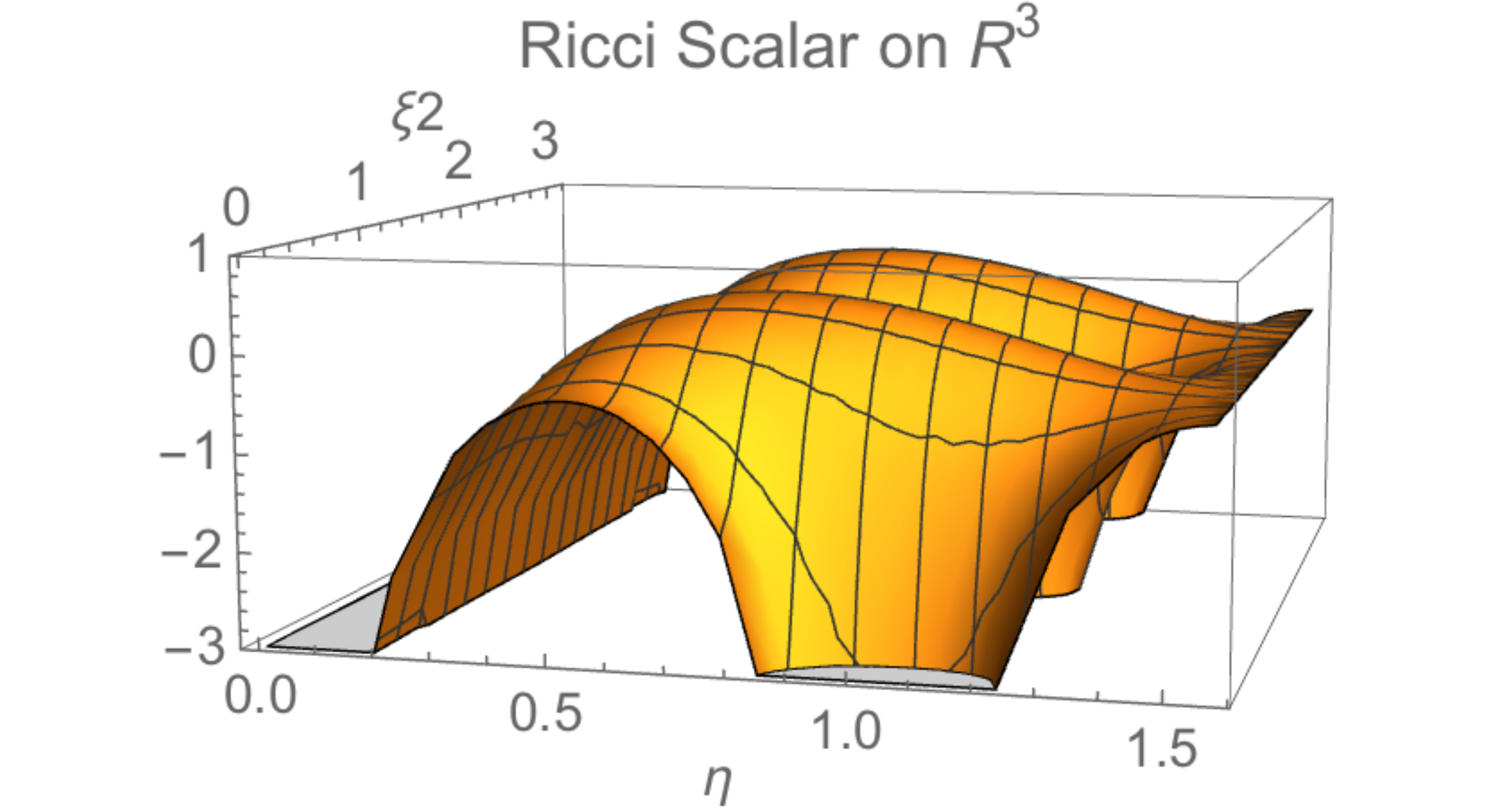}
		\caption{\small }
	\end{subfigure}
\caption{\small Ricci scalar $R$ for zero energy and equal masses on $\mC^2$ and $\mR^3$ for the Newtonian potential (in units of ${1/G m^{5/2} r}$). $R$ on $\mC^2$ is strictly negative while that on $\mR^3$ can have either sign.}
\label{f:scalar-curvature-inv-r}
\end{figure}

Though we do not have a submersion to the shape sphere, the quantity $U(\eta, \xi_2)$ in the conformal factor may be regarded as a function on a $2$-sphere of radius one-half. This allows us to express the scalar curvatures as 
	\beqs
	\label{e:scalar-curvatures-coulomb}
	R_{\mC^2} &=& \frac{3}{2 G m^{5/2} r U ^3}
 \left(3 U ^2+ |\grad U|^2-2 U \D U \right) \quad \text{and} \cr
	R_{\mR^3} &=& \frac{1}{4 Gm^{5/2} r U^3} \left(30 U^2+6|\grad U|^2 -8  U  \D U  \right)
	\eeqs
where $\D U$ is the Laplacian and $\grad U$ the gradient relative to the round metric on a $2$-sphere of radius $1/2$. Evidently, both the scalar curvatures vanish in the limit $r \to \infty$ of large moment of inertia $I_{\rm CM} = r^2$; they are plotted in Fig. \ref{f:scalar-curvature-inv-r}. Numerically, we find that for any fixed $r$, $R_{\mC^2}$ is strictly negative and reaches its global maximum $-3/(2 Gm^{5/2} r)$ at the Lagrange configurations $L_{4,5}$, while $R_{\mR^3}$ has a positive global maximum $1/(2 Gm^{5/2} r)$ at the same locations. Note that $R_{\mR^3} = 2 R_{\mC^2}/3 + (9U^2 + |\grad U|^2 )/(2 Gm^{5/2} r U^3)$. As argued in Eq. (\ref{e:scalar-curv-c2-compare-s2}), the second term is strictly positive and vanishes only when $r \to \infty$. Using the negativity of $R_{\mC^2}$, it follows that $R_{\mR^3} > R_{\mC^2}$ with $(R_{\mR^3} - R_{\mC^2})$ attaining its minimum $2/(Gm^{5/2} r)$ at $L_{4,5}$. Thus in a sense, the geodesic dynamics on $\mC^2$ is more linearly unstable than on shape space.  Like the Ricci scalars, sectional curvatures on coordinate $2$-planes are $(1/r) \times $ a function of $\eta$ and $\xi_2$. We find that sectional curvatures are largely negative and often go to $\pm \infty$ at collision points (see Eq. (\ref{e:sec-curv-near-collision-coulomb-pot})).

\subsection{Near-collision geometry and geodesic incompleteness}
\label{s:geodesic-incompleteness-newtonian-pot}

Unlike for the inverse-square potential, the scalar curvatures on $\mC^2$ and $\mR^3$ (\ref{e:scalar-curvatures-coulomb}) diverge at binary and triple collisions. To examine the geometry near pairwise collisions of equal masses, it suffices to study the geometry near $C_3$ ($\eta = 0$, $r \ne 0$, $\xi_{1,2}$ arbitrary) which represents a collision of $m_1$ and $m_2$. We do so by retaining only those terms in the expansion of the zero-energy metrics around $\eta=0$:
	\beqs
	\label{e:near-collision-1/r-c2}
	ds_{\mC^2}^2 &\approx& \left( \fr{G m^{5/2}}{\sqrt{2}\eta r} \right) \left( dr^2+ r^2  \left(d\eta^2+d\xi_1^2-2(1-2\eta^2) d\xi_1 d\xi_2+d\xi_2^2\right)\right) 
	\quad \text{and} \cr
	ds_{\mR^3}^2 &\approx& \left( \fr{G m^{5/2}}{r}  \right) \left( \ov{\sqrt{2} \eta} + 2\sqrt{\fr{2}{3}}\right) \left( dr^2+ r^2  \left(d\eta^2+4\eta^2d\xi_2^2\right)\right),
	\eeqs
that are necessary to arrive at the following curvatures to leading order in $\eta$:
	\beqs
	\label{e:sec-curv-near-collision-coulomb-pot}
	\text{on $\mC^2$:} && R= \fr{-3}{\varrho} \; \text{and} \; 
	K(\pdr_\eta,\pdr_{r,\xi_{1,2}}) = 2 K(\pdr_r,\pdr_{\xi_{1,2}}) = - 2K(\pdr_{\xi_1},\pdr_{\xi_2}) =\fr{-1}{\varrho } \cr
	\text{on $\mR^3$:} && R = \fr{-1}{\varrho}, \;
\; K(\pdr_\eta,\pdr_r) = -2 K(\pdr_r,\pdr_{\xi_2}) = \fr{-1}{\varrho } \cr
&&\text{and} \quad  K(\pdr_\eta,\pdr_{\xi_2}) = -\fr{2\sqrt{{2}/{3}}}{Gm^{5/2}}
	\eeqs
where $\varrho = \sqrt{2} G m^{5/2}\eta r$. The curvature singularity at $\eta = 0$ is evident in the simple poles in the Ricci scalars and all but one of the sectional curvatures in coordinate planes.

We use the near-collision JM metric of Eq. (\ref{e:near-collision-1/r-c2}) to show that a pairwise collision point lies at finite geodesic distance from another point in its neighborhood. Thus, unlike for the inverse-square potential, the geodesic reformulation {\it does not} regularize the gravitational three-body problem. Consider a point $P$ near $\eta = 0$ with coordinates $(r,\eta_0,\xi_1, \xi_2)$. We estimate its distance to the collision point $C_3$ $(r,0,\xi_1, \xi_2)$. To do so, we consider a curve $\g$ of constant $r$, $\xi_1$ and $\xi_2$ running from $P$ to $C_3$ parametrized by $\eta_0 \geq \eta \geq 0$. We will show that $\g$ has finite length so that the geodesic distance to $C_3$ must be finite. In fact, from (\ref{e:near-collision-1/r-c2}):
	\beq
	\text{Length}(\g) = \int_{\eta_0}^{0} \sqrt{\fr{G r m^{5/2}}{\sqrt{2}}}  \fr{d\eta}{\sqrt{\eta}} = - 2 \sqrt{\fr{G r m^{5/2}}{\sqrt{2}}} \sqrt{\eta_0} < \infty.
	\eeq
Furthermore, the image of $\g$ under the Riemannian submersion to shape space $\mR^3$ is a curve of even shorter length ending at a collision point. Thus geodesics on $\mC^2$ and $\mR^3$ can reach binary collisions in finite time, where the scalar curvature is singular. It is therefore interesting to study regularizations of collisions in the three body problem and their geometric interpretation.

\chapter[Instabilities, chaos and ergodicity in the classical three-rotor problem]{Instabilities, chaos and ergodicity in the classical three-rotor problem}
\chaptermark{Classical three--rotor problem}
\label{chapter:three-rotor}

In this chapter, we investigate periodic orbits, instabilities and onset of chaos in the system of three coupled rotors. Furthermore, we investigate ergodicity, mixing and recurrence time statistics in a band of energies. This chapter is based on \cite{gskhs-cnsd-3rotor}, \cite{gskhs-3rotor} and \cite{gskhs-3rotor-ergodicity}.

\section{Three coupled classical rotors}
\label{s:three-rotor-setup}

We study a periodic chain of three identical rotors of mass $m$ interacting via attractive cosine potentials. The Lagrangian is 
	\beq
	L = \sum_{i=1}^3  \left\{ \half { m r^2} \dot\tht_i^2 - g [1 - \cos\left(\tht_i-\tht_{i+1} \right) ] \right\}
	\eeq
with $\tht_4 \equiv \tht_1$. Here, $\tht_i$ are $2\pi$-periodic coordinates on a circle of radius $r$. Though we only have nearest neighbor  interactions, each pair interacts as there are only three rotors. We consider the `ferromagnetic' case where the coupling $g > 0$ so that the rotors attract each other. Unlike in the gravitational three-body problem, the inter-rotor forces vanish when a pair of them coincide so that rotors can `pass' through each other: this is physically reasonable since they occupy distinct sites. The equations of motion for $i = 1$, $2$ and $3$ (with $\tht_0 \equiv \tht_3$ and $\tht_1 \equiv \tht_4$) are
	\beq
	m r^2 \ddot \tht_i = g \sin (\tht_{i-1}-\tht_i) - g \sin (\tht_i-\tht_{i+1}).
	\eeq
This is a system with three degrees of freedom, the configuration space is a 3-torus $0 \leq \tht_i \leq 2\pi$. The conjugate angular momenta are $\pi_i = m r^2 \dot \tht_i$ and the Hamiltonian is
	\beq
	H = \sum_{i =1}^3  \left\{ \fr{\pi_i^2}{2 m r^2} + g [1- \cos\left(\tht_i-\tht_{i+1} \right) ] \right\}.
	\eeq
Hamilton's equations 
	\beq
	\dot \tht_i = \fr{\pi_i}{mr^2} \quad \text{and} \quad \dot \pi_i = g [\sin (\tht_{i-1}-\tht_i) - \sin (\tht_i-\tht_{i+1})]
	\label{e:hamilton-eom-theta}
	\eeq
define a smooth Hamiltonian vector field on the 6d phase space of the three-rotor problem. The additive constant in $H$ is chosen so that the minimal value of energy is zero. This system has three independent dimensionful physical parameters $m$, $r$ and $g$ that can be scaled to one by a choice of units. However, once such a choice of units has been made, all other physical quantities (such as $\hbar$) have definite numerical values. This circumstance is similar to that in the Toda model \cite{gutzwiller-book}. As discussed in Appendix \ref{a:n-rotor-from-xy}, the quantum $n$-rotor problem, which models a chain of Josephson junctions, also arises by Wick-rotating a partial continuum limit of the XY model on a lattice with nearest neighbor ferromagnetic coupling $J$, $n$ horizontal sites and horizontal and vertical spacings $a$ and $b$ (\ref{e:classical-action-intermediate-problem}). The above parameters are related to those of the Wick-rotated XY model via $m = J/c^2$, $r = \sqrt{L b^2/a}$ and $g = J L/a$ where $L = n a$ and $c$ is a speed associated to the Wick rotation to time.

The Hamiltonian vector field (\ref{e:hamilton-eom-theta}) is non-singular everywhere on the phase space. In particular, particles may pass through one another without encountering collisional singularities. Though the phase space is not compact, the constant energy $(H = E)$ hypersurfaces are compact 5d submanifolds without boundaries. Indeed, $0 \leq \tht_i \leq 2\pi$ are periodic coordinates on the compact configuration space $T^3$. Moreover, the potential energy is non-negative so that $\pi_i^2 \leq 2 m r^2 E$. Thus, the angular momenta too have finite ranges. Consequently, we cannot have `non-collisional singularities' where the (angular) momentum or position diverges in finite time. Solutions to the initial value problem (IVP) are therefore expected to exist and be unique for all time. 

Alternatively, the Hamiltonian vector field is globally Lipschitz since it is everywhere differentiable and its differential bounded in magnitude on account of energy conservation. This means that there is a common Lipschitz constant on the energy hypersurface, so that a unique solution to the IVP is guaranteed to exist for $0 \leq t \leq t_0$ where $t_0 > 0$ is independent of initial condition (IC). Repeating this argument, the solution can be extended for $t_0 \leq t \leq 2t_0$ and thus can be prolonged indefinitely in time for any IC, implying global existence and uniqueness \cite{global-existence-ode}. 

In \S \ref{s:JM-approach}, we will reformulate the dynamics as geodesic flow on a two-torus (or three-torus upon including center of mass motion, see below), which must be geodesically complete as a consequence. For $E > 4.5g$, this is expected on account of compactness and lack of boundary of the energetically allowed Hill region. For $E < 4.5 g$, though the trajectories can (in finite time) reach the Hill boundary, they simply turn around. Examples of such trajectories are provided by the $\vf_1 = 0$ pendulum solutions described in \S \ref{s:pendulum-soln}.

\subsection{Center of mass and relative coordinates}

 It is convenient to define the center of mass (CM) and relative angles
	\beq
	\vf_0 = \fr{\tht_1 + \tht_2 + \tht_3}{3}, 
	\;\; \vf_1 = \tht_1 - \tht_2 \;\; \text{and} \;\;  \vf_2 = \tht_2 - \tht_3
	\label{e:phi0-cm-relative-coords}
	\eeq
or equivalently,
	\beq
	\tht_1 = \vf_0 + \fr{2\vf_1}{3} + \fr{\vf_2}{3}, \quad \tht_2 = \vf_0 - \fr{\vf_1}{3} + \fr{\vf_2}{3} \quad \text{and} \quad
	 \tht_3 = \vf_0 - \fr{\vf_1}{3} - \fr{2\vf_2}{3}.
	\eeq
As a consequence of the $2\pi$-periodicity of the $\tht$s, $\vf_0$ is $2\pi$-periodic while $\vf_{1,2}$ are $6\pi$-periodic. However, the cuboid ($0 \leq \vf_0 \leq 2\pi$, $0 \leq \vf_{1,2} \leq 6\pi$) is a nine-fold cover of the fundamental cuboid $0 \leq \tht_{1,2,3} \leq 2\pi$. In fact, since the configurations $(\vf_0, \vf_1 - 2\pi, \vf_2)$, $(\vf_0, \vf_1, \vf_2 + 2\pi)$ and $(\vf_0 +2\pi/3, \vf_1, \vf_2)$ are physically identical, we may restrict $\vf_{1,2}$ to lie in $[0,2\pi]$. Here, the $\vf_i$ are not quite periodic coordinates on $T^3 \equiv [0,2\pi]^3$. Rather, when $\vf_1 \mapsto \vf_1 \pm 2\pi$ or $\vf_2 \mapsto \vf_2 \mp 2\pi$, the CM variable $\vf_0 \mapsto \vf_0 \pm 2\pi/3$. In these coordinates, the  Lagrangian becomes $L = \T - \V$ where
	\beq
	\T =  \fr{3}{2} m r^2 \dot \vf_0^2 + \fr{1}{3} m r^2 \left[  \dot \vf_1^2 + \dot \vf_2^2 + \dot \vf_1\dot \vf_2 \right] \;\; \text{and} \;\;
 \V = g \left[ 3 - \cos \vf_1  - \cos \vf_2  - \cos( \vf_1 +  \vf_2) \right],  
 	\label{e:lagrangian-phi-coords-3rotors}
 	\eeq
with the equations of motion (EOM) $3 m r^2 \ddot \vf_0 = 0$, 
	\beq
	m r^2 \left(2 \ddot \vf_1 + \ddot \vf_2 \right) = - 3 g \left[ \sin \vf_1 + \sin ( \vf_1 + \vf_2) \right] \;\; \text{and} \;\; 1 \leftrightarrow 2. 
	\label{e:3rotors-cm-rel-coords-lagrangian-eom-2nd-order}
	\eeq
The evolution equations for $\vf_1$ (and $\vf_2$ with $1 \leftrightarrow 2$) may be rewritten as
	\beq
	 m r^2 \ddot \vf_1 =  - g \left[ 2 \sin \vf_1 - \sin \vf_2 + \sin(\vf_1 + \vf_2)\right].	
	\label{e:3rotors-EOM-ph1ph2}
	\eeq
Notice that when written this way, the `force' on the RHS isn't the gradient of any potential, as the equality of mixed partials would be violated. The (angular) momenta conjugate to $\vf_{0,1,2}$ are $p_0 = 3 m r^2 \dot \vf_0$,
	\beq
	p_1 = \fr{m r^2}{3} (2 \dot \vf_1 + \dot \vf_2) \;\;\; \text{and}
	\;\;\; p_2 = \fr{m r^2}{3} ( \dot \vf_1 + 2 \dot \vf_2).
	\eeq
The remaining three EOM on phase space are $\dot p_0 = 0$ (conserved due to rotation invariance),
	\beq
	\dot p_1 = - g \left[ \sin \vf_1 + \sin (\vf_1 + \vf_2) \right] \quad \text{and} \quad
	\dot p_2 = - g \left[ \sin \vf_2 + \sin (\vf_1 + \vf_2) \right].
	\label{e:3rotors-EOM-p0p1p2}
	\eeq
The EOM admit a conserved energy which is a sum of CM, relative kinetic and potential energies:
	\beq
	E =   \fr{3}{2} m r^2 \dot \vf_0^2 + \fr{1}{3} m r^2 \left[  \dot \vf_1^2 + \dot \vf_2^2 + \dot \vf_1\dot \vf_2 \right] + \V(\vf_1, \vf_2).
 	\label{e:egy-3rotors-phi1-phi2-coords}
	\eeq
The above EOM are Hamilton's equations $\dot f = \{f, H\}$ for canonical Poisson brackets (PBs) $\{ \vf_i, p_j \} = \del_{ij}$ with the Hamiltonian
	\beq
	H = \fr{p_0^2}{6m r^2} + \fr{p_1^2 + p_2^2 - p_1 p_2}{m r^2} + \V(\vf_1, \vf_2).
	\eeq

\subsection{Analogue of Jacobi coordinates}
\label{s:jacobi-coordinates}

We define the Jacobi coordinates for the three-rotor problem to be $\vf_0$ (\ref{e:phi0-cm-relative-coords}) and
	\beq
	\vf_+ = (\vf_1 + \vf_2)/2 = (\tht_1-\tht_3)/2 \quad \text{and} \quad \vf_- = (\vf_1 - \vf_2)/2 = (\tht_1+\tht_3)/2 - \tht_2.
	\eeq
Up to a change in order, these are analogous to the Jacobi vectors of the three-body problem (see Fig.~\ref{f:jacobi-vectors}): $\vf_0$ is the center of mass of the three rotors, $2\vf_+$ is the angle of the first rotor relative to the third and $-\vf_-$ is the angle of the second rotor with respect to the center of mass of the first and the third rotors. Unlike in the CM and relative coordinates and as in the three-body problem, the kinetic energy as a quadratic form in velocities is diagonal. Indeed, $L = \T - \V$ where
	\beq
	\T = \fr{3}{2} m r^2 \dot \vf_0^2 + m r^2 \dot \vf_+^2  + \fr{1}{3} m r^2 \dot \vf_-^2 \quad \text{and} \quad
	\V = g \left(3 - 2 \cos \vf_- \cos \vf_+ - \cos 2\vf_+  \right).
	\eeq
The conjugate momenta $p_0$ and $p_\pm =  p_1 \pm p_2$ are proportional to the velocities and the EOM are
	\beqs
	\dot p_0 = 0, \quad \dot p_+ = -2g \sin \vf_+ \left( \cos \vf_- + 2 \cos \vf_+ \right) \quad
	 \text{and} \quad
	\dot p_- = - 2g \cos \vf_+ \sin \vf_-.
	\eeqs
The fundamental domain which was the cube $0 \leq \vf_{0,1,2} \leq 2\pi$ now becomes the cuboid ($0 \leq \vf_0 \leq 2\pi$, $0 \leq \vf_+ \leq 2\pi$, $0 \leq \vf_- \leq \pi$). As before, though $\vf_\pm$ are periodic coordinates on a 2-torus, $\vf_{0,\pm}$ are not quite periodic coordinates on $T^3$. The transformation of the CM variable $\vf_0$ under $2\pi$-shifts of $\vf_{1,2}$ discussed above may be reformulated as follows. When crossing the segments $\vf_+ + \vf_- = 2\pi$ from left to right or $\vf_+ - \vf_- = 0$ from right to left, $\vf_0$ increases by $2\pi/3$ [and $\vf_0 \mapsto \vf_0 - 2\pi/3$ when the segments are crossed in the opposite direction].

\section[Dynamics on the $\vf_1$-$\vf_2$ torus]{Dynamics on the \texorpdfstring{$\vf_1$-$\vf_2$}{phi1-phi2} torus \sectionmark{Dynamics on the $\vf_1$--$\vf_2$ torus}}
\sectionmark{Dynamics on the $\vf_1$--$\vf_2$ torus}
\label{s:dynamics-on-2torus}

The dynamics of $\vf_{1}$ and $\vf_2$  (or equivalently that of $\vf_\pm$) decouples from that of the CM coordinate $\vf_0$. The former may be regarded as periodic coordinates on the 2-torus $[0,2\pi] \times [0,2\pi]$. On the other hand, $\vf_0$, which may be regarded as a fibre coordinate over the $\vf_{1,2}$ base torus, evolves according to 
	\beq
	\vf_0 = \fr{p_0 t}{3mr^2} + \vf_0(0) + \fr{2 \pi}{3} (n_2 - n_1) \quad \text{mod}\; 2\pi.
	\eeq
Here, $n_{1,2}$ are the `greatest integer winding numbers' of the trajectory around the cycles of the base torus. If a trajectory goes continuously from $\vf^i_{1,2}$ to $\vf^f_{1,2}$ (regarded as real rather than modulo $2\pi$), then the greatest integer winding numbers are defined as $n_{1,2} = [(\vf^f_{1,2}-\vf^i_{1,2})/2\pi]$. 

Consequently, we may restrict attention to the dynamics of $\vf_1$ and $\vf_2$. The equations of motion on the corresponding 4d phase space (the cotangent bundle of the 2-torus) are
	\beq
	\dot \vf_{1} = (2 p_{1} - p_{2})/m r^2, \;\;\;
	\dot p_1 = - g \left[ \sin \vf_1 + \sin ( \vf_1 + \vf_2) \right]
	\label{e:3rotors-CM-EOM-phasespace}
	\eeq
and $1\leftrightarrow 2$. These equations define a singularity-free vector field on the phase space. They follow from the canonical PBs with Hamiltonian given by the relative energy
	\beq
	H_\text{rel} = \fr{p_1^2 + p_2^2 - p_1 p_2}{m r^2} + \V(\vf_1, \vf_2).
	\label{e:hamiltonian-3rotor-phi1phi2space-full}
	\eeq
These equations and Hamiltonian are reminiscent of those of the planar double pendulum with the Hamiltonian
	\beq
	H_{\rm dp} = \fr{p_1^2 - 2 c_{12}\, p_1\, p_2 + 2 p_2^2}{2 m l^2 (2 -c_{12}^2)} - m g l ( 2 \cos \tht_1 + \cos \tht_2)
	\eeq
where $\tht_{1,2}$ are the angles between the upper and the lower rods (each of length $l$) and the vertical and $c_{12} = \cos (\tht_1-\tht_2)$.

\subsection{Static solutions and their stability}
\label{s:static-sol-and-stability}

\begin{figure}
	\centering
	\begin{subfigure}[t]{6.5cm}
		\centering
		\includegraphics[width=6.5cm]{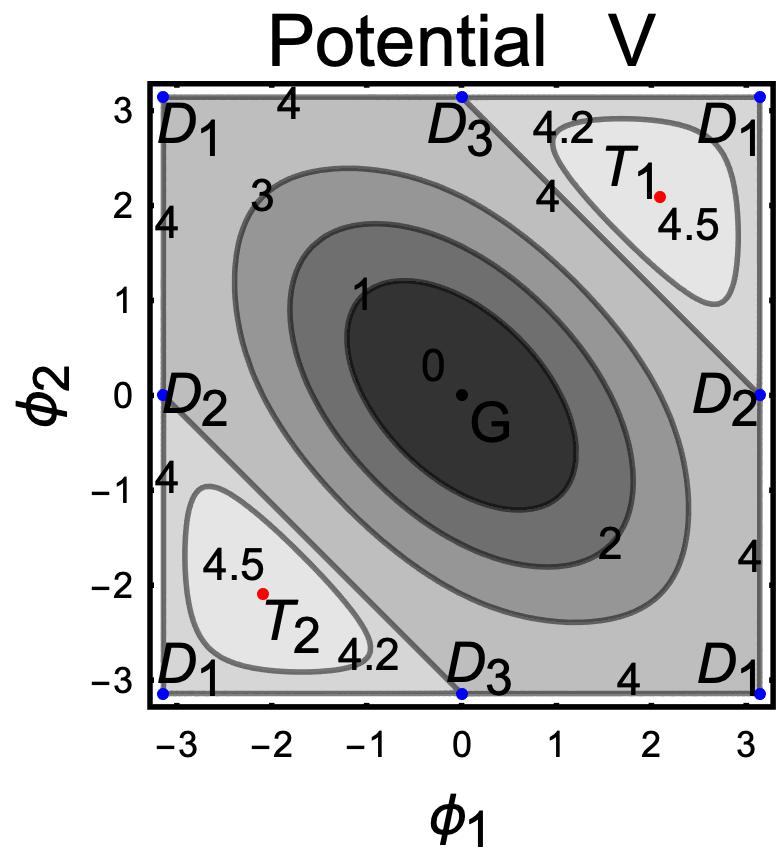}
		\caption{\small Contours of $\V$.}
		\label{f:potential-contour-plot}
	\end{subfigure}
	\qquad
	\begin{minipage}[b]{0.45\textwidth}
	\begin{subfigure}[t]{3.6cm}
		\centering
		\includegraphics[width=3cm]{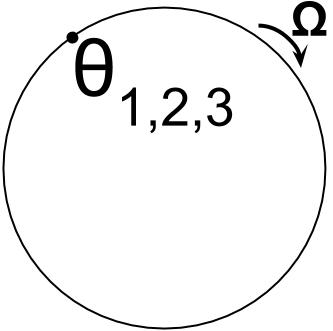}
		\caption{\small  Ground state G.}
		\label{f:3rotors-ground-state}	
	\end{subfigure}	
	\begin{subfigure}[t]{3.6cm}
		\centering
		\includegraphics[width=3cm]{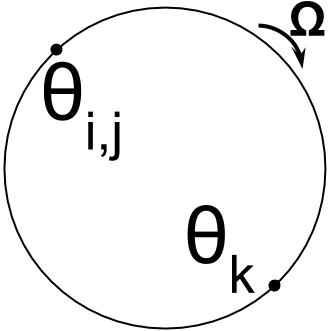}
		\caption{\small Diagonal states D.}
		\label{f:3rotors-first-excited}	
	\end{subfigure}	
	\begin{subfigure}[t]{3.6cm}
		\centering
		\includegraphics[width=3cm]{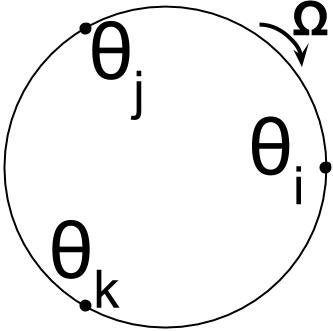}
		\caption{\small Triangle states T.}
		\label{f:3rotors-second-excited}
	\end{subfigure}
	\end{minipage}	
	\caption{\small (a) Potential energy $\V$ in units of $g$ on the $\vf_1$-$\vf_2$ configuration torus with its extrema (locations of static solutions G, D and T) indicated. The contours also encode changes in topology of the Hill region ($\V \leq E$) when $E$ crosses $E_{\rm G} = 0$, $E_{\rm D} = 4g$ and $E_{\rm T} = 4.5g$. (b, c, d) Uniformly rotating three-rotor solutions obtained from G, D and T. Here, $i,j$ and $k$ denote any permutation of the numerals $1$, $2$ and $3$. (b) and (d) are the simplest examples of choreographies discussed in \S \ref{s:choreographies}.}
	\label{f:static-solutions-3rotors}
\end{figure}

Static solutions for the relative motion correspond to zeros of the vector field where the force components in (\ref{e:3rotors-CM-EOM-phasespace}) vanish: $p_1 = p_2 = 0$ and
	\beq
	\sin \vf_1 + \sin (\vf_1+\vf_2) = \sin \vf_2 + \sin (\vf_1 + \vf_2) = 0.
	\eeq 
In particular, we must have $\vf_1 = \vf_2$ or $\vf_1 = \pi - \vf_2$. When $\vf_1 = \vf_2$, the force components are both equal to $\sin \vf_1 (1 + 2\cos \vf_1)$ which vanishes at the following configurations:
	\beq
	(\vf_1, \vf_2)\; = \;(0,0), \; \left(\pi, \pi\right) \; \text{and} \; \left(\pm {2\pi}/{3}, \pm {2\pi}/{3}\right).
	\eeq
On the other hand, if $\vf_1 = \pi - \vf_2$, we must have $\sin \vf_1 = 0$  leading to two more static configurations $(0, \pi)$ and $(\pi, 0)$. Thus we have six static solutions which we list in increasing order of (relative) energy:
	\beqs
 E=0 :&& G(0,0), \cr
 E=4g :&& D_1(\pi, \pi), D_2(\pi, 0), D_3(0, \pi) \cr
 \text{and} \quad E = 9g/2 :&& T_{1,2}(\pm {2\pi}/{3},\pm{2\pi}/{3}). 
	\eeqs
Below, we clarify their physical meaning by viewing them as uniformly rotating three body configurations.

\subsubsection{Uniformly rotating three-rotor solutions from G, D and T} 
\label{s:rotating-static-sol}

If we include the uniform rotation of the CM angle ($\dot \vf_0 = \Om$ is arbitrary), these six solutions correspond to the following uniformly rotating rigid configurations of three-rotors (see Fig.~\ref{f:static-solutions-3rotors}):  (a) the ferromagnetic ground state G where the three particles coalesce ($\tht_1 = \tht_2 = \tht_3$), (b) the three `diagonal' `anti ferromagnetic N\'eel'  states D where two particles coincide and the third is diametrically opposite ($\tht_1 = \tht_2 = \tht_3 + \pi$ and cyclic permutations thereof) and (c) the two `triangle' `spin wave' states T where the three bodies are equally separated ($\tht_1 = \tht_2 + 2\pi/3 = \tht_3 + 4\pi/3$ and $\tht_2 \leftrightarrow \tht_3$). 

\subsubsection{Stability of static solutions}

 The linearization of the EOM (\ref{e:3rotors-EOM-ph1ph2}) for perturbations to G, D and T ($\vf_{1,2} = \bar \vf_{1,2} + \delta\vf_{1,2}(t)$) take the form
	\beqs
	 &mr^2 \fr{d^2}{dt^2} \colvec{2}{\delta\vf_1} {\delta\vf_2} = -g A \colvec{2}{\delta\vf_1} {\delta\vf_2}
	\quad \text{where} \quad
	A_{\text{G}} = 3 I, \cr 
	&A_{\text{D$_3$}(0,\pi)}=\colvec{2}{1 & 0}{-2 & -3}, \quad
	 A_{\text{D$_2$}(\pi,0)} = \colvec{2}{-3 & -2}{0 & 1}, \cr
	& A_{\text{D$_1$}(\pi, \pi)}=\colvec{2}{-1 & 2}{2 & -1}\quad \text{and}
	\quad
	A_{\rm T}= -3I/2.
	\label{e:EOM-ph1ph2-perturbation-static-solution}
	\eeqs
Here $I$ is the $2 \times 2$ identity matrix. Perturbations to G are stable and lead to small oscillations with equal frequencies $\om_0 = \sqrt{3g/mr^2}$. The saddles D have one stable direction with frequency $\om_0/\sqrt{3}$ and one unstable eigendirection with growth rate $\om_0$. On the other hand, both eigendirections around T are unstable with growth rate $\om_0/\sqrt{2}$.

\subsection{Changes in topology of Hill region with growing energy}
\label{s:topology-hill-region}

\begin{figure}
	\centering
	\begin{subfigure}[c]{10cm}
		\centering
		\includegraphics[width=10cm]{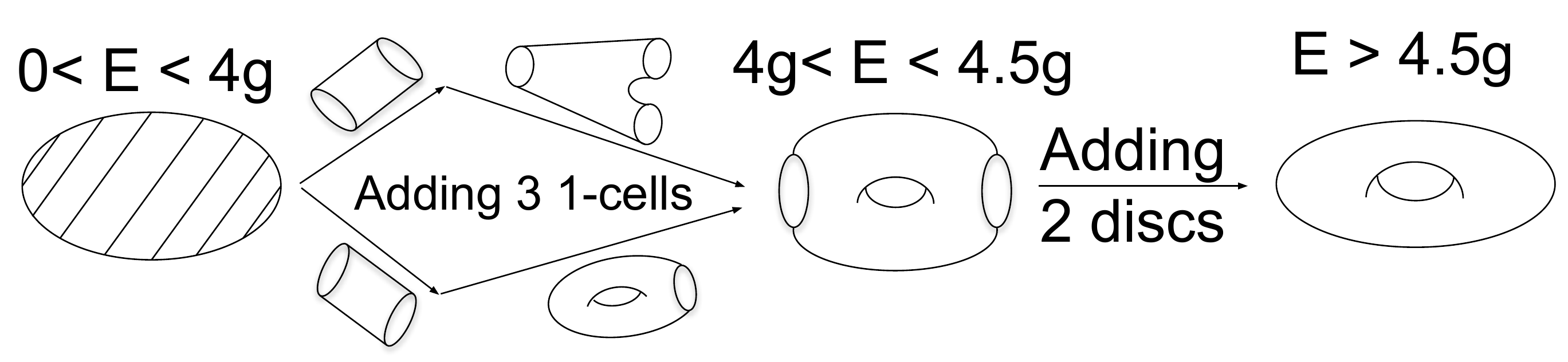}
		\caption{\small  }
		\label{f:topology-hill-region-a}
	\end{subfigure}	
	\quad 
	\begin{subfigure}[c]{5cm}
		\centering
		\includegraphics[width=5cm]{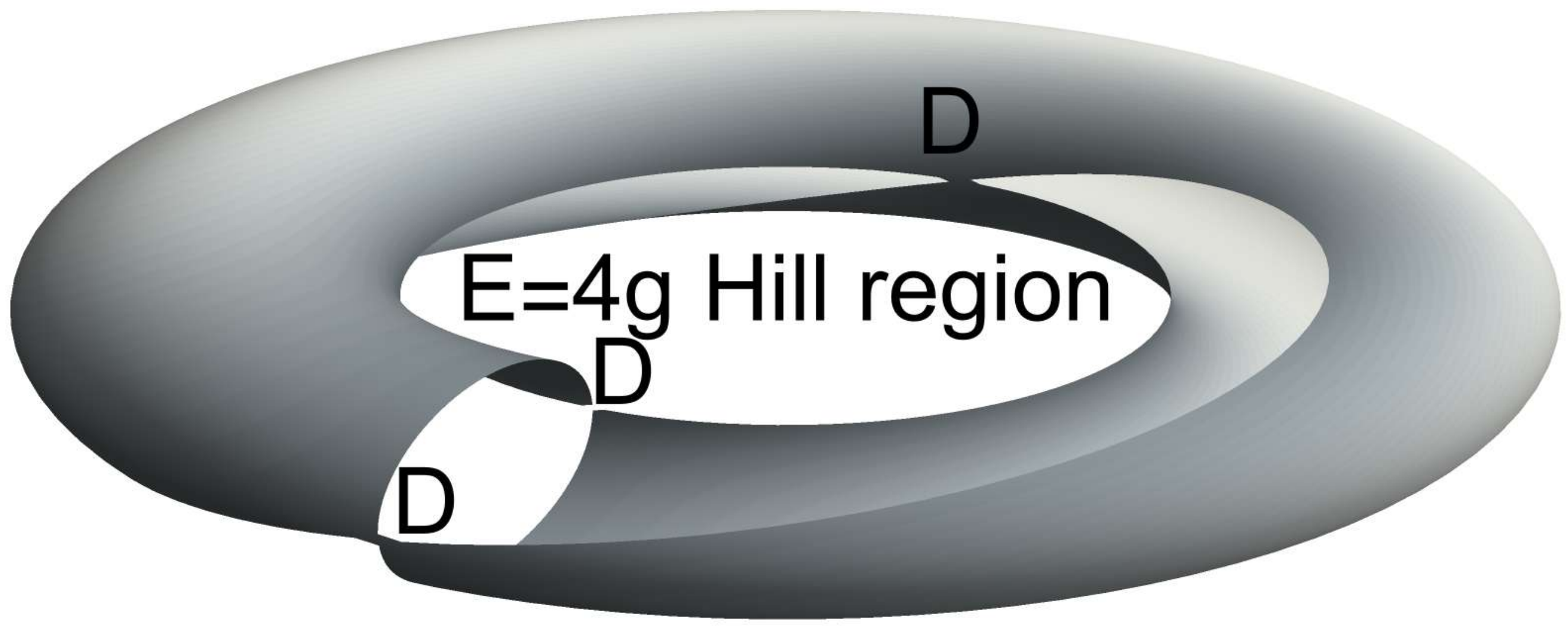}
		\caption{\small  }
		\label{f:topology-hill-region-b}
	\end{subfigure}	
	\caption{\small  (a) Topology of Hill region of configuration space $(\V(\vf_1, \vf_2) \leq E)$ showing transitions at $E = 4g$ and $4.5g$ as implied by Morse theory (see \S \ref{s:dynamics-on-2torus}). (b) The Hill region for $E = 4g$ is not quite a manifold; its boundary consists of 3 non-contractible closed curves on the torus meeting at the D configurations.} 
	\label{f:topology-hill-region}
\end{figure}

The Hill region of possible motions ${\cal H}_E$ at energy $E$ is the subset $\V(\vf_1, \vf_2) \leq E$ of the $\vf_1$-$\vf_2$ configuration torus. The topology of the Hill region for various energies can be read-off from Fig.~\ref{f:potential-contour-plot}. For instance, for $0 < E < 4g$, ${\cal H}_E$ is a disc while it is the whole torus for $E > 4.5 g$. For $4g < E < 4.5g$, it has the topology of a torus with a pair of discs (around T$_1$ and T$_2$) excised (see also Fig. \ref{f:curvature-3rotors-phi1-phi2-torus}). These changes in topology are confirmed by Morse theory \cite{milnor} if we treat $\V$ as a real-valued Morse function, since its critical points are non-degenerate (non-singular Hessian). In fact, the critical points of $\V$ are located at G (minimum with index 0), D$_{1,2,3}$ (saddles with indices 1) and T$_{1,2}$ (maxima with indices 2). Thus, the topology of ${\cal H}_E$ can change only at the critical values $E_G = 0, E_D = 4g$ and $E_T = 4.5g$ (see Fig. \ref{f:topology-hill-region-a}). The topological transition from ${\cal H}_{E < 4g}$ (disc) to ${\cal H}_{4g < E < 4.5g}$ (torus with two discs excised) can be achieved by the successive addition of three 1-cells to the disc (proceeding either via a cylinder and a pair of pants or a cylinder and a torus with one disc excised). Similarly, one arrives at the toroidal Hill region for $E > 4.5 g$ by sewing two 2-cells to cover the excised discs of ${\cal H}_{4g < E < 4.5g}$ as depicted in  Fig. \ref{f:topology-hill-region-a}. At the critical value $E = 0$, ${\cal H}_E$ shrinks to a point while at $E = 4.5 g$, it is a twice-punctured torus. Fig.~\ref{f:topology-hill-region-b} illustrates the Hill region at the critical value $E = 4g$ where alone it is not quite a manifold.


\subsection{Low and high energy limits}
\label{s:low-and-high-egy-limit}

In the CM frame, the three-rotor problem (\ref{e:3rotors-CM-EOM-phasespace}) has a 4-dimensional phase space but possesses only one known conserved quantity (\ref{e:hamiltonian-3rotor-phi1phi2space-full}). However, an extra conserved quantity emerges at zero and infinitely high energies:

(a) For $E \gg g$, the  kinetic energy dominates and $H \approx (p_1^2 - p_1 p_2 + p_2^2)/mr^2$. Here $\vf_{1,2}$ become cyclic coordinates and  $p_{1,2}$ are both approximately conserved.

(b) For $E \ll g$, the system executes small oscillations around the ground state G $(\vf_{1,2} \equiv 0)$. The quadratic approximation to the Lagrangian (\ref{e:lagrangian-phi-coords-3rotors}) for relative motion is
	\beq
	L_{\rm low} = \fr{mr^2}{3} \left[\dot \vf_1^2 + \dot \vf_2^2 + \dot \vf_1 \dot \vf_2 \right] - g \left(\vf_1^2 + \vf_2^2 + 
\vf_1 \vf_2 \right).
	\label{e:Lagrangian-low-egy}
	\eeq
The linear equations of motion for $\vf_1$ and $\vf_2$ decouple,
	\beq
	m r^2 \ddot \vf_{1} = - 3 g \vf_{1} \quad \text{and} \quad m r^2 \ddot \vf_{2} = - 3 g \vf_{2}
	\label{e:decoupled-sho-eqn-3rotors-low-egy-limit}
	\eeq
leading to the separately conserved normal mode energies $E_{1,2} = \left( m r^2 \dot \vf_{1,2}^2 + 3 g \vf_{1,2}^2 \right)/2$. The equality of frequencies implies that any pair of independent linear combinations of $\vf_1$ and $\vf_2$ are also normal modes. Of particular significance are the Jacobi-like variables $\vf_\pm = (\vf_1 \pm \vf_2)/2$ that diagonalize the kinetic and potential energy quadratic forms:
	\beq
	L_{\rm low} = m r^2 \dot \vf_+^2 - 3g \vf_+^2 + m r^2 \dot \vf_-^2/3 - g \vf_-^2.
	\eeq
Though (\ref{e:decoupled-sho-eqn-3rotors-low-egy-limit}) are simply the EOM for a pair of decoupled oscillators, the Lagrangian and Poisson brackets $\{ \cdot , \cdot \}$ inherited from the non-linear theory are different from the standard ones. With conjugate momenta $p_{1,2} = (mr^2/3)  (2 \dot \vf_{1,2} + \dot \vf_{2,1})$, the Hamiltonian corresponding to (\ref{e:Lagrangian-low-egy}) is
	\beq
	H_{\rm low} = \frac{p_1^2 - p_1 p_2 + p_2^2}{ mr^2} +  g \left(\vf_1^2 + \vf_2^2 + \vf_1 \vf_2 \right).
	\label{e:low-egy-hamiltonian-phi1-phi2-3rotors}
	\eeq
Note that $p_{1,2}$ differ from the standard momenta $p_{1,2}^{\rm s} = m r^2 \dot \vf_{1,2}$ whose PBs are now non-canonical,  $\{\vf_i, p_j^{\rm s}\} = -1 + 3 \delta_{ij}$.

\subsubsection{Three low-energy constants of motion}

$H_{\rm low}$ and the normal mode energies
	\beq
	H_{1,2} = \left( 2 p_{1,2} - p_{2,1} \right)^2/2 m r^2 + {3g}\vf_{1,2}^2/2
	\label{e:low-egy-normal-modes-3rotors-phi1-phi2}
	\eeq
are three independent constants of motion in the sense that the corresponding 1-forms $dH$, $dH_1$ and $dH_2$ are generically linearly independent ($dH \wedge dH_1 \wedge dH_2 \not \equiv 0$ on the 4d phase space). On the other hand, we also have a conserved `angular momentum' $L_z = m r^2 (\vf_1 \dot \vf_2 - \vf_2 \dot \vf_1)$ corresponding to the rotation invariance of the decoupled oscillators in (\ref{e:decoupled-sho-eqn-3rotors-low-egy-limit}). It turns out that $H_{\rm low}$ may be expressed as
	\beq
	H_{\rm low} = \fr{2}{3}\left[ H_1 + H_2 +\sqrt{ H_1 H_2 - (3 g/4 m r^2) L_z^2} \right].
	\eeq
The low energy phase trajectories lie on the common level curves of $H_{\rm low}$, $H_1$ and $H_2$. Though $H_1$ and $H_2$ are conserved energies of the normal modes, they do not Poisson commute. In fact, the Poisson algebra of conserved quantities is $\{H_{1,2}, H_{\rm low}\} = \{L_z, H_{\rm low} \} = 0$,
	\beqs
	\{H_1, H_2\} = -3 g L_z/m r^2 \quad \text{and} \quad 
	 \{L_z, H_{1,2}\} = \pm 2(3H_{\rm low} - 2H_{1,2} - H_{2,1}).
	\eeqs
It is also noteworthy that the integrals $H_1 + H_2$ and $H_1 H_2 - 3g L_z^2/(4mr^2)$ are in involution.

\section{Reductions to one degree of freedom} 
\label{s:reduction-one-dof}

\begin{figure}	
	\centering
	\begin{subfigure}[t]{5cm}
		\centering
		\includegraphics[width=5cm]{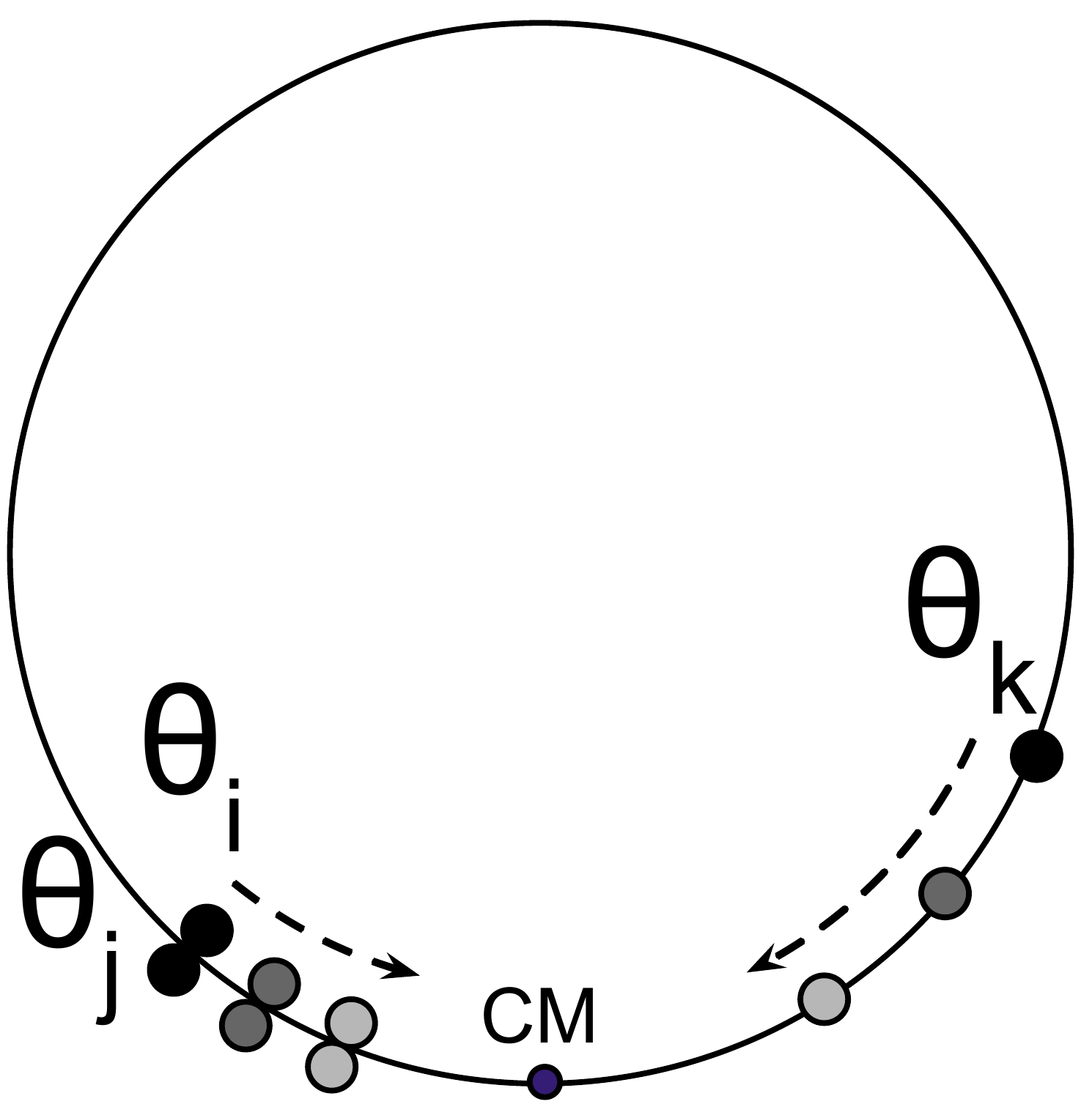}
	\caption{\small  Pendula}
	\label{f:pendula}
	\end{subfigure}
\qquad
	\begin{subfigure}[t]{5cm}
		\centering
		\includegraphics[width=5cm]{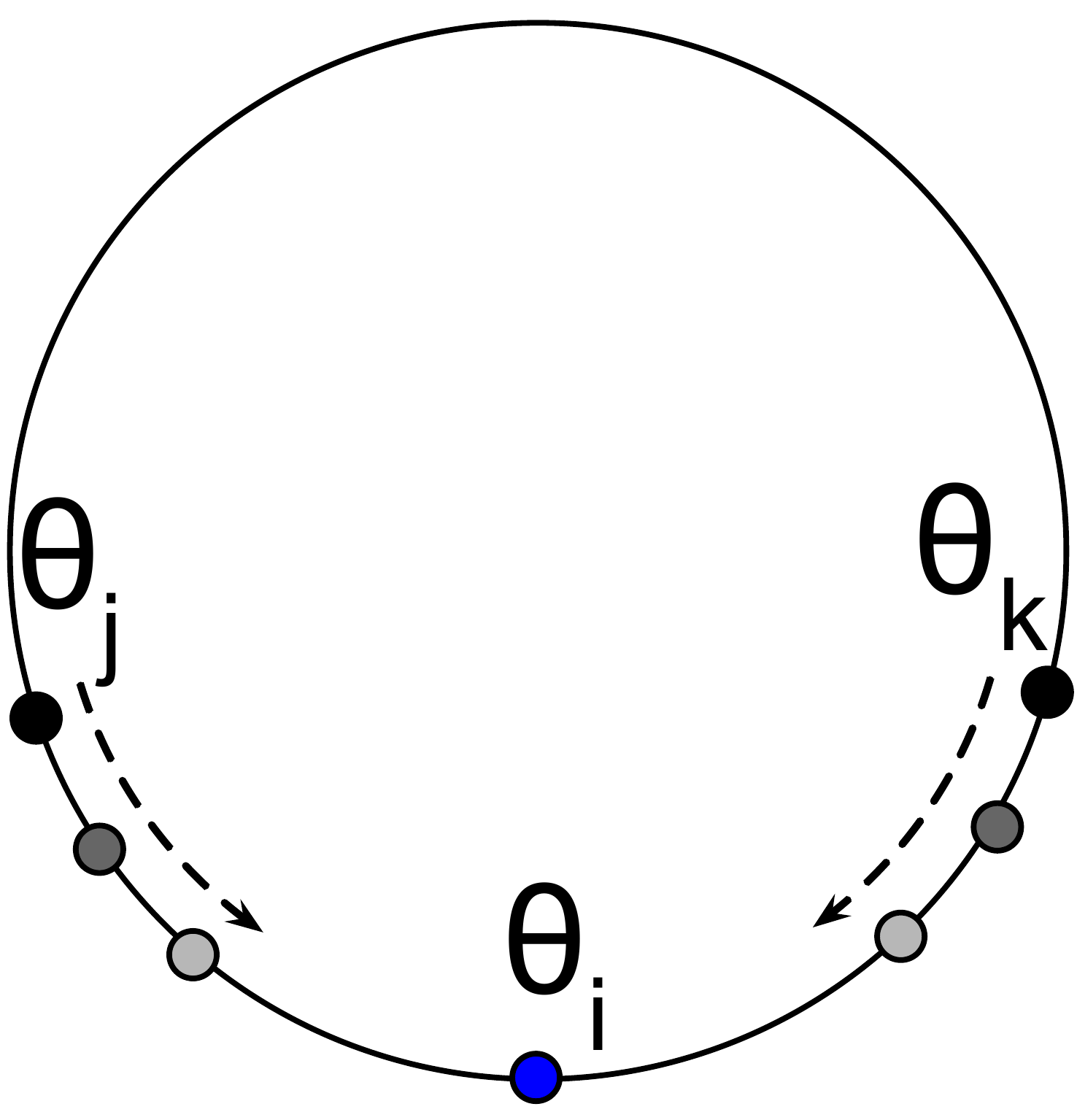}
	\caption{\small  Isosceles `breathers'}
	\label{f:breathers}
	\end{subfigure}	
	\caption{\small   In pendula, $\tht_i$ and $\tht_j$ form a molecule that along with $\tht_k$ oscillates about their common CM. In breathers, $\tht_i$ is at rest at the CM with $\tht_j$ and $\tht_k$ oscillating symmetrically about the CM. Here, $i,j$ and $k$ denote any permutation of the numerals $1$, $2$ and $3$.}
	\label{f:periodic-soln}
\end{figure}

Recall that the Euler and Lagrange solutions of the planar three-body problem arise through a reduction to the two body Kepler problem. We find an analogue of this construction for three rotors, where pendulum-like systems play the role of the Kepler problem. We find two such families of periodic orbits, the pendula and isosceles breathers (see Fig. \ref{f:periodic-soln}). They exist at all energies and go from librational to rotational motion as $E$ increases. They turn out to have remarkable stability properties which we deduce via their monodromy matrices.

\subsection{Periodic pendulum solutions} 
\label{s:pendulum-soln}

We seek solutions where one pair of rotors form a `bound state' with their angular separation remaining constant in time. We show that consistency requires this separation to vanish, so that the two behave like a single rotor and the equations reduce to that of a two-rotor problem. There are three such families of `pendulum' solutions depending on which pair is bound together (see Fig. \ref{f:pendula}). For definiteness, we suppose that the first two particles have a fixed separation $\zeta$ ($\tht_1 = \tht_2 + \zeta$ or $\vf_1 = \zeta$). Putting this in (\ref{e:3rotors-EOM-ph1ph2}), we get a consistency condition and an evolution equation for $\vf_2$:
	\beqs
	 2 \sin \zeta - \sin \vf_2 + \sin(\zeta + \vf_2) = 0 \quad  \text{and} \quad
	 m r^2 \ddot \vf_2 = - g \left[ 2 \sin \vf_2 - \sin \zeta + \sin(\zeta + \vf_2)\right].
	\eeqs
The consistency condition is satisfied only when the separation $\zeta = 0$, i.e., rotors 1 and 2 must coincide so that $\vf_1 = 0$ and $\dot \vf_1 = 0$ (or $p_2 = 2 p_1$) at all times (the other two families are defined by $\vf_2 = \dot \vf_2 = 0$ and $\vf_1+\vf_2 = \dot\vf_1 + \dot \vf_2 = 0$). The evolution equation for $\vf_2$ reduces to that for a simple pendulum: 
	\beq
	m r^2 \ddot \vf_2 =  - 3 g \sin \vf_2 \;\; \text{with} \;\; E = \fr{m r^2 \dot \vf_2^2}{3}  + 2g (1 - \cos \vf_2)
	\eeq
being the conserved energy. The periodic solutions are either librational (for $0 \leq E < 4g$) or rotational (for $E>4g$) and may be expressed in terms of the Jacobi elliptic function sn:  
	\beq
	\bar \vf_2(t) =
	\begin{cases}
	2 \arcsin (k \,{\rm sn} (\om_0 t, k))
	& \text{for} \quad 0 \leq E \leq 4g, \\
	 2 \arcsin ({\rm sn}(\om_0 t/\kappa, \kappa))	& \text{for} \quad E \geq 4g. 	\end{cases}
	\label{e:pendulum-solutions}
	\eeq
Here, $\om_0 = \sqrt{3g/mr^2}$ and the elliptic modulus $k=\sqrt{E/4g}$ with $\kappa = 1/k$. Thus $0 \leq k < 1$ for libration and $0 \leq \kappa < 1$ for rotation. The corresponding periods are $ \tau_{\rm lib} = 4 K(k)/\om_0$ and $\tau_{\rm rot} = 2\kappa K(\kappa)/\om_0$, where $K$ is the complete elliptic integral of the first kind. As $E \to 4g^\pm$, the period diverges and we have the separatrix $\bar \vf_2(t) =  2 \arcsin (\tanh (\om_0 t))$. The conditions $\vf_1 = 0$ and $p_2 = 2p_1$ define a 2d `pendulum submanifold' of the 4d phase space foliated by the above pendulum orbits. Upon including the CM motion of $\vf_0$, each of these periodic solutions may be promoted to a quasi-periodic orbit of the three-rotor problem. There is a two-parameter family of such orbits, labelled for instance, by the relative energy $E$ and the CM angular momentum $p_0$.

\subsubsection{Stability of pendulum solutions via monodromy matrix}
\label{s:pendulum-stability}

Introducing the dimensionless variables 
	\beq
	\tilde p_{1,2} = p_{1,2}/\sqrt{m r^2 g} \quad \text{and} \quad \tilde t =  t\sqrt{g/mr^2},
	\eeq 
the equations for small perturbations 
	\beq
	\vf_1 = \delta \vf_1, \;\; \vf_2 = \bar \vf_2 + \delta \vf_2 \;\; \text{and} \;\; p_{1 ,2} = \bar p_{1,2} + \delta p_{1,2}
	\eeq
to the above pendulum solutions (\ref{e:pendulum-solutions}) to (\ref{e:3rotors-CM-EOM-phasespace}) are
	\beq
	\fr{d^2}{d\tilde t^2} \colvec{2}{\delta \vf_1}{\delta \vf_2} = - \colvec{2}{2 + \cos \bar \vf_2 & 0}{\cos \bar \vf_2 - 1 & 3 \cos \bar \vf_2} \colvec{2}{\delta  \vf_1}{\delta \vf_2}.
	\eeq
This is a pair of coupled Lam\'e-type equations since $\bar \vf_2$ is an elliptic function. The analogous equation in the 2d anharmonic oscillator reduces to a single Lam\'e equation\cite{yoshida84, brack-mehta-tanaka}. Our case is a bit more involved and we will resort to a numerical approach here. To do so, it is convenient to consider the first order formulation
	\beq
	\fr{d}{d\tilde t}\colvec{4}{\delta  \vf_1}{\delta \vf_2}{\delta {\tilde{p}}_1}{\delta {\tilde{p}}_2} = -\colvec{4}{0 & 0 & -2 & 1}{0 & 0 & 1 & -2}{1 + \cos \bar \vf_2 & \cos \bar \vf_2 & 0 & 0}{\cos \bar \vf_2 & 2 \cos \bar \vf_2 & 0 & 0} \colvec{4}{\delta \vf_1}{\delta \vf_2}{\delta \tilde p_1}{\delta \tilde p_2}.
	\label{e:coeff-matrix}	
	\eeq
Since $m, g$ and $r$ have been scaled out, there is no loss of generality in working in units where $m = g = r = 1$, as we do in the rest of this section. The solution is $\psi(t) = U(t,0) \; \psi(0)$ where the real time-evolution matrix is given by a time-ordered exponential $U(t,0) = T\exp\{ \int_0^{t} A(t)\, dt\}$ where $A(t)$ is the coefficient matrix in (\ref{e:coeff-matrix}) and $T$ denotes time ordering. The tracelessness of $A(t)$ implies $\det U(t,0) = 1$ and preservation of phase volume. Though $A(t)$ is $\tau$-periodic, $\psi(t + \tau) = M(\tau) \psi(t)$ where the monodromy matrix $M(\tau) = U(t+\tau,t)$ is independent of $t$. Thus, $\psi(t + n \tau) = M^n \psi(t)$ for $n = 1,2,\ldots$, so that the long-time behavior of the perturbed solution may be determined by studying the spectrum of $M$. In fact, the eigenvalues $\la$ of $M$ may be related to the Lyapunov exponents associated to the pendulum solutions 
	\beq
	\mu = \lim_{t \to \infty} \fr{1}{t} \ln \fr{|\psi(t)|}{|\psi(0)|} \quad \text{via} \quad \mu = \frac{\log |\la|}{\tau}.
	\eeq
Since ours is a Hamiltonian system with 2 degrees of freedom, two of the eigenvalues of $M$ must equal one and the other two must be reciprocals \cite{hadjime}. On account of the reality of $M$, the latter two $(\la_3, \la_4)$ must be of the form $(e^{i \phi}, e^{- i \phi})$ or $(\la, 1/\la)$ where $\phi$ and $\la$ are real. It follows that two of the Lyapunov exponents must vanish while the other two must add up to zero. The stability of the pendulum orbit is governed by the stability index $\sigma = \tr M - 2$. We have stability if $|\sigma| \leq 2$ which corresponds to the eigenvalues $e^{\pm i \phi}$ and instability if $|\sigma| = |\la + 1/\la| > 2$.

\begin{figure}	
	\centering
	\begin{subfigure}[t]{7cm}
		\centering
		\includegraphics[width=7cm]{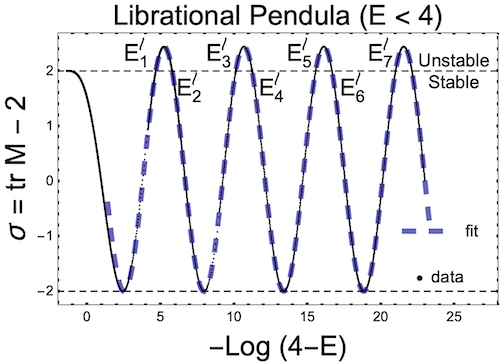}
	\end{subfigure}
\quad
	\begin{subfigure}[t]{7cm}
		\centering
		\includegraphics[width=7cm]{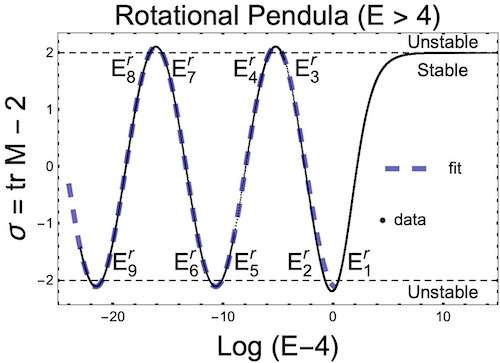}
	\end{subfigure}	
	\caption{\small   Numerically obtained stability index of pendulum solutions showing approach to periodic oscillations between stable and unstable phases as $E \to 4^\pm$. Equations (\ref{e:fit-trace-monodromy-lib}) and (\ref{e:fit-trace-monodromy-rot}) are seen to fit the data as $E \to 4^\pm$.}
	\label{f:monodromy-evals}
\end{figure}

We now discuss the energy dependence of the stability index for pendula. In the limit of zero energy, (\ref{e:pendulum-solutions}) reduces to the ground state G and $A(t)$ becomes time-independent and similar to $2\pi i \times {\rm diag}(1, 1, -1, -1)$. Consequently, $M = \exp (A \tau)$ is the $4 \times 4$ identity $I$. Thus G is stable and small perturbations around it are periodic with period $\tau = 2\pi/\om_0$, as we know from Eq. (\ref{e:EOM-ph1ph2-perturbation-static-solution}). For $E > 0$, we evaluate $M$ numerically. We find it more efficient to regard $M$ as the fundamental matrix solution to $\dot \psi = A(t) \psi$ rather than as a path ordered exponential or as a product of infinitesimal time-evolution matrices. Remarkably, as discussed below, we find that while the system is stable for low energies $0 \leq E \leq E^\ell_1 \approx 3.99$ and high energies $E \geq E^\err_1 \approx 5.60$, the neighborhood of $E = 4$ consists of a doubly infinite sequence of intervals where the behavior alternates between stable and unstable (see Fig. \ref{f:monodromy-evals}). This is similar to the infinite sequence of transition energies for certain periodic orbits of a class of Hamiltonians studied in \cite{churchill} and to the singly infinite sequence of transitions in the 2d anharmonic oscillator as the coupling $\al$ goes from zero to infinity \cite{yoshida84}:
	\beq 
	H_{\rm anharm} = \half \left( p_1^2 + p_2^2 \right) + \ov{4} \left( q_1^4 + q_2^4 \right) + \al \: q_1^2 q_2^2.
	\label{e:anharmonic-oscillator-yoshida}
	\eeq
This accumulation of stable-to-unstable transition energies at the threshold for librational `bound' trajectories is also reminiscent of the quantum mechanical energy spectrum of Efimov trimers that form a geometric sequence accumulating at the zero-energy threshold corresponding to arbitrarily weak two-body bound states with diverging S-wave scattering length \cite{efimov}.

\subsubsection{Stability of librational pendula \texorpdfstring{$(E < 4)$}{(E < 4)}}

In the first stable phase $0 \leq E \leq E^\ell_1$, $\phi = \arg \la_3$ monotonically increases from $0$ to $2\pi$ with growing energy and $\la_4 = e^{-i \phi}$ goes round the unit circle once clockwise. There is a stable to unstable phase transition at $E_1^\ell$. In the unstable phase $E_1^\ell < E  < E_2^\ell$, $\sig > 2$ corresponding to  real positive $\la_4$ increasing from $1$ to $1.9$ and then dropping to $1$ (see Fig. \ref{f:monodromy-evals}). There is then an unstable to stable transition at $E^\ell_2$. This pattern repeats so that the librational regime $0 < E < 4$ is divided into an infinite succession of progressively narrower stable and unstable phases. Remarkably, we find that the stable phases asymptotically have equal widths on a logarithmic energy scale just as the unstable ones do. Indeed, if we let $E_{2n +1}^\ell$ and $E_{2n}^\ell$ denote the energies of the stable to unstable and unstable to stable transitions for $n = 1,2,3, \ldots$, then the widths $w^{lu}_n$ and $w^{ls}_{n+1}$ of the $n^{\rm th}$ unstable and $n+1^{\rm st}$ stable phases are 
	\beqs 
	w^{lu}_n &=& E^\ell_{2n} - E^\ell_{2n-1} \approx (E^\ell_{2} - E^\ell_{1}) \times e^{- \Lambda \, (n-1)} \quad \text{and} \cr
	w^{ls}_{n+1} &=& E^\ell_{2n+1} - E^\ell_{2n} \approx (E^\ell_{3} - E^\ell_{2}) \times e^{- \Lambda \, (n-1)}.
	\eeqs 
Here, $E^\ell_{2} - E^\ell_{1} \approx e^{-4.67} (1 - e^{-1.11})$ and $E^\ell_{3} - E^\ell_{2} \approx e^{-5.78} (1 - e^{-4.34})$ are the lengths of the first unstable and second stable intervals while $\Lambda \approx 1.11 + 4.34 = 5.45$ is the combined period on a log scale. The first stable phase has a width $E^\ell_1 - 0 \approx 4 - e^{-4.67}$ that does not scale like the rest. Our numerically obtained stability index (see Fig. \ref{f:monodromy-evals}) is well approximated by
	\beq
	\sigma \approx
	2.22 \cos \left[ \fr{2}{\sqrt{3}}\log (4 - E) + .24 \right] + .22 \; \text{as} \; E \to 4^-.
	\label{e:fit-trace-monodromy-lib}
	\eeq
On the other hand, $\sig(E) \sim 2 - {\cal O} (E^3)$ when $E \to 0$.

\subsubsection{Stability of rotational pendula \texorpdfstring{$(E > 4)$}{(E > 4)}}

For sufficiently high energies $E \geq E_1^\err$, the rotational pendulum solutions are stable. In fact, as $E$ decreases from $\infty$ to $E^\err_1$, $\la_4 = e^{-i \phi}$ goes counterclockwise around the unit circle from 1 to $-1$. There is a stable to unstable transition at $E^\err_1$. As $E$ decreases from $E^\err_1$ to $E^\err_2$, $\la_4$ is real and negative, decreasing from $-1$ to $-1.5$ and then returning to $-1$ (see Fig. \ref{f:monodromy-evals}). This is followed by a stable phase for $E_2^\err \geq E \geq E_3^\err$ where $\la_4$ completes its passage counterclockwise around the unit circle reaching $1$ at $E_3^\err$. The last phase of this first cycle consists of an unstable phase between $E^\err_3$ and $E^\err_4$ where $\la_4$ is real and positive, increasing from $1$ to $1.4$ and then going down to $1$. The structure of this cycle is to be contrasted with those in the librational regime where $\la_4$ made complete revolutions around the unit circle in each stable phase and was always positive in unstable phases. This is reflected in the stability index overshooting both $2$ and $-2$ for rotational solutions but only exceeding 2 in the librational case. Furthermore, as in the librational case, there is an infinite sequence of alternating stable and unstable phases accumulating from above at $E = 4$, given by
	\beq
	\text{stable energies} = \left [ E^\err_1 ,\infty \right ) \bigcup_{n = 1}^\infty \left[ E^\err_{2n+1}, E^\err_{2n} \right ] 
	\text{ and unstable energies} = \bigcup_{n = 1}^\infty \left( E^\err_{2n}, E^\err_{2n-1} \right ).
	\eeq
As before, with the exception of the two stable and one unstable intervals of highest energy, the widths of the stable and unstable energy intervals are approximately constant on a log scale:
	\beqs
	w^{ru}_n &=&  E^\err_{2n-1} - E^\err_{2n} \approx (E^\err_{3} - E^\err_{4}) \times e^{- \Lambda \, (n-2)} \quad \text{and} \cr
	w^{rs}_{n+1} &=& E^\err_{2n} - E^\err_{2n+1} \approx  (E^\err_{4} - E^\err_{5}) \times e^{- \Lambda \, (n-2)}
	\eeqs
for $n = 2,3,4 \cdots$. Here, $E^\err_{3} - E^\err_{4} \approx e^{-4.7} (1 - e^{-1.1})$ and $E^\err_{4} - E^\err_{5} \approx e^{-5.8} (1 - e^{-4.3})$ are the lengths of the second unstable and third stable intervals while $\Lambda \approx 1.1+4.3 = 5.4$ is the combined period. The three highest energy phases are anomalous: (a) $E \geq E_1^\err \approx 5.60$ is a stable phase of infinite width, (b) the unstable phase $E_1^\err > E > E_2^\err \approx 4.48$ has width $1.2 > 1.1$ on a log scale and manifests more acute instability and (c) the stable phase $E_2^\err \geq E \geq E_3^\err \approx 4.01$ has a less than typical width $3.9 < 4.3$ (see Fig. \ref{f:monodromy-evals}). As before, we obtain the fit
	\beq
	\sigma \approx - 2.11 \cos\left[ \fr{1}{\sqrt{3}} \log (E - 4) - .12 \right] \;\;  \text{as} \;\; E \to 4^+
	\label{e:fit-trace-monodromy-rot}
	\eeq
while $\sig(E) \sim 2- {\cal O} (1/E)$ when $E \to \infty$.
\subsubsection{Energy dependence of eigenvectors}

 Since the pendulum solutions form a one parameter family of periodic orbits $(0, \vf_2, p_1, 2p_1)$ with continuously varying time periods, a perturbation tangent to this family takes a pendulum trajectory to a neighboring pendulum trajectory and is therefore neutrally stable. These perturbations span the 1-eigenspace $\span (v_1,v_2)$ of the monodromy matrix, where $v_1 = (0, 1, 0, 0) = \pdr_{\vf_2}$ and $v_2 = (0, 0, 1, 2) = \pdr_{p_1} + 2 \pdr_{p_2}$. The other two eigenvectors of $M$ have a simple dependence on energy and thus help in ordering the eigenvalues $\la_3$ and $\la_4$ away from transitions. In the `unstable' energy intervals
	\beq
	( E^\ell_1, E^\ell_2 ) \cup ( E^\err_2, E^\err_1 ) \cup ( E^\ell_3, E^\ell_4 ) \cup ( E^\err_4, E^\err_3 ) \cup \ldots,
	\eeq 
$M = \text{diag}(1,1,\la_3, 1/\la_3)$ in the basis $(v_1, v_2, v_+, v_-)$ where $v_\pm = (2a(E), -a(E), \pm b (E), 0)$. In the same basis, $M = \text{diag} (1, 1, R_\phi)$ in the complementary `stable' energy intervals $( 0, E^\ell_1 ) \cup ( E^\err_1, \infty ) \cup \cdots$. Here, $R_\phi$ is the $2 \times 2$ rotation matrix $(\cos \phi, \sin \phi | - \sin \phi, \cos \phi)$. At the transition energies, either $a$ or $b$ vanishes so that $v_+$ and $v_-$ become collinear and continuity of eigenvectors with $E$ cannot be used to unambiguously order the corresponding eigenvalues across transitions. For instance, the eigenvalue that went counterclockwise around the unit circle for $E < E_1^\ell$ could be chosen to continue as the real eigenvalue of magnitude either greater or lesser than one when $E$ exceeds $E_1^\ell$.

\vspace{.2cm}

{\bf \fl Pitfall in trigonometric and quadratic approximation at low energies:} Interestingly, if for low energies ($0 \leq E \ll g$), we use the simple harmonic/trigonometric approximation to (\ref{e:pendulum-solutions}), $\bar \vf_2 \approx \sqrt{E/g} \sin \omega_0 t$ with $\om_0 = \sqrt{3g/mr^2}$ and $E \approx (mr^2/3) \dot {\bar \vf}_2^2 + g \bar\vf_2^2$ and approximate $\cos \bar \vf_2$ by $1 - \bar\vf_2^2/2$ in (\ref{e:coeff-matrix}), we find that the eigenvalues of the monodromy matrix are of the form $e^{\pm i \tht}$ and $e^{\pm i \phi}$ where $\tht$ and $\phi$ monotonically increase from zero with energy up to moderate energies. By contrast, as we saw above, two of the eigenvalues $\la_{1,2}$ are always equal to one, a fact which is not captured by this approximation. 

\subsection{Periodic isosceles `breather' solutions}
\label{s:iso-periodic-sol}

We seek solutions where two of the separations remain equal at all times: $\tht_i - \tht_j = \tht_j - \tht_k$ where ($i, j, k$) is any permutation of (1,2,3). Loosely, these are `breathers' where one rotor is always at rest midway between the other two (see Fig. \ref{f:breathers}). For definiteness, suppose $\tht_1 - \tht_2 = \tht_2 - \tht_3$ or equivalently $\vf_1 = \vf_2$. Putting this  in Eq. (\ref{e:3rotors-EOM-ph1ph2}), we get a single evolution equation for $\vf_1 = \vf_2 = \vf$,
	\beq
	m r^2 \ddot \vf = - g (\sin \vf + \sin 2\vf),
	\label{e:isosceles-EOM}
	\eeq
which may be interpreted as a simple pendulum with an additional periodic force. As before, each periodic solution of this equation may, upon inclusion of CM motion, be used to obtain quasi-periodic solutions of the three-rotor problem.

  \begin{figure} \centering
  \includegraphics[width=6cm]{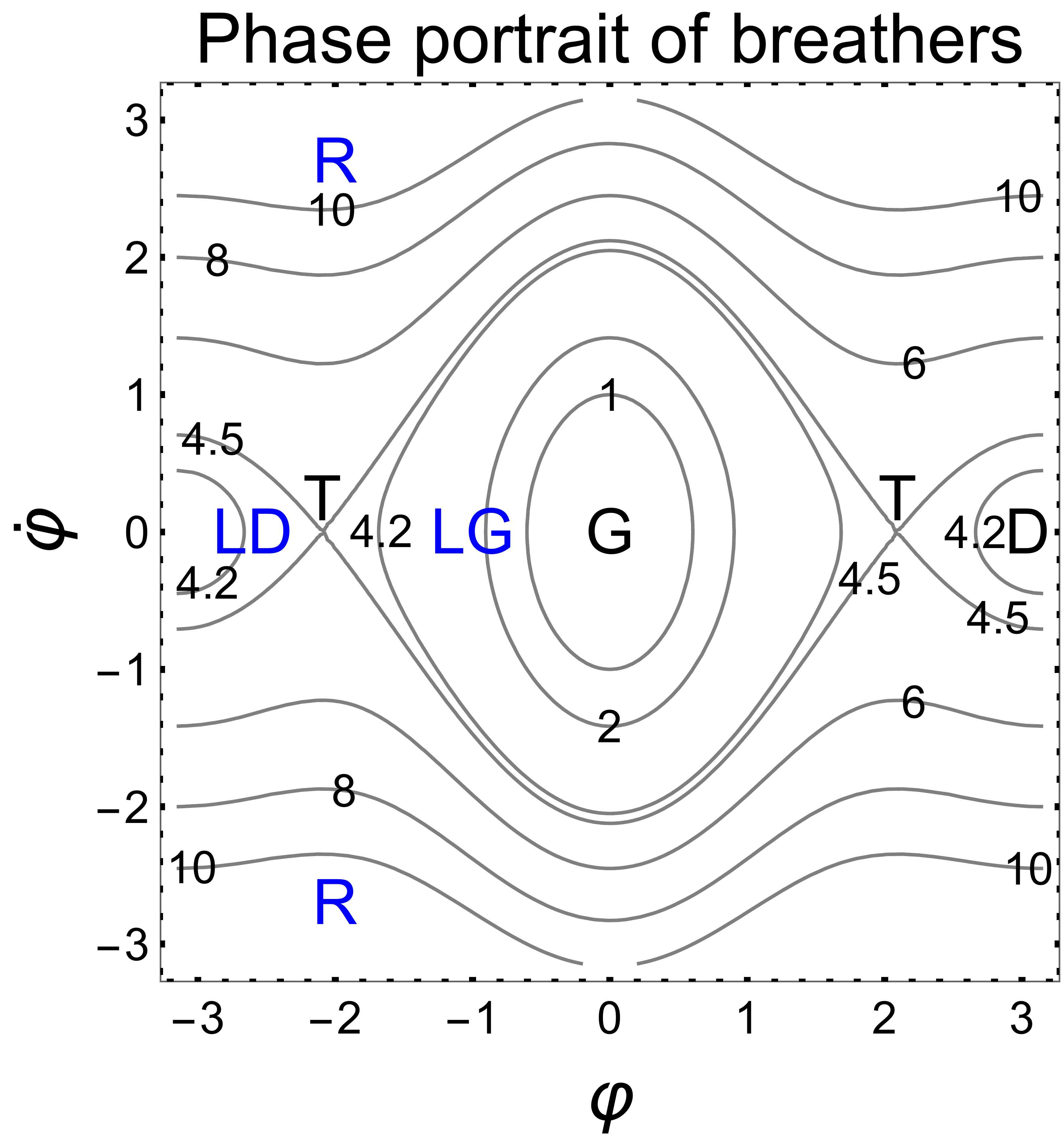}
  \caption{\small \label{f:isosceles-phaseportrait} Level contours of $E$ on a phase portrait of the LG, LD and R families of isosceles periodic solutions.}
  \end{figure}

 At $E = 0$, the isosceles solutions reduce to the ground state G. More generally, there are two families of librational breathers. With $E$ denoting energy in units of $g$, they are LG (oscillations around G $(\vf = 0)$ for $0 \leq E \leq 9/2$) and LD (oscillations around D $(\vf = \pi)$ for $4 \leq E \leq 9/2$) with monotonically growing time period which diverges at the separatrix at $E = 9/2$ (see Fig.~\ref{f:isosceles-phaseportrait}). For $E > 9/2$, we have rotational modes R with time period diminishing with energy ($\tau^{\rm rot}(E) \sim 2\pi/\sqrt{E}$ as $E \to \infty$). At very high energies, one rotor is at rest while the other two rotate rapidly in opposite directions. Eq. (\ref{e:isosceles-EOM}) may be reduced to quadrature by use of the conserved relative energy (\ref{e:egy-3rotors-phi1-phi2-coords}):
	\beq
	E = m r^2 \dot \vf^2 + g (3 - 2 \cos \vf - \cos 2 \vf).
	\eeq
For instance, in the case of the LG family,
	\beq
	\fr{\om_0 t}{\sqrt{3}} 
	= \fr{1}{\sqrt{2}} \int_0^{u} \fr{du}{\sqrt{u(2-u)(u^2 - 3u + E/2)}}
	\eeq
where $u = 1- \cos \vf$. The relative angle $\vf$ may be expressed in terms of Jacobi elliptic functions. Putting $\eps = \sqrt{9-2E}$,
	\beqs
	\vf(t) &=& \arccos \left( 1- \fr{E \eta^2}{2 \eps + (3-\eps) \eta^2} \right) \quad \text{where} \cr
	 \eta(t) &=& \text{sn} \left(\fr{\sqrt{\eps}\om_0 t}{\sqrt{3}} , \sqrt{\fr{(\eps - 1)(3 - \eps)}{8 \eps}} \right).
	\eeqs
It turns out that the periods of both LG and LD families are given by a common expression,
	\beq
	\tau^{\rm lib}(E) = \fr{4\sqrt{3}}{\om_0 \sqrt{\eps}} K\left( \sqrt{\ov{2} - \fr{6 - E}{4 \eps}} \right) \;\; \text{for} \;\; 0 \leq E \leq 4.5.
	\eeq
As $E \to 4.5$, $\tau^{\rm lib}$ diverges as $2 \sqrt{2/3} \log (4.5 - E)$. The time period of rotational solutions (for $E \geq 4.5$) is
	\beq
	\tau^{\rm rot}(E) =\fr{4\sqrt{3}}{\om_0} (E^2 - 4 E)^{-1/4} K \left(\sqrt{ \half + \fr{6 - E}{2\sqrt{E^2 - 4E}}} \right).
	\eeq

\subsubsection{Linear stability of breathers}
\label{s:stability-of-breathers}

\begin{figure}	\centering		\includegraphics[width=7cm]{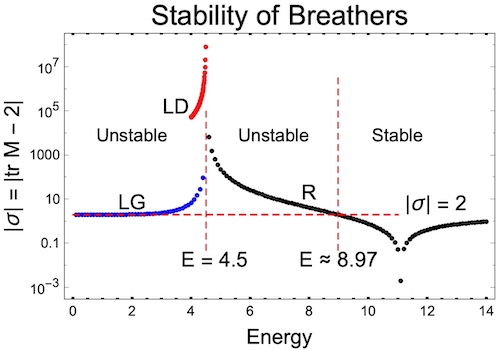}
		\caption{\small \label{f:isosceles-stability-index} Absolute value of the stability index of the isosceles breathers as a function of energy.}
\end{figure}

The stability of isosceles solutions as encoded in the stability index $(\sigma = \tr M -2)$ is qualitatively different from that of the pendulum solutions. In particular, there is only one unstable to stable transition occurring at $E \approx 8.97$ (see Fig. \ref{f:isosceles-stability-index}). Indeed, by computing the monodromies, we find that both families LG and LD of librational solutions are unstable. The stability index $\sig_{\rm LG}$ grows monotonically from $2$ to $\infty$ as the energy increases from $0$ to $4.5$. In particular, even though arbitrarily low energy breathers are small oscillations around the stable ground state G, they are themselves unstable to small perturbations. By contrast, we recall that low energy pendulum solutions around G are stable. On the other hand, the LD family of breathers are much more unstable, indeed, we find that $\sig_{\rm LD}$ increases from $\approx 5.3\times10^4$ to $\infty$ for $4 < E < 4.5$. This is perhaps not unexpected, given that they are oscillations around the unstable static solution D. The rotational breathers are unstable for $4.5 < E < 8.97$ with $\sig_{\rm R}$ growing from $-\infty$ to $-2$. These divergences of $\sigma$ indicate that isosceles solutions around $E = 4.5$ suffer severe instabilities not seen in the pendulum solutions. Beyond $E = 8.97$, the rotational breathers are stable with $\sig_{\rm R}$ growing from $-2$ to $2$ as $E \to \infty$. This stability of the breathers is also evident from the Poincar\'e sections of \S \ref{s:poincare-section}. In fact, the isosceles solutions go from intersecting the Poincar\'e surface `$\vf_1 = 0$' at  hyperbolic to elliptic fixed points as the energy is increased beyond $E \approx 8.97$ (see Fig. \ref{f:psec-egy=2-3}-\ref{f:psec-egy=4.5-to-18}).

\section[Jacobi-Maupertuis metric and curvature]{Jacobi-Maupertuis metric and curvature \sectionmark{Jacobi--Maupertuis metric and curvature}}
\sectionmark{Jacobi--Maupertuis metric and curvature}
\label{s:JM-approach}

We now consider a geometric reformulation of the classical three-rotor problem, that suggests the emergence of widespread instabilities for $E > 4$ from a largely stable phase at lower energies and a return to regularity as $E \to \infty$.  This indicates the presence of an `order-chaos-order' transition which will be confirmed in \S \ref{s:poincare-section}.

As discussed in \S \ref{s:traj-as-geodesics}, configuration space trajectories of the Lagrangian $L = (1/2) m_{ij} \dot q_i \dot q_j - \V(q)$ may be regarded as reparametrized geodesics of the Jacobi-Maupertuis (JM) metric $g^{\rm JM}_{ij} = (E - \V) m_{ij}$ which is conformal to the mass/kinetic metric $m_{ij}(q)$. The sectional curvatures of this metric have information on the behavior of nearby trajectories with positive/negative curvature associated to (linear) stability/instability. For the three-rotor problem, the JM metric on the $\vf_1$-$\vf_2$ configuration torus is given by
	\beq
	ds_{\rm JM}^2 = \fr{2 m r^2}{3} (E - \V) (d\vf_1^2 + d\vf_1 d\vf_2 + d\vf_2^2),
	\eeq
where $\V = g[3 - \cos \vf_1 - \cos \vf_2 - \cos( \vf_1 + \vf_2) ]$. Letting $f$ denote the conformal factor $E - \V$ and using the gradient and Laplacian defined with respect to the flat kinetic metric, the corresponding scalar curvature ($2 \times$ the Gaussian curvature) is
	\beqs
	R &=& \fr{|\grad f|^2 - f \Delta f}{f^3} 
	 = \fr{g^2}{m r^2 (E-\V)^3} \cr && \times \left[6 +  \left(\fr{2E}{g} - 3\right)\left(3 - \fr{\V}{g}\right) + \cos(\vf_1-\vf_2)   + \cos(2\vf_1+\vf_2) + \cos(\vf_1+2\vf_2) \right]. \;\;\qquad
 \label{e:JM-scalar-curv}
	\eeqs

\subsection{Behavior of JM curvature}

\begin{figure}	
	\begin{subfigure}[t]{5cm}
		\centering
		\includegraphics[width=5cm]{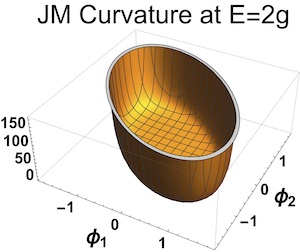}
	\end{subfigure}	
	\begin{subfigure}[t]{5cm}
		\centering
		\includegraphics[width=5cm]{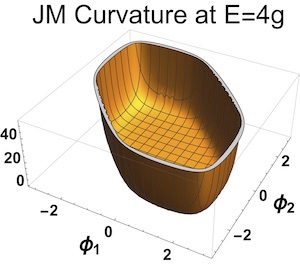}
	\end{subfigure}	
	\begin{subfigure}[t]{5cm}
		\centering
		\includegraphics[width=5cm]{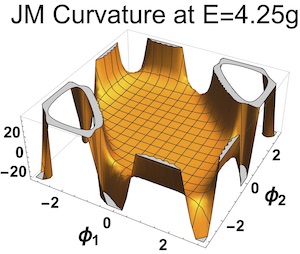}
	\end{subfigure}	
	\begin{subfigure}[t]{5cm}
		\centering
		\includegraphics[width=5cm]{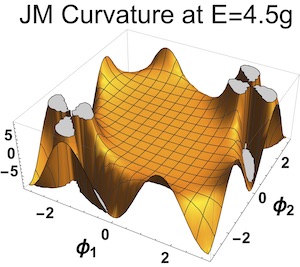}
	\end{subfigure}
	\begin{subfigure}[t]{5cm}
		\centering
		\includegraphics[width=5cm]{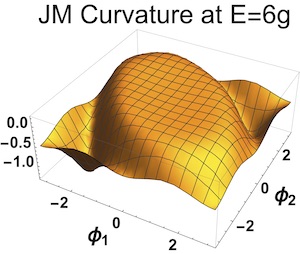}
	\end{subfigure}	
	\caption{\small Scalar curvature $R$ of the JM metric on the Hill region of the $\vf_1$-$\vf_2$ torus. In the regions shaded grey, $|R|$ is very large. We see that $R > 0$ for $E \leq 4g$ but has both signs for $E > 4g$ indicating instabilities.}
	\label{f:curvature-3rotors-phi1-phi2-torus}
\end{figure}

 For $0 \leq E \leq 4g$, $R$ is strictly positive  in the classically allowed Hill region $(\V < E)$ and diverges on the Hill boundary $\V = E$ where the conformal factor vanishes (see Appendix \ref{a:positivity-of-JM-curvature} for a proof and the first two `bath-tub' plots of $R$ in Fig.~\ref{f:curvature-3rotors-phi1-phi2-torus}). Thus the geodesic flow should be stable for these energies. Remarkably, we also find a near absence of chaos in all Poincar\'e sections for $E \lesssim 3.8g$ (see Fig. \ref{f:psec-egy=2-3} and  \ref{f:chaos-vs-egy}). We will see that Poincar\'e surfaces show significant chaotic regions for $E > 4g$. This is perhaps related to the instabilities associated with $R$ acquiring both signs above this energy. Indeed, for $4g < E \leq 9g/2$, the above `bath-tub' develops sinks around the saddles $D(0,\pi)$, $D(\pi,0)$ and $D(\pi,\pi)$  where $R$ becomes negative, though it continues to diverge on the Hill boundary which is a union of two closed curves encircling the local maxima T$(\pm 2\pi/3, \pm 2\pi/3)$. For $E > 9g/2$, the Hill region expands to cover the whole torus. Here, though bounded, $R$ takes either sign while ensuring that the total curvature $\int_{T^2} R \, \sqrt{\det g_{ij}}\; d\vf_1 \, d\vf_2$ vanishes. For asymptotically high energies, the JM metric tends to the flat metric $E \,m_{ij}$ and $R \sim 1/E^2 \to 0$ everywhere indicating a return to regularity.

\subsection{Geodesic stability of static solutions}

 The static solutions G, D and T lie on the boundary of the Hill regions corresponding to the energies $E_{\rm G, D, T} = 0$, $4g$ and $4.5 g$. We define the curvatures at G, D and T by letting $E$ approach the appropriate limiting values in the following formulae: 
	\beqs
	R_{(0,0)} = \fr{6 g}{m r^2 E^2}, \quad R_{(0,\pi),(\pi,0),(\pi,\pi)} = \fr{- 2 g/ m r^2}{(E - 4g)^2} \quad
	 \text{and} \quad R_{\left(\pm \fr{2\pi}{3},\pm \fr{2\pi}{3}\right)} = \fr{- 12 g/m r^2}{(2E -9g)^2}.
	\eeqs
Thus $R_{\rm G} = \infty$ while $R_{\rm D} = R_{\rm T} = - \infty$ indicating that G is stable while D and T are unstable. These results on geodesic stability are similar to those obtained from (\ref{e:EOM-ph1ph2-perturbation-static-solution}). Note that we do not define the curvatures at G, D and T by approaching these points from within the Hill regions as these limits are not defined for G and T and gives $+\infty$  for D. On the other hand, it is physically forbidden to approach the Hill boundary from the outside. Thus we approach G, D and T by varying the energy while holding the location on the torus fixed.

\section{Poincar\'e sections: periodic orbits and chaos}
\label{s:poincare-section}

To study the transitions from integrability to chaos in the three-rotor problem, we use the method of Poincar\'e sections. Phase trajectories are constrained to lie on energy level sets which are compact 3d submanifolds of the 4d phase space parametrized by $\vf_1$, $\vf_2$, $p_1$ and $p_2$ (cotangent bundle of the 2-torus). By the Poincar\'e surface `$\vf_1 = 0$' at energy $E$ (in units of $g$), we mean the 2d surface $\vf_1 = 0$ contained in a level-manifold of energy. It may be parametrized by $\vf_2$ and $p_2$ with the two possible values of $p_1(\vf_2,p_2;E)$ determined by the conservation of energy. Similarly, we may consider other Poincar\'e surfaces such as the ones defined by $\vf_2 = 0$, $p_1 = 0$, $p_2 = 0$ etc. We record the points on the Poincar\'e surface where a trajectory that begins on it returns to it under the Poincar\'e return map, thus obtaining a Poincar\'e section for the given initial condition (IC). For transversal intersections, a periodic trajectory leads to a Poincar\'e section consisting of finitely many points while quasi-periodic trajectories produce sections supported on a finite union of 1d curves. We call these two types of sections `regular'. By a chaotic section, we mean one that is not supported on such curves but explores a 2d region. In practice, deciding whether a numerically obtained Poincar\'e section is regular or chaotic can be a bit ambiguous in borderline cases when it is supported on a thickened curve (see Fig. \ref{f:psec-egy=18} and around $I$ in Fig. \ref{f:psec-egy=2-3}). We define the chaotic region of a Poincar\'e surface at energy $E$ to be the union of all chaotic sections at that energy.

\begin{figure}[ht]
	\centering
	\begin{subfigure}{7.2cm}
		\centering
		\includegraphics[width=7.2cm]{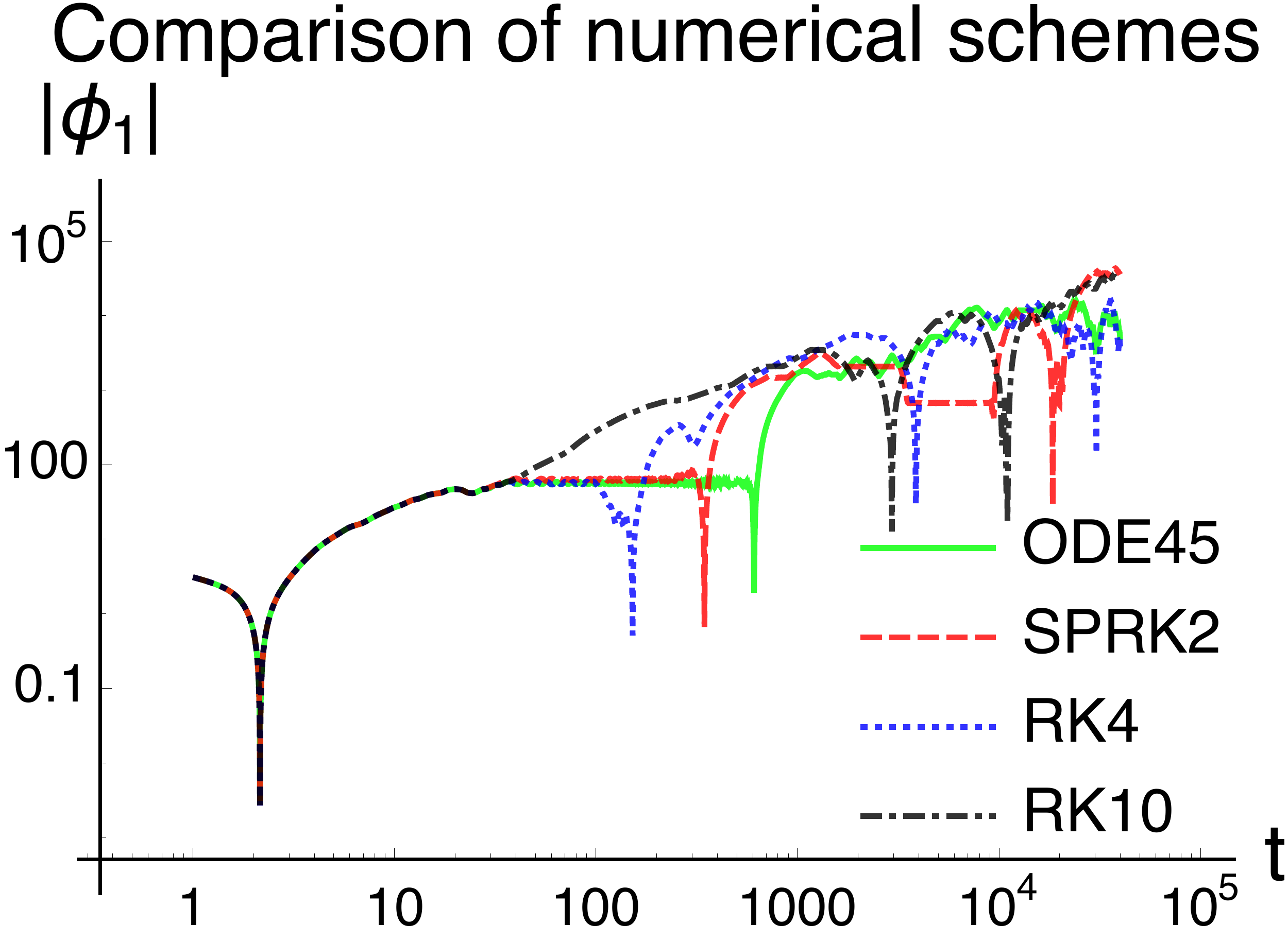}
		\caption{\small }
	\end{subfigure}\qquad 
	\begin{minipage}{0.5\textwidth}	
	\begin{subfigure}[t]{3.2cm}
		\centering
		\includegraphics[width=3.2cm]{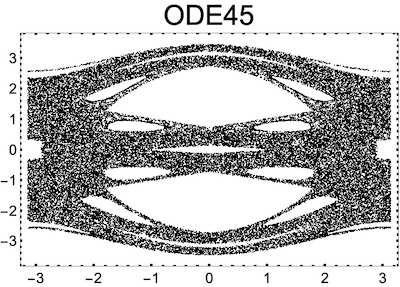}
		\caption{\small }
	\end{subfigure}
	\begin{subfigure}[t]{3.2cm}
		\centering
		\includegraphics[width=3.2cm]{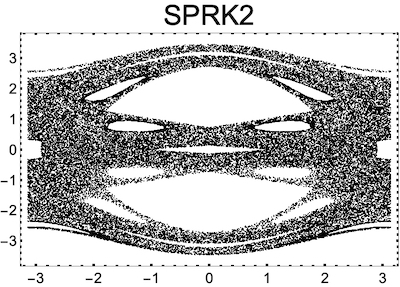}
		\caption{\small }
	\end{subfigure} \vfill
	\begin{subfigure}[t]{3.2cm}
		\centering
		\includegraphics[width=3.2cm]{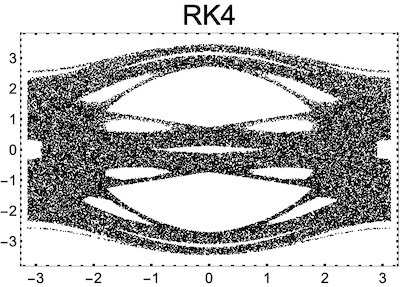}
		\caption{\small }
	\end{subfigure}
	\begin{subfigure}[t]{3.2cm}
		\centering
		\includegraphics[width=3.2cm]{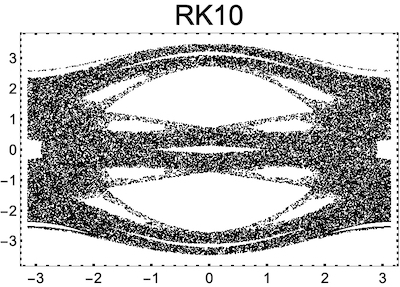}
		\caption{\small }
	\end{subfigure}
	\end{minipage}
	\caption{\small   (a) The trajectories (e.g., $|\vf_{1}|$) obtained via different numerical schemes cease to agree after $t \sim 10^2$ for the IC $\vf_1 = 6.23$, $\vf_2 = 3.00$, $p_1 = -.90$ and $p_2 = 1.87$ with $E=9.98$. (b, c, d, e) However, Poincar\'e sections (with $\approx 5 \times 10^4$ points) obtained via different schemes are seen to explore qualitatively similar regions when evolved till $t = 10^5$ (though {\em not} for shorter times $\sim 10^3$).}	
	\label{f:scheme-dependent-trajectories}
\end{figure}

\subsection{Transition to chaos and global chaos}
\label{s:transition-to-chaos-global-chaos}

\subsubsection{Numerical schemes and robustness of Poincar\'e sections}

 To obtain Poincar\'e sections, we implement the following numerical schemes: ODE45: explicit Runge-Kutta with difference order 5; RK4 and RK10: implicit Runge-Kutta with difference orders 4 and 10 and SPRK2: symplectic partitioned Runge-Kutta with difference order 2. Due to the accumulation of errors, different numerical schemes (for the same ICs) sometimes produce trajectories that cease to agree after some time, thus reflecting the sensitivity to initial conditions. Despite this difference in trajectories, we find that the corresponding Poincar\'e sections from all schemes are roughly the same when evolved for sufficiently long times (see Fig.~\ref{f:scheme-dependent-trajectories}). Moreover, we find a strong correlation between the degree to which different schemes produce the same trajectory and the degree of chaos as manifested in Poincar\'e sections. As the agreement in trajectories between different schemes improves, the Poincar\'e sections go from being spread over 2d regions to being concentrated on a finite union of 1d curves. Since ODE45 is computationally faster than the other schemes, the results presented below are obtained using it. Furthermore, we find that for all ICs studied, all four Poincar\'e sections on surfaces defined by $\vf_1 = 0$, $\vf_2 = 0$, $p_1 = 0$ and $p_2 = 0$ are qualitatively similar with regard to the degree of regularity or chaos. Thus, in the sequel, we restrict to the Poincar\'e surface defined by $\vf_1 = 0$.

\begin{figure}	
	\centering
	\begin{subfigure}[t]{5cm}
		\centering
		\includegraphics[width=5cm]{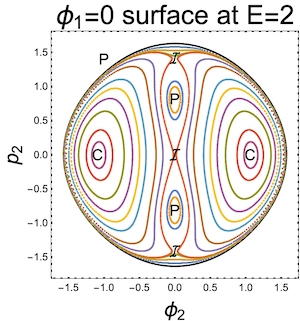}
	\end{subfigure}	
\qquad
	\begin{subfigure}[t]{5cm}
		\centering
		\includegraphics[width=5cm]{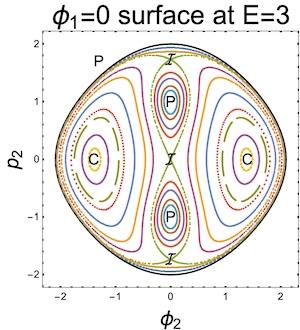}
	\end{subfigure}	
	\caption{\small   Several Poincar\'e sections in the energetically allowed `Hill' region on the `$\vf_1 = 0$' surface for $E = 2$ and $3$. All sections (indicated by distinct colors online) are largely regular and possess up-down and left-right symmetries. The Hill boundary is the librational pendulum solution $\vf_1 = 0$. P, ${\cal I}$ and C indicate pendulum, isosceles and choreography periodic solutions. More careful examination of the vicinity of the $\cal I$s shows small chaotic sections.}
	\label{f:psec-egy=2-3}
\end{figure}

\subsubsection{Symmetry breaking accompanying  onset of chaos}

\begin{figure}
	\centering
	\begin{subfigure}[t]{5.4cm}
		\includegraphics[width=5.4cm]{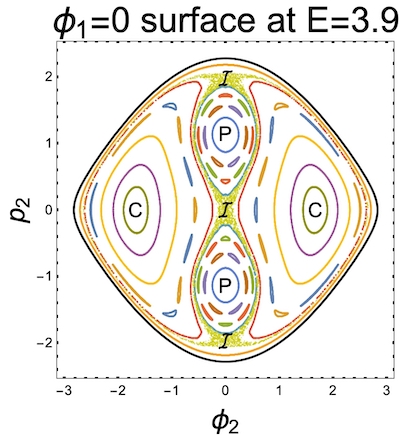}
		\caption{\small }
		\label{f:psec-egy=3.9}
	\end{subfigure}	
	\begin{subfigure}[t]{5.4cm}
		\includegraphics[width=5.4cm]{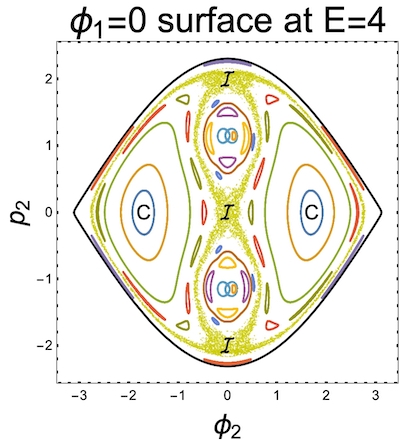}
		\caption{\small }
		\label{f:psec-egy=4}
	\end{subfigure}	
	\begin{subfigure}[t]{5.4cm}
		\includegraphics[width=5.4cm]{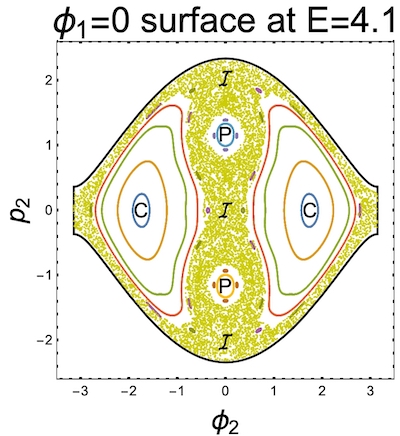}
		\caption{\small }
		\label{f:psec-egy=4.1}
	\end{subfigure}	
	\caption{\small   Several Poincar\'e sections on the `$\vf_1 = 0$' surface in the vicinity of $E = 4$ where the chaotic region (shaded, yellow  online) makes its first significant appearance. Distinct sections have different colors online. On each surface, one sees breaking of both up-down and left-right symmetries. Aside from a couple of exceptions on the $E=4$ surface, the set of ICs is left-right and up-down symmetric. The boundary of the Hill region on the `$\vf_1 = 0$' Poincar\'e surface is the $\vf_1 = 0$ pendulum solution. It becomes disconnected for $E > 4$ owing to the bifurcation of the librational pendula into clockwise and counterclockwise rotational pendula.}
	\label{f:psec-egy-near-4}
\end{figure}

We find that for $E \lesssim 4$, all Poincar\'e sections (on the surface `$\vf_1 = 0$') are nearly regular and display left-right $(\vf_2 \to - \vf_2)$ and up-down $(p_2 \to - p_2)$ symmetries  (see Fig. \ref{f:psec-egy=2-3}). Though there are indications of chaos even at these energies along the periphery of the four stable lobes (e.g., near the unstable isosceles fixed points $\cal I$), chaotic sections occupy a negligible portion of the Hill region. Chaotic sections make their first significant appearance at $E \approx 4$ along the figure-8 shaped separatrix and along the outer periphery of the regular `lobes' that flank it (see Fig. \ref{f:psec-egy-near-4}). This transition to chaos is accompanied by a spontaneous breaking of both the above symmetries. Interestingly, the $\vf_2 \to - \vf_2$ symmetry (though not $p_2 \to - p_2$) seems to be restored when $E \gtrsim 4.4$. The lack of $p_2 \to - p_2$ symmetry at high energies is not unexpected: rotors at high energies either rotate clockwise or counter-clockwise.

\begin{figure}[ht]	
	\begin{subfigure}[t]{5cm}
		\centering
		\includegraphics[width=5cm]{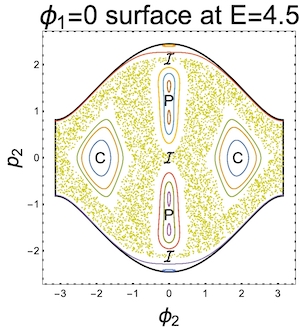}
		\caption{\small }
		\label{f:psec-egy=4.5}
	\end{subfigure}		
	\begin{subfigure}[t]{5cm}
		\centering
		\includegraphics[width=5cm]{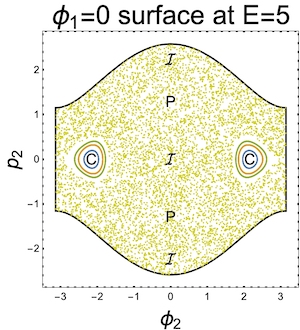}
		\caption{\small }
		\label{f:psec-egy=5}
	\end{subfigure}	
	\begin{subfigure}[t]{5cm}
		\centering
		\includegraphics[width=5cm]{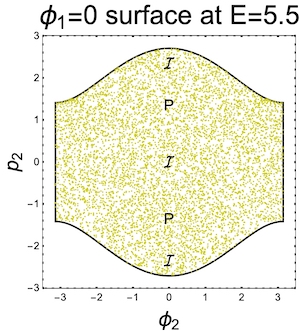}
		\caption{\small }
		\label{f:psec-egy=5.5}
	\end{subfigure}	
	\begin{subfigure}[t]{5cm}
		\centering
		\includegraphics[width=5cm]{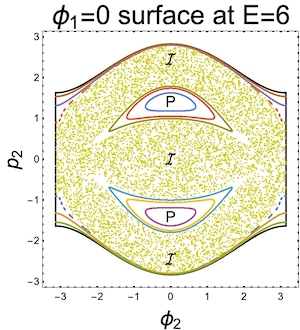}
		\caption{\small }
		\label{f:psec-egy=6}
	\end{subfigure} \quad
	\begin{subfigure}[t]{5cm}
		\centering
		\includegraphics[width=5cm]{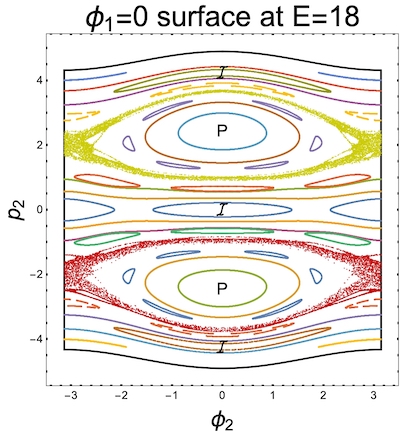}
		\caption{\small }
		\label{f:psec-egy=18}
	\end{subfigure}	
	\caption{\small   The up-down symmetry remains broken, though the left-right symmetry is restored on Poincar\'e plots at higher energies. The periodic orbits corresponding to points marked C are choreographies for $E \lesssim 5.33$.}
	\label{f:psec-egy=4.5-to-18}
\end{figure}

At moderate energies $E \gtrsim 4$, we observe that all chaotic sections (irrespective of the ICs) occupy essentially the same region, as typified by the examples in Fig. \ref{f:psec-egy=4.5-to-18}. At somewhat higher energies (e.g. $E = 14$), we find chaotic sections that fill up both the entire chaotic region and portions thereof when trajectories are evolved up to $t = 10^5$. At yet higher energies (e.g. $E = 18$, Fig. \ref{f:psec-egy=18}), there is no single chaotic section that occupies the entire chaotic region as the $p_2 \to - p_2$ symmetry is broken. 

\subsubsection{Fraction of chaos and global chaos} 

For a range of energies beyond $4$, we find that the area of the chaotic region increases with $E$ (see Fig.~\ref{f:psec-egy-near-4} and \ref{f:psec-egy=4.5-to-18}). At $E \approx 5.5$, the chaotic region coincides with the energetically allowed portion of the Poincar\'e surface (see Fig.~\ref{f:psec-egy=5.5}). Beyond this energy, chaotic sections are supported on increasingly narrow bands (see Fig. \ref{f:psec-egy=18}). This progression towards regular sections is expected since the system acquires an additional conserved quantity in the limit $E \to \infty$. To quantify these observations, we find the `fraction of chaos' $f$ by exploiting the feature that the density of points in chaotic sections is roughly uniform for {\it all energies} on the `$\vf_1 = 0$' surface (this is not true for most other Poincar\'e surfaces). Thus $f$ is estimated by calculating the fraction of the area of the Hill region covered by chaotic sections (see Appendix \ref{a:estimate-f} and Fig. \ref{f:chaos-vs-egy}).

The near absence of chaos is reflected in $f$ approximately vanishing for $E \lesssim 3.8$. There is a rather sharp transition to chaos around $E \approx 4$ ($f \approx 4\%$, $20\%$ and $40\%$ at $E = 3.85$, $4$ and $4.1$; see lower inset of Fig. \ref{f:chaos-vs-egy}). This is a bit unexpected from the viewpoint of KAM theory and might encode a novel mechanism by which KAM tori break down in this system. Thereafter, $f$ rapidly rises and reaches the maximal value $f \approx 1$ at $E \approx 5.33$. As illustrated in the upper inset of Fig. \ref{f:chaos-vs-egy}, this `fully chaotic' phase persists up to $E \approx 5.6$. Interestingly, we find that for this range of energies, $f \approx 1$ on a variety of Poincar\'e surfaces examined (see Fig. \ref{f:global-chaos}), so that this may be regarded as a phase of `global chaos'. Furthermore, chaotic sections fill up Poincar\'e surfaces in a roughly uniform manner, resulting in uniform density of points on all Poincar\'e surfaces in this phase of global chaos indicating some sort of ergodicity (see \S \ref{s:ergodicity}). Additionally, the pendula and breathers are unstable in this phase (see \S \ref{s:reduction-one-dof}) and it would be interesting to know whether this is the case with all periodic solutions. Remarkably, the cessation of the band of global chaos happens to coincide with the energy $E_1^\err \approx 5.6$ above which pendulum solutions are always stable (see Fig. \ref{f:monodromy-evals}). Beyond $E \approx 5.6$, $f$ decreases gradually to zero as $E \to \infty$. Interestingly, the sharp transition to chaos at $E \approx 4$ is also reflected in the JM curvature of \S \ref{s:JM-approach} going from being positive for $E < 4$ to admitting both signs for $E>4$. It is noteworthy that the stable to unstable transition energies in pendula also accumulate from both sides at $E=4$ (see Fig. \ref{f:monodromy-evals}).

\begin{figure}
	\begin{subfigure}[t]{8cm}
		\centering
		\includegraphics[height=6.3cm]{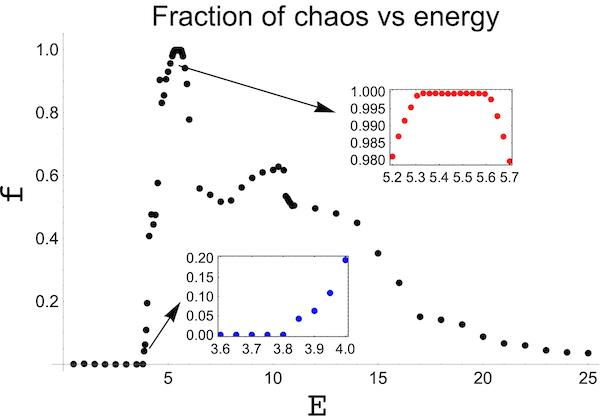}
		\caption{\small }
		\label{f:chaos-vs-egy}
	\end{subfigure}	
	\quad
	\begin{subfigure}[t]{8cm}
		\centering
		\includegraphics[height=6.3cm]{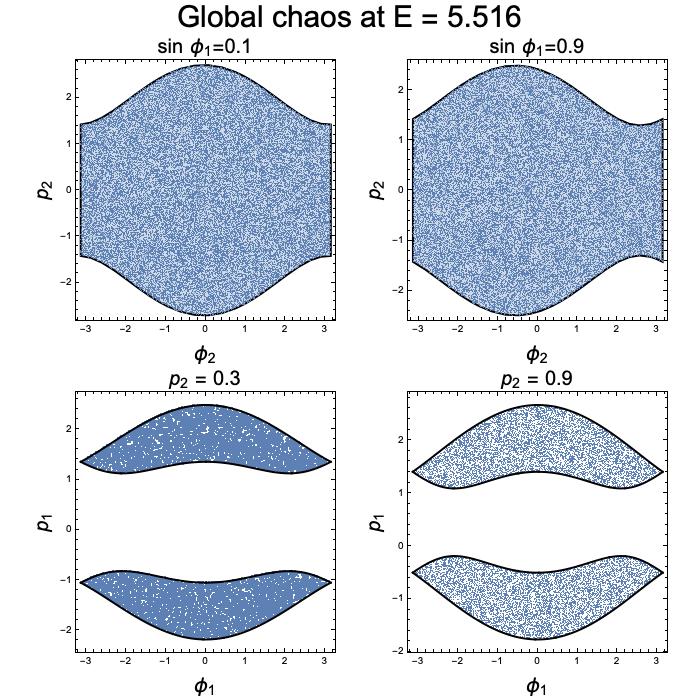}
		\caption{\small }
		\label{f:global-chaos}
	\end{subfigure}		
	\caption{\small (a) Energy dependence of the area of the chaotic region on the `$\vf_1 = 0$' Poincar\'e surface as a fraction of the area of the Hill region. (b) Various Poincar\'e surfaces showing global chaos at $E = 5.516$.}
\end{figure}


\subsection{Periodic solutions on the \texorpdfstring{`$\vf_1 = 0$'}{`phi1 = 0'} Poincar\'e surface}
\label{s:periodic-sol-on-Psec}

Here, we identify the points on the Poincar\'e surface corresponding to the periodic pendulum and isosceles solutions. Remarkably, careful examination of the Poincar\'e sections also leads us to a new family of periodic `choreography' solutions which are defined and discussed further in \S \ref{s:choreographies}.

\subsubsection{Pendula}

 The $\vf_1 = 0$ pendulum solutions are everywhere tangent to the Poincar\'e surface `$\vf_1 = 0$' and interestingly constitute the `Hill' energy boundary (see Fig. \ref{f:psec-egy=2-3}-\ref{f:psec-egy=4.5-to-18}). [Nb. This connection between pendulum solutions and the Hill boundary is special to the surfaces `$\vf_1 = 0$' and `$\vf_2 = 0$'.] By contrast, the other two classes of pendulum trajectories ($\vf_2 = 0$ and $\vf_1 + \vf_2 = 0$) are transversal to this surface, meeting it at the pendulum points P($0, \pm \sqrt{E/3}$) halfway to the boundary from the origin. These are period-2 and period-1 fixed points for  librational and rotational solutions respectively. Examination of the Poincar\'e sections indicates that pendulum solutions must be stable for $E \lesssim 3.9$ and $E \gtrsim 5.6$ leaving open the question of their stability at intermediate energies. As discussed in \S \ref{s:pendulum-soln}, the pendulua go from being stable to unstable infinitely often as $E \to 4^\pm$. Additionally, by considering initial conditions near the pendulum points, we find that the pendulum solutions lie within the large chaotic section only between $E \approx 4.6$ and the cessation of global chaos at $E \approx 5.6$.

\subsubsection{Breathers}

 Unlike pendula, all isosceles periodic orbits intersect the `$\vf_1 = 0$' surface transversally at points on the vertical axis. Indeed, the breathers defined by $\vf_1 = \vf_2$ and $\vf_2 + 2\vf_1 = 0$ intersect the surface at the isosceles points ${\cal I}(0, \pm \sqrt{E})$ which form a pair of period-2 fixed points for $E < 4.5$ and become period-1 in the rotational regime (see Fig. \ref{f:psec-egy=2-3}-\ref{f:psec-egy=4.5-to-18}). The breathers defined by $\vf_1 + 2\vf_2 = 0$ intersect the surface at the period-1 fixed point at the origin. In agreement with the conclusions of \S \ref{s:stability-of-breathers}, the Poincar\'e sections show that all three isosceles points are unstable at low energies, lie in the large chaotic section for $3.9 \lesssim E \lesssim 8.97$ and are stable at higher energies.

\subsubsection{A new family of periodic solutions}

 The period-2 fixed points C at the centers of the right and left lobes on the Poincar\'e surfaces of Fig. \ref{f:psec-egy=2-3} and \ref{f:psec-egy-near-4} correspond to a new family of periodic solutions. Evidently, they go from being stable to unstable as the energy crosses $E \approx 5.33$. We argue in \S \ref{s:choreographies} that they are choreographies for $E \lesssim 5.33$.

\section{Choreographies} 
\label{s:choreographies}

\begin{figure}	
	\centering
	\begin{subfigure}[t]{8cm}
		\centering
		\includegraphics[width=8cm]{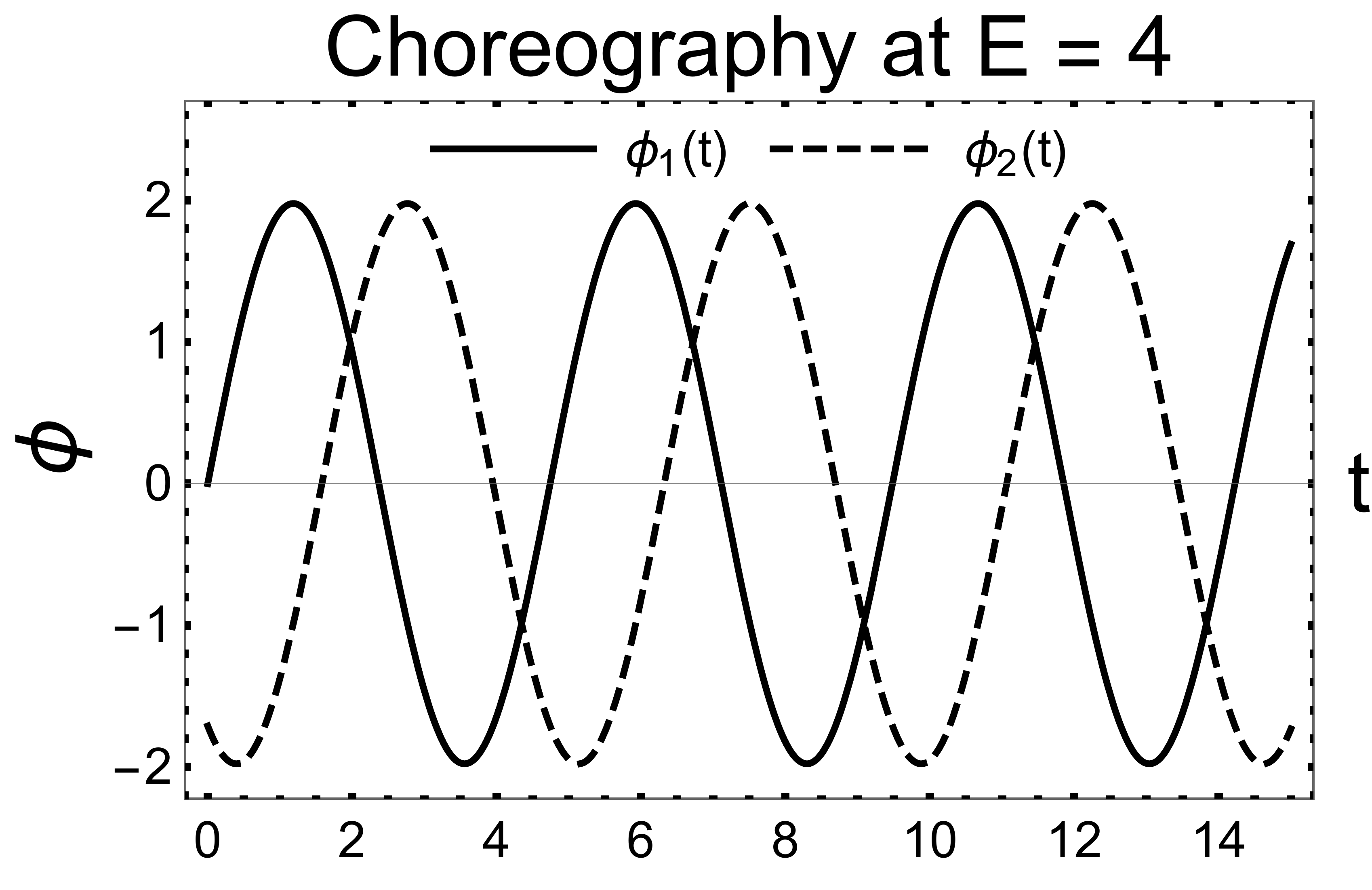}
		\caption{\small }
		\label{f:choreo-phi1-phi2-plot}
	\end{subfigure}
\qquad
	\begin{subfigure}[t]{7cm}
		\centering
		\includegraphics[width=7cm]{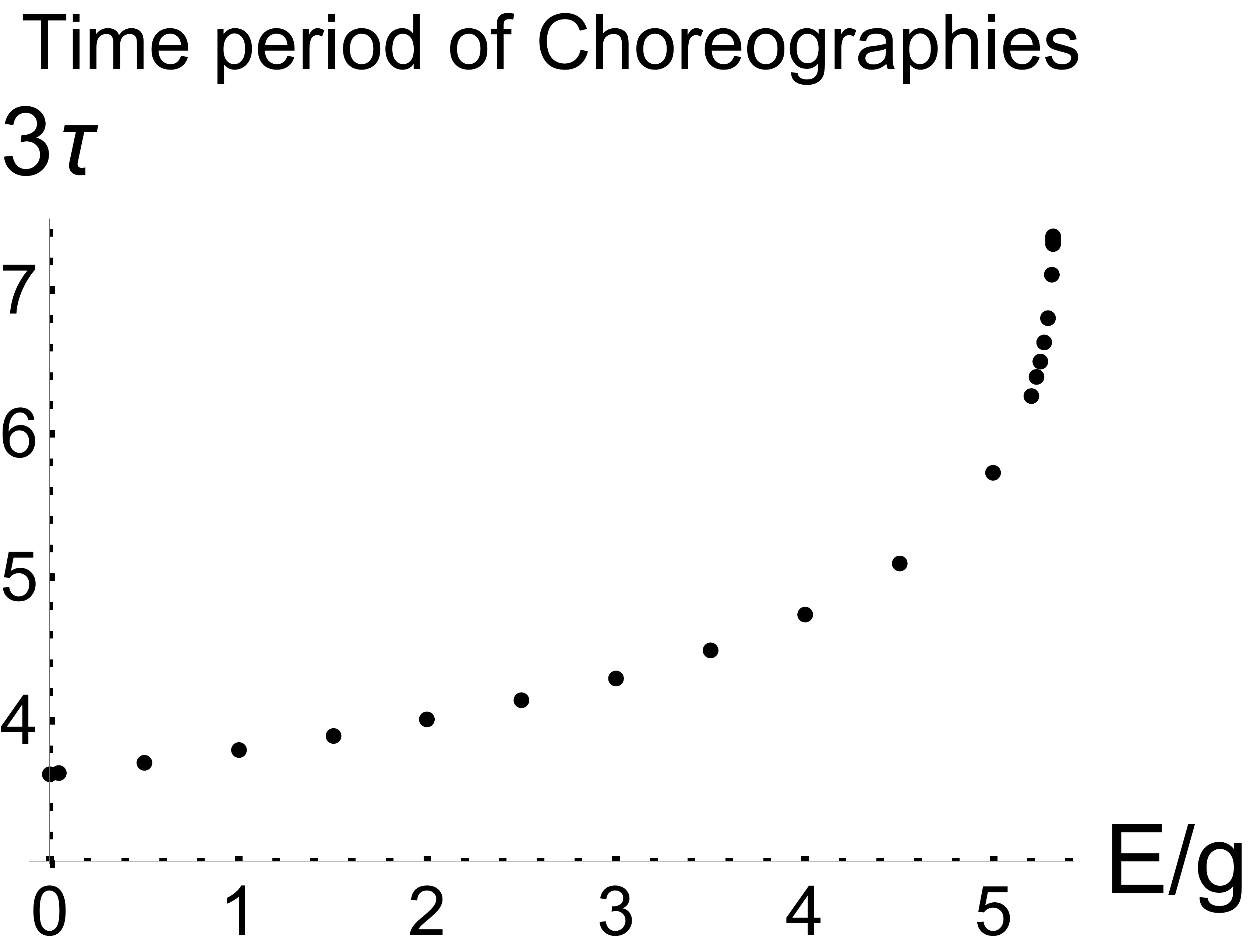}
		\caption{\small }
		\label{f:choreo-time-vs-egy}
	\end{subfigure}
	\caption{\small   (a) A non-rotating choreography at $E = 4g$ showing that the time lag between $\vf_1$ and $\vf_2$ is one-third the period. (b) The time period $3\tau$ of non-rotating choreographies as a function of energy indicating divergence at $E \approx 5.33g$.}
	\label{f:choreo-time-vs-egy-phi1-phi2-plot}
\end{figure}

Choreographies are an interesting class of periodic solutions of the $n$-body problem where all particles follow the same closed curve equally separated in time \cite{montgomery-choreographies}. The Lagrange equilateral solution where three equal masses move on a common circle and the stable zero-angular momentum figure-8 solution discovered by C. Moore \cite{c-moore-braid-group} (see also \cite{chenciner-montgomery}) are perhaps the simplest examples of choreographies in the equal mass gravitational three-body problem. Here, we consider choreographies in the three-rotor problem where the angles $\tht_i(t)$ of the three rotors may be expressed in terms of a single $3\tau$-periodic function, say $\tht_1(t)$:
	\beq
	\label{e:defn-choreography}
	\tht_2(t) = \tht_1(t+\tau) \;\; 
	\text{and} 
	\;\; \tht_3(t) = \tht_1(t + 2\tau).
	\eeq
This implies that the CM and relative coordinates $\vf_0$, $\vf_1(t)$ and $\vf_2(t) = \vf_1(t+\tau)$ must be $3\tau$ periodic (see Fig. \ref{f:choreo-phi1-phi2-plot}) and satisfy the delay algebraic equation
	\beqs
	\vf_1(t) + \vf_1(t+\tau) + \vf_1(t+2\tau) =  
	\tht_1 - \tht_2 + \tht_2 - \tht_3 + \tht_3 - \tht_1 \equiv 0 \mod 2\pi.
	\label{e:choreography-delay-algebraic}
	\eeqs
The EOM (\ref{e:3rotors-EOM-ph1ph2}) become $3 m r^2 \ddot \vf_0 = 0$ and the pair of delay differential equations
	\beqs
	mr^2 \ddot \vf_1(t) &=& - g\big[ 2 \sin \vf_1(t) - \sin \vf_1(t+\tau)  + \sin (\vf_1(t) + \vf_1(t+\tau)) \big] \quad \text{and} \cr 
	mr^2 \ddot \vf_2(t) &=& mr^2 \ddot \vf_1(t+\tau) \cr
	&=& -g \big[ 2 \sin \vf_1(t+\tau) 
	- \sin \vf_1(t) + \sin (\vf_1(t) + \vf_1(t+\tau)) \big]. \qquad
	\label{e:choreography-phi-1-2}
	\eeqs
In fact, the second equation in (\ref{e:choreography-phi-1-2}) follows from the first by use of the delay algebraic equation (\ref{e:choreography-delay-algebraic}). Moreover, using the definition of $\vf_0$, the constant angular velocity of the CM
	\beq
	\dot \vf_0 = \ov{\tau} \left[\vf_0(t+\tau) - \vf_0(t)\right] 
	= - \fr{1}{3\tau} \left[ \vf_1(t) + \vf_1(t+\tau) + \vf_1(t+2\tau)\right]. 
	\label{e:choreography-CM-angular-velocity}
	\eeq
It is verified that any $3\tau$ periodic triple $\vf_{0,1,2}$ satisfying (\ref{e:choreography-delay-algebraic}), (\ref{e:choreography-phi-1-2}) and (\ref{e:choreography-CM-angular-velocity}) leads to a choreography of the three-rotor system. Thus, to discover a choreography we only need to find a $3\tau$-periodic function $\vf_1$ satisfying (\ref{e:choreography-delay-algebraic}) and the first of the delay differential equations (\ref{e:choreography-phi-1-2}) with the period $3 \tau$ self-consistently determined. Now, it is easy to show that choreographies cannot exist at asymptotically high (relative) energies. In fact, at high energies, we may ignore the interaction terms ($\propto g$) in (\ref{e:choreography-phi-1-2}) to get $\vf_1(t) \approx \om t + \vf_1(0)$ for $|\om| \gg 1$. However, this is inconsistent with (\ref{e:choreography-delay-algebraic}) which requires $3 \om t \equiv 0 \mod 2\pi$ at all times. On the other hand, as discussed below, we {\it do} find examples of choreographies at low and moderate relative energies.

\subsection{Examples of choreographies}

 Uniformly rotating (at angular speed $\Om$) versions of the static solutions G and T (but not D) (see \S \ref{s:rotating-static-sol} and Fig.~\ref{f:static-solutions-3rotors}) provide the simplest examples of choreographies with $\tht_1(t) = \Om t$ and $\tau = 2\pi/\Om$ for G and $\tau = 2\pi/3\Om$ for T where $\Om$ is arbitrary. In the case of G, though all particles coincide, they may also be regarded as separated by $\tau$. The energies (\ref{e:egy-3rotors-phi1-phi2-coords}) of these two families of choreographies come from the uniform CM motion and a constant relative energy:
	\beq
	E^\text{(G)}_{\rm tot} = \fr{3}{2} m r^2 \Om^2 \quad \text{and} \quad E^\text{(T)}_{\rm tot} = \fr{3}{2} m r^2 \Om^2 + \fr{9g}{2}.
	\eeq
These two families of choreographies have the scaling property: if $\tht(t)$ with period $3\tau$ describes a choreography in the sense of (\ref{e:defn-choreography}), then $\tht(a t)$ with period $|3\tau/a|$ also describes a choreography for any real $a$. It turns out that the above two are the only such `scaling' families of choreographies. To see this, we note that both $\tht(t)$ and $\tht(at)$ must satisfy the delay differential equation
	\beq
	\ddot \tht(t+\tau) - \ddot \tht(t) = \fr{-g}{mr^2} \left[ 2 \sin (\tht(t+\tau) - \tht(t))  - \sin (\tht(t) - \tht(t-\tau)) + \sin (\tht(t+\tau) - \tht(t-\tau)) \right]
	\eeq
implying that either $a^2 = 1$ or $\ddot \tht(t + \tau) = \ddot \tht(t)$. However, the latter implies that $\dot \tht(t+\tau) - \dot \tht(t) = -\dot \vf_1(t)$ is a constant which must vanish for the delay algebraic equation (\ref{e:choreography-delay-algebraic}) to be satisfied. Consequently, $\dot \vf_2$ must also vanish implying that the choreography is a uniformly rotating version of G or T.

\subsection{Non-rotating choreographies}

 Remarkably, we have found another 1-parameter family of choreographies (e.g., Fig. \ref{f:choreo-phi1-phi2-plot}) that start out as small oscillations around G. At low energies, they have a period $3 \tau = 2\pi/\om_0$ and reduce to 
	\beq
	\vf_1(t) \approx \sqrt{\fr{2E}{3g}} \sin (\om_0 (t - t_0)) \quad \text{for} \quad E \ll g
	\label{e:choreography-low-egy}
	\eeq
where $\om_0 = \sqrt{3 g/mr^2}$. It is easily verified that (\ref{e:choreography-delay-algebraic}) is identically satisfied while  (\ref{e:choreography-phi-1-2}) is satisfied for $E \ll g$. Moreover, using (\ref{e:choreography-CM-angular-velocity}), we find that the angular speed $\dot \vf_0$ of the CM must vanish for (\ref{e:choreography-low-egy}) so that the energy is purely from the relative motion. The phase trajectory corresponding to (\ref{e:choreography-low-egy}) intersects the $\vf_1 = 0$ Poincar\'e surface at the pair of period-2 fixed points C$(\pm \sqrt{E/2g},0)$ which lie at the centers of the left and right stable `lobes' pictured in Fig. \ref{f:psec-egy=2-3} at $E = 2g$ and $3g$. 

More generally, we numerically find that when the ICs are chosen at the stable fixed points at the centers of these lobes, the trajectories are a one-parameter family of choreographies $\vf_1(t;E)$ varying continuously with $E$ up to $E \approx 5.33$. It can be argued that these choreographies are non-rotating (involve no CM motion). Indeed, from (\ref{e:choreography-CM-angular-velocity}) and (\ref{e:choreography-delay-algebraic}), we must have  $3 \tau \dot \vf_0 \equiv 0 \mod 2\pi$, implying that $\dot\vf_0$ cannot jump discontinuously. Since, $3 \tau \dot \vf_0 = 0$ as $E \to 0$ (\ref{e:choreography-low-egy}), it must remain zero when $E$ is continuously increased from $0$ to $5.33$. Though we do not study their stability here by the monodromy approach, the Poincar\'e sections (see Fig.~\ref{f:psec-egy=2-3} and \ref{f:psec-egy-near-4}) indicate that they are stable. As shown in Fig. \ref{f:choreo-time-vs-egy}, the time period $3 \tau$ grows monotonically with $E$ and appears to diverge at $E \approx 5.33$, which coincides with the beginning of the band of `global chaos' (see \S \ref{s:poincare-section}). For $E \gtrsim 5.33$, the period-2 choreography points C on the `$\vf_1 = 0$' Poincar\'e surface become unstable   and lie in a chaotic region (see Fig. \ref{f:psec-egy=4.5-to-18}), preventing us from finding such a choreography, if it exists, using the above numerical technique. As argued before, choreographies are forbidden at very high energies. For instance, on the `$\vf_1 = 0$' Poincar\'e surface at $E = 18$ (see Fig. \ref{f:psec-egy=18}), the analogues of the C points correspond to unstable periodic orbits which are {\it not} choreographies. In fact, we conjecture that this family of periodic solutions ceases to be a choreography beyond $E \approx 5.33$.

\section{Ergodicity in the band of global chaos}
\label{s:ergodicity}

\begin{figure}	
\centering
	\begin{subfigure}[t]{4cm}
		\centering
		\includegraphics[width=4cm]{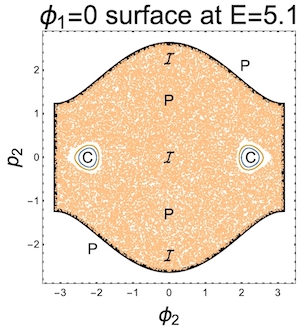}
	\end{subfigure}
	\begin{subfigure}[t]{4cm}
		\centering
		\includegraphics[width=4cm]{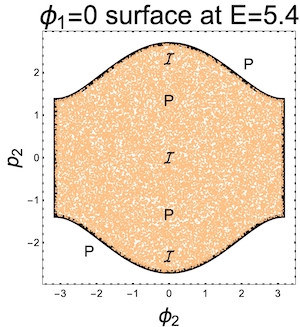}
	\end{subfigure}
	\begin{subfigure}[t]{4cm}
		\centering
		\includegraphics[width=4cm]{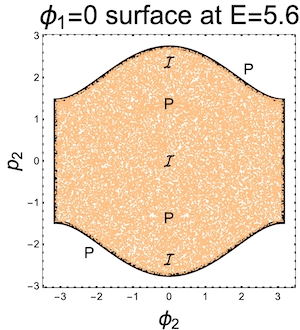}
	\end{subfigure}	
	\begin{subfigure}[t]{4cm}
		\centering
		\includegraphics[width=4cm]{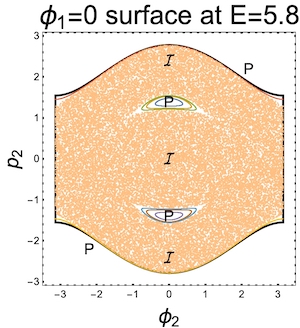}
	\end{subfigure}	
	\caption{\small  Approach to the band of global chaos ($5.33g \leq E \leq 5.6g$) on the Poincar\'e surface $\vf_1 = 0$. The last elliptic islands to cease to exist (as $E \to 5.33g^-$) are around choreographies (C) and the first elliptic islands to open up (when $E$ exceeds $5.6g$) are around pendula (P) which also occur along the Hill boundary. Isosceles solutions intersect this surface at the points marked $\cal I$.}
	\label{f:band-of-global-chaos-choreography-pendula}
\end{figure}

In \S \ref{s:transition-to-chaos-global-chaos}, we found a band of global chaos ($5.33g \leq E \leq 5.6g$) and conjectured ergodic behavior. Intriguingly, the beginning of this band coincides with the divergence in the period of the non-rotating choreographies which additionally cease to exist above this energy (see Fig.~\ref{f:band-of-global-chaos-choreography-pendula}). Similarly, the cessation of this band coincides with the energy at which pendula become stable. In this section, we provide evidence for ergodicity in this band by comparing distributions of $\vf_{1,2}$ and $p_{1,2}$ on constant energy hypersurfaces (weighted by the Liouville measure) with their distributions along generic (chaotic) numerically determined trajectories. For ergodicity, the distribution along a generic trajectory (over sufficiently long times) should be independent of initial condition and tend to the corresponding distribution over the energy hypersurface \cite{arnold-avez,gutzwiller-book}. We also examine the rate of approach to ergodicity in time and deviations from ergodicity outside the band of global chaos. Our numerical and analytical results, while indicative of ergodic behavior, are nonetheless not sufficient to establish it, since we examine only a restricted set of observables.

\subsection{Distributions along trajectories and over energy hypersurfaces}
\label{s:ergodicity-time-ensemble-avgs}

\begin{figure}[!ht]	
\centering
	\includegraphics[width=16cm]{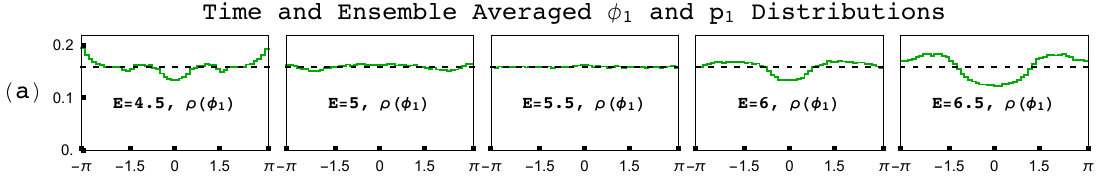}
	\includegraphics[width=16cm]{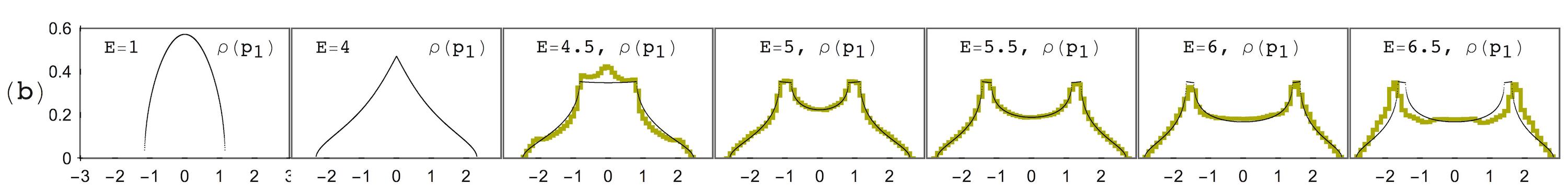}
	\caption{\small Distribution along generic trajectories (yellow, lighter) and distribution over constant energy hypersurface (black, darker) of (a) relative angle ($\vf_1$) and (b) relative momentum ($p_1$) for a range of increasing energies with $m = r = g = 1$. The horizontal axis is $\vf_1$ in (a) and $p_1$ in (b). Note that $\vf_1$ and $\vf_2$ have the same distributions as do $p_1$ and $p_2$. The  distribution along a generic (chaotic) trajectory is found to be insensitive to the IC chosen. The  momentum distribution over constant energy hypersurfaces transitions from a Wigner semi-circle to a bimodal distribution with increasing energy. The two distributions agree only in the band of global chaos $(5.33 \leq E \leq 5.6)$ consistent with ergodicity in this band.}
	\label{f:collage-ensemble-time-avg-dist}
\end{figure}

{\fl \bf Distribution along generic trajectories:} By the {\it distribution function} of a dynamical variable $F(p,\vf)$ (such as $p_1$ or $\vf_1$) along a {\it given trajectory} parametrized by time $t$, we mean
	\beq
	\varrho_F(f) = \lim_{T \to \infty} \ov{T} \int_{0}^T \del(F(p(t),\varphi(t)) - f) \: dt.
	\eeq
The time average of $F$ along the trajectory is then given by the first moment $\bra F \ket_{\rm t} = \int f \:\varrho_F(f) df$. In practice, to find the distribution of $F$, we numerically evolve a trajectory starting from a random initial condition (IC) and record the values $f$ of $F$ at equally spaced intervals of time (say, $\Delta t = .25$) up to $t_{\rm max} = 3 \times 10^5$ in units where $g = m = r = 1$. For such $t_{\rm max}$ and for energies in the globally chaotic band, we find that the histograms of recorded values  approach asymptotic distributions (see Fig.~\ref{f:collage-ensemble-time-avg-dist}) that are largely independent of the choice of $\Delta t$ and ICs.

{\fl \bf Distributions over energy hypersurfaces:} The ensemble average $\bra \cdot \ket_{\rm e}$ of a dynamical variable $F(p,\vf)$ at energy $E$ is defined with respect to the Liouville volume measure on phase space. Since $\vf_i$ and $p_j$ are canonically conjugate, we have
	\beq
	\bra F \ket_{\rm e}  = \ov{V_E} \int F \;\del(H - E) \; d\vf_1 \, d\vf_2 \, dp_1 \, dp_2 
	\quad \text{where} \quad	
	V_E = \int \del(H - E) d\vf_1 d\vf_2 dp_1 dp_2
	\label{e:ensemble-avg-distr-vol-egy-surf}
	\eeq
is the volume of the $H = E$ energy hypersurface $M_E$. More generally, the {\it distribution} of $F(p,\vf)$ over the energy $E$ hypersurface weighted by the Liouville measure is the following phase space integral:
 	\beq
	\rho_{F,E}(f) = \ov{V_E} \int \del(F(p,\vf) - f) \del(H - E) d\vf_1 d\vf_2  dp_1 dp_2.
	\eeq
Loosely, it is like the Maxwell distribution of speeds in a gas. We will often omit the subscripts $F$ and/or $E$ when the observable and/or the energy are clear from the context. By definition, the above distribution is a probability density: $\int \rho(f) df = 1$. The ensemble average $\bra F \ket_{\rm e}$ is its first moment:
	\beq 
	\bra F \ket_{\rm e} = \int f \: \rho_{F,E}(f) \: df.
	\eeq
To find distributions over an energy hypersurface $M_E$, we need to integrate over it. For instance, to find the volume $V_E$ of the energy hypersurface, we observe that the Hamiltonian $H = \T + \V$ is quadratic in $p_2$ where
	\beq
	\T =  \fr{p_1^2 + p_2^2 - p_1 p_2}{m r^2} 
	\quad \text{and} \quad
	\V(\vf_1, \vf_2) = g \left[3 - \cos \vf_1  - \cos \vf_2  - \cos(\vf_1 + \vf_2) \right]. \quad
	\label{e:hamiltonian}
	\eeq
Hence, we cover $M_E$ by two coordinate patches parametrized by $\vf_1, \vf_2$ and $p_1$ with
	\beq
	p_2^\pm = \half \left( p_1 \pm \sqrt{ 4 m r^2 (E-\V(\vf_1, \vf_2)) - 3 p_1^2} \right).
	\label{e:p2PM}
	\eeq
Using the factorization $H - E = (p_2 - p_2^+) (p_2 - p_2^-)$, we evaluate the integral over $p_2$ in Eq. (\ref{e:ensemble-avg-distr-vol-egy-surf}) to arrive at
	\beq
V_E =   \iint\displaylimits_{(\vf_1, \vf_2) \in {\cal H}_E} d\vf_1 \, d\vf_2 \int\displaylimits_{- p_{\rm max}}^{p_{\rm max}} \fr{dp_1}{(p_2^+ - p_2^-)}
	\eeq
where $p_{\rm max} = \sqrt{4mr^2 (E - \V)/{3}}$. Here, $\vf_{1,2}$ are restricted to lie in the Hill region ${\cal H}_E$ ($\V \leq E$). Interestingly, the integral over $p_1$ is independent of $\vf_1$ and $\vf_2$ as well as $E$ so that
	\beq
	\int\displaylimits_{- p_{\rm max}}^{p_{\rm max}} \fr{dp_1}{(p_2^+ - p_2^-)} = \fr{\pi}{\sqrt{3}} 
	\;\; \imply \;\;
	V_E = \fr{\pi}{\sqrt{3}} \times \text{Area}({\cal H}_E).
	\label{e:p1-integral}
	\eeq
Here, Area(${\cal H}_E$) is the area of the Hill region with respect to the measure $d\vf_1 d\vf_2$. It is a monotonically increasing function of $E$ and saturates at the value $4\pi^2$ for $E \geq 4.5$ when the Hill region includes the entire $\vf_1$-$\vf_2$ torus. We now derive formulae for distributions over energy hypersurfaces.

{\fl \bf Distribution of angles:} The {\it joint distribution function} of $\vf_1$ and $\vf_2$ is given by ($p_2^\pm$ are as in (\ref{e:p2PM}))
	\beq
	\rho_E(\vf_1, \vf_2) = \ov{V_E} \int \del(H - E) \; dp_1 \, dp_2 
	 = \ov{V_E} \int\displaylimits_{- p_{\rm max}}^{p_{\rm max}} \fr{dp_1}{(p_2^+ - p_2^-)} = \fr{\pi}{V_E\sqrt{3}},
	\eeq
since from (\ref{e:p1-integral}), the integral over $p_1$ is $\pi/\sqrt{3}$ for all $E$ and $\vf_{1}$. In other words, ($\vf_1,\vf_2$) is uniformly distributed on the Hill region. Furthermore, for $E \geq 4.5$, the Hill region is the whole torus and $\rho_E(\vf_1, \vf_2) = 1/4\pi^2$. Thus, $\vf_1$ and $\vf_2$ are each uniformly distributed on $[0,2\pi]$ for $E \geq 4.5$. Fig.~\ref{f:collage-ensemble-time-avg-dist}a shows that the distributions of $\vf_1$ and $\vf_2$ along a trajectory with energy $E = 5.5$ in the band of global chaos agrees with this uniform phase space distribution (the fractional deviation is at most .2 \% across all angles).

\begin{figure}	
	\centering
	\begin{subfigure}[t]{16cm}
		\centering
		\includegraphics[width=16cm]{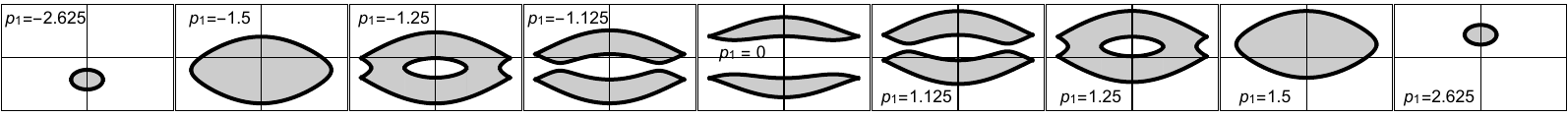}
	\end{subfigure}
	\caption{\small The energetically allowed portion (shaded gray) of the $\vf_2$-$p_2$ Poincar\'e surface for a sequence of increasing values of $p_1$ at $E = 5.5$ in the band of global chaos for $m = r = g = 1$.  On each plot, the horizontal axis is $\vf_2 \in [-\pi, \pi]$ and the vertical axis is $p_2 \in [-3, 3]$. The value of the distribution function $\rho_E(p_1)$ is the Liouville area of the shaded region. It is plausible that $\rho_E(p_1)$ is even and that as $p_1$ goes from $0$ to $p_{\rm max} = \sqrt{4 m r^2 E/3} \approx 2.71$, $\rho_E(p_1)$ initially increases from a non-zero local minimum, reaches a maximum and then drops to zero as shown in the $E = 5.5$ subfigure of Fig. \ref{f:collage-ensemble-time-avg-dist}b.}
	\label{f:collage-psec}
\end{figure}

{\fl \bf Distribution of momenta:} The momentum distribution functions turn out to be more intricate. Due to the $1\leftrightarrow 2$ symmetry of the Hamiltonian (\ref{e:hamiltonian}), the 1-particle momentum distribution functions $\rho_E(p_1)$ and $\rho_E(p_2)$ are equal and given by the marginal distribution
	\beq
	\rho_E(p_1) = \ov{V_E} \int \del(H - E) \; d\vf_1 \, d\vf_2 \, dp_2 
	= \ov{V_E} \iint\displaylimits_{(\vf_1, \vf_2) \in {\cal H}_{E,p_1}} \fr{d\vf_1 \, d\vf_2}{p_2^+ - p_2^-}.
	\eeq
Here, ${\cal H}_{E,p_1}$ is the portion of the $\vf_1$-$\vf_2$ torus allowed for the given values of $E$ and $p_1$. Since $p_2^\pm$ must be real, from (\ref{e:p2PM}) we see that $4mr^2 (E - \V) - 3 p_1^2 \geq 0$ or $\V \leq E - 3 p_1^2/4 mr^2$. Thus, $\vf_1$ and $\vf_2$ must lie in the Hill region for the modified energy $E' = E - 3 p_1^2/4 mr^2$. For this Hill region to be non-empty, we must have $E' \geq 0$. Thus, the distribution function $\rho_E(p_1)$ is supported on the interval $[-\sqrt{4 m r^2 E/3}, \sqrt{4 m r^2 E/3}]$ and is given by 
	\beq
	\rho_E(p_1) = \ov{V_E} \iint_{{\cal H}_{E'}} \fr{d\vf_1 \, d\vf_2}{ \sqrt{4 m r^2 (E'(p_1) - \V)}}.
	\label{e:rhoe-p1-general}
	\eeq
On account of $E'(p_1)$ being even, $\rho_E(p_1) = \rho_E(-p_1)$. Upon going to Jacobi coordinates $\vf_\pm = (\vf_1 \pm \vf_2)/2$, the integral over $\vf_-$ can be expressed in terms of an incomplete elliptic integral of the first kind. Though the resulting formulae are lengthy in general, for low energies $\rho_E(p_1)$ turns out to be the Wigner semi-circular distribution (see Fig. \ref{f:collage-ensemble-time-avg-dist}b). Indeed, upon going to Jacobi coordinates and using the quadratic approximation for the potential $\V_{\rm low} = 3 g \vf_+^2 + g \vf_-^2$, we find that at low energies, the Hill region ${\cal H}_{E'}$ is the elliptical disk $ 3 g \vf_+^2 + g \vf_-^2 \leq E'(p_1)$. Thus, 
	\beq
	V_E = \fr{\pi}{\sqrt{3}} \times {\rm Area}({\cal H}_E) = \fr{2 \pi^2 E}{3 g} \quad \text{for} \quad E \ll g
	\eeq
leading to the Wigner semi-circular distribution
	\beq
	\rho_E (p_1) \approx \ov{V_E} \iint_{{\cal H}_{E'}} \fr{2 d\vf_+ d\vf_-}{\sqrt{4 m r^2(E'(p_1) - \V_{\rm low})}} 
	= \fr{3}{2 \pi m r^2 E} \sqrt{\fr{4}{3} m r^2 E - p_1^2} \quad \text{for} \quad E \ll g.
	\eeq
For larger $E$, we perform the integral (\ref{e:rhoe-p1-general}) numerically. Fig. \ref{f:collage-ensemble-time-avg-dist}b shows that the distribution goes from being semi-circular to bimodal as $E$ crosses $4g$. Loosely, $\rho_E(p_1)$ is the analogue of the Maxwell distribution for the relative momenta of the three-rotor problem. Fig. \ref{f:collage-psec} provides a qualitative explanation of the bimodal shape of $\rho_E(p_1)$ for an energy in the band of global chaos. Fig. \ref{f:collage-ensemble-time-avg-dist}b shows that the distribution of $p_1$ along a generic trajectory closely matches its distribution $\rho_E(p_1)$ over the constant energy hypersurface in the band of global chaos $(5.33 \leq E \leq 5.6)$ but deviates at other energies, providing evidence for ergodic behavior in this band.

\subsection{Approach to ergodicity}
\label{s:approach-to-ergodicity}

\begin{figure}	
	\centering
	\begin{subfigure}[t]{5.3cm}
		\centering
		\includegraphics[width=5.3cm]{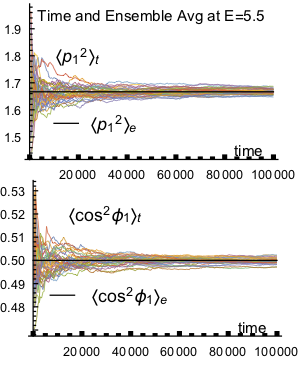}
		\caption{\small }
	\end{subfigure}
\quad
	\begin{subfigure}[t]{10cm}
		\centering
		\includegraphics[width=10cm]{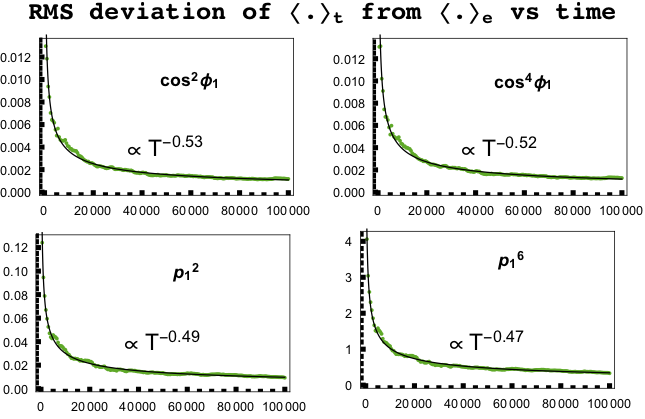}
		\caption{\small }
	\end{subfigure}	
	\caption{\small (a) Time averages $\bra p_1^2 \ket_{\rm t}$ and $\bra \cos^2 \vf_1 \ket_{\rm t}$ as a function of time $T$ for 35 randomly chosen trajectories at $E = 5.5$. They are seen to approach the corresponding ensemble averages ($\bra \cdot \ket_{\rm e}$ indicated by thick black lines) as time grows. (b) Root mean square deviation (over 35 chaotic initial conditions) of time averages from the corresponding ensemble average as a function of time $T$ for $E = 5.5$ in the band of global chaos for the observables $\cos^2 \vf_1$, $\cos^4 \vf_1$, $p_1^2$ and $p_1^6$. The fits show a $T^{-1/2}$ approach to ergodicity.}
	\label{f:mean-vs-time}
\end{figure}

To examine the rate of approach to ergodicity for energies in the band of global chaos, we compare ensemble averages of variables such as $\cos^2 \vf_1$ and $p_1^2$ with their time averages over increasingly long times.

{\fl \bf Ensemble average:} The ensemble average $\bra \cdot \ket_{\rm e}$ of a variable $F$ at energy $E$ defined in (\ref{e:ensemble-avg-distr-vol-egy-surf}) reduces to 
	\beq
	\bra F \ket_{\rm e} = \ov{V_E} \iint\displaylimits_{(\vf_1, \vf_2) \in {\cal H}_E} d\vf_1 \, d\vf_2 \int\displaylimits_{- p_{\rm max}}^{p_{\rm max}} dp_1 \: \fr{F(\vf_1, \vf_2, p_1, p_2^+) + F(\vf_1, \vf_2, p_1, p_2^-)}{2(p_2^+ - p_2^-)}
	\eeq
upon using the factorization $H - E = (p_2 - p_2^+) (p_2 - p_2^-)$ to evaluate the integral over $p_2$. Since for $E \geq 4.5$, $\vf_1$ and $\vf_2$ are independently uniformly distributed on $[0,2\pi]$, we have
	\beq
	\bra \cos^{m} \vf_1 \cos^{n} \vf_2 \ket_{\rm e} = \bra \cos^{m} \vf_1 \ket_{\rm e} \bra \cos^{n} \vf_2 \ket_{\rm e}
	\eeq
with $\bra \cos^{2n} \vf_1 \ket_{\rm e} = \fr{(2n)!}{2^{2n} (n!)^2}$ and the odd moments vanishing. Remarkably, the phase space averages of momentum observables are also exactly calculable for $E \geq 4.5$:
	\beqs
	\bra p_1^2 \ket_{\rm e} &=& 2E/3 - 2, \quad
	\bra p_1^4 \ket_{\rm e} = 2E^2/3  - 4 E + 7, \cr
	\bra p_1^2 p_2^2 \ket_{\rm e} &=& E^2/3 - 2 E + 7/2 
	\;\; \text{and} \;\;
	\bra p_1^6 \ket_{\rm e} = 20 E^3/27 - 20E^2/3 + {70E}/{3} - {260}/{9}. \;\;
	\eeqs
Though we restrict to $E \geq 4.5$ to obtain simple formulae for ensemble averages, this includes the band of global chaos $5.33 \leq E \leq 5.6$ where {\it alone} we can expect ergodic behavior. 

To compare with time averages, for each energy, we pick $N_{\rm traj} = 35$ random ICs (on the $\vf_1 = 0$ surface) and evolve them forward. As Fig. \ref{f:mean-vs-time}a indicates, though the time averages ($\ov{T} \int_0^T F \: dt$) display significant fluctuations at early times, they have approached their asymptotic values by $T = 10^5$. To estimate the rate of approach to ergodicity, we compute the root mean square deviation $\sigma(T)$ of the time average from the ensemble average as a function of time:
	\beq
	\sigma^2(T) = \ov{N_{\rm traj}} \sum_a \left( \bra F \ket_{t, a}(T)  - \bra F \ket_{\rm e} \right)^2 \quad
	\text{where} \quad \bra F \ket_{t, a}(T) = \ov{T} \int_0^T F(t'_a) \: dt'_a
	\eeq
is the time average over the $a^{\rm th}$ trajectory. Fig. \ref{f:mean-vs-time}b shows that for several variables $F = \cos^2 \vf_1, p_1^2$ etc., the mean square deviation decays roughly as the reciprocal of time, $\sig \sim 1/\sqrt{T}$, as expected of an ergodic system where correlations decay sufficiently fast as shown in Appendix \ref{a:approach-to-ergodicity} (see also \cite{prl-dechant} for a stochastic formulation).

\begin{figure}	
	\centering
	\begin{subfigure}[t]{8cm}
		\centering
		\includegraphics[width=7.5cm]{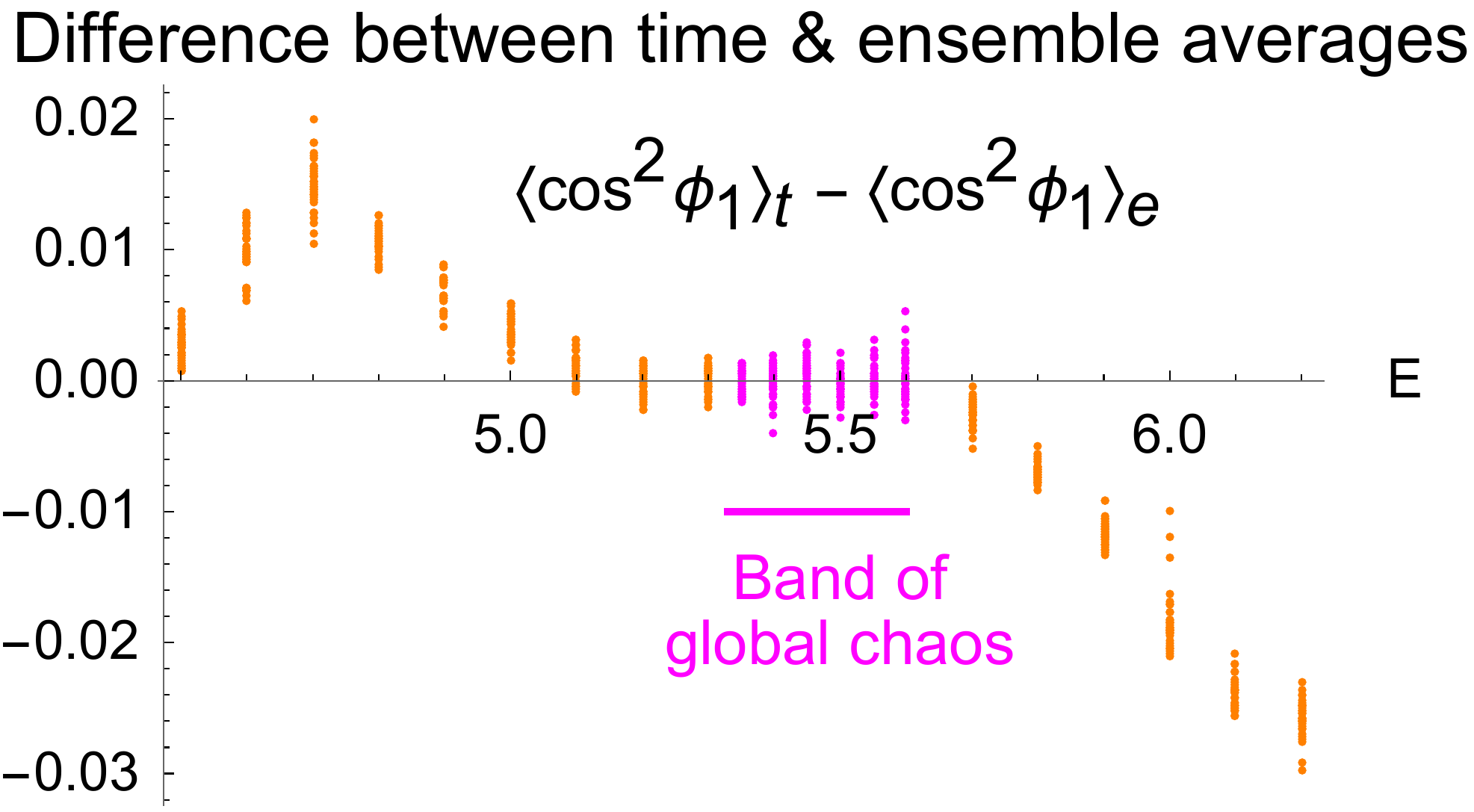}
	\end{subfigure}
	\begin{subfigure}[t]{8cm}
		\centering
		\includegraphics[width=7.5cm]{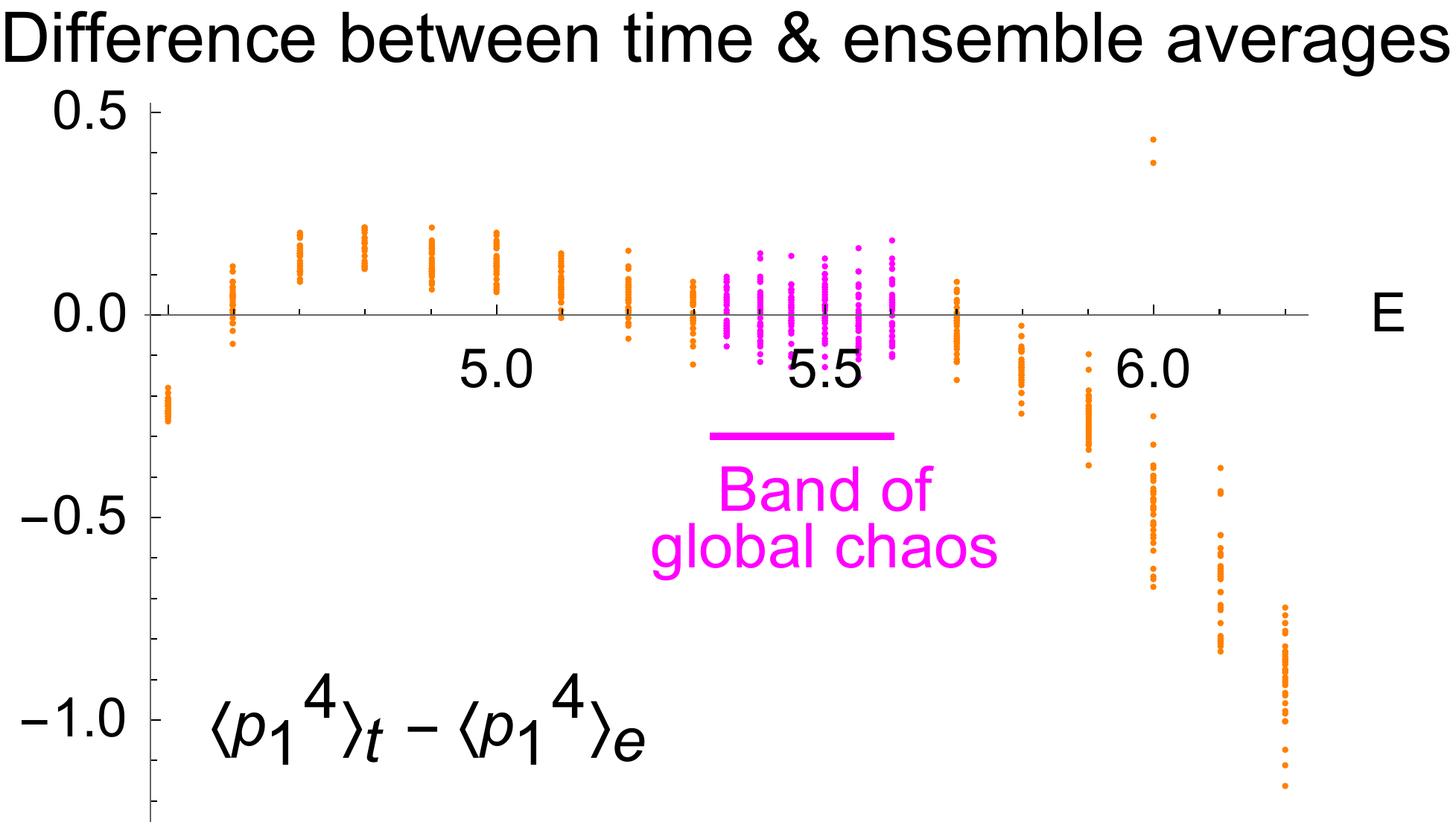}
	\end{subfigure}	
	\caption{ \small  Difference between time averages $\bra \cdot \ket_{\rm t}$ over a time $T = 10^5$ (for 35 randomly chosen chaotic trajectories) and ensemble average $\bra \cdot \ket_{\rm e}$ for $\cos^2 \vf_1$ and $p_1^2$ indicating ergodicity in the band of global chaos $5.33 \leq E \leq 5.6$ (magenta) and discernible departures outside this band (orange). The spread in $\bra \cdot \ket_{\rm t} - \bra \cdot \ket_{\rm e}$  at a fixed energy is due to the finiteness of $T$. However, this spread is small compared to  the average values $\bra \cos^2 \vf_1 \ket_{\rm e} = .5$ and $\bra p_1^4 \ket_{\rm e} = 2E^2/3 - 4E + 7$ demonstrating that time averages over distinct chaotic trajectories converge to a common value. Note that the spread in $\bra p_1^4 \ket_{\rm t} - \bra p_1^4 \ket_{\rm e}$ increases with $E$ as the average values themselves increase with $E$.}
	\label{f:mean-vs-egy}
\end{figure}

Finally, we examine the approach to ergodicity as the energy approaches the band of global chaos $5.3 \lesssim E \lesssim 5.6$. To this end, we compare the  ensemble averages of a few variables with their time averages for 35 randomly chosen {\it chaotic trajectories} over a range of energies. Fig. \ref{f:mean-vs-egy} shows that the time averages of $\cos^2 \vf_1$ and $p_1^2$ agree reasonably well with their ensemble averages in the band of global chaos. At lower and higher energies, there are discernible deviations from the ensemble averages, showing ergodicity breaking. (a) For $E$ slightly outside the band of global chaos, we find that there is a single chaotic region (see Fig. \ref{f:band-of-global-chaos-choreography-pendula}), and time averages along trajectories from this region converge to a common value which however differs from the ensemble average over the whole energy hypersurface (see Fig. \ref{f:mean-vs-egy}). (b) At energies significantly outside the band of global chaos, there can be several distinct chaotic regions (see Fig. \ref{f:psec-egy=18}). We find that time averages of an observable along chaotic trajectories from these distinct regions generally converge to different values, none of which typically agrees with the ensemble average over the whole energy hypersurface.


\section{Mixing in the band of global chaos}
\label{s:mixing}

\begin{figure}	
	\centering
	\begin{subfigure}[t]{8cm}
		\centering
		\includegraphics[width=8cm]{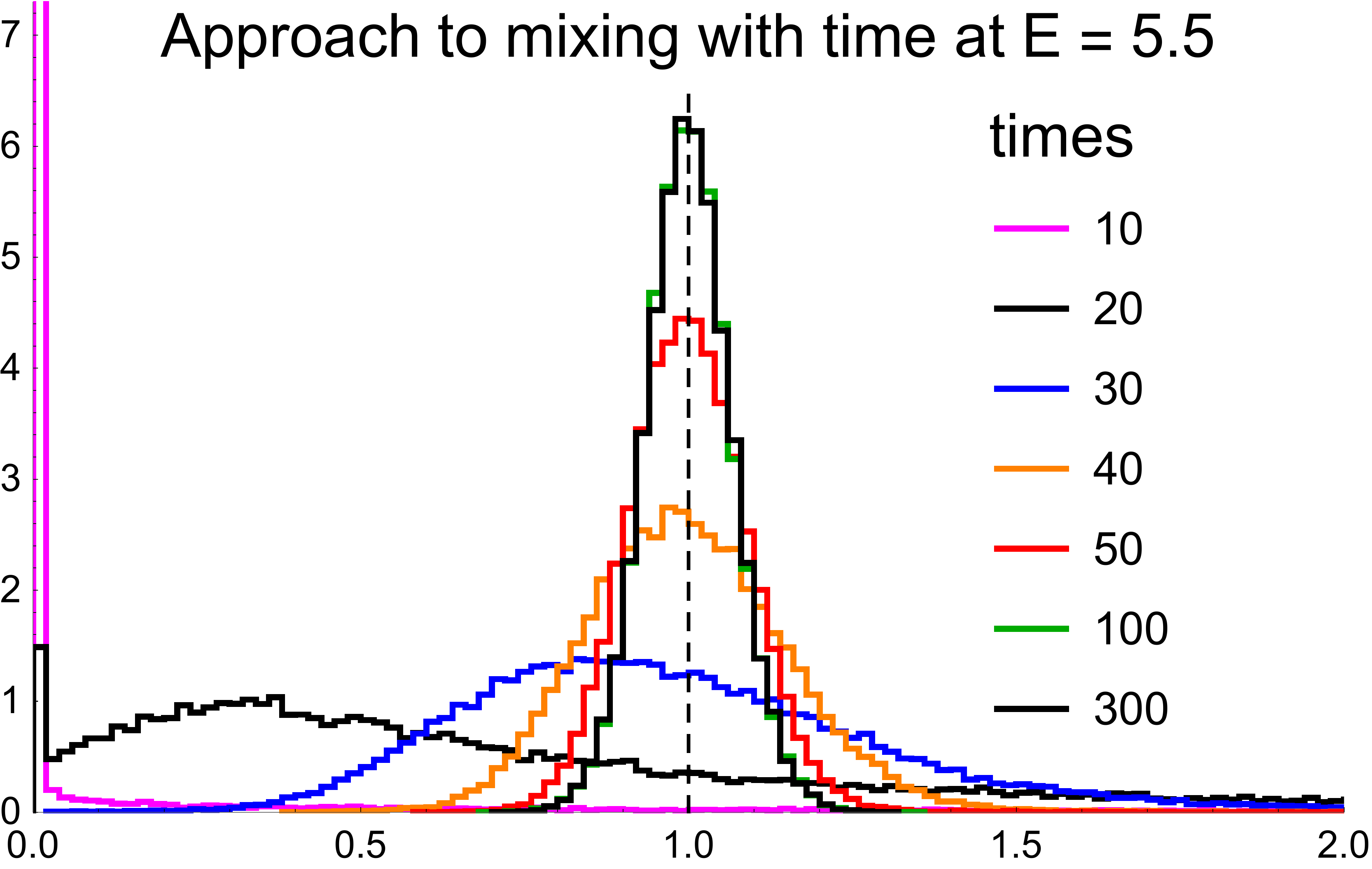}
		\caption{ \small  }
		\label{f:approach-to-mixing-in-time}
	\end{subfigure}
\quad
	\begin{subfigure}[t]{7.3cm}
		\centering
		\includegraphics[width=7.3cm]{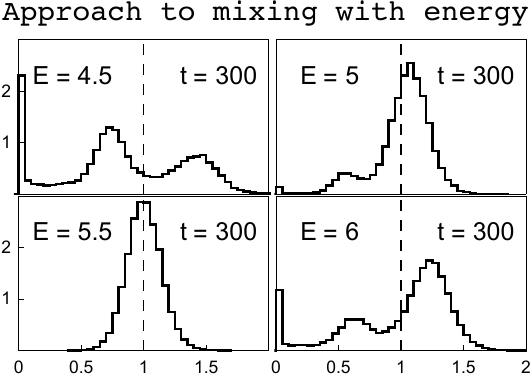}
		\caption{ \small  }
		\label{f:approach-to-mixing-in-energy}
	\end{subfigure}		
	\caption{ \small Histograms of number of trajectories $n_i(t)$ in each cell $i$ of an energy hypersurface. To facilitate comparison across energies and numbers of ICs considered, the histograms of $\tilde n_i(t) = (n_i(t) V_E) / (\mu_i N)$ (see Eq. \ref{e:mean-counts-mixing}) are displayed. For the flow to be mixing, the histograms should strongly peak around $\tilde n_i(t) = 1$. Fig. (a) shows the approach to mixing in time at an energy $E = 5.5$ in the band of global chaos. The histogram is seen to migrate from peaking at zero to 1 with advancing time. Fig. (b) shows these histograms at reasonably late times ($t=300$) showing how the flow becomes mixing as we approach the band of global chaos (represented here by $E = 5.5$).}
	\label{f:approach-to-mixing}
\end{figure}

In \S \ref{s:ergodicity}, we provided numerical evidence for ergodicity in the three-rotor problem for energies in the band of global chaos. We now investigate whether the dynamics is mixing in this regime. A flow $\phi_t$ on the energy hypersurface $M_E$ of the phase space is said to be strongly mixing if for all subsets $A, B \subseteq M_E$ with positive measures ($\mu(A)>0$ and $\mu(B) >0$), we have
	\beq
	\lim_{t \to \infty} \mu(\phi_t(B) \cap A) = \mu(B) \times \mu(A)/\mu(M_E)
	\eeq	
where $\mu$ is the Liouville volume measure on $M_E$ \cite{arnold-avez,gutzwiller-book}. To numerically examine whether the dynamics of three-rotors is mixing in the band of global chaos, we work in units where $m=r=g=1$ and consider a large number $N$ ($= 1.3 \times 10^7$) of random ICs with energy $E$ in a small initial region of phase space (e.g., $|\vf_{1,2}|, |p_1| < .05$ with $p_2 = p_2^+$ (\ref{e:p2PM}) determined by $E$). The trajectories are numerically evolved forward in time and their locations recorded at discrete time intervals (e.g., $t = 10$, 20, $\cdots$, 300). If the dynamics is mixing, then in the limit $N \to \infty$ and $t \to \infty$, the number of trajectories located at time $t$ in a Liouville volume $V$ must equal $N V/V_E$ where $V_E$ is the Liouville volume of the energy hypersurface. Poincar\'e sections (see Fig.~\ref{f:band-of-global-chaos-choreography-pendula}) as well as investigations of ergodicity in \S \ref{s:ergodicity} rule out the possibility of mixing for energies outside the regime of global chaos. Thus, we restrict to $5.33 \leq E \leq 5.6$ where $V_E = {4\pi^3}/\sqrt{3}$, a formula that holds for any $E \geq 4.5$ (\ref{e:p1-integral}). Now, for convenience, we divide the 3d energy hypersurface into cuboid-shaped cells of equal geometric volume $V^g$. The Liouville volumes of these cells are not equal, so we denote by $\mu_i$ the Liouville volume of the $i^{\rm th}$ cell. In practice, we take cells of linear dimensions ${2\pi}/{d}$ each in $\vf_1$ and $\vf_2$ and ${2p_1^{\rm max}}/{d}$ in $p_1$ where $d = 40$ is the number of subdivisions and $p_1^{\rm max}$ the maximal value of $p_1$ corresponding to energy $E$. Though we compute $\mu_i$ exactly, it is approximately $V^g \times$ the Liouville density at the center of the $i^{\rm th}$ cell:
	\beq
	\mu_i \approx \ov{2(p_2^+ - p_2^-)} \times \fr{2\pi}{d} \times \fr{2\pi}{d} \times \fr{2p_1^{\rm max}}{d}
	\eeq 
where $p_2^\pm$ (\ref{e:p2PM}) are evaluated at the center of the cell. Cells that lie outside or straddle the boundary of the energy hypersurface are not considered. At various times, we record the instantaneous locations of the trajectories and count the number $n_i(t)$ of trajectories that lie in the cell $i$. If the dynamics is mixing, the number of trajectories in the $i^{\rm th}$ cell should be
	\beq
	n_i = N \times \fr{\mu_i}{V_E}.
	\label{e:mean-counts-mixing}
	\eeq 

\begin{figure}[!ht]	
	\centering
		\includegraphics[width=9cm]{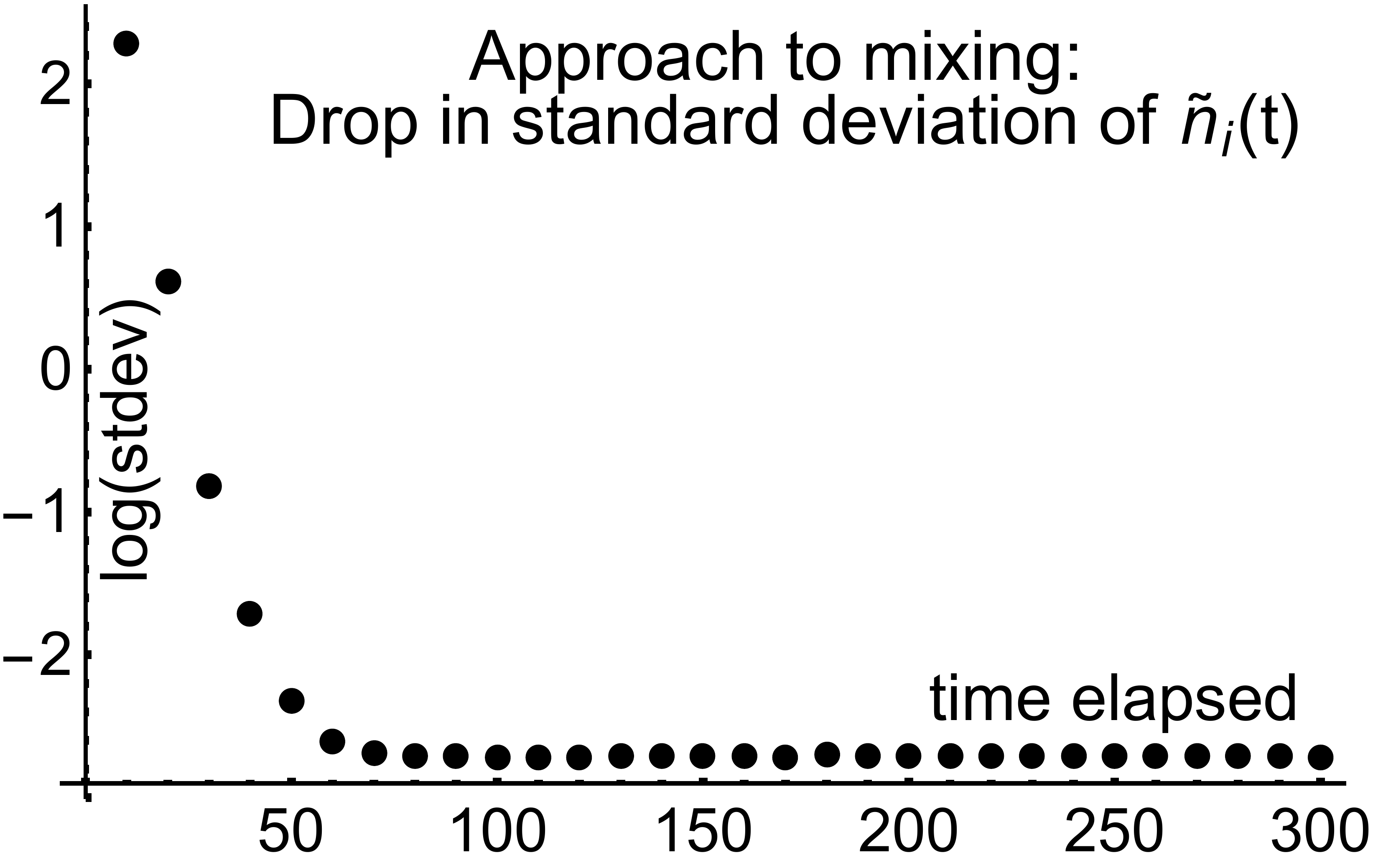}
	\caption{\small Drop with time of the standard deviation of the distribution (see Fig.~\ref{f:approach-to-mixing-in-time}) of the scaled number of trajectories $\tilde n_i(t)$ in each cell of the energy $E=5.5$ hypersurface.  The latter is partitioned into $N_{\rm cells} \approx 4 \times 10^4$ cells and $N = 1.3 \times 10^7$ trajectories have been considered. The plot shows that the standard deviation has dropped to $0.066$ at $t = 300$. This is close to the expected standard deviation $0.055$ if the $N$ trajectories were distributed uniformly among the $N_{\rm cells}$ cells at the instant considered.}
	\label{f:stdev-of-ntilde}
\end{figure}

\begin{figure}[!ht]	
	\centering
	\begin{subfigure}[t]{7cm}
		\centering
		\includegraphics[width=7cm]{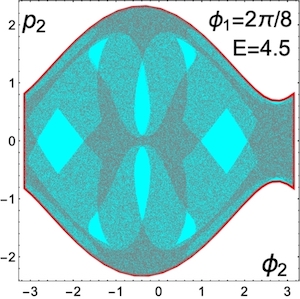}
	\end{subfigure}
\qquad \quad
	\begin{subfigure}[t]{7cm}
		\centering
		\includegraphics[width=7cm]{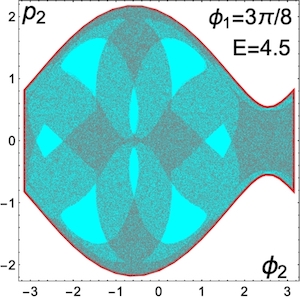}
	\end{subfigure}
\caption{\small Two distinct densities (shaded dark and light) of points (from trajectories for $0 \leq t \leq 10^5$) on chaotic sections of Poincar\'e surfaces at $E = 4.5$ corresponding to the two non-zero peaks in the histogram of Fig.~\ref{f:approach-to-mixing-in-energy} showing characteristic departure from mixing. The unshaded regions are energetically allowed but are not visited by these chaotic trajectories and correspond to the peak around zero in the same histogram.}
\label{f:psec-outside-globalchaos}
\end{figure}

To test the mixing hypothesis and the rate of approach to mixing, we plot in Fig.~\ref{f:approach-to-mixing} at various times $t = 10, 20, \cdots, 300$, a histogram of $n_i(t)$. To be more precise, we plot a histogram of $\tilde n_i(t) = n_i(t) V_E / (\mu_i N)$ so that the expected mean is 1, to facilitate comparison across energies, times and numbers of ICs considered. At very early times ($t\lesssim 10$), most cells have not been visited by trajectories, so that the histogram is strongly peaked around zero counts. As $t$ increases, we observe from Fig.~\ref{f:approach-to-mixing-in-time} that the histograms shift, and become progressively narrower, peaking around the expected value of 1 with the expected width (see Fig.~\ref{f:stdev-of-ntilde}). This provides evidence for mixing in the regime of global chaos. In Fig.~\ref{f:approach-to-mixing-in-energy}, we compare these histograms at sufficiently late times ($t = 300$) for a range of energies and observe significant departures from mixing for energies outside the band of global chaos. In fact, for energies such as $E = 4.5$ and $E = 6$, the histograms in Fig.~\ref{f:approach-to-mixing-in-energy} show three distinct peaks corresponding to cells that are never visited and two other types of cells (in chaotic regions) that are visited with unequal frequencies  (see Fig.~\ref{f:psec-outside-globalchaos}). This characteristic departure from mixing with respect to the Liouville measure (even when restricted to chaotic regions) is also reflected in the two distinct densities of points in Poincar\'e plots at such energies, as seen in Fig.~\ref{f:psec-outside-globalchaos}.

\section{Recurrence time statistics}
\label{s:recurrenc-time-dist}

\begin{wrapfigure}[10]{R}{6cm}
\centering
\includegraphics[width=5cm]{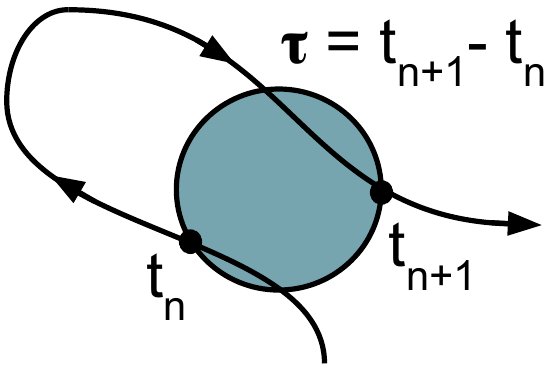}
\caption{\small Recurrence time $\tau$.}
\label{f:rec-time-def}
\end{wrapfigure}
Here, we study the statistics of Poincar\'e recurrence times to a three-dimensional cell in an energy-$E$ hypersurface of the phase space. For convenience, we choose the cell to be a cuboid of width $w$, e.g, $-w/2 \leq \vf_1, \vf_2, p_1 \leq w/2$ with $p_2 = p_2^+$ (\ref{e:p2PM}) determined by energy for a cell centered at the origin. We choose a large number ($\sim 3 \times 10^4$) of initial conditions distributed uniformly randomly within the cell and numerically evolve them forward in time. The recurrence time $\tau$ for a given trajectory is defined as the time from the first exit to the next exit from the cell (see Fig. \ref{f:rec-time-def}) \cite{zaslavsky}. Evidently, starting from the instant the trajectory first exits the cell, $\tau$ is the sum of the times it spends outside the cell and while traversing the cell. A histogram of the recurrence times (normalized to be a probability distribution) is then plotted as in Fig. \ref{f:recurrence-time-distribution-a}.

\subsection{Exponential law}

\begin{figure}[!ht]	
	\centering
	\begin{subfigure}[t]{8cm}
		\centering
		\includegraphics[width=8cm]{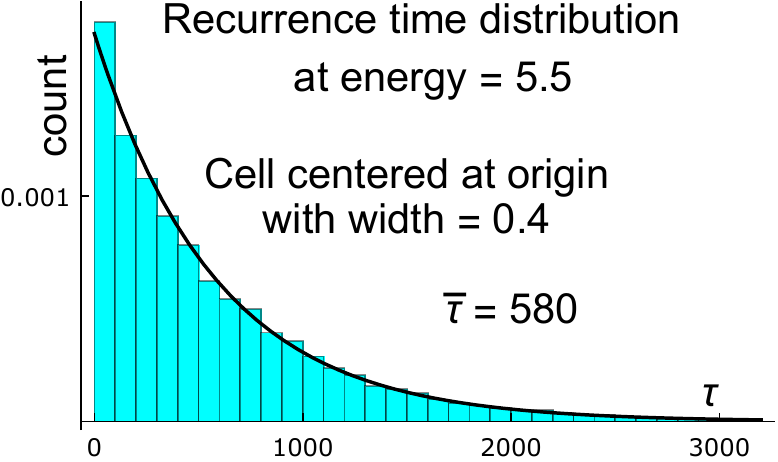}
		\caption{ \small   }
		\label{f:recurrence-time-distribution-a}
	\end{subfigure}
\quad
	\begin{subfigure}[t]{8cm}
		\centering
		\includegraphics[width=8cm]{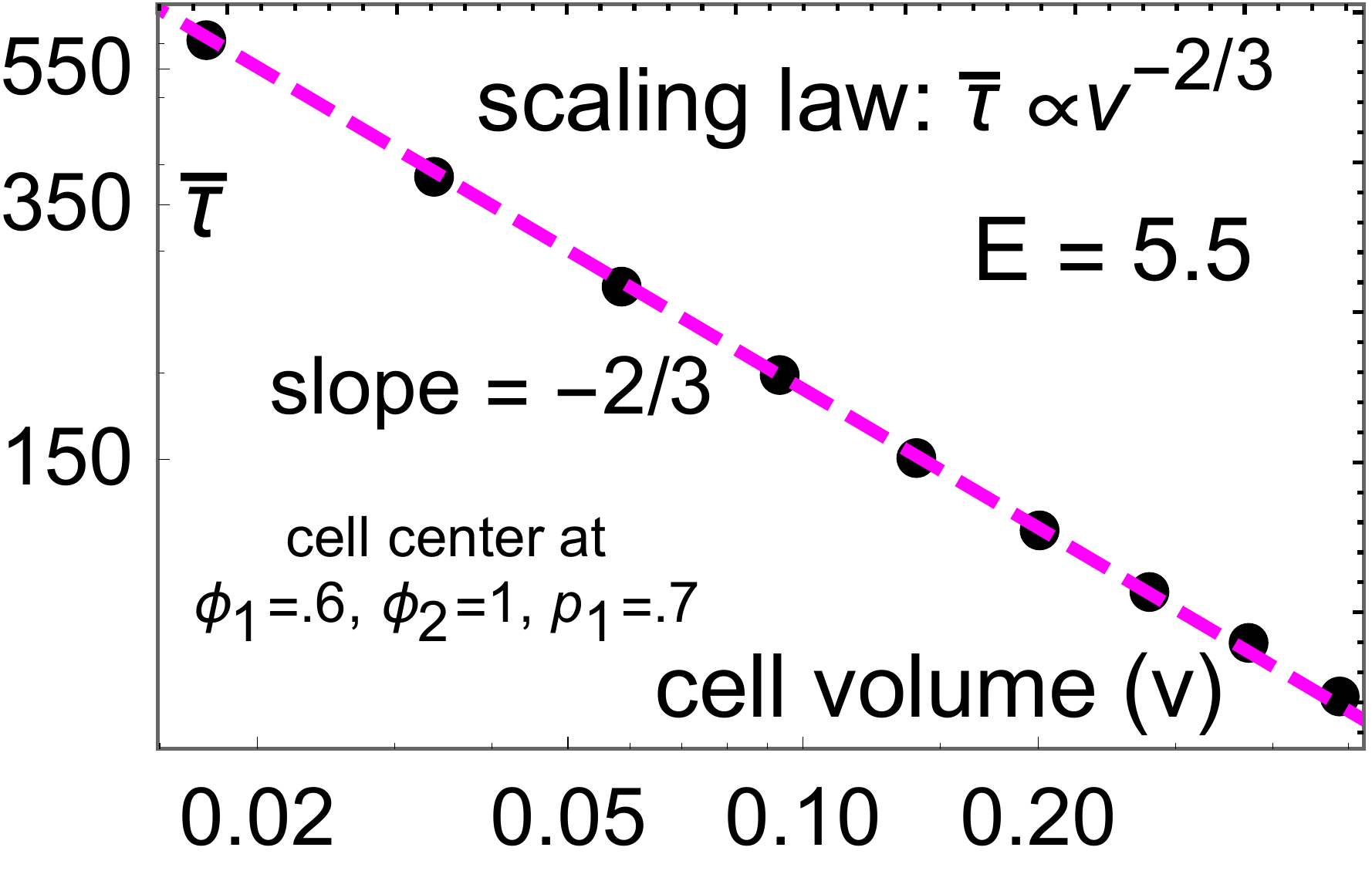}
		\caption{ \small   }
		\label{f:recurrence-time-distribution-b}
	\end{subfigure}
	\begin{subfigure}[t]{8cm}
		\centering
		\includegraphics[width=8cm]{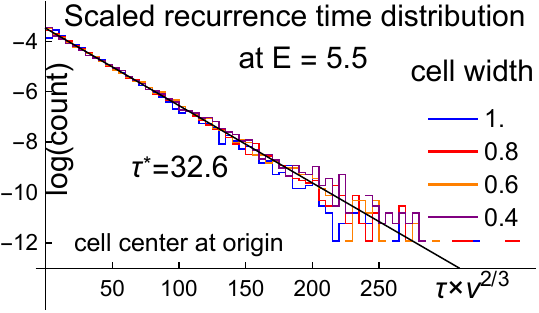}
		\caption{ \small   }
		\label{f:recurrence-time-distribution-c}
	\end{subfigure}
\quad
	\begin{subfigure}[t]{8cm}
		\centering
		\includegraphics[width=8cm]{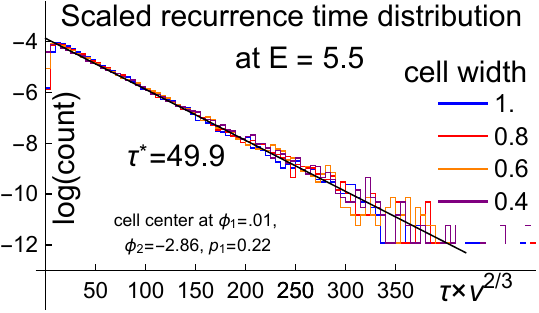}
		\caption{ \small   }
		\label{f:recurrence-time-distribution-d}
	\end{subfigure}
	\caption{ \small   (a) Histogram of recurrence times (normalized to be a probability distribution) for a cubical cell centered at the origin ($p_1=\vf_1=\vf_2= 0$) of the globally chaotic energy-$5.5$ (in units where $m = r = g = 1$) hypersurface showing an exponential law $(1/\bar\tau)\exp(-\tau/\bar\tau)$ where $\bar\tau$ is the fitted mean recurrence time. Note that $\bar \tau \approx 580$ is much larger than the time scale of the linearized system ($1/\om_0 =\sqrt{mr^2/3g}$). (b) At any cell location, $\bar \tau$ scales as the minus two-thirds power of the Liouville volume of the cell, consistent with ergodicity. (c, d) Normalized histogram of (recurrence times) $\times$ (cell volume)$^{2/3}$ plotted on a log-linear scale for cells of various widths, showing a universal exponential distribution $(1/\tau^*){\rm exp}(-\tau/\tau^*)$ away from very small $\tau$. The larger spread at large $\tau \times {\rm v}^{2/3}$ is due to lower statistics. The rescaled fitted mean recurrence time $\tau^*$ varies with cell location but only weakly depends on energy within the band of global chaos.}
	\label{f:recurrence-time-distribution}
\end{figure}

 For uniformly mixing dynamics, it is expected that this normalized distribution follows an exponential law $(1/\bar \tau) e^{-\tau/\bar\tau}$ where $\bar \tau$ is the mean recurrence or relaxation time \cite{zaslavsky}. As shown in Fig. \ref{f:recurrence-time-distribution}, this exponential law for recurrence times holds for energies in the band of global chaos though there can be (sometimes significant) deviations for very small values of $\tau$ (e.g., $\tau \lesssim 25 \ll \bar \tau \approx 250$ for $w$ = .6 in Fig.~\ref{f:recurrence-time-distribution-d}). These deviations could be attributed to a memory effect, the finite time that the system takes before the dynamics displays mixing (see Fig. \ref{f:approach-to-mixing-in-time}). Thus, $\bar \tau$ is to be interpreted as the time constant in the above exponential law that best fits the distribution away from very small $\tau$.

A heuristic argument for the exponential law follows; for a more detailed treatment, see \cite{kac, balakrishnan-pre} and references therein. We pick a large number $N$ of ICs uniformly from a region $\Om$ of volume $V_\Om$ in an energy-$E$ hypersurface of volume $V_E$. They are evolved in time and their locations sampled at a temporal frequency $\Delta$. At each such instant, the probability of returning to $\Om$ is $p = V_\Om/V_E$ provided a sufficiently long time $T$ has elapsed for correlations to have died out. Suppose a fraction $f$ of trajectories have {\it not} returned to $\Om$ by this time $T$. Then, the probability that the {\it first return time} $\tau$ equals $T + \Delta$ is $P(\tau = T + \Delta) = f p$ (leaving aside possible returns that the sampling at frequency $\Delta$ does not detect). If $\Delta$ is chosen large enough ($\gtrsim$ transit time across $\Om$), we also have $P(\tau = T + 2\Delta) = f (1-p) p$ and similarly $P(\tau = T + n \Delta) = f (1-p)^{n-1} p$ for $n = 1,2,\cdots$. In the limit $N \to \infty$, $\Delta \to 0$ and $V_\Om \to 0$ holding $\Delta/p = \bar \tau$ fixed, and omitting prefactors (independent of $t$) that go into the normalization,
		\beq
		P( t \leq \tau \leq t+dt) \propto \lim_{\Delta \to 0} (1 - p)^{t/\Delta} = e^{-t/\bar \tau}.
		\eeq

\subsection{Scale invariance}

Though $\bar\tau$ varies with the width $w$, we find that when rescaled by the two-third power of the Liouville volume ${\rm v}$ of the cell, it becomes independent of cell size within the band of global chaos\footnote{For ergodic systems defined by the iterations of a map, Kac's Lemma implies that the mean first return time to a cell is inversely proportional to the measure of the cell \cite{Kac-paper}. What we observe here is a continuous time version of it.}. In other words, $\bar\tau \times {\rm v}^{2/3} = \tau^*$ is constant for cells centered at a given location (see Fig. \ref{f:recurrence-time-distribution-b}). Thus, as shown in Figs.  \ref{f:recurrence-time-distribution-c} and \ref{f:recurrence-time-distribution-d}, the rescaled recurrence time distributions for various cell sizes, all follow the {\it same} exponential law for a given energy and cell center. This scaling law may be viewed as a 3d energy hypersurface analogue of the 2d phase space version given in Eq. (36) of \cite{balakrishnan-scaling-law} as well as of the scaling law for the mean recurrence time of the second type in \cite{gao}. Heuristically, the mean recurrence time $\bar \tau$ is inversely proportional to the surface area ($\sim {\rm v}^{2/3}$) of the cell and allows us to view the `attractor' as being three dimensional, which is consistent with global chaos and ergodicity. On the other hand, we find that the scaling exponent deviates from two-thirds in chaotic regions outside this band. This is to be expected since the dynamics at such energies is not mixing in such chaotic regions, as shown in Figs.~\ref{f:approach-to-mixing-in-energy} and \ref{f:psec-outside-globalchaos}.

The above scaling law defines for us the  {\it scaled} mean recurrence time $\tau^*$ for cells centered at a given location of an energy hypersurface. We find that $\tau^*$ varies with location. For instance, for cells centered along an isosceles trajectory (see \S \ref{s:iso-periodic-sol}), we find that the values of $\tau^*$ display a reflection symmetry about the triple collision configuration and vary over the range $31 \lesssim \tau^* \lesssim 56$. On the other hand, within the band of global chaos, $\tau^*$ hardly varies with energy for a given cell location.

\subsection{Loss of memory} 

We also observe the absence of memory in the sense that the gaps between successive recurrence times are uncorrelated. For instance, let us denote by $\tau_1$ and $\tau_2$ the first recurrence time and the gap between second and first recurrence times for a given trajectory and cell, and define the the correlation coefficient
	\beq
	\scripty{r} = 
	\left[ { \bra \tau_1 \tau_2 \ket - \bra \tau_1 \ket \bra \tau_2 \ket} \right] / ({\sig_1 \sig_2}).
	\eeq
The averages here are performed with respect to a random collection of trajectories and $\sig_{1,2}$ denote the standard deviations of $\tau_{1,2}$. We find that $|\scripty{r}| \approx 10^{-3}-10^{-5} \ll 1$ for cells of widths $0.4-1.2$ centered at the origin of the energy $E = 5.5$ hypersurface, indicating uncorrelated recurrences.

\chapter{Discussion}
\label{c:discussion}

In the first part of this thesis (Chapter \ref{chapter:three-body}), we investigate the planar three-body problem with Newtonian and inverse-square potentials from a geometric viewpoint where trajectories are reparametrized geodesics of the Jacobi-Maupertuis metric on the configuration space. Symmetries are used to pass to quotients of the configuration space using the method of Riemannian submersions. We study the near-collision dynamics and show that the geodesic formulation regularizes collisions in the inverse-square potential, though not for the Newtonian potential. Explicit calculations are facilitated by a good choice of coordinates in which Killing vector fields point along coordinate vector fields. By estimating scalar and sectional curvatures, we establish the presence of widespread geodesic instabilities. The results are summarized in \S \ref{s:intro-3body}. An interesting direction for further research is to study the dynamical consequences of sectional curvatures of the JM metric possessing either sign and to relate the local geodesic instabilities to medium- and long-time behavior as well as to chaos. Though this remains an open issue in the three-body problem, we have been able to establish closer connections of this sort in the problem of three coupled rotors, which turns out to be a very interesting problem in its own right.

In the second part (Chapter \ref{chapter:three-rotor}), we propose and study the three-rotor problem which can be viewed as arising as the classical limit of a model for chains of coupled Josephson junctions. We find that it displays an order-chaos-order transition with increasing energy and discover novel signatures of its  transition to chaos. We also uncover `pendulum' and `isosceles-breather' periodic solutions as well as choreographies and discuss their stability properties. Moreover, we discover a band of energies where the dynamics is globally chaotic and provide evidence for ergodicity and mixing in this band.  \S \ref{s:intro-3rotor} contains a concise summary of our results. Here, we discuss some open questions arising from our work. 

The classical three-rotor problem and the planar restricted three-body problem are similar in the sense that both have essentially two degrees of freedom and only one known conserved quantity. In the latter, Bruns and Poincar\'e \cite{whittaker} proved the non-existence of additional conserved quantities of certain types (analytic in small mass ratios and orbital elements). It would be reassuring to obtain a similar result for the three-rotor problem. Analogously, the extension to our system, of Ziglin's \cite{ziglin} and Melnikov's \cite{melnikov} arguments for non-integrability is also of interest.

While we found the trace of the monodromy for periodic `pendulum' solutions numerically, it would be interesting to prove the accumulation of stability transitions at $E = 4g$  as in  \cite{churchill} and establish its asymptotic periodicity on a log scale, for instance by finding an analytical expression for the stability index as Yoshida \cite{yoshida84} does in the 2d anharmonic oscillator of Eq. (\ref{e:anharmonic-oscillator-yoshida}). This accumulation at the threshold for bound librational trajectories with diverging time periods and the periodicity on a log scale is reminiscent of the quantum energy spectrum of Efimov trimers that accumulate via a geometric sequence at the two-body bound state threshold with diverging S-wave scattering length \cite{efimov}. It would also be interesting to explore a possible connection between this accumulation of transitions and the accumulation of homoclinic points at a hyperbolic fixed point in a chaotic system. The nature of bifurcations \cite{brack-mehta-tanaka} and local scaling properties \cite{LSS} at these transitions are also of interest. In another direction, one would like to understand if there is any connection between the accumulation of transition energies and the change in topology of the Hill region $(\V \leq E)$ of the configuration torus as $E$ crosses the value $4g$ at the three critical points (saddles D) of the Morse function $\V$ (see \S \ref{s:topology-hill-region}). One would also like to analyze the onset of widespread chaos in this system using methods such as those of Chirikov \cite{chirikov-overlap-resonance} and Greene \cite{green}. 

We have argued that the three-rotor system is integrable at $E = 0$ and $\infty$ ($g = \infty, 0$), where  additional conserved quantities emerge. One wonders whether it is `integrable' at any other energy. In other words, is there any non-trivial energy hypersurface in phase space on which all trajectories are periodic or quasi-periodic so that the corresponding Poincar\'e sections are regular? Our estimate of the fraction of chaos on the `$\vf_1 = 0$' Poincar\'e surface strongly suggests that any integrable energy $E_{\rm I}$ is either isolated or $E_{\rm I} \lesssim 3.8g$. However, even for low energies, we expect chaotic sections in the neighborhood of the isosceles points $\cal I$ (see Fig. \ref{f:psec-egy=2-3}). In fact, we conjecture that the three-rotor problem has no non-trivial integrable energies unlike the 2d anharmonic oscillator\cite{yoshida84}. 

 While we have provided a qualitative explanation for the shape of the momentum distribution over energy hypersurfaces in \S \ref{s:ergodicity-time-ensemble-avgs}, it would be nice to understand the mechanisms underlying the phase transitions observed in $\rho(p_1)$ as the energy is varied. In another direction, outside the band of global chaos, it would be interesting to determine whether the dynamics, when restricted to a chaotic region, is ergodic and/or mixing with respect to a suitable measure. In fact, Figs. \ref{f:approach-to-mixing-in-energy} and \ref{f:psec-outside-globalchaos} suggest that this measure cannot be the Liouville measure. In \S \ref{s:recurrenc-time-dist}, the scaled mean recurrence time $\tau^*$ to cells at a given location is found to vary with the location on the energy hypersurface. It would be of interest to study the nature of this variation and its physical implications. We also wonder whether global chaos and ergodicity are to be found in the problems of four or more rotors.

Unlike billiards and kicked rotors, the equations of the three-rotor system do not involve impulses/singularities. It would be interesting to identify other such continuous time autonomous Hamiltonian systems that display global chaos and ergodicity. As noted, the three-rotor problem may also be formulated as geodesic flow on a two-torus of non-constant Jacobi-Maupertuis curvature. A challenging problem would be to try to extend the analytic treatments of ergodicity in geodesic flows on constant curvature Riemann surfaces to the three-rotor problem.

Finally, a deeper understanding of the physical mechanisms underlying the onset of chaos in this system would be desirable, along with an examination of quantum manifestations of the classical chaos and an exploration of ergodicity and recurrence in the quantum three-rotor system.


\appendix

\chapter[Three-body problem]{Three-body problem}
\chaptermark{Three--body problem}

\section[Some landmarks in the history of the three-body problem]{Some landmarks in the history of the three-body problem \sectionmark{Some landmarks in the history of the three--body problem}}
\sectionmark{Some landmarks in the history of the three--body problem}
\label{a:history-three-body}

We consider the problem of three point masses ($m_a$ with position vectors $\bfr_a$ for $a = 1,2,3$) moving under their mutual gravitational attraction. The importance of the three-body problem lies in part in the developments that arose from attempts to solve it \cite{diacu-holmes,musielak-quarles}. These have had an impact all over astronomy, physics and mathematics. The system has 9 degrees of freedom, whose dynamics is determined by 9 coupled second order nonlinear ODEs:
	\beq 
	m_a \fr{d^2\bfr_a}{dt^2} = \sum_{b \neq a} G m_a m_b \fr{\bfr_b-\bfr_a}{|\bfr_b-\bfr_a |^3} \quad \text{for} \quad a = 1, 2, 3.
	\label{e:newtonian-3body-ODE}
	\eeq
The three components of momentum $\bfP = \sum_a m_a \dot \bfr_a$, three components of angular momentum $\bfL = \sum_a \bfr_a \times \bfp_a$ and energy 
	\beq
	E = \half \sum_{a=1}^3 m_a \dot \bfr_a^2 - \sum_{a < b} \frac{G m_a m_b}{|\bfr_a - \bfr_b|}  \equiv T + V
	\eeq
furnish $7$ independent conserved quantities. Joseph-Louis Lagrange used these conserved quantities to reduce the above equations of motion to 7 first order ODEs.

\begin{figure}	
	\centering
	\begin{subfigure}[t]{5cm}
		\centering
		\includegraphics[width=5cm]{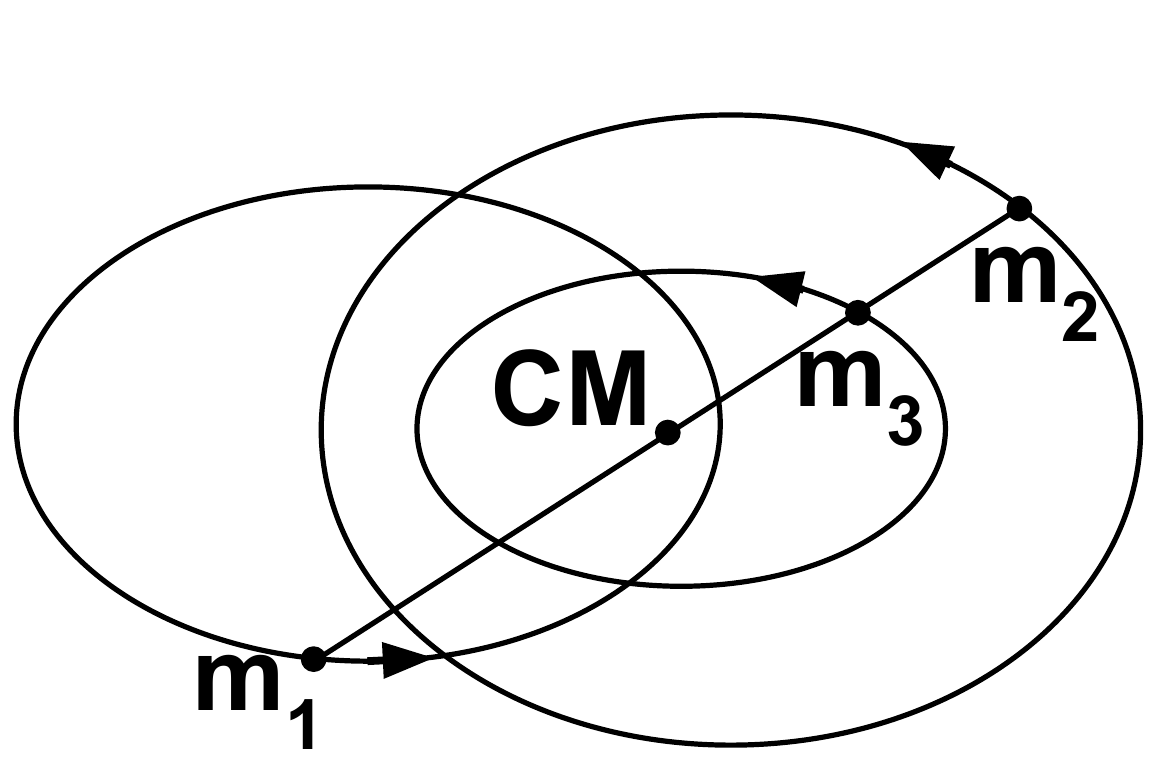}
		\caption{\small Euler collinear solution}
		\label{f:euler-periodic}		
	\end{subfigure}
	\;
	\begin{subfigure}[t]{5cm}
		\centering
		\includegraphics[width=3cm]{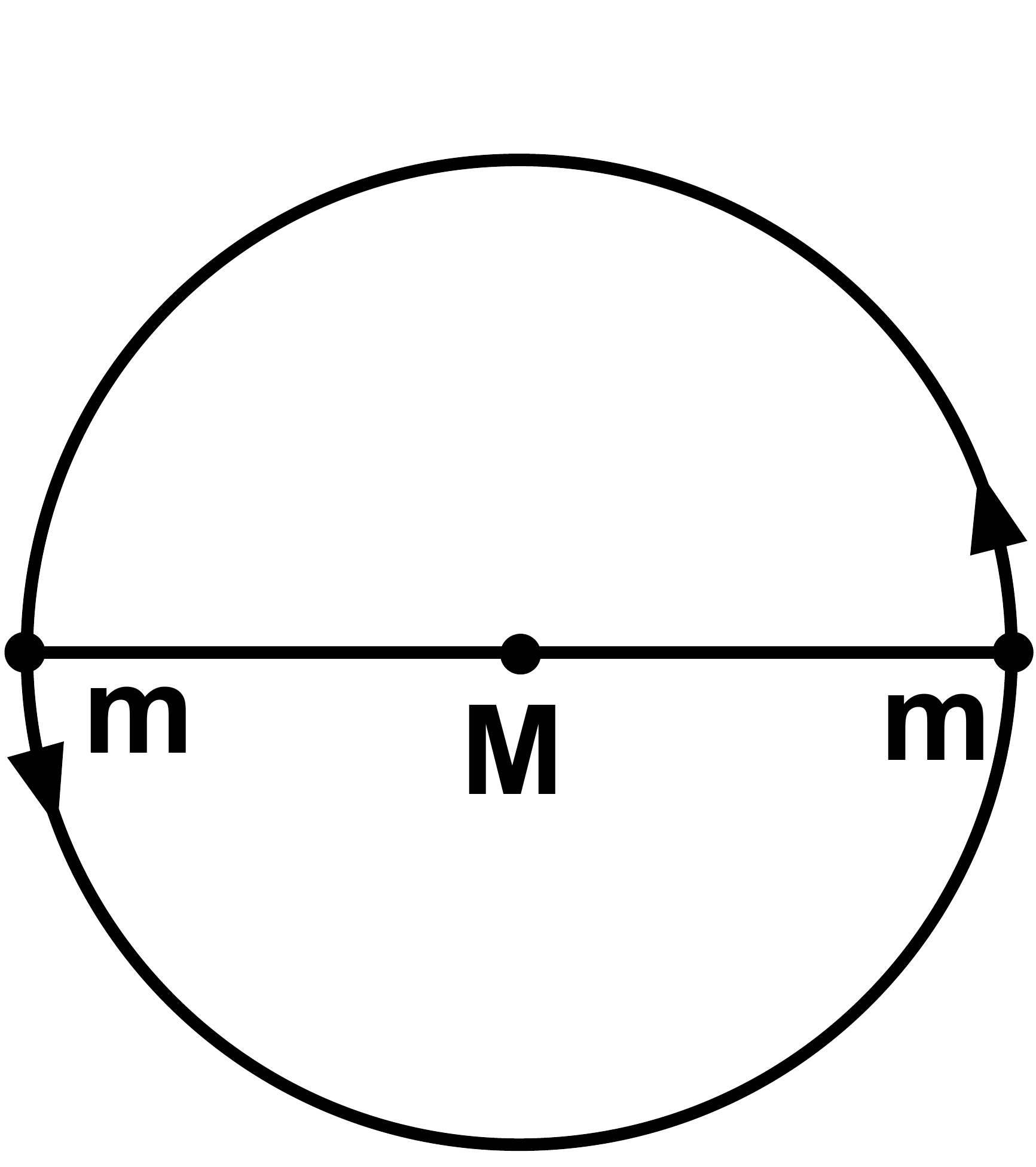}
		\caption{\small Euler collinear solution}
		\label{f:euler-eq-mass}		
	\end{subfigure}	
	\;
	\begin{subfigure}[t]{5.5cm}
		\centering
		\includegraphics[width=4.5cm]{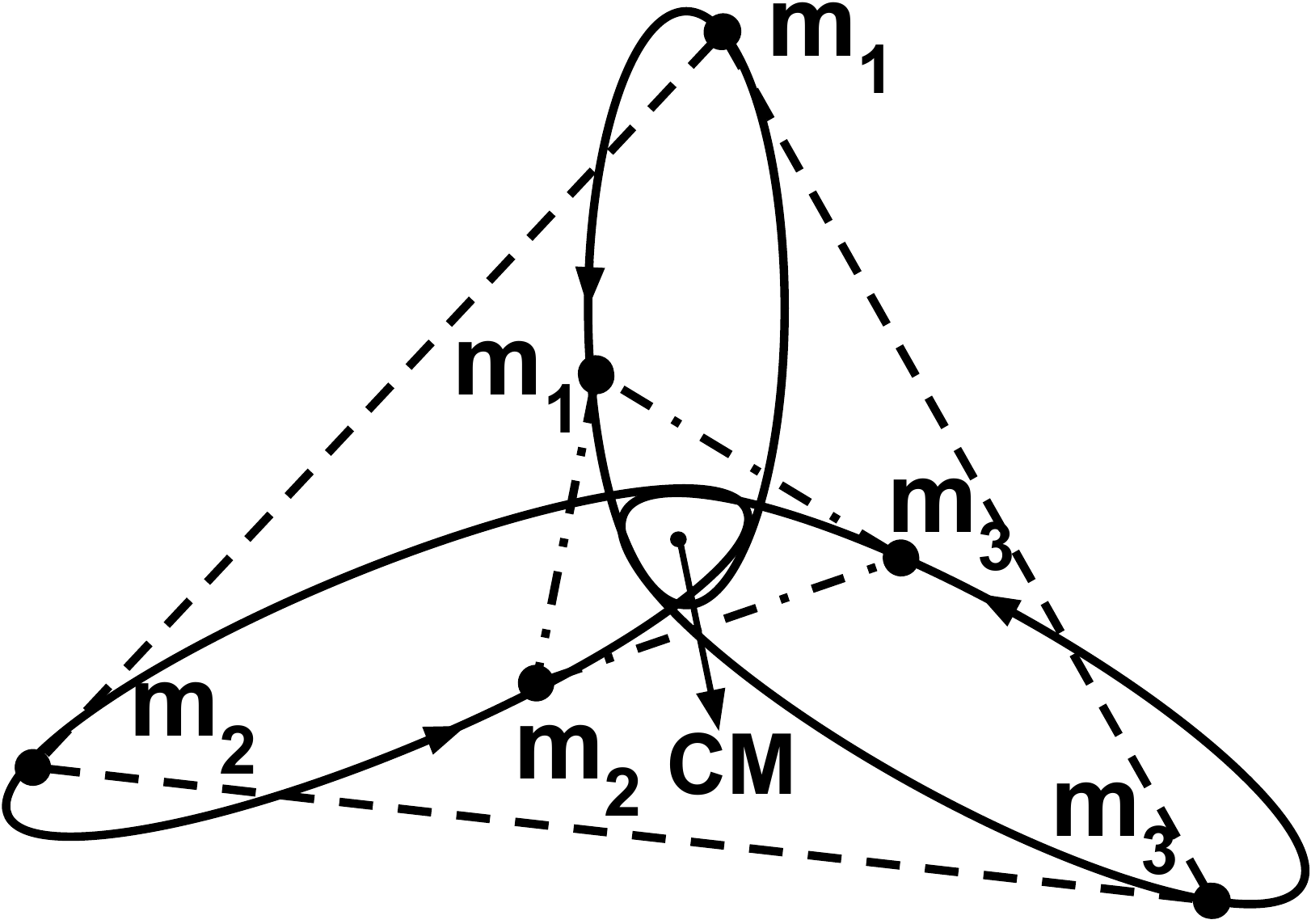}
		\caption{\small Lagrange equilateral solution}
		\label{f:lagrange-periodic}		
	\end{subfigure}	
	\caption{\small Euler's and Lagrange's periodic solutions of the three-body problem. The constant ratios of separations are functions of the mass ratios alone. (a) Euler collinear solution where masses traverse Keplerian ellipses with one focus at the CM. (b) Euler's solution where two equal masses $m$ are in a circular orbit around a third mass $M$ at their CM. (c) Lagrange's periodic solution with three bodies at vertices of equilateral triangles.}
	\label{f:three-body-periodic}
\end{figure}

The planar three-body problem is the special case where the masses always lie on a fixed plane. For instance, this happens when the center of mass (CM) is at rest ($\dot \bfJ_3 = 0$) and the angular momentum about the CM vanishes ($\bfL_{\rm CM} = M_1 \bfJ_1 \times \dot \bfJ_1 + M_2 \bfJ_2 \times \dot \bfJ_2 = 0$). In 1767, Leonhard Euler discovered simple periodic solutions to the planar three-body problem where the masses are always collinear, with each body traversing a Keplerian orbit about their common CM. The line through the masses rotates about the CM with the ratio of separations remaining constant (see Figs.~\ref{f:euler-periodic} and \ref{f:euler-eq-mass}). Lagrange rediscovered Euler's solution in 1772 and also found new periodic solutions where the masses are always at the vertices of equilateral triangles whose size and angular orientation may change with time (see Fig.~\ref{f:lagrange-periodic}). In the limiting case of zero angular momentum, the three bodies move toward/away from their CM along straight lines. These implosion/explosion solutions are called Lagrange homotheties. Euler collinear and Lagrange equilateral configurations are the only central configurations\footnote{Three-body configurations in which the acceleration of each particle points towards the CM and is proportional to its distance from the CM (${\bf a}_b= \om^2 (\bfR_{\rm CM} - \bfr_b)$ for $b = 1,2,3$) are called `central configurations'.} in the three-body problem. In 1912, Karl Sundmann showed that triple collisions are asymptotically central configurations. 

Can planets collide, be ejected from the solar system or suffer significant deviations from their Keplerian orbits? This is the question of the stability of the solar system. In the $18^{\rm th}$ century, Lagrange and Pierre-Simon Laplace obtained the first significant results on stability. They showed that to first order in the ratio of planetary to solar masses ($M_p/M_S$), there is no unbounded variation in the semi-major axes of the orbits, indicating stability of the solar system. Their compatriot Sim\'eon Denis Poisson extended this result to second order in $M_p/M_S$. However, in what came as a surprise, Spiru Haretu (1878) overcame significant technical challenges to find secular terms (growing linearly and quadratically in time) in the semi-major axes at third order!  Haretu's result did not prove instability as the effects of his secular terms could cancel out. However, it effectively put an end to the hope of proving the stability/instability of the solar system using such a perturbative approach.

The development of Hamilton's mechanics and its refinement in the hands of Carl Jacobi was still fresh when the dynamical astronomer Charles Delaunay (1846) began the first extensive use of canonical transformations in perturbation theory \cite{gutzwiller-three-body}. The scale of his hand calculations is staggering: he applied a succession of 505 canonical transformations to a $7^{\rm th}$ order perturbative treatment of the three-dimensional elliptical restricted three-body problem\footnote{The restricted three-body problem is a simplified version of the three-body problem where one of the masses is assumed much smaller than the two primaries.}. He arrived at the equation of motion for the small mass in Hamiltonian form using $3$ pairs of canonically conjugate orbital variables (3 angular momentum components, the true anomaly, longitude of the ascending node and distance of the ascending node from perigee). He obtained the latitude and longitude of the moon in trigonometric series of about $450$ terms with secular terms  eliminated. It wasn't till 1970-71 that Delaunay's heroic calculations were checked and extended using computers at the Boeing Scientific Laboratories \cite{gutzwiller-three-body}!

Anders Lindstedt (1883) developed a systematic method to approximate solutions to nonlinear ODEs when naive perturbation series fail due to secular terms. The technique was further developed by Poincar\'e. Lindstedt assumed the series to be generally convergent, but Poincar\'e soon showed that they are divergent in most cases. Remarkably, nearly 70 years later, Kolmogorov, Arnold and Moser showed that in many of the cases where Poincar\'e's arguments were inconclusive, the series are in fact convergent, leading to the celebrated KAM theory of integrable systems subject to small perturbations.

George William Hill was motivated by discrepancies in lunar perigee calculations. His celebrated paper on this topic was published in 1877 while working with Simon Newcomb at the American Ephemeris and Nautical Almanac\footnote{Simon Newcomb's project of revising all the orbital data in the solar system established the missing $42''$ in the $566''$ centennial precession of Mercury's perihelion. This played an important role in validating Einstein's general theory of relativity.}. He found a new family of periodic orbits in the circular restricted (Sun-Earth-Moon) three-body problem by using a frame rotating with the Sun's angular velocity instead of that of the Moon. The solar perturbation to lunar motion around the Earth results in differential equations with periodic coefficients. He used Fourier series to convert these ODEs to an infinite system of linear algebraic equations and developed a theory of infinite determinants to solve them and obtain a rapidly converging series solution for lunar motion. He also discovered new `tight binary' solutions to the three-body problem where two nearby masses are in nearly circular orbits around their center of mass CM$_{12}$, while CM$_{12}$ and the far away third mass in turn orbit each other in nearly circular trajectories.

\begin{figure}[!ht]
\centering
	\begin{subfigure}[t]{8cm}
	\center
	\raisebox{.7cm}{\includegraphics[width=6cm]{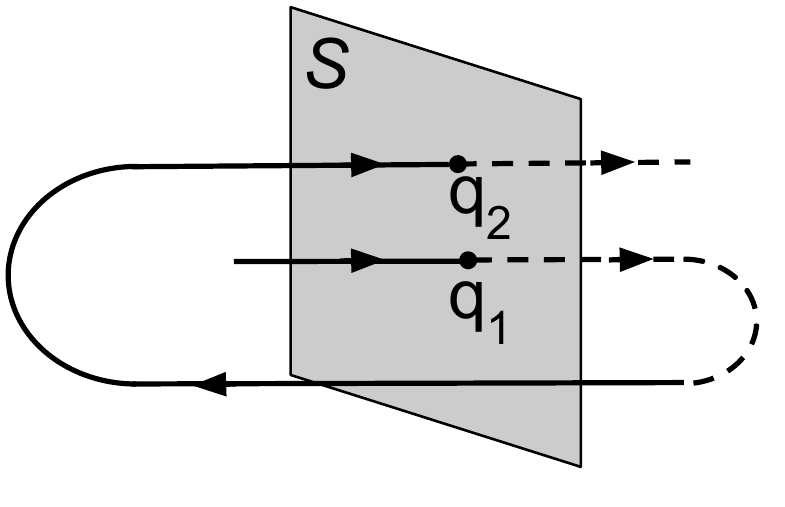}}
	\caption{\small \label{f:poincare-return-map}}
	\end{subfigure}
	\quad
	\begin{subfigure}[t]{8cm}
	\center
	\includegraphics[width=6cm]{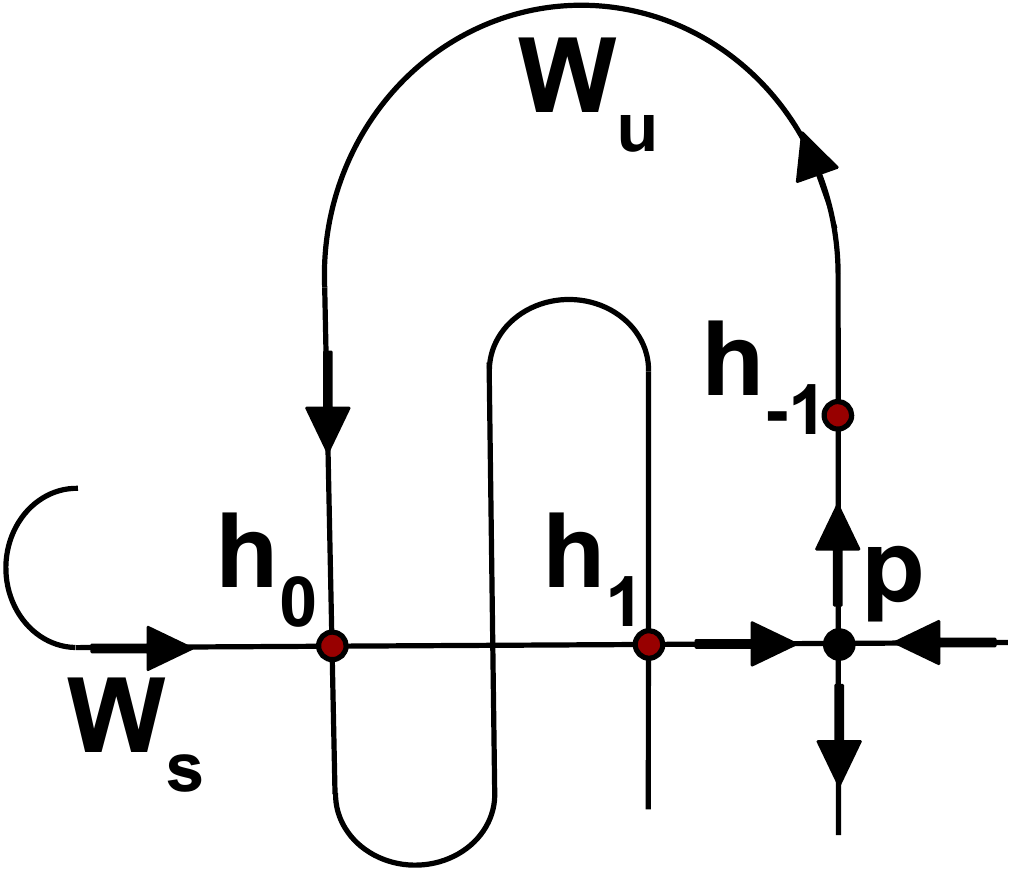}
	\caption{\small \label{f:homoclinic-points}}
	\end{subfigure}
	\caption{\small (a) A Poincare surface $S$ transversal to a trajectory is shown. The trajectory through $q_1$ on $S$ intersects $S$ again at $q_2$. The map taking $q_1$ to $q_2$ is called Poincar\'e's first return map. (b)  The saddle point $p$ and its stable and unstable spaces $W_s$ and $W_u$ are shown on a Poincar\'e surface through $p$. The points at which $W_s$ and $W_u$ intersect are called homoclinic points, e.g., $h_0,$ $h_1$ and $h_{-1}$. Points on $W_s$ (or $W_u$) remain on $W_s$ (or $W_u$) under forward and backward iterations of the return map. Thus, the forward and backward images of a homoclinic point under the return map are also homoclinic points. In the figure, $h_0$ is a homoclinic point whose image is $h_1$ on the segment $[h_0,p]$ of $W_s$. Thus, $W_u$ must fold back to intersect $W_s$ at $h_1$. Similarly, if $h_{-1}$ is the backward image of $h_0$ on $W_u$, then $W_s$ must fold back to intersect $W_u$ at $h_{-1}$. Further iterations produce an infinite number of homoclinic points accumulating at $p$. The first example of a homoclinic tangle was discovered by Poincar\'e in the restricted three-body problem and is a signature of its chaotic nature.}
\end{figure}

The mathematician/physicist/engineer Henri Poincar\'e began by developing a qualitative theory of differential equations from a global geometric viewpoint of the dynamics on phase space. This included a classification of the types of equilibria on the phase plane (nodes, saddles, foci/spiral and centers). His 1890 memoir on the three-body problem was the prize-winning entry in King Oscar II's $60^{\rm th}$ birthday competition (for a detailed account see \cite{barrow-green-poincare-three-body}). He proved the divergence of series solutions for the three-body problem developed by Delaunay, Hugo Gyld\'en and Lindstedt (in many cases) and convergence of Hill's infinite determinants. To investigate the stability of three-body motions, Poincar\'e defined his `surfaces of section' and a discrete-time dynamics via the `return map' (see Fig.~\ref{f:poincare-return-map}). A Poincar\'e surface $S$ is a two-dimensional surface in phase space transversal to trajectories. The first return map takes a point $q_1$ on $S$ to $q_2$, which is the next intersection of the trajectory through $q_1$ with $S$. Given a hyperbolic fixed point (e.g., a saddle point) $p$ on a surface $S$, he defined its stable and unstable spaces $W_s$ and $W_u$ as points on $S$ that tend to $p$ upon repeated forward or backward applications of the return map (see Fig.~\ref{f:homoclinic-points}). He initially assumed that $W_s$ and $W_u$ on a surface could not intersect and used this to argue that the solar system is stable. This assumption turned out to be false, as he discovered with the help of Lars Phragm\'en. In fact, $W_s$ and $W_u$ can intersect transversally on a surface at a homoclinic point\footnote{Homoclinic refers to the property of being `inclined' both forward and backward in time to the same point.} if the state space of the underlying continuous dynamics is at least three-dimensional. What is more, he showed that if there is one homoclinic point, then there must be infinitely many of them accumulating at $p$ (see Fig.~\ref{f:homoclinic-points}). Moreover, $W_s$ and $W_u$ fold and intersect in a very complicated `homoclinic tangle' in the vicinity of $p$. This was the first example of what we now call chaos. 

\begin{figure}	
	\centering
	\begin{subfigure}[t]{7.5cm}
		\centering
		\includegraphics[height=4cm]{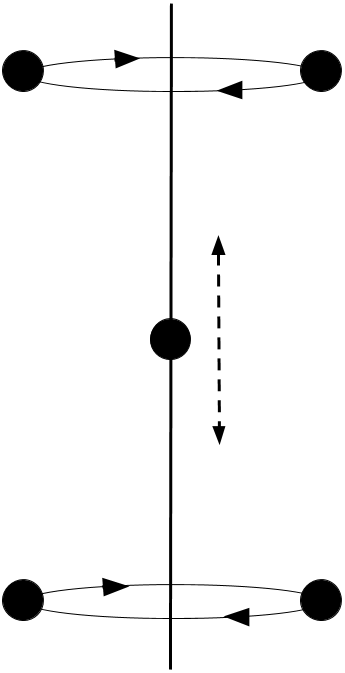}
		\caption{\small Xia's example}
		\label{f:xia-example}
	\end{subfigure}
\quad
	\begin{subfigure}[t]{7.5cm}
		\centering
		\raisebox{.4cm}{\includegraphics[height=3cm]{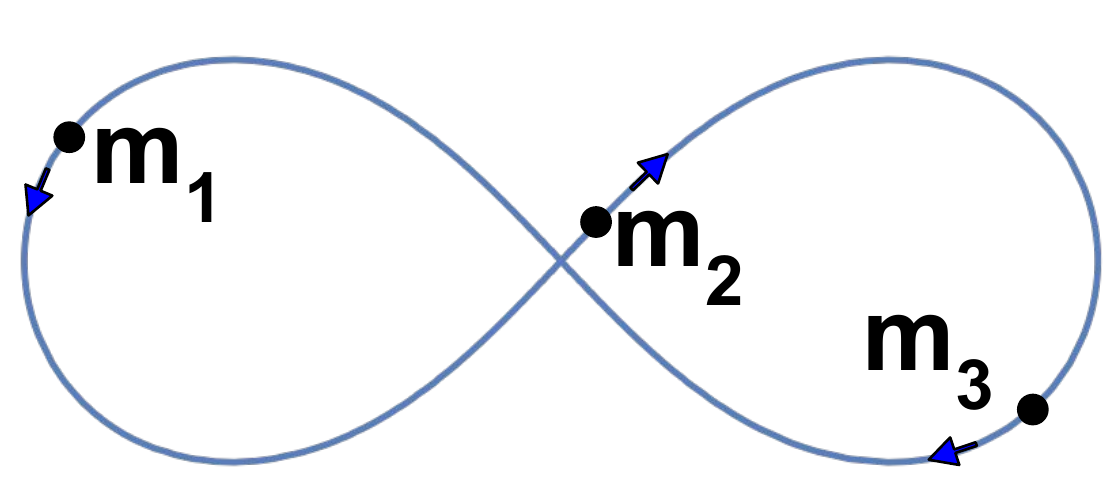}}
		\caption{\small Figure-8 solution}
		\label{f:figure-8}
	\end{subfigure}
	\caption{\small (a) An example due to Xia leading to a non-collisional singularity in the 5-body problem where a mass oscillates with ever-growing frequency and amplitude between two pairs of collapsing tight binaries that escape to infinity in finite time. (b) Equal-mass zero-angular momentum figure-8 choreography solution to the three-body problem. A choreography is a periodic solution where all masses traverse the same orbit separated equally in time.}
\end{figure}

When two gravitating point masses collide, their relative speed diverges and solutions to the equations of motion become singular at the collision time $t_c$. More generally, a singularity occurs when either a position or velocity diverges in finite time. Paul Painlev\'e (1895) showed that binary and triple collisions are the only possible singularities in the three-body problem. However, he conjectured that non-collisional singularities (e.g. where the separation between a pair of bodies goes to infinity in finite time) are possible for four or more bodies. It took nearly a century for this conjecture to be proven, culminating in the work of Donald Saari and Zhihong Xia (1992) and Joseph Gerver (1991) who found explicit examples of non-collisional singularities in the $5$-body and $3n$-body problems for $n$ sufficiently large \cite{saari}. In Xia's example, a particle oscillates with ever-growing frequency and amplitude between two pairs of tight binaries (see Fig. \ref{f:xia-example}). The separation between the binaries diverges in finite time, as does the velocity of the oscillating particle. 

Tulio Levi-Civita (1901) attempted to avoid singularities and thereby `regularize' collisions in the three-body problem by a change of variables in the differential equations. For example, the ODE for the one-dimensional Kepler problem $\ddot x = - k/x^2$ is singular at the collision point $x=0$. This singularity can be regularized\footnote{Solutions which could be smoothly extended beyond collision time (e.g., the bodies elastically collide) were called regularizable. Those that could not were said to have an essential or transcendent singularity at the collision.} by introducing a new coordinate $x = u^2$ and a reparametrized time $ds = dt/u^2$, which satisfy the nonsingular oscillator equation $u''(s) = E u/2$ with conserved energy $E = (2 \dot u^2 - k)/u^2$. Such regularizations could shed light on near-collisional trajectories (`near misses') provided the differential equations remain physically valid\footnote{Note that the point particle approximation to the equations for celestial bodies of non-zero size breaks down due to tidal effects when the bodies get very close}. 

Karl Sundman (1912) began by showing that binary collisional singularities in the three-body problem could be regularized by a repararmetrization of time, $s = |t_1-t|^{1/3}$ where $t_1$ is the binary collision time \cite{siegel-moser}. He used this to find a {\it convergent} series representation (in powers of $s$) of the general solution of the three-body problem in the absence of triple collisions\footnote{Sundman showed that for non-zero angular momentum, there are no triple collisions in the three-body problem.}. The possibility of such a convergent series had been anticipated by Karl Weierstrass in proposing the three-body problem for King Oscar's 60th birthday competition. However, Sundman's series converges exceptionally slowly and has not been of much practical or qualitative use.

The advent of computers in the $20^{\rm th}$ century allowed numerical investigations into the three-body (and more generally the $n$-body) problem. Such numerical simulations have made possible the accurate placement of satellites in near-Earth orbits as well as our missions to the Moon, Mars and the outer planets. They have also facilitated theoretical explorations of the three-body problem including chaotic behavior, the possibility for ejection of one body at high velocity (seen in hypervelocity stars \cite{hypervelocity-stars}) and quite remarkably, the discovery of new periodic solutions. For instance, in 1993, Chris Moore discovered the zero angular momentum figure-8 `choreography' solution. It is a stable periodic solution with bodies of equal masses chasing each other on an $\infty$-shaped trajectory while separated equally in time (see Fig.~\ref{f:figure-8}). Alain Chenciner and Richard Montgomery \cite{chenciner-montgomery} proved its existence using an elegant geometric reformulation of Newtonian dynamics that relies on the variational principle of Euler and Maupertuis. In fact, we use the associated Jacobi-Maupertuis metric formulation in our geometric approach to the planar three-body problem in Chapter \ref{chapter:three-body}.

\section{Proof of an upper bound for the scalar curvature}
\label{a:upper-bound-scalar-curvature}
 
Here we establish a strict lower bound on the quantity that appears in the relation (\ref{e:scalar-curv-c2-compare-s2}) between Ricci scalars on $\mC^2$ and $\mS^2$. Since Montgomery has shown that $R_{\mS^2} \leq 0$, this helps us  establish strictly negative upper bounds for the scalar curvatures on $\mC^2$, $\mR^3$ and $\mS^3$. We will show here that
	\beq
	\label{e:inequality-zeta}
	 12 h^2 + |\grad h|^2  > \zeta h^3 \quad \text{where} \quad  \zeta = {55}/{27} \approx 2.04.
	\eeq 
The best possible $\zeta$ is estimated numerically to be $\zeta = 8/3$ and the minimum occurs at the Euler points $E_{1,2,3}$. We define the power sum symmetric functions $u_{2n} = \sum_{i=1}^3 v_i^n$ in terms of which the pre-factor in the JM metric (\ref{e:conformal-prefactor-h}) is $h = v_1 + v_2 + v_3 = u_2$. In \cite{montgomery-pants} Montgomery shows that $|\grad h|^2 = 4s$ where the symmetric polynomial
	\beq
	s = (1/2) \left(-2 u_2^2 + 4 u_2 u_4 - 3u_4^2 + 3 u_8 \right).
	\eeq
This gives 
	\beq
	\label{e:A-B}
	12 h^2 + |\grad h|^2 = u_2^3 \left( 8 A + 6 B\right) \quad \text{where} \quad 	A = \fr{u_2 + u_4}{ u_2^2} \quad \text{and} \quad B = \fr{u_8- u_4^2}{u_2^3} .
	\eeq
We will show below that	$ A \geq 17/27$ and $B > -1/2$, from which Eq. (\ref{e:inequality-zeta}) follows (numerically we find that $B \geq -32/81$ which leads to the above-mentioned optimal value $\zeta = 8/3$). To prove the inequality for $B$, we define $c = \cos 2\eta$ and $s = \sin 2\eta \cos 2\xi_2$ which lie in the interval $[-1,1]$. Then  
	\beq
	\fr{u_8 - u_4^2}{u_2^3} > -\half \quad \Leftrightarrow \quad
u_8 - u_4^2 + \fr{u_2^3}{2} > 0 \quad	\Leftrightarrow \quad
	 \frac{3}{8} \left(20 - 3 (c^2 + s^2)^2 - 8 c^3 + 24 c s^2 \right) > 0.
	\eeq
For the latter to hold it is sufficient that $17 - 8 c^3 + 24 c s^2 > 0$ which is clearly true for $0 \leq c \le 1$. For $-1 \leq c < 0$ put $c = -d$. Then it is enough to show that $17 + 8 d^3 - 24 d (1-d^2) > 0$ since $s^2 \leq 1 - d^2$. This holds as the LHS is positive at its boundary points $d = 0, 1$ as well as at its local extremum $d = 1/2$.

\begin{figure}[h]
	\centering
	\includegraphics[width=7cm]{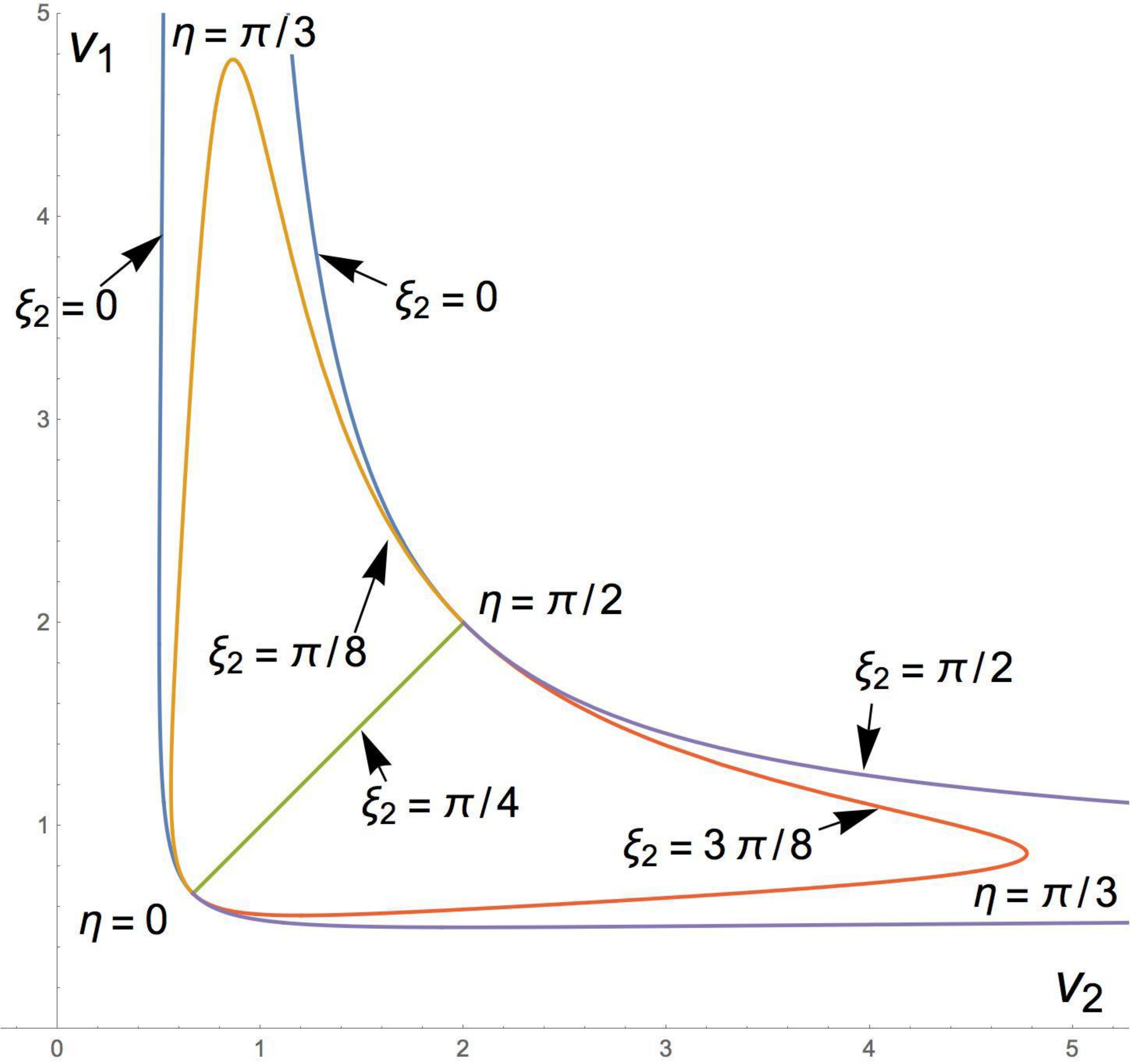}
	\caption{\small The boundary $\pdr D$ of the region $D$ in the $v_1$-$v_2$ plane is given by the level curves $\xi_2 = 0, \pi/2$. These level curves run from the collision point $\eta = 0$ to the Euler point $\eta = \pi/2$, passing through the collision points at $v_1 = \infty$ or $v_2 = \infty$ (where $\eta = \pi/3$). The level curves $\xi_2 = \pi/8, \pi/4, 3\pi/8$ in the interior $D$ are also shown. Note that $D$ lies within the quadrant $v_{1,2} \geq 1/2$.}
	\label{f:boundary}
	\end{figure}

The quantity $A$ defined in Eq. (\ref{e:A-B}) is a symmetric function of $v_1, v_2$ and $v_3$ which in turn are functions of $\eta$ and $\xi_2$ (\ref{e:conformal-prefactor-h}) for $0 \leq \eta \leq \pi/2$ and $0 \leq \xi_2 \leq \pi$. Since $\sum_i 1/v_i = 3$, we may regard $A$ as a function of any pair, say $v_1$ and $v_2$. The allowed values of $\eta$ and $\xi_2$ define a domain $\bar D = D \amalg \pdr D$ in the $v_1$-$v_2$ plane. To show that $A \geq 17/27$, we seek its global minimum, which must lie either at a local extremum in the interior $D$ or on the boundary $\pdr D$. $\pdr D$ is defined by the curves $\xi_2 = 0$ and $\xi_2 = \pi/2$ which meet at $\eta = 0$ and $\eta = \pi/2$ and include the points $(v_1 = \infty, v_2 = 2/3)$ and $(v_1 = 2/3, v_2 = \infty)$ (see Fig. \ref{f:boundary}). This is because, for any fixed $\eta$, $v_1$ and $v_2$ (\ref{e:conformal-prefactor-h}) are monotonic functions of $\xi_2$ for $0 \leq \xi_2 \leq \pi/2$ and symmetric under reflection about $\xi_2 = \pi/2$.
Along $\pdr D$, $A = (5 \cos 6 \eta + 22)/27$ is independent of $\xi_2$ and minimal at the Euler configurations $\eta = \pi/6$ and $\pi/2$ with the common minimum value $17/27$, which turns out to be the global minimum of $A$. This is because its only local extremum in $D$ is at the Lagrange configuration $v_1 = v_2 = v_3 = 1$ where $A = 2/3$. To see this, we note that local extrema of $A$ in $D$ must lie at the intersections of $\pdr A/\pdr v_1 = 0$ and $\pdr A/\pdr v_2 = 0$. Now $\pdr A/ \pdr v_1 = (v_1 - v_3)F(v_1, v_2)/v_1^2 u_2^3$ where
	\beq
	F(v_1,v_2)=  u_2 \left\{ v_1 + v_3 + 2 \left( v_1^2 + v_1 v_3 + v_3^2 \right) \right\} - 2 (v_1 + v_3) ( u_2 + u_4).
	\eeq
For $\pdr A/\pdr v_1$ to vanish, either $v_1 = v_3$ or $F(v_1,v_2) = 0$ or one of the $v_i = \infty$. The collision points $v_i = \infty$ do not lie in $D$. The conditions for $\pdr A / \pdr v_2$ to vanish are obtained via the exchange $v_1 \leftrightarrow v_2$. The intersection of the conditions $v_1 = v_3$ and $v_2 = v_3$ lies at the Lagrange configurations $v_i = 1$ where $A = 2/3$. It turns out that the only intersection of $v_1 = v_3$ with $F(v_2,v_1) = 0$ or of $v_2 = v_3$ with $F(v_1, v_2) = 0$ lying in $D$ occurs at the above Lagrange configuration. For instance, when $v_1 = v_3 = v$, $F(v_2,v_1) = -3 v^2 (4v-1)(v-1)/(3v - 2)^2$ vanishes when $v = 1$ or $v = 1/4$ (which violates $v \geq 1/2$). Finally, we account for extrema lying on the zero loci of both $F(v_1,v_2)$ and $F(v_2,v_1)$, which using $u_{-2} = 3$, must satisfy 
	\beq
	F(v_1,v_2) - F(v_2,v_1) = (v_1 - v_2) \left[12 v_1 v_2 v_3 - (v_1 + v_2 + v_3) \right] = 0.
	\eeq
So either $v_1 = v_2$ or $12 v_1 v_2 v_3 = u_2$. Now, we have shown above that the only extrema of $A$ on $v_1 = v_3$ in $D$ lie at the Lagrange configurations. Since $A$ is a symmetric function of the $v_i$, it follows that its only extrema on $v_1 = v_2$ also lies at the Lagrange configurations. On the other hand, $12 v_1 v_2 v_3 - (v_1 + v_2 + v_3) \geq 0$ for $v_i \geq 1/2$, with equality only at $v_i = 1/2$ which is not in $D$. Thus the only extremum of $A$ in $D$ is at the Lagrange configurations (where $A = 2/3$) and hence its global minimum occurs on $\pdr D$ at the Euler configurations  (where $A = 17/27$).

\chapter[Three-rotor problem]{Three-rotor problem}
\chaptermark{Three--rotor problem}

\section[Quantum $N$-rotor problem from  XY model]{Quantum \texorpdfstring{$N$-rotor}{N-rotor} problem from  XY model \sectionmark{Quantum $N$-rotor problem from  XY model}}
\sectionmark{Quantum $N$-rotor problem from  XY model} 
\label{a:n-rotor-from-xy}

\begin{figure}	
	\centering
	\begin{subfigure}[t]{7.5cm}
		\centering
		\includegraphics[width=7.5cm]{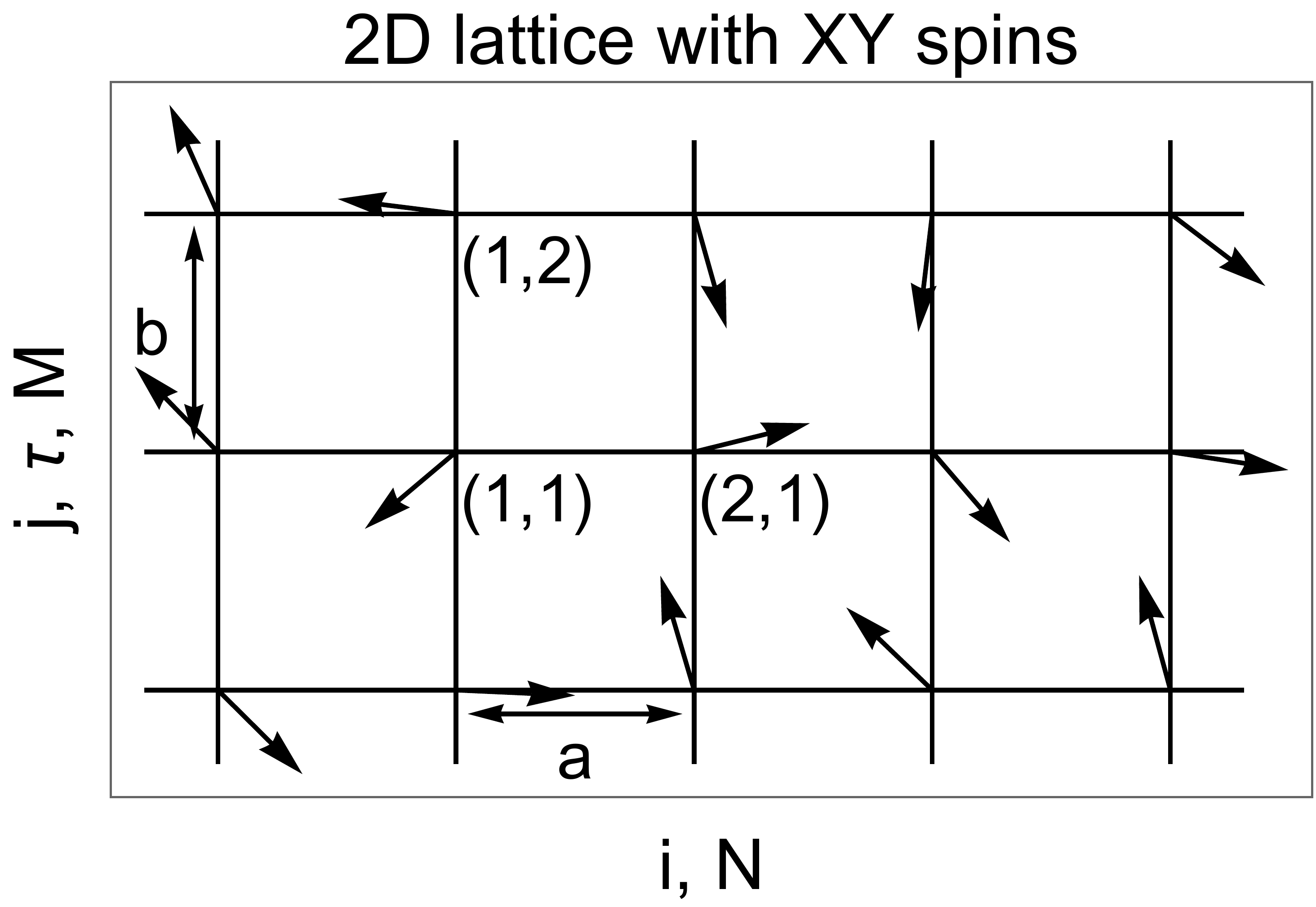}
	\end{subfigure}
	\caption{\small The quantum $N$-rotor problem arises from a partial continuum limit of the Wick-rotated XY model of classical statistical mechanics.}
	\label{f:3rotor-xy-model}
\end{figure}

The quantum $N$-rotor problem may be related to the 2d XY model of classical statistical mechanics which displays the celebrated Kosterlitz-Thouless topological phase transition \cite{sachdev}.
The dynamical variables of the XY model are 2d unit-vector spins ${\bf S}_\al$ (or phases $e^{i \tht_\al}$) at each site $\al$ of an $N \times M$ rectangular lattice with horizontal and vertical spacings $a$ and $b$ and nearest neighbor  ferromagnetic interaction energies $- J \: {\bf S}_\al \cdot {\bf S}_\beta = - J \cos (\tht_\al - \tht_\beta)$ with $J > 0$ (see Fig. \ref{f:3rotor-xy-model}). One often considers $a=b$ and assumes that $\tht$ varies gradually so that in the continuum limit $a \to 0$ and $N, M \to \infty$ holding $aN$ and $aM$ fixed, the Hamiltonian becomes $H = \fr{J}{2} \int |\grad \tht|^2 \: d^2{\bf r}$. This defines the 1+1 dimensional $O(2)$ principal chiral model.

Here, we approximately reformulate the XY model as an interacting quantum $N$-rotor problem by taking a partial continuum limit in the vertical direction followed by a Wick rotation. The resulting quantum system has been used to model a 1d array of coupled Josephson junctions and is known to be related to the XY model in a Villain approximation \cite{sondhi-girvin,wallin-1994}. With $i$ and $j$ labelling the columns and rows of the lattice, the XY model Hamiltonian is 
	\beq
	H = - J \sum_{i,j} \left[ \cos(\tht_{i,j+1}-\tht_{i,j}) + \cos(\tht_{i+1,j}-\tht_{i,j}) \right]  
	\label{e:Hamiltonian-XY}
	\eeq
with $J > 0$. In the first term, the sum is over $1\leq i \leq N$ and $1\leq j \leq M-1$ while for the second term, we have $1 \leq i \leq N-1$ and $1 \leq j \leq M$. We will impose periodic boundary conditions (BCs) in the horizontal but not in the vertical direction (open BCs are also of interest). We will take a continuum limit in two steps. We first make the spacing between rows small by introducing a continuous vertical coordinate $\tau$ in place of $j$ such that $\tau(j+1) - \tau(j) = \del \tau = b$. Next, we approximate $\cos(\tht_{i,j+1}-\tht_{i,j})$ by
	\beqs
	\cos(\tht_{i}(\tau+\del \tau)-\tht_i(\tau)) \approx 1-\half (\tht_i(\tau + \del \tau)-\tht_i(\tau))^2
	\approx 1- \half \tht_i'(\tau)^2 \, b \, d\tau.
	\eeqs 
Here, we have chosen to write $(\del\tau)^2$ as $b \: d \tau$ in anticipation of taking $b \to 0$ in the second step. 
Within this approximation, the Hamiltonian (\ref{e:Hamiltonian-XY}) up to an additive constant becomes 
	\beq
	H = J \sum_i \int \left\{ \fr{b}{2} \tht_i'(\tau)^2 - \ov{b} \cos\left[\tht_{i+1}(\tau)-\tht_{i}(\tau) \right] \right\} \: d\tau
	\eeq
using the prescription $\sum_j b f(\tau_j) \to \int f(\tau) d\tau$. The resulting partition function
	\beq
	Z = \int \prod_{k=1}^N D[\tht_k] \exp\left[-\beta H \right]
	\eeq
after a Wick rotation $\tau = i c t$, may be written as	
	\beqs
	Z = \int D[\tht] e^{i S/\hbar} \quad \text{where} \quad 
	\fr{S}{\hbar} = \beta J c \sum_i \int dt \left[ \fr{b}{2 c^2} \dot \tht_i(t)^2 + \ov{b} \cos\left[\tht_{i+1}(t)-\tht_{i}(t) \right] \right].
	\label{e:partition-fn-cnt-vertical-discrete-horizontal}
	\eeqs
We introduced a parameter $c > 0$ with dimensions of speed so that $t$ has dimensions of time. We may take a second continuum limit, this time in the horizontal direction by replacing $\sum_i$ by $\int \fr{dx}{a}$ by taking $a \to 0$ and $N \to \infty$ while holding $aN$ and $a/b$ fixed to get
	\beqs
	\fr{S}{\hbar} &\approx& \beta J c \int \fr{dx}{a}  \int dt \left\{ \fr{b}{2c^2} \left( \fr{\pdr \tht}{\pdr t} \right)^2 + \ov{b} \cos\left( a \fr{\pdr \tht}{\pdr x} \right) \right\} \cr
	&\approx& \half \beta J c \int dx \; dt \left\{ \fr{b}{a}\ov{c^2} \dot \theta^2 - \fr{a}{b} \tht'^2 \right\}.
	\eeqs
The path integral $\int D[\tht] e^{i S/\hbar}$ is what we would have obtained if we had taken the conventional continuum limit ($a,b \to 0$) of the XY model partition function and then performed a Wick rotation. Our two-step continuum limit has allowed us to approximately identify the quantum $N$-rotor problem (\ref{e:partition-fn-cnt-vertical-discrete-horizontal}) where $b$ has not yet been taken to zero.

For fixed $N, a$ and $b$, the physical interpretation of (\ref{e:partition-fn-cnt-vertical-discrete-horizontal}) is facilitated by letting $Lb/ac \beta$ play the role of $\hbar$ where $L$ is a length that remains finite in the limit $a, b \to 0$. $L$ could be the horizontal linear dimension of the system. This $\hbar$ has dimensions of action and tends to $0$ at low temperatures where quantum fluctuations in the Wick rotated theory should be small. With this identification of $\hbar$, we read off the classical action
	\beq
	\label{e:classical-action-intermediate-problem}
	S[\tht] = \sum_{i} \int  \left\{ \fr{ J L b^2}{2 a c^2} \dot\tht_i^2 + \fr{J L}{a} \cos\left[\tht_i-\tht_{i+1} \right] \right\} \; dt. 
	\eeq
Letting $m = J/c^2$, $r = \sqrt{L b^2/a}$ and $g = J L/a$, the corresponding Hamiltonian (with $\tht_{N+1} \equiv \tht_1$)
	\beq
	H = \sum_{i=1}^N  \left\{ \half { m r^2} \dot\tht_i^2 + g [1 - \cos\left(\tht_i-\tht_{i+1} \right) ] \right\}
	\eeq
describes the equal mass $N$-rotor problem. The rotor angles $\tht_i$ parametrize $N$ circles whose product  is the $N$-torus configuration space. Though the rotors are identical, each is associated to a specific site and thus are distinguishable. In particular, the wavefunction $\psi(\tht_1, \tht_2, \cdots \tht_N)$ need not be symmetric or antisymmetric under exchanges. We may also visualize the motion by identifying all the circles  but allowing the rotors/particles to remember their order from the chain. So particles $i$ and $j$ interact only if $i-j = \pm 1$. In particular, particles with coordinates $\tht_1$ and $\tht_3$ can freely `pass through' each other! Furthermore, on account of the potential, particles $i$ and $i+1$ can also cross without encountering singularities. Finally, we note that the quantum Hamiltonian corresponding to (\ref{e:classical-action-intermediate-problem}),
	\beq
	\hat H 
	= \sum_i - \fr{\hbar^2}{2 m r^2} \fr{\pdr^2}{\pdr\tht_i^2} - g \cos(\tht_i - \tht_{i+1})
	\eeq
has been used to model a 1d array of coupled Josephson junctions (see Fig. \ref{f:3rotor-JJchain}) with the capacitive charging and Josephson coupling energies given by $E_C = \hbar^2/m r^2 = L/a \beta^2 J$ and $E_J = g = JL/a$ \cite{sondhi-girvin}.

\section{Positivity of the JM curvature for \texorpdfstring{$0 \leq E \leq 4g$}{0 < E < 4g}} 
\label{a:positivity-of-JM-curvature}
 
Here, we prove that for $0 \leq E \leq 4g$, the JM curvature $R$ of \S \ref{s:JM-approach} is strictly positive in the Hill region ($E > \V$) of the $\vf_1$-$\vf_2$ configuration torus. It is negative outside and approaches $\pm \infty$ on the Hill boundary $E = \V$.  It is convenient to work in Jacobi coordinates $\vf_\pm = (\vf_1 \pm \vf_2)/2$ introduced in \S \ref{s:jacobi-coordinates} and define $P = \cos \vf_+$ and $Q = \cos \vf_-$. In these variables,
 	\beq
	R = \fr{g^2 {\cal N}_E(P,Q)}{mr^2 (E - \V)^3}  \quad \text{where} \quad {\cal N}_E = 5 + 2Q^2 - 6 P Q + 8 P^3 Q  
	+  \left[\fr{2E}{g} - 3\right] (2P^2 + 2 P Q - 1).
	\label{e:N}
	\eeq
Since $E - \V > 0$ in the Hill region, it suffices to show that ${\cal N}_E \geq 0$ on the whole torus and strictly positive in the Hill region. It turns out that (a) ${\cal N}_E \geq 0$ for $E = 0$ and $4g$ and (b) for $E = 0$, ${\cal N}_E$ vanishes only at the ground state G while for $E = 4g$, it vanishes only at the saddles D, with both G and the Ds lying on the Hill boundary. Since G is distinct from the Ds, linearity of ${\cal N}_E$  then implies that ${\cal N}_E > 0$ on the entire torus for $0 < E < 4g$. It only remains to prove (a) and (b).

To proceed, we regard ${\cal N}_E$ as a function on the $[-1,1]\times[-1,1]$ $PQ$-square. (i) When $E=0$, ${\cal N}_0$ has only one local extremum in the interior of the $PQ$-square at $(0,0)$ where ${\cal N}_0(0,0) = 8$. On the boundaries of the $PQ$-square,
	\beqs
	{\cal N}_0(\pm 1,Q) = 2(1 \mp  Q)^2 \geq 0 
	\quad \text{and} \quad
	{\cal N}_0(P, \pm 1) = 2(P\mp1)^2 (5 \pm 4 P) \geq 0
	\eeqs
with ${\cal N}_0$ vanishing only at $(1,1)$ and $(-1,-1)$ both of which correspond to G. Thus, ${\cal N}_0 \geq 0$ on the whole torus and vanishes only at G which lies on the Hill boundary. (ii) When $E = 4g$, the local extrema in the interior of the $PQ$-square are at $(0,0)$ and $(\pm 1, \mp 5/3)/\sqrt{3}$ where ${\cal N}_{4g}$ takes the values $0$ and $40/27$. On the boundaries of the $PQ$-square,
	\beq
	{\cal N}_{4g}(\pm 1,Q) = 2 (1 \pm Q) (5 \pm Q) \geq 0
	\quad \text{and} \quad
	{\cal N}_{4g}(P,\pm 1) = 2 (1 \pm P)(1 \pm P + 4 P^2) \geq 0
	\eeq
with ${\cal N}_{4g}$ vanishing only at $(1, -1)$ and $(-1, 1)$. Hence, for $E = 4g$,  ${\cal N}_{4g} \geq 0$ on the whole torus and vanishes only at the three saddle points (Ds) all of which lie on the Hill boundary.

\section[Measuring area of chaotic region on the `$\vf_1 = 0$' Poincar\'e surface]{Measuring area of chaotic region on the `$\vf_1 = 0$' Poincar\'e surface \sectionmark{Measuring area of chaotic region on Poincar\'e surfaces}}
\sectionmark{Measuring area of chaotic region on Poincar\'e surfaces}
\label{a:estimate-f}

\begin{figure}	
	\centering
	\begin{subfigure}[t]{7.5cm}
		\centering
		\includegraphics[width=7.5cm]{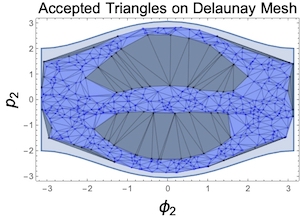}
		\caption{\small }
		\label{f:accepted-triangles-mesh}
	\end{subfigure}
\qquad
	\begin{subfigure}[t]{7.5cm}
		\centering
		\raisebox{.2cm}{\includegraphics[width=7.5cm]{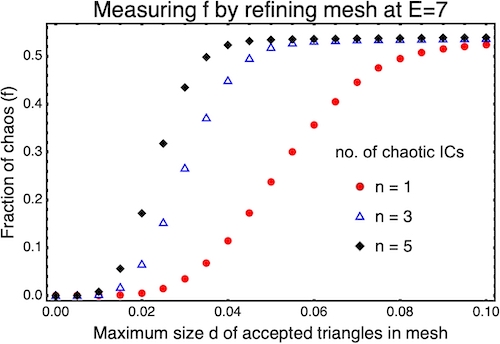}}
		\caption{\small }
		\label{f:refine-mesh}
	\end{subfigure}
	\caption{\small (a) Accepted (chaotic, shaded lighter/blue) and rejected (regular, shaded darker/grey) triangles on Delaunay Mesh for a {\it sample} chaotic region on the `$\vf_1=0$' Poincar\'e surface at $E = 7$ for maximal edge length $d=1$. The light colored region on the periphery inside the Hill region consists of regular sections. (b)  Estimates of the fraction of chaos (area of accepted region/area of Hill region) for various choices of  $d$. An optimal estimate for $f$ is obtained by picking $d$ where $f$ saturates. The three data sets displayed have $n=1,3,5$ chaotic ICs, each evolved for the same duration $t = 10^5$.}
\end{figure}

To estimate the fraction of the area of the Hill region (at a given $E$) occupied by the chaotic sections on the `$\vf_1=0$' Poincar\'e surface, we need to assign an area to the corresponding scatter plot (e.g., see Fig \ref{f:psec-egy=4.5}). We use the DelaunayMesh routine in Mathematica to triangulate the scatter plot so that every point in the chaotic region lies at the vertex of one or more triangles (see Fig. \ref{f:accepted-triangles-mesh}). For such a triangulation and a given $d > 0$, the $d$-area of the chaotic region is defined as the sum of the areas of those triangles with maximal edge length $\leq d$ (accepted triangles in Fig. \ref{f:accepted-triangles-mesh}). Fig.~\ref{f:refine-mesh} shows that the area initially grows rapidly with $d$, and then saturates for a range of $d$. Our best estimate for the area of the chaotic region is obtained by picking $d$ in this range. Increasing $d$ beyond this admits triangles that are outside the chaotic region. Increasing the number of points in the scatter plot (either by evolving each IC for a longer time or by including more chaotic ICs, which is computationally more efficient) reduces errors and decreases the threshold value of $d$ as illustrated in Fig. \ref{f:refine-mesh}.

\section[Power-law approach to ergodicity in time]{Power-law approach to ergodicity in time \sectionmark{Power--law approach to ergodicity in time}}
\sectionmark{Power--law approach to ergodicity in time}
\label{a:approach-to-ergodicity}

Assuming correlations decay sufficiently fast, as expected for a chaotic system, we give here a heuristic explanation for our observed (see \S \ref{s:approach-to-ergodicity}) power-law approach to ergodicity in time (see also \cite{prl-dechant} for a discussion based on a stochastic framework). Let $F(p, \vf)$ be a dynamical variable with ensemble average at energy $E$ denoted $\bar F = \bra F \ket_{\rm e}$  (\ref{e:ensemble-avg-distr-vol-egy-surf}). Its time average, over the interval $[0,T]$, along an energy-$E$ phase trajectory $( \vec p_i(t), \vec \vf_i(t))$ labelled $i$, is denoted
	\beq
	\tilde F_{i}(T) = \ov{T} \int_0^T F_i(t) \: dt \equiv \ov{T} \int_0^T F(\vec p_i(t), \vec \vf_i(t)) \: dt.
	\eeq
To examine the rate at which time averages along different trajectories $i$ approach the ensemble average, we define the mean square deviation of $\tilde F_{i}(T)$ from $\bar F$ for a family $\cal I$ of trajectories:
	\beq
	{\rm var}_F(T) = \Bra \left( \tilde F_{i}(T) - \bar F \right)^2 \Ket \equiv \ov{\# (\cal I)} \sum_{i \in \cal I} \left( \tilde F_{i}(T) - \bar F \right)^2.
	\eeq
Expanding, we write the mean square deviation as
	\beq
	{\rm var}_F(T) = \Bra \tilde F_{i}(T)^2 \Ket  + \bar F^2 - 2 \bar F \Bra \tilde F_{i}(T) \Ket.
	\eeq
We now assume that the ICs for the trajectories in ${\cal I}$ are distributed uniformly with respect to the Liouville measure on the energy-$E$ hypersurface. Since the dynamics is Hamiltonian, by Liouville's theorem the trajectories remain uniformly distributed at all times $T$, so that as ${\# (\cal I)} \to \infty$,
	\beq
	\Bra \tilde F_{i}(T) \Ket = \bar F.
	\eeq
Thus, the mean square deviation becomes
	\beqs
	{\rm var}_F(T) &=& \Bra \tilde F_{i}(T)^2 \Ket - \bar F^2 =  \Bra \tilde F_{i}(T)^2 - \bar F^2 \Ket \cr
	&=& \Bra \ov{T^2} \int_0^T \int_0^T  [F_i(t_1) F_i(t_2) - \bar F^2] dt_1 dt_2 \Ket \cr
	&=& \ov{T^2} \int_0^T \int_0^T  \Bra F_i(t_1) F_i(t_2) - \bar F^2 \Ket
 dt_1 dt_2. 	\quad
 	\eeqs
We now assume that $F_i(t_1)$ and $F_i(t_2)$ are practically uncorrelated if $|t_1 - t_2| > \eps$ for some time $\eps$, i.e.,
	\beq
	\Bra F(t_1) F(t_2) - \bar F^2 \Ket \approx \begin{cases}0 & {\rm if} \quad |t_1 - t_2| > \eps  \quad \text{and} \\
	{\cal C}(t_1-t_2) & {\rm otherwise} \end{cases}
	\eeq
by time-translation invariance, for some (2nd cumulant) function ${\cal C}(t_1-t_2)$. We now change integration variables from $t_{1,2}$ to $u = t_1 - t_2$ and $v = (t_1+t_2)/2$ with $dt_1 dt_2 = du \: dv$ and assume $T \gg \eps$ to get
	\beqs
	{\rm var}_F(T) 
	&\approx& \ov{T^2} \int_{0}^T dv \int_{-\eps}^\eps  du \: {\cal C}(u)    
	= \ov{T}  \int_{-\eps}^\eps  {\cal C}(u) du.  \qquad
	\eeqs
Thus, the RMS deviation of time averages from the ensemble average vanishes like $1/\sqrt{T}$ as $T \to \infty$.

\end{document}